\documentclass[12pt,titlepage]{article}%
\usepackage{heck}
\usepackage{graphicx}
\usepackage{cite}
\usepackage{multicol}
\usepackage{geometry}
\usepackage{amsfonts}
\usepackage{mathrsfs}
\usepackage{amssymb}
\usepackage{amsmath}%
\providecommand{\U}[1]{\protect\rule{.1in}{.1in}}
\numberwithin{equation}{section}

\def\^{{\wedge}}
\def\*{{\star}}

\def\bar{\overline}

\begin{document}

\date{May, 2008}

\preprint{arXiv:0806.0102}

\institution{HarvardU}{Jefferson Physical Laboratory, Harvard University, Cambridge,
MA 02138, USA}%

\title
{GUTs and Exceptional Branes in \\[-0.25cm]F-theory - II: \\[1cm]Experimental Predictions}%
%

\authors{Chris Beasley\footnote{e-mail: {\tt beasley@physics.harvard.edu}},
Jonathan J. Heckman\footnote{e-mail: {\tt jheckman@fas.harvard.edu}} and
Cumrun Vafa\footnote{e-mail: {\tt vafa@physics.harvard.edu}}}%

\longabstract{We consider realizations of GUT models in F-theory.  Adopting a bottom
up approach, the assumption that the dynamics
of the GUT model can in principle decouple from Planck scale physics leads to a surprisingly
predictive framework.  An internal $U(1)$ hypercharge flux Higgses the GUT group
directly to the MSSM or to a flipped GUT model, a mechanism
unavailable in heterotic models.  This new ingredient automatically addresses a number of puzzles
present in traditional GUT models.  The internal $U(1)$ hyperflux allows us
to solve the doublet-triplet splitting problem, and explains the qualitative features
of the distorted GUT mass relations for lighter generations due to
the Aharanov-Bohm effect.  These models typically come with nearly exact global symmetries which
prevent bare $\mu$ terms and also forbid dangerous baryon number violating operators.
Strong curvature around our brane leads to a repulsion mechanism for Landau
wave functions for neutral fields.   This leads to large hierarchies of the form $\text{exp}(-c/\varepsilon^{2\gamma})$ where $c$ and $\gamma$ are order one parameters and $\varepsilon \sim \alpha_{GUT}^{-1}M_{GUT}/M_{pl}$.  This
effect can simultaneously generate a viably small $\mu$ term as well as an acceptable Dirac
neutrino mass on the order of $0.5\times 10^{-2 \pm 0.5}$ eV.  In another scenario, we find a
modified seesaw mechanism which predicts that the light neutrinos have masses in the expected range
while the Majorana mass term for the heavy neutrinos is $\sim 3\times 10^{12 \pm 1.5}$ GeV.
Communicating supersymmetry breaking to the MSSM can be elegantly
realized through gauge mediation.  In one scenario, the same repulsion mechanism also leads to messenger
masses which are naturally much lighter than the GUT scale.}%

\maketitle

\tableofcontents

\pagebreak

\section{Introduction}

Despite many theoretical advances in our understanding of string theory, this
progress has not produced a single verifiable prediction which can be tested
against available experiments. \ Part of the problem is that in its current
formulation, string theory admits a vast landscape of consistent low energy
vacua which look more or less like the real world.

Reinforcing this gloomy state of affairs is the fact that the particle content
of the Standard Model is generically of the type encountered in string theory.
\ Indeed, the gauge group of the Standard Model is of the form $\prod
_{i}U(N_{i})$ and the chiral matter content corresponds to bi-fundamental
fields transforming in representations such as $(N_{i},\overline{N}_{j})$.
\ While this may reinforce the idea that string theory is on the right track,
precisely because this appears to be such a generic feature of string
constructions, this also unfortunately limits the predictivity of the theory.
\ To rectify this situation, we must impose additional criteria to narrow down
the search in the vast landscape.

From a top down approach, one idea is to further incorporate some specifically
stringy principles. \ For instance, we have learned that the large $N$ limit
of many $U(N)$ gauge theories causes the gauge system to `melt' into a dual
gravitational background \cite{juanAdS}. \ Moreover, this large $N$ gauge
theory can undergo a duality cascade to a small $N$ gauge theory
\cite{KlebanovStrassler}. \ Indeed, the Standard Model could potentially
emerge at the end of such a process. \ In the string theory literature, this
idea has been explored in
\cite{CascalesCascade,SMCASCADE,FrancoVerlindeCascade}. Interesting as this
idea is, it does not incorporate the idea of grand unification of the gauge
forces into one gauge factor in any way.

From a bottom up approach, it is natural to ask whether there is some way to
incorporate the important fact that the gauge coupling constants of:%
\begin{equation}
G_{std}\equiv SU(3)_{C}\times SU(2)_{L}\times U(1)_{Y}%
\end{equation}
seem to unify in the minimal supersymmetric extension of the Standard Model
(MSSM). \ This not only supports the idea that supersymmetry is realized at
low energies, but also suggests that the multiple gauge group factors of the
Standard Model unify into a single simple group such as $SU(5)$ or $SO(10)$.
\ Moreover, the fact that the matter content of the Standard Model
economically organizes into representations of the groups $SU(5)$ and $SO(10)$
provides a strong hint that the basic idea of grand unified theories (GUTs) is
correct. For example, it is quite intriguing that all of the chiral matter of
a single generation \textit{precisely} organizes into the spinor
representation $16$ of $SO(10)$. \ Hence, we ask whether the principle of
grand unification can narrow down the large list of candidate vacua in the
landscape to a more tractable, and predictive subset.

Despite the many attractive features of the basic GUT framework, the simplest
implementations of this idea in four-dimensional models suffer from some
serious drawbacks. \ For example, the minimal four-dimensional
supersymmetric\ $SU(5)$ GUT with standard Higgs content seems to be
inconsistent with present bounds on proton decay \cite{MurayamaPierce}. \ In
the absence of higher dimensional representations of $SU(5)$ or somewhat
elaborate higher dimension operator contributions to the effective
superpotential, this model also leads to mass relations and over-simplified
mixing matrices which are generically too strong to be correct. \ This
presents an opportunity for string theory to intervene: Can string theory
preserve the nice features of GUT\ models while avoiding their drawbacks?

Indeed, the $E_{8}\times E_{8}$ heterotic string seems very successful in this
regard because the usual GUT\ groups $SU(5)$ and $SO(10)$ can naturally embed
in one of the $E_{8}$ factors. \ See \cite{GSW} for an early review on how
GUT\ models could potentially originate from compactifying the heterotic
string on a Calabi-Yau threefold. \ Moreover, because no appropriate
four-dimensional GUT Higgs field is typically available to break the
GUT\ group to the Standard Model gauge group, it is necessary to employ a
higher-dimensional breaking mechanism. \ When the internal space has
non-trivial fundamental group, the gauge group can break via a discrete Wilson
line. \ In this way, the gauge group in four dimensions is always the Standard
Model gauge group but the matter content and gauge couplings still unify.
\ Moreover, such higher dimensional GUTs provide natural mechanisms to
suppress proton decay and avoid unwanted mass relations. \ See
\cite{BouchardDonagi,OvrutHET,RabyHET,AndreasCurioRecentI,AndreasCurioRecentII}
for some recent attempts in this direction.

However, the heterotic string has its own drawbacks simply because it is
rather difficult to break the gauge symmetry down to $G_{std}$.\footnote{At a
pragmatic level, the perturbative regime of the heterotic string also seems to
be inconsistent with the relation between the GUT scale $M_{GUT}$ and the
four-dimensional Planck scale $M_{pl}$. \ A discussion of this discrepancy and
related issues may be found in \cite{DienesPaths}. \ One potential way to
bypass this problem requires going to the regime of strong coupling
\cite{WittenStrong}.} \ One popular method is to use internal Wilson lines to
directly break the gauge symmetry to that of the MSSM. \ This requires that
the fundamental group of the Calabi-Yau must be non-trivial. \ Although this
can certainly be arranged, the generic Calabi-Yau threefold is simply
connected and this mechanism is unavailable. \ Moreover, when the GUT\ group
has rank five or higher, gauge group breaking by Wilson lines can also leave
behind additional massless $U(1)$ gauge bosons besides $U(1)$ hypercharge.
\ Present constraints on additional long rang forces are quite stringent, and
in many cases it is not always clear how to remove these unwanted states from
the low energy spectrum. \ In the absence of a basic principle which naturally
favors a non-trivial fundamental group, it therefore seems reasonable to look
for other potential realizations of the GUT paradigm in string theory.

There are two other natural ways that GUTs can appear in string theory.
\ These possibilities correspond to non-perturbatively realized
four-dimensional $\mathcal{N}=1$ compactifications of type IIA\ and
IIB\ string theory. \ In the type IIA case, the GUT\ models originate from the
compactification of M-theory on manifolds with $G_{2}$ holonomy. \ For type
IIB\ theories, the corresponding vacua are realized as compactifications of
F-theory on Calabi-Yau fourfolds. In the latter case, the gauge theory degrees
of freedom of the GUT localize on the worldvolume of a non-perturbative
seven-brane. \ The $ADE$ gauge group on the seven-brane corresponds to the
discriminant locus of the elliptic model where the degeneration is locally of
$ADE$ type. \ Of these two possibilities, the holomorphic geometry of
Calabi-Yau manifolds provides a more tractable starting point for addressing
detailed model building issues. \ It was with this aim that we initiated an
analysis of how GUT models can be realized in F-theory \cite{BHV}. \ See
\cite{WijnholtDonagi,WatariTATARHETF} for related discussions in the context
of F-theory/heterotic duality.

Even so, there is a certain tension between string theory and the
GUT\ paradigm. From a top down perspective, it is a priori unclear why there
should be any distinction between the Planck scale $M_{pl}$ and the GUT\ scale
$M_{GUT}$. \ In the bottom up approach, the situation is completely reversed.
\ Indeed, insofar as effective field theory is valid at the GUT\ scale, it is
quite important that $M_{GUT}/M_{pl}$ is small and \textit{not} an order one
number. \ For example, in the extreme situation where the only chiral matter
content of a four-dimensional GUT\ model originates from the MSSM, the
resulting theory is asymptotically free.

In geometrically engineered gauge theories in string theory, asymptotic
freedom translates to the existence of a consistent decompactification limit.
\ It is therefore quite natural to ask if \textit{at least in principle} we
could have decoupled the two scales $M_{GUT}$ and $M_{pl}$. \ This is also in
accord with the bottom up approach to string phenomenology
\cite{KiritsisBottomUp,UrangaBottomUp,WijnholtVerlindeBottomUp,GrayJejjalaBottomUp}%
. \ In the present paper our main focus will therefore be to search for vacua
which at least in principle admit a limit where $M_{pl}\rightarrow\infty$
while $M_{GUT}$ remains finite. \ Of course, in realistic applications
$M_{pl}$ should also remain finite. \ For completeness, we shall also present
some examples of models where $M_{GUT}$ and $M_{pl}$ cannot be decoupled. \ In
such cases, we note that it is not a priori clear whether the correct value of
$M_{GUT}$ can be achieved.

Nevertheless, the mere \textit{existence} of a decoupling limit turns out to
endow the resulting candidate models with surprising predictive power.\ \ It
turns out that the only way to achieve such a decoupling limit requires that
the spacetime filling seven-brane must wrap a del Pezzo surface. \ The fact
that the relevant part of the internal geometry in this setup is limited to
just ten distinct topological types is very welcome! \ In a certain sense,
there is a unique choice corresponding to the del Pezzo 8 surface because all
of the other del Pezzo surfaces can be obtained from this one by blowing down
various two-cycles.

At the next level of analysis, we must determine what kind of seven-brane
should wrap the del Pezzo surface. \ As explained in \cite{BHV}, realizing the
primary ingredients of GUT\ models requires that the singularity type
associated with the seven-brane should correspond to a subgroup of the
exceptional group $E_{8}$. \ Because the Standard Model gauge group has rank
four, this determines a lower bound on the rank of any putative GUT\ group. At
rank four, $SU(5)$ is the only available GUT\ group. \ Hence, the most
`minimal' choice is to have an $SU(5)$ seven-brane wrapping the del Pezzo 8
surface. We will indeed find that this minimal scenario is viable. \ The upper
bound on the rank of a candidate GUT\ group is six. \ This bound comes about
from the fact that if the rank is any higher, the model will generically
contain localized light degrees of freedom at points on the del Pezzo surface
which do not appear to admit a standard interpretation in gauge theory
\cite{MorrisonFlorea,BHV}. \ This is because on complex codimension one
subspaces, the rank of the gauge group goes up by one, and on complex
codimension two subspaces, i.e. points, the rank goes up by two. Hence, if the
rank is greater than six, the compactification contains points on the del
Pezzo with singularities of rank nine and higher which do not admit a standard
gauge theoretic interpretation because $E_{8}$ is the maximal compact
exceptional group.

In the minimal scenario where the seven-brane has gauge group $SU(5)$, we find
that there is an essentially unique mechanism by which the GUT\ group can
break to a four-dimensional model with gauge group $G_{std}$. \ This breaking
pattern occurs in vacua where the $U(1)$ hypercharge flux in the internal
directions of the seven-brane is non-trivial. \ This mechanism is unavailable
in heterotic compactifications because the $U(1)$ hypercharge always develops
a string scale mass via the Green-Schwarz mechanism \cite{WittenU(1)}. As
noted for example in \cite{WittenU(1)}, in order to preserve a massless $U(1)$
hypercharge gauge boson, additional $U(1)$ factors must mix non-trivially with
this direction, which runs somewhat counter to the idea of grand unification.
Nevertheless, for suitable values of the gauge coupling constants for these
other factors, a semblance of unification can be maintained. See
\cite{BlumenhagenWeigenadLineBundles,BlumenhagenMosterWeigandI,BlumenhagenMosterWeigandII,TatarWatariU1}
for further discussion on vacua of this type.

In F-theory, we show that there is no such generic obstruction. \ This is a consequence of the fact that
while the cohomology class of the flux on the seven-brane can be non-trivial,
it can nevertheless represent a trivial class in the base of the F-theory
compactification. \ This topological condition is necessary and also
sufficient for the corresponding four-dimensional $U(1)$ gauge boson to remain
massless. \ An important consequence of this fact is that these F-theory vacua
do not possess a heterotic dual.

The particular choice of internal $U(1)$ flux which breaks the GUT group is
also unique. \ To see how this comes about, we first recall that the middle
cohomology of the del Pezzo 8 surface splits as the span of the canonical
class and the collection of two-cycles orthogonal to this one-dimensional
lattice. \ With respect to the intersection form on two-cycles, this
orthogonal subspace corresponds to the root lattice of $E_{8}$. \ Moreover,
the admissible fluxes of the $U(1)$ hypercharge are in one to one
correspondence with the roots of $E_{8}$. \ This restriction occurs because
for more generic choices of $U(1)$ flux, the low energy spectrum contains
exotic matter which if present would ruin the unification of the gauge
coupling constants. \ In keeping with the general philosophy outlined in
\cite{BHV}, we always specify the appropriate line bundle first and only then
determine whether an appropriate K\"{a}hler class exists so that the vacuum is
supersymmetric. \ In this sense, there is a unique choice of flux because the
Weyl group of $E_{8}$ acts transitively on the roots of $E_{8}$. On general
grounds, this internal flux will also induce a small threshold correction near
the GUT scale. Determining the size and sign of this correction would clearly
be of interest to study.\footnote{After our work appeared, 
this question has been studied in \cite{DonagiWijnholtII,BlumenhagenThreshold}.}

The matter and Higgs fields localize on Riemann surfaces in the del Pezzo
surface. \ In F-theory, these Riemann surfaces are located at the intersection
between the GUT model seven-brane and additional seven-branes in the full
compactification. \ Along these intersections, the rank of the singularity
type increases by one. \ This severely limits the available representation
content so that the matter fields can only transform in the $5$ or
$\overline{5}$ along an enhancement to $SU(6)$ and the $10$ or $\overline{10}$
for local enhancement to $SO(10)$.

The internal hypercharge flux automatically distinguishes the Higgs fields
from the other chiral matter content of the MSSM. \ The Higgs fields localize
on matter curves where the $U(1)$ hypercharge flux is non-vanishing, and the
chiral matter of the MSSM\ localizes on Riemann surfaces where the net flux
vanishes. \ In other words, the two-cycles for the Higgs curves intersect the
root corresponding to this internal flux while all the other chiral matter of
the MSSM\ localizes on two-cycles orthogonal to this choice of flux. \ This
internal choice of flux implies that the chiral matter content will always
fill out complete representations of $SU(5)$, while the Higgs doublets can
never complete to full GUT\ multiplets. \ Moreover, by a suitable choice of
flux on the other seven-branes, the spectrum \ will contain no extraneous
Higgs triplets, thus solving the doublet-triplet splitting problem. \ In
certain cases, superheavy Higgs triplets can still cause the proton to decay
too quickly. \ In traditional four-dimensional GUT\ models the missing partner
mechanism is often invoked to avoid generating dangerous dimension five
operators which violate baryon number. \ Here, this condition translates into
the simple geometric condition that the Higgs up and down fields must localize
on distinct matter curves.

In our study of Yukawa couplings, we shall occasionally encounter situations
involving two fields charged under the GUT group and one neutral field (for
example a $1\times5 \times\overline{5}$ interaction). In such cases, the
neutral field lives on a matter curve normal to the del Pezzo which intersects
this surface at a point. In order to determine the strength of the Yukawa
couplings, we need to estimate the strength of the corresponding zero mode
wave functions at the intersection point. It turns out that since the del
Pezzo is strongly positively curved ($\mathcal{R}\sim M_{GUT}^{2}$), the
normal geometry is negatively curved. Moreover, this leads to the wave
function being either attracted to, or repelled away from our brane, depending
on the choice of the gauge flux on the normal intersecting seven-branes. In
one case the wave function is attracted to our seven-brane, making it behave
as if the wave function is localized inside the brane. In another case the
wave function is repelled away from our brane, leading to an exponentially
small amplitude at our brane. The exponential hierarchy is given by
$\exp(-cR_{\bot}^{2}/R_{GUT}^{2})$ where $c$ is a positive order one constant,
$R_{\bot}$ is the radius of the normal geometry to the brane, and $R_{GUT}$ is
the length associated to GUT. The estimate for $R_{\bot}$ depends on
assumptions about how the geometry normal to our brane looks, and in
particular to what extent it is tubular. We find that:
\begin{equation}
\frac{R_{\bot}}{R_{GUT}}=\varepsilon^{-\gamma}%
\end{equation}
where $1/3\lesssim\gamma\lesssim1$ is a measure of the normal eccentricity and
$\varepsilon$ is a small parameter:%
\begin{equation}
\varepsilon\sim\frac{M_{GUT}}{\alpha_{GUT}M_{pl}}\sim7.5\times10^{-2}\text{.}%
\end{equation}
This leads to a natural hierarchy given by
\begin{equation}
\exp\left(  -c\frac{R_{\bot}^{2}}{R_{GUT}^{2}}\right)  \sim\exp\left(
-c\frac{1}{\varepsilon^{2\gamma}}\right)  \text{.}%
\end{equation}
There are various vector-like pairs which can only develop a mass through a
cubic Yukawa coupling with a third field coming from a neutral normal wave
function. This suppression mechanism will be useful in many such cases,
including solving the $\mu$ problem and also obtaining a small Dirac neutrino
mass leading to realistic light neutrino masses without using the seesaw mechanism.

There are two ways we can solve the $\mu$ problem. \ Perhaps most simply, we
can consider geometries where the Higgs up and down fields localize on
distinct matter curves which do not intersect. \ In this case, the $\mu$ term
is identically zero. When these curves do intersect, the value of the $\mu$
term depends on the details of a gauge singlet wave function which localizes
on a matter curve normal to the del Pezzo surface. \ In the case of
attraction, the $\mu$ term is near the GUT scale, which is untenable. \ In the
repulsive case, the $\mu$ term is suppressed to a much lower value:
\begin{equation}
\frac{\mu}{M_{GUT}}\sim\exp\left(  -c\frac{1}{\varepsilon^{2\gamma}}\right)
\text{,}%
\end{equation}
so that the resulting value of $\mu$ can then naturally fall in a
phenomenologically viable range.

In fact, a similar exponential suppression in the wave functions of the
right-handed neutrinos can generate small Dirac neutrino masses of the form:%
\begin{equation}
m_{\nu}^{D}\sim\frac{\mu\varepsilon^{-\gamma}}{\langle H_{u}\rangle}%
\times\frac{\langle H_{u}\rangle^{2}}{M_{GUT}}\sim0.5\times10^{-2\pm0.5}\text{
eV}%
\end{equation}
which differs by a factor of $\mu\varepsilon^{-\gamma}/\left\langle
H_{u}\right\rangle $ from the value predicted by the simplest type of seesaw
mechanisms with Majorana masses at the GUT scale. \ We note that the value we
obtain is in reasonable agreement with recent experimental results on neutrino
oscillations. \ In this case, the Majorana mass term must identically vanish
to remain in accord with observation.

A variant of the standard seesaw mechanism is also available when the
right-handed neutrino wave functions are attracted to the del Pezzo surface.
\ In this case, the Majorana mass terms in the neutrino sector are suppressed
by some overall volume factors. Although the standard seesaw mechanism again
generates naturally light neutrino masses $\sim2\times10^{-1\pm1.5}$ eV, we
find that the Majorana mass term is naturally somewhat lighter than the GUT
scale and is on the order of $\sim3\times10^{12\pm1.5}$ GeV. \ It is
interesting that the numerical values we obtain in either scenario are both in
a range of values consistent with leptogenesis, as well as the observed light
neutrino masses.

Non-trivial flavor structures can potentially arise in a number of ways in
this class of models. \ For example, one common approach in the model building
literature is to use a discrete symmetry to induce additional structure in the
form of the Yukawa couplings. \ The Weyl group symmetries of the exceptional
groups naturally act on the del Pezzo surfaces. \ This symmetry can be
partially broken by the choice of the K\"{a}hler classes of two-cycles. This
may potentially lead to a model of flavor based on the discrete symmetry
groups $S_{3}$, $A_{4}$ or $S_{4}$. \ Indeed, these are all subgroups of the
Weyl group of $E_{8}$.

One of the main conceptual issues with the usual GUT framework is to explain
why $m_{b}\sim m_{\tau}$ at the GUT scale while the lighter generations do not
satisfy such a simple mass relation. \ At a qualitative level, the behavior of
the omnipresent internal $U(1)$ hypercharge flux again plays a central role in
the resolution of this issue. \ Although the net hypercharge flux vanishes on
curves which support full GUT\ multiplets, in general it will not vanish
pointwise. \ Hence, the hypercharge flux can still leave behind an important
imprint on the wave functions of the fields in the MSSM. \ Indeed, because the
individual components of a GUT\ multiplet have different hypercharge, the
Aharonov-Bohm effect will alter the distinct components of a GUT\ multiplet
differently, leading to violations in the most naive mass relations. \ In
fact, because the mass of a generation is higher the smaller the volume of the
matter curve, the amount of flux which can pierce the curve also decreases.
\ In this way, the most naive mass relations remain approximately intact for
the heaviest generation but will in general receive corrections for the
lighter generations.

In the next to minimal GUT\ scenario, we can consider seven-branes where the
bulk gauge group has rank five. \ In this case there are three choices
corresponding to $SO(10)$, $SO(11)$ and $SU(6)$. \ In this paper we mainly
focus on the $SO(10)=E_{5}$ case because it fits most closely with our general
philosophy that the exceptional groups play a distinguished role in
GUT\ models. \ It turns out that this model can only descend to the MSSM by a
sequence of breaking patterns where the eight-dimensional theory first breaks
to a four-dimensional flipped $SU(5)$ model with gauge group $SU(5)\times
U(1)$. \ The model then operates as a traditional four-dimensional flipped
$SU(5)$ GUT which breaks to the Standard Model gauge group when a field in the
$10_{-1}$ of $SU(5)\times U(1)$ develops a suitable vev. \ Indeed, direct
breaking of $SO(10)$ to the Standard Model gauge group via fluxes taking
values in a $U(1)\times U(1)$ subgroup always generates exotic matter which
would ruin the unification of the gauge coupling constants. \ Many of the more
refined features of these models such as textures and our solution to the
$\mu$ problem share a common origin to those studied in the minimal $SU(5)$ model.

Even though our main emphasis in this paper is on models which admit a
decoupling limit, we also consider models where such a limit does not exist.
\ In such cases the problem of engineering a GUT model becomes more flexible
because the local model is incomplete. \ We study examples of this situation
because there are well-known difficulties in heterotic models in realizing
traditional four-dimensional GUT\ group breaking via fields in the adjoint
representation. \ This is due to the fact that in many cases, the requisite
adjoint-valued fields do not exist. \ Indeed, gauge group breaking by Wilson
lines is not so much an elegant ingredient in heterotic constructions as much
as it is a necessary element of any construction.\footnote{It is also possible
to avoid this constraint in heterotic models which descend to a
four-dimensional flipped $SU(5)$ GUT. \ See
\cite{HagelinEllisFlipped,NanopDbranesFlipped,BlumenhagenMosterWeigandI} for further
details on this approach. \ We also note that in certain cases, chiral
superfields transforming in other representations can arise from higher
Kac-Moody levels of the heterotic string.} \ Gauge group breaking via Wilson
lines can also occur in F-theory when the surface wrapped by the seven-brane
has non-trivial fundamental group. \ For example, a well-studied surface with
$\pi_{1}(S)\neq0$ is the Enriques surface which can be viewed as the $%
\mathbb{Z}
_{2}$ quotient of a $K3$ surface.

Given the large proliferation of four-dimensional GUT\ models which exist in
the model building literature, it is also natural to ask whether there exist
purely four-dimensional GUT\ models in F-theory with adjoint-valued GUT\ Higgs
fields. \ We find that this can be done provided the surface wrapped by the
seven-brane has non-zero Hodge number $h^{2,0}(S)\neq0$. \ But in contrast to
the usual approach to four-dimensional effective field theories where it is
common to assume that Planck scale physics can in principle be decoupled, here
we see that the traditional four-dimensional GUT \textit{cannot} be decoupled
from Planck scale physics.

We also briefly consider supersymmetry breaking in our setup. \ This is
surprisingly simple to accommodate because extra messenger fields can
naturally arise from additional matter curves which do not intersect any of
the other curves on which the matter content of the MSSM\ localizes.
\ Supersymmetry breaking can then communicate to the MSSM\ via the usual gauge
mediation mechanism. \ We note that because the $\mu$ term naturally develops
a value around the electroweak scale independently of any supersymmetry
breaking mechanism, we can retain many of the best features of gauge mediation
such as the absence of additional flavor changing neutral currents (FCNCs)
while avoiding some of the problematic elements of this scenario which are
related to generating appropriate values for the $\mu$ and $B\mu$ terms.
\ Depending on the local behavior of the wave functions which propagate in
directions normal to the del Pezzo surface, the messenger scale can quite
flexibly range from values slightly below the GUT scale to much lower but
still phenomenologically viable mass scales.

The organization of this paper is as follows. In Section
\ref{LowEnergyConstraints}, we formulate what we wish to achieve in our GUT
constructions. \ In Section \ref{BasicSetup} we review and slightly extend our
previous work on realizing GUT\ models in F-theory. \ To this end, we describe
many of the necessary ingredients for an analysis of the matter content and
interaction terms of any potential model.\ \ Before proceeding to any
particular class of models, in Section \ref{MASSSCALES} we discuss the various
mass scales which will generically appear throughout this paper. \ In Section
\ref{GeneralOverview}, we give a general overview of the class of GUT\ models
in F-theory we shall study. These models intrinsically divide based on how the
GUT breaks to the MSSM. \ We first study models where the GUT\ scale cannot be
decoupled from the Planck scale. \ In Section \ref{GeneralType} we discuss
models where GUT\ breaking proceeds just as in four-dimensional models.
\ Next, we discuss GUT\ breaking via discrete Wilson lines in Section
\ref{DiscreteWilson}. \ In the remainder of the paper we focus on the primary
case of interest where a decoupling limit exists. \ Section
\ref{GeomdelPezzReview} reviews some relevant geometrical facts about del
Pezzo surfaces. \ This is followed in Section \ref{GUTBreakviaFlux} by a study
of GUT\ breaking to the MSSM\ via an internal $U(1)$ hypercharge flux. In
Section \ref{Exotica} we determine which bulk gauge groups can break directly
to the Standard Model gauge group via internal fluxes. \ We also explain in
greater detail how to obtain the exact spectrum of the MSSM\ from such models.
In Section \ref{R-parity} we discuss a geometric realization of matter parity,
and in Section \ref{DTS} we study the interrelation between proton decay and
doublet triplet splitting in our models. After giving a simple criterion for
avoiding the simplest dimension five operators responsible for proton decay,
in Section \ref{ExtraU1} we explain how extra global $U(1)$ symmetries in the
low energy effective theory are encoded geometrically in F-theory, and in
particular, how these symmetries can forbid potentially dangerous higher
dimension operators. In Section \ref{TowardsYukawas} we discuss some coarse
properties of Yukawa couplings and also speculate on how further details of
flavor physics could in principle be incorporated. \ In this same Section we
also provide a qualitative explanation for why the usual mass relations of GUT
models become increasingly distorted as the mass of a generation decreases.
\ In Section \ref{Suppression} we show that interaction terms involving matter
fields which localize on Riemann surfaces outside of the surface can generate
hierarchically small values for both the $\mu$ term as well as Dirac neutrino
masses. \ We also study a variant on the usual seesaw mechanism which
generates the expected mass scale for the light neutrinos. \ Intriguingly, the
Majorana mass of the right-handed neutrinos is somewhat lower than the value
expected in typical GUT\ models. \ In Section \ref{SUSYBREAK} we propose how
supersymmetry breaking could communicate to the MSSM, and in Section
\ref{SU5MODEL} we present an $SU(5)$ model which incorporates some (but not
all!) of the mechanisms developed in previous sections. \ Our expectation is
that further refinements are possible which are potentially more realistic.
\ In a similar vein, in Section \ref{FLIPPED} we present a flipped $SU(5)$
model. \ Section \ref{NUMEROLOGY} collects various numerical estimates
obtained throughout the paper, and Section \ref{CONCLUSIONS} presents our
conclusions. \ The Appendices contain further background material used in the
main body of the paper and which may also be of use in future model building
efforts.
\begin{figure}
[ptb]
\begin{center}
\includegraphics[
height=4.3708in,
width=4.721in
]%
{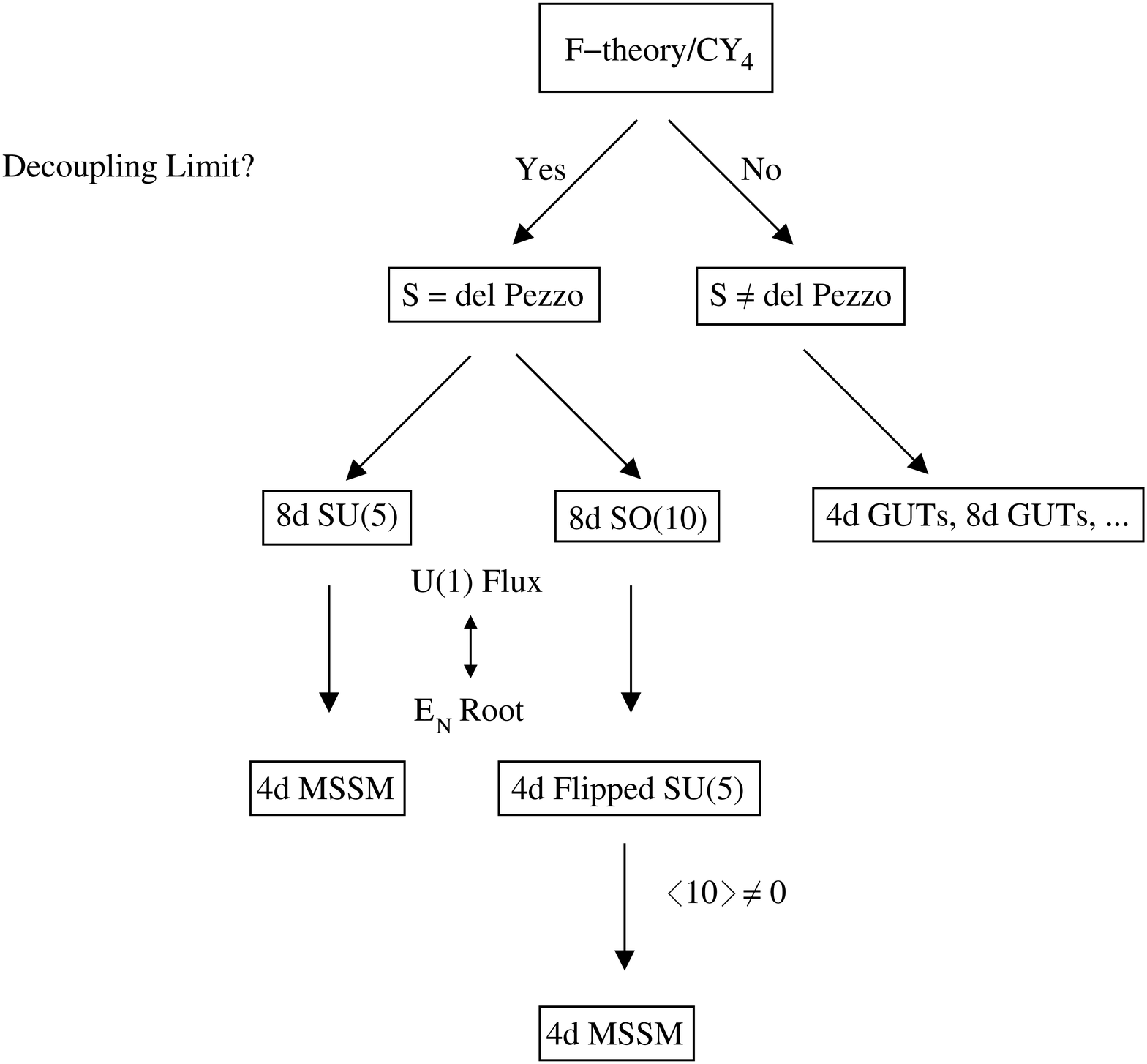}%
\caption{General overview of how GUT\ breaking constrains the type of
GUT\ model.}%
\label{flowchart}%
\end{center}
\end{figure}

\section{Constraints From Low Energy Physics \label{LowEnergyConstraints}}

In this Section we define the criteria by which we shall evaluate how
successfully our models reproduce features of low energy physics obtained by a
minimal extrapolation of experimental data to the MSSM. \ There are a number
of open questions in both phenomenology and string theory which must
ultimately be addressed in any approach. \ See
\cite{DienesTwentyQuestionsPheno,KaneLykkenTwentyFive} for an expanded
discussion of some of the issues we briefly address here.

At the crudest level, we require that any viable model contain precisely three
generations of chiral matter. \ It is an experimental fact that the chiral
matter content of the Standard Model organizes into $SU(5)$ and $SO(10)$
GUT\ multiplets. \ Coupled with the fact that the gauge couplings of the MSSM
appear to unify at an energy scale $M_{GUT}\sim3\times10^{16}$ GeV, we shall
aim to reproduce these features in all of the models we shall consider. \ For
all of these reasons, we require that the low energy content of all of our
models must match to the matter content of the MSSM. \ By this we mean that in
addition to achieving the correct chiral matter content and Higgs content of
the MSSM, all additional matter charged under the gauge groups must at the
very least fit into vector-like pairs of complete GUT\ multiplets in order to
retain gauge coupling unification.\footnote{While it is in principle possible
to consider models where vector-like exotics preserve gauge coupling
unification, we believe this runs contrary to the spirit of GUT\ models.
\ Although we shall not entertain this possibility here, see
\cite{MartinExotica,RabyExotica} for further discussion of this possibility.}
\ In the minimal incarnation of GUT\ models considered here, we shall further
require that the low energy spectrum of particles charged under the Standard
Model gauge group must exactly match to the matter content of the MSSM. \ We
note that historically, even this qualitative requirement has been difficult
to achieve in Calabi-Yau compactifications of the perturbative heterotic string.

Although the correct particle content is a necessary step in achieving a
realistic model, it is certainly not sufficient because we must also reproduce
the superpotential of the MSSM:%
\begin{equation}
W=\mu H_{u}H_{d}+\lambda_{ij}^{u}Q^{i}U^{j}H_{u}+\lambda_{ij}^{d}Q^{i}%
D^{j}H_{d}+\lambda_{ij}^{l}L^{i}E^{j}H_{d}+\lambda_{ij}^{\nu}L^{i}N_{R}%
^{j}H_{u}+... \label{WMSSM}%
\end{equation}
where the indices $i$ and $j$ label the three generations. \ While the precise
form of the Yukawa matrices labeled by the $\lambda$'s will lead to masses and
mixing terms between the generations, a necessary first step is that there are
in principle non-zero contributions to the above superpotential! \ As a first
approximation, we require that the tree level superpotential of the theory at
high energy scales generate a non-trivial interaction term for the third
generation so that there is a rough hierarchy in mass scales. \ In the context
of GUT\ models, it is well-known that because the particle content of the
Standard Model organizes into complete GUT\ multiplets, the Yukawa couplings
couple universally to fields organized in such multiplets. \ One attractive
feature of the tree level superpotential in most GUT\ models is that the third
generation obeys a simple mass relation of the form $m_{b}/m_{\tau}\sim1$ at
the GUT scale. \ Evolving this relation under the renormalization group to the
weak scale yields the relation $m_{b}/m_{\tau}\sim3$ which is roughly in
agreement with experiment. \ Unfortunately, this relation is violated for the
lighter generations. \ Ideally, it would be of interest to find models which
naturally preserve the mass relations of the third generation while modifying
the relations of the first two generations.

At the next level of approximation, any model should be consistent with
current experimental bounds on the lifetime of the proton ($\geq
10^{31}-10^{33}$ yrs \cite{PDG}). \ This requires that certain operators must
be absent or sufficiently suppressed in the low energy superpotential.
\ Indeed, note that in equation (\ref{WMSSM}), we have implicitly only
included renormalizable R-parity invariant couplings because if present, the
interaction terms $\lambda_{ijk}U^{i}D^{j}D^{k}$ and $\lambda_{ijk}^{\prime
}L^{i}L^{j}E^{k}$ will cause the proton to decay too rapidly. We shall
consider models with and without R-parity. \ In the latter case, we therefore
must present alternative reasons to expect renormalizable operators
responsible for R-parity to vanish.

Proton decay is a hallmark of GUT models. \ Aside from renormalizable
interaction terms, the dominant contribution to proton decay in the simplest
GUT\ models comes from the dimension five operator
\cite{DimFIVEWeinberg,DimFIVESakaiYanagida}:\footnote{There is an additional
contribution to the superpotential given by $UUDE$. \ At the level of
discussion in this paper, it is sufficient to only deal with the term $QQQL$.}%
\begin{equation}
O_{5}=\frac{c_{5}}{M_{GUT}}\int d^{2}\theta QQQL
\end{equation}
and the dimension six operator:%
\begin{equation}
O_{6}=\frac{c_{6}}{M_{GUT}^{2}}\int d^{4}\theta QQU^{\dag}E^{\dag}\text{.}%
\end{equation}
The operator $O_{5}$ can originate from the exchange of heavy Higgs triplets
and can cause the decay $p\rightarrow K^{+}\overline{\nu}$. \ The operator
$O_{6}$ can originate from the exchange of heavy off-diagonal GUT\ group gauge
bosons and can cause the decay $p\rightarrow e^{+}\pi_{0}$. \ To remain in
accord with current bounds on nucleon decay, $c_{6}$ can typically be an order
one coefficient whereas $c_{5}$ must be suppressed at least to the order of
$10^{-7}$. \ See \cite{RabyProtonDecay} for further discussion on proton decay
in GUT\ models.

In four-dimensional GUT\ models, this issue is closely related to the
mechanism responsible for removing the Higgs triplets from the low energy
spectrum. \ One common approach is to invoke some continuous or discrete
symmetry to sufficiently suppress this operator. \ The use of discrete
symmetries in compactifications of M-theory on manifolds with $G_{2}$ holonomy
has been studied in \cite{WittenDecon}. \ Note that while the Higgs triplet
must develop a sufficiently large mass in order to reproduce the particle
content of the MSSM, we must also require that the supersymmetric Higgs mass
$\mu$ should be on the order of the weak scale.

While the above problems are necessary requirements for any potentially viable
model, there are many additional phenomenological constraints which must be
satisfied in a fully realistic compactification. In principle, a complete
model should also naturally accommodate hierarchical masses for the quarks and
leptons. \ For example, in conventional GUT\ models, the seesaw mechanism
allows the neutrino masses in the Standard Model to be much lighter than the
electroweak scale. \ At a more refined level, a full model should explain why
the CKM\ matrix is nearly equal to the identity matrix whereas the MNS\ matrix
contains nearly maximal mixing between the neutrinos.

A fully realistic model must of course specify how supersymmetry is broken and
provide a mechanism for communicating this breaking to the MSSM. \ Our
expectation is that this issue can be treated independently from the
supersymmetric models which shall be our primary focus here. \ We note that
for general string compactifications, supersymmetry breaking is closely
entangled with moduli stabilization. \ While we will not specify a method for
stabilizing moduli, we note that F-theory provides a natural arena for further
study of this issue. \ See \cite{DenefFixingAllModuli} for a particular
example of moduli stabilization in F-theory and \cite{DenefReview} for a
review of this active area of research.

\section{Basic Setup\label{BasicSetup}}

In this Section we review the basic properties of exceptional seven-branes in
F-theory. \ In particular, we explain how to compute the low energy matter
spectrum as well as the effective superpotential of the four-dimensional
theory. \ Further details may be found in \cite{BHV}.

F-theory compactified on an elliptically fibered Calabi-Yau fourfold preserves
$\mathcal{N}=1$ supersymmetry in the four uncompactified spacetime dimensions.
\ Letting $B_{3}$ denote the base of the Calabi-Yau fourfold, the discriminant
locus of the elliptic fibration determines a subvariety $\Delta$ of complex
codimension one in the base $B_{3}$. Denoting by $S$ the K\"{a}hler surface
defined by an irreducible component of $\Delta$, when this degeneration locus
is a singularity of $ADE$ type, the resulting eight-dimensional theory defines
the worldvolume of an exceptional seven-brane with gauge group $G_{S}$ of
$ADE$ type. \ This singularity type can enhance along complex codimension one
curves in $S$ to a singularity of type $G_{\Sigma}$ and can further enhance at
complex codimension two points in $S$ to a singularity of type $G_{p}$. \ Such
points correspond to the triple intersection of three matter curves. \ Because
the Cartan subalgebra of each singularity type is visible to the geometry
\cite{KatzMorrison,KatzVafa}, these enhancements satisfy the containment
relations:%
\begin{equation}
G_{S}\times U(1)\times U(1)\subset G_{\Sigma}\times U(1)\subset G_{p}\text{.}%
\end{equation}
As argued in \cite{BHV}, many necessary features of even semi-realistic
GUT\ models require that $G_{p}\subset E_{8}$. \ In particular, this implies
that the rank of the bulk gauge group $G_{S}$ is at most six. \ This
significantly limits the available bulk gauge groups because the rank of
$G_{S}$ must be at least four in order to contain the Standard Model gauge group.

In this paper we shall assume that given a choice of matter curves, there
exists a Calabi-Yau fourfold which contains the corresponding local
enhancement in singularity type. \ While this assumption is clearly not fully
justified for compact models, in the context of local models this can always
be done. \ As an example, we now engineer a local model where the bulk gauge
group $E_{6}$ enhances along a matter curve $\Sigma$ in $S$ to an $E_{7}$
singularity. \ A local elliptic model of this type is:%
\begin{equation}
y^{2}=x^{3}+fxz^{3}+q^{2}z^{4}\text{.} \label{ellipticexample}%
\end{equation}
In the above, $q$ is a section of $\mathcal{O}_{S}(\Sigma)$, $f$ is a section
of $L\otimes K_{S}^{-3}$ and the coordinates $(x,y,z)$ transform as a section
of \cite{BHV}:%
\begin{equation}
L^{2}\oplus L^{3}\oplus L\otimes K_{S}%
\end{equation}
where $K_{S}$ denotes the canonical bundle on $S$ and $L$ is a line bundle
which can be expressed in terms of $K_{S}$ and $\mathcal{O}_{S}(\Sigma)$.
\ The essential point of this example is that in a local model, there always
exists a line bundle $L$ such that the resulting local model is well-defined.
\ For example, in this case we have:%
\begin{equation}
L=\mathcal{O}_{S}(\Sigma)\otimes K_{S}^{2}\text{.}%
\end{equation}
Further, we shall make the additional assumption that there is no mathematical
obstruction to various twofold enhancements in the rank of the singularity at points of the surface $S$. \
It would certainly be of interest to study this issue.

We now describe in greater detail the effective action of exceptional
seven-branes. \ In terms of four-dimensional $\mathcal{N}=1$ superfields, the
matter content of the theory consists of an $\mathcal{N}=1$ vector multiplet
which transforms as a scalar on $S$, a collection of chiral superfields
$\mathbb{A}_{\overline{i}}$ which transform as a $(0,1)$ form on $S$ (the bulk
gauge bosons) and a collection of chiral superfields $\Phi$ which transform as
a holomorphic $(2,0)$ form on $S$. \ The bulk modes couple through the
superpotential term:%
\begin{equation}
W_{S}=\underset{S}{\int}Tr\left[  \left(  \overline{\partial}\mathbb{A}%
+\mathbb{A}\wedge\mathbb{A}\right)  \wedge\Phi\right]  \text{.} \label{WBULK}%
\end{equation}

When two irreducible components $S$ and $S^{\prime}$ of $\Delta$ intersect on
a Riemann surface $\Sigma$, the singularity type enhances further. \ In this
case, additional six-dimensional hypermultiplets localize along $\Sigma$. \ As
in \cite{KatzVafa}, the representation content of these fields is given by
decomposing the adjoint representation of the enhanced singularity to the
product $G_{S}\times G_{S^{\prime}}$ associated with the gauge groups on $S$
and $S^{\prime}$. \ In terms of four-dimensional $\mathcal{N}=1$ superfields,
the matter content localized on a curve consists of chiral superfields
$\Lambda$ and $\Lambda^{c}$ which transform as spinors on $\Sigma$. \ The bulk
modes couple to matter fields localized on the curve via the superpotential
term:%
\begin{equation}
W_{\Sigma}=\underset{\Sigma}{\int}\left\langle \Lambda^{c},(\overline
{\partial}+\mathbb{A}+\mathbb{A}^{\prime})\Lambda\right\rangle \label{WCURVE}%
\end{equation}
where $\left\langle \cdot,\cdot\right\rangle $ denotes the natural pairing
which is independent of any metric data.

Finally, when three irreducible components of $\Delta$ intersect at a point
$p$, the singularity type can enhance even further. \ Evaluating the overlap
of three $\Lambda$'s for three matter curves yields a further contribution to
the four-dimensional effective superpotential:%
\begin{equation}
W_{p}=\Lambda_{1}\Lambda_{2}\Lambda_{3}|_{p}\text{.} \label{WPOINT}%
\end{equation}
An analysis similar to that given below equation (\ref{ellipticexample}) shows
that given three matter curves which form a triple intersection, so long as
the resulting interaction term is consistent with group theoretic
considerations, there exists a local Calabi-Yau fourfold with the desired
twofold enhancement in singularity type.

Having specified the individual contributions to the quasi-topological
eight-dimensional theory, the superpotential is:%
\begin{equation}
W[\Phi,A,\Lambda]=W_{S_{1}}+...+W_{S_{l}}+W_{\Sigma_{1}}+...+W_{\Sigma
_{m}}+W_{p_{1}}+...+W_{p_{n}}+W_{flux}+W_{np}\text{.} \label{WFULL}%
\end{equation}
In the above, the corresponding fields entering the above expression are to be
viewed as a large collection of four-dimensional chiral superfields labeled by
points of the complex surfaces $S_{i}$ and the Riemann surfaces $\Sigma_{i}$.
\ We have also included the contribution from the flux-induced superpotential
which couples to the various $(2,0)$ forms of the seven-branes and indirectly
to matter fields localized on curves. \ As explained in \cite{BHV}, the vevs
for the $(2,0)$ form and fields localized on matter curves correspond to
complex deformations of the Calabi-Yau fourfold. \ Because the flux-induced
superpotential couples to the complex structure moduli of the Calabi-Yau
fourfold, such terms will generically be present. \ In equation (\ref{WFULL}),
we have also included the term $W_{np}$ which denotes all non-perturbative
contributions from wrapped Euclidean three-branes. \ These terms are
proportional to $\exp(-aVol(S))\sim\exp(-c/\alpha_{GUT})$ where $c$ is an
order one positive constant. \ In a GUT model where the gauge coupling
constants unify perturbatively, such contributions are negligible.

The fields of the four-dimensional effective theory correspond to zero mode
solutions in the presence of a background field configuration. \ As in
\cite{BHV}, we shall confine our analysis of the matter spectrum to
backgrounds where all fields other than the bulk gauge field are expanded
about zero. \ In the presence of a non-trivial background gauge field
configuration, the chiral matter content of the four-dimensional effective
theory descends from bulk modes on $S$ and Riemann surfaces which we denote by
the generic label $\Sigma$. \ An instanton taking values in a subgroup $H_{S}$
will break $G_{S}$ to the commutant subgroup. \ Decomposing the adjoint
representation of $G_{S}$ to the maximal subgroup of the form $\Gamma
_{S}\times H_{S}$, the chiral matter transforming in a representation $\tau$
of $\Gamma_{S}$ descends from the bundle-valued cohomology groups:%
\begin{equation}
\tau\in\text{ }H_{\overline{\partial}}^{0}(S,\mathcal{T}^{\ast})^{\ast}\oplus
H_{\overline{\partial}}^{1}(S,\mathcal{T})\oplus H_{\overline{\partial}}%
^{2}(S,\mathcal{T}^{\ast})^{\ast}%
\end{equation}
where $\mathcal{T}$ denotes a bundle transforming in the representation $T$ of
$H_{S}$ obtained by the decomposition of the adjoint representation of the
associated principle $G_{S}$ bundle on $S$. \ When $S$ is a del Pezzo surface,
the cohomology groups $H_{\overline{\partial}}^{0}$ and $H_{\overline
{\partial}}^{2}$ vanish for supersymmetric gauge field configurations so that
the number of zero modes transforming in the representation $\tau$ is given by
an index:%
\begin{equation}
n_{\tau}=\chi(S,\mathcal{T})=-\left(  1+\frac{1}{2}c_{1}(S)\cdot
c_{1}(\mathcal{T})+\frac{1}{2}c_{1}(\mathcal{T})\cdot c_{1}(\mathcal{T}%
)\right)  \text{.}%
\end{equation}

An analogous computation holds for the zero mode content localized on a
Riemann surface transforming in a representation $\nu\times\nu^{\prime}$ of
$H_{S}\times H_{S^{\prime}}$:%
\begin{equation}
\nu\times\nu^{\prime}\in H_{\overline{\partial}}^{0}(\Sigma,K_{\Sigma}%
^{1/2}\otimes\mathcal{V}\otimes\mathcal{V}^{\prime}) \label{mattercurvecohom}%
\end{equation}
so that the net number of zero modes is given by the index:%
\begin{equation}
n_{\nu\times\nu^{\prime}}-n_{\overline{\nu\times\nu^{\prime}}}=\deg\left(
\mathcal{V}\otimes\mathcal{V}^{\prime}\right)  \text{.} \label{curveIndex}%
\end{equation}
In many cases we shall compute the relevant cohomology groups in equation
(\ref{mattercurvecohom}) by assuming a canonical choice of spin structure.
\ As argued in \cite{BHV}, this can always be done when the curve corresponds
to the vanishing locus of the holomorphic $(2,0)$ form in the
eight-dimensional theory.

When $\pi_{1}(S)\neq0$, it is also possible to consider vacua with non-trivial
Wilson lines. \ In order to avoid complications from the reduction of
additional supergravity modes, we shall always assume that $\pi_{1}(S)$ is a
finite group. \ The discussion closely parallels a similar analysis in
heterotic compactifications (see for example
\cite{DonagiOvrutWilsonLineSpectra}). \ Recall that admissible Wilson lines
are specified by a choice of element $\rho_{S}\in Hom(\pi_{1}(S),G_{S})$. \ In
order to maintain continuity with the discussion reviewed above, we shall
require that the non-trivial portion of the discrete Wilson line takes values
in the subgroup $\Gamma_{S}\subset G_{S}$ defined above. \ More generally,
this restriction can be lifted and may allow additional possibilities for
projecting out phenomenologically unviable representations from the low energy
spectrum. \ Under these restrictions, the unbroken four-dimensional gauge
group is given by the commutant subgroup of $\rho_{S}(\pi_{1}(S))\times H_{S}$
in $G_{S}$.

We now determine the zero mode content of the theory in the presence of a
non-trivial discrete Wilson line. \ As in Calabi-Yau compactifications of the
heterotic string, our strategy will be to lift all computations to a covering
theory. \ Because $\pi_{1}(S)$ is finite, the universal cover of $S$ denoted
by $\widetilde{S}$ is a compact K\"{a}hler surface. \ Letting $p:\widetilde
{S}\rightarrow S$ denote the covering map, the bundle $\mathcal{T}$ on $S$ now
lifts to a bundle $\widetilde{\mathcal{T}}=p^{\ast}(\mathcal{T})$ on
$\widetilde{S}$. \ Under the present restrictions, the Wilson line corresponds
to a flat $\Gamma_{S}$-bundle induced from the covering map from
$\widetilde{S}$ to $S$. \ The deck transformation defined by the action of
$\pi_{1}(S)$ on $\widetilde{S}$ also determines a group action of $\pi_{1}(S)$
on the cohomology groups $H_{\overline{\partial}}^{i}(\widetilde{S}%
,\widetilde{\mathcal{T}})$. \ Treating $H_{\overline{\partial}}^{i}%
(\widetilde{S},\widetilde{\mathcal{T}})$ as a complex vector space, the
eigenspace decomposition of $H_{\overline{\partial}}^{i}(\widetilde
{S},\widetilde{\mathcal{T}})$ is of the form:%
\begin{equation}
H_{\overline{\partial}}^{i}(\widetilde{S},\widetilde{\mathcal{T}}%
)\simeq\underset{\lambda}{\oplus}%
\mathbb{C}
_{\lambda}%
\end{equation}
in the obvious notation. \ The irreducible representation of $\Gamma_{S}$
defined by $\tau$ decomposes into irreducible representations of the maximal
subgroup $\Gamma\times\rho_{S}(\pi_{1}(S))\subset\Gamma_{S}$ as:%
\begin{equation}
\tau\simeq\underset{i}{\oplus}\tau_{i}\otimes R_{i}\text{.}%
\end{equation}
The zero modes transforming in the representation $\tau_{i}$ are therefore
specified by the $\rho_{S}$ invariant subspaces:%
\begin{equation}
\tau_{i}:\left[  H_{\overline{\partial}}^{0}(\widetilde{S},\widetilde
{\mathcal{T}}^{\ast})^{\ast}\otimes R_{i}\right]  ^{\rho_{S}}\oplus\left[
H_{\overline{\partial}}^{1}(\widetilde{S},\widetilde{\mathcal{T}}^{\ast
})^{\ast}\otimes R_{i}\right]  ^{\rho_{S}}\oplus\left[  H_{\overline{\partial
}}^{2}(\widetilde{S},\widetilde{\mathcal{T}}^{\ast})^{\ast}\otimes
R_{i}\right]  ^{\rho_{S}}\text{.}%
\end{equation}

Having specified the zero mode content of the theory, we can now in principle
determine the full superpotential of the low energy effective theory by
integrating out all Kaluza-Klein modes from equation (\ref{WFULL}). \ This is
similar to the treatment of Chern-Simons gauge theory as a string theory
\cite{ChernSimonsStringWitten}. \ For quiver gauge theories defined by D-brane
probes of Calabi-Yau threefolds, the higher order terms of the effective
superpotential are given by integrating out all higher Kaluza-Klein modes from
the associated holomorphic Chern-Simons theory for B-branes
\cite{DouglasTomasso}.

In the present context, we can follow the procedure outlined in
\cite{DijkgraafVafaDecon} to determine the full expression for the effective
superpotential. \ This is given by a bosonic partition function with action
given by the superpotential of equation (\ref{WFULL}). \ \ Viewing the
higher-dimensional fields as a collection of four-dimensional chiral
superfields labeled by points of the internal space, the effective
superpotential is now given by the bosonic path integral:%
\begin{equation}
\exp\left(  -W_{eff}[\Phi_{0},A_{0},\Lambda_{0}]\right)  =\underset{1PI}{\int
}[d\Phi][dA][d\Lambda]\exp\left(  -W[\Phi+\Phi_{0},A+A_{0}%
,\Lambda+\Lambda_{0}]\right)  \label{partitionfunction}%
\end{equation}
where the zero subscript denotes the zero mode, and the path integral is over
all one particle irreducible Feynman diagrams. \ In this expression,
$W_{tree}$ should be viewed as a bosonic action with functional dependence
identical to that of equation (\ref{WFULL}). \ The complete four-dimensional
effective superpotential for the zero modes is then determined by the
partition function of the quasi-topological theory. \ We emphasize that this
partition function is well-defined without any reference to metric data. \ A
very similar procedure for extracting the superpotential by integrating out
Kaluza-Klein modes in heterotic compactifications has been given in
\cite{WittenU(1)}. \ Some examples of similar computations for quiver gauge
theories can be found in \cite{AspinwallKatz}. \ To conclude this Section, we
note that any symmetry of the full eight-dimensional theory descends to the
four-dimensional effective superpotential for the zero modes. \ Neglecting the
contribution due to non-perturbative effects in equation (\ref{WFULL}), the
extra $U(1)$ factors which are always present when the singularity type
enhances will provide additional global symmetries in the effective theory
which will typically forbid some higher dimension operators from being
generated. \ Although non-perturbative effects can violate these symmetries,
the corresponding contribution to $W_{eff}[\Phi_{0},A_{0},\Lambda_{0}]$ will
typically be small enough that we may safely neglect such contributions.

These general considerations already constrain the matter content of any
candidate theory.\ \ Modes propagating in the bulk of the surface $S$ must
transform in the adjoint representation of the bulk gauge group. \ Moreover,
although matter fields can localize on a curve $\Sigma$ inside of $S$, these
fields must descend from the adjoint representation of $G_{\Sigma}$. \ For
example, for $SU(N)$ gauge group factors which do not embed in $E_{8}$, the
only available local enhancements are to higher $A$ or $D$ type singularities.
\ In such cases, the decomposition of the adjoint representation only contains
two index representations. \ Similar restrictions apply for $SO(N)$ gauge
group factors which do not embed in $E_{8}$. \ In particular, the spinor
representation never appears in such cases. \ In a sense, this is to be
expected because these are precisely the types of configurations which can be
realized within perturbative type IIB\ vacua.

For $SO(N)\subset E_{8}$ gauge groups, the available representations are the
vector, spinor or adjoint representations, and for $SU(N)\subset E_{8}$ gauge
groups, the only available representations are the one, two or three index
anti-symmetric and the adjoint representations.\footnote{Strictly speaking
there are additional possibilities if the rank of the bulk singularity
enhances by more than one rank. \ If one allows more general breaking patterns
involving higher $SU(N)$ and $SO(2N)$ type enhancements, it is also possible
to achieve two index symmetric representations of $SU(N)$ theories. \ For
example, letting $A_{2N}$ denote the two index anti-symmetric representation
of $SU(2N)$, $A_{2N}$ decomposes to $SU(N)\times SU(N)$ as $A_{2N}\rightarrow
A_{N}\otimes1+1\otimes A_{N}+F_{N}\otimes F_{N}$. \ Higgsing this to the
diagonal $SU(N)$ subgroup, we note that the product $F_{N}\otimes F_{N}$
contains two index symmetric representations. This is a rather exotic
possibility and we shall therefore not consider it further in this paper.} For
example, when $G_{S}=SO(10)$, this implies that all of the matter fields
transform in the $10,16,\overline{16}$ or $45$, while for $G_{S}=SU(5)$, the
only available representations are the $5$, $\overline{5}$, $10$,
$\overline{10}$ or $24$. \ in the specific case of del Pezzo models, this
matter content is even more constrained. \ Indeed, as explained in \cite{BHV},
the bulk zero mode content for del Pezzo models never contains chiral
superfields which transform in the adjoint representation of the unbroken
gauge group in four dimensions.

In fact, the type of twofold enhancement strongly determines the qualitative
behavior of the associated triple intersection of matter curves. \ For
example, the possible rank two enhancements of $SU(5)$ are $E_{6}$, $SO(12)$,
and $SU(7)$. \ In the case of $E_{6}$ and $SO(12)$, the associated curves
which form a triple intersection all live inside of $S$. \ Indeed, by group
theory considerations, the matter fields localized on each curve transform in
non-trivial representations of $SU(5)$ \cite{BHV}. \ On the other hand, this
is qualitatively different from a local enhancement to $SU(7)$. \ In this
case, two of the curves of the triple intersection support matter in the
fundamental and anti-fundamental of $SU(5)$ and therefore live in $S$, while
the third curve of the intersection supports matter in the singlet representation.

More generally, we note that as opposed to a generic field theory, in
F-theory, vector-like pairs of the bulk gauge group can only interact through
cubic superpotential terms involving a field localized on a curve which
intersects $S$ at a point. \ While the vev of this gauge singlet can induce a
mass term for the vector-like pair, the dynamics of this field in the
threefold base $B_{3}$ is qualitatively different from fields which localize
inside of $S$.

\section{Mass Scales and Decoupling Limits\label{MASSSCALES}}

\begin{figure}
[ptb]
\begin{center}
\includegraphics[
height=3.10295in,
width=5.7843in
]%
{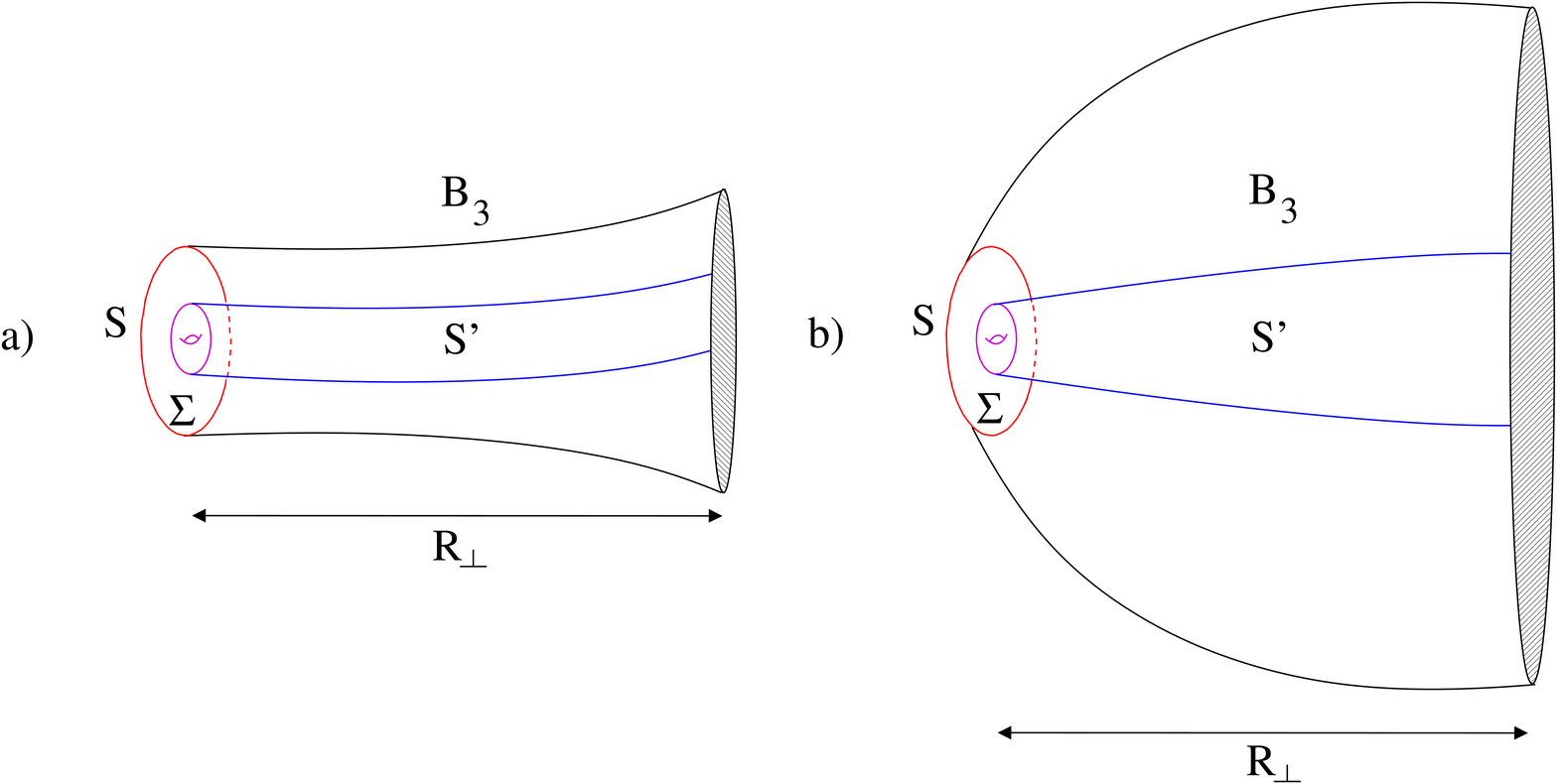}%
\caption{Depiction of F-theory compactified on a local model of a Calabi-Yau
fourfold with non-compact base threefold $B_{3}$. \ The diagram shows the
behavior of the geometry in the neighborhood of a compact K\"{a}hler surface
$S$ on which the gauge degrees of freedom of the GUT\ model can localize in
the cases where $B_{3}$ is given by a roughly tubular geometry, as in case a),
as well as geometries where $B_{3}$ is more homogeneous, as in case b). \ In
both cases, the directions orthogonal to $S$ in $B_{3}$ are large compared to
$S$, but not warped. \ To regulate the geometry of the local model it is
necessary to introduce a cutoff length scale which we denote by $R_{\bot}$.
\ The intersection locus between the compact surface $S$ and a non-compact
surface $S^{\prime}$ appears as a curve $\Sigma$ in $S$. \ When seven-branes
wrap both surfaces, additional light states will localize on this matter
curve.}%
\label{massscales}%
\end{center}
\end{figure}

Before proceeding to specific models, we first present a general analysis of
the relevant mass scales in the local models we treat in this paper. \ Rather
than specify one particular profile for the threefold base $B_{3}$, we
consider both geometries where $B_{3}$ is roughly tubular so that it
decomposes as the product of~$S$ with two non-compact directions orthogonal to
$S$ in $B_{3}$, as well as more homogeneous profiles. To parameterize our
ignorance of the details of the geometry, we define the length scales:%
\begin{align}
R_{S}  &  \equiv Vol(S)^{1/4}\\
R_{B}  &  \equiv Vol(B_{3})^{1/6}%
\end{align}
as well as a cutoff length scale $R_{\bot}$ which measures the radius normal
to $S$:%
\begin{equation}
R_{\bot}\equiv R_{B}\times\left(  \frac{R_{B}}{R_{S}}\right)  ^{\nu}%
\end{equation}
so that the exponent $\nu$ ranges from $\nu=0$ when $B_{3}$ is homogeneous, to
the value $\nu=2$ when $B_{3}$ is the product of $S$ with two non-compact
directions. \ Indeed, the approximations we consider in this paper are valid
in the regime $0\lesssim\nu\lesssim2$. \ Note that under the assumption
$R_{B}>R_{S}$, the three length scales are related by:%
\begin{equation}
R_{\bot}>R_{B}>R_{S}\text{.} \label{RADIIINEQU}%
\end{equation}
See figure \ref{massscales} for a comparison of the local behavior of $B_{3}$
for $\nu\sim0$ and $\nu\sim2$. \ To clarify, although the directions normal to
$S$ are \textquotedblleft non-compact\textquotedblright\ in our local model,
in a globally consistent compactification of F-theory they will still be quite
small, and all on the order of the GUT scale, as will be discussed below.
\ Indeed, this is quite different from models based on large extra dimensions
which can be either flat, but still compact \cite{ArkaniHamedEXTRADIM}, or
potentially highly warped and of infinite extent \cite{RSTWO}.%

Compactifying on a threefold base $B_{3}$, the ten-dimensional
Einstein-Hilbert action is:%
\begin{equation}
S_{EH}\sim M_{\ast}^{8}\underset{%
\mathbb{R}
^{3,1}\times B_{3}}{\int}R\sqrt{-g}d^{10}x
\end{equation}
where $M_{\ast}$ is a particular mass scale associated with the supergravity
limit of the F-theory compactification. \ In perturbative type IIB\ string
theory, the parameter $M_{\ast}$ is given in string frame by the relation
$M_{\ast}^{8}=M_{s}^{8}/g_{s}^{2}$. \ Upon reduction to four dimensions, the
four-dimensional Planck scale $M_{pl}$ satisfies the relation:%
\begin{equation}
M_{pl}^{2}\sim M_{\ast}^{8}Vol(B_{3})\text{.} \label{4dMPLANCK}%
\end{equation}
The tension of a seven-brane wrapping a K\"{a}hler surface $S$ in $B_{3}$
determines the gauge coupling constant of the four-dimensional effective
theory. \ More precisely, the coefficient of the kinetic term for the gauge
field strength is of the form:\footnote{The astute reader will notice a
difference in sign between the gauge kinetic term used here, and the
convention adopted in \cite{BHV}. \ In \cite{BHV}, we adopted an
anti-hermitian basis of Lie algebra generators in order to conform to
conventions typically used in topological gauge theory. \ Because our emphasis
here is on the four-dimensional effective field theory, in this paper we have
reverted back to the standard sign convention in the physics literature so
that all Lie group generators are hermitian.}%
\begin{equation}
S_{kin}\sim-M_{\ast}^{4}\underset{%
\mathbb{R}
^{3,1}\times S}{\int}Tr\left(  F\wedge\ast_{8}F\right)  \text{.}%
\end{equation}
The value of the gauge coupling constant at the scale of unification is
therefore:%
\begin{equation}
\alpha_{GUT}^{-1}\sim M_{\ast}^{4}Vol(S)\text{.} \label{GUTCOUP}%
\end{equation}
Equations (\ref{4dMPLANCK}) and (\ref{GUTCOUP}) now imply:%
\begin{equation}
Vol(B_{3})\sim\left(  \alpha_{GUT}M_{pl}Vol(S)\right)  ^{2}
\label{basicscalerelation}%
\end{equation}
or:%
\begin{equation}
R_{B}^{6}\sim\left(  \alpha_{GUT}M_{pl}R_{S}^{4}\right)  ^{2}\text{.}%
\end{equation}

We now convert these geometric scales into mass scales in the low energy
effective theory. \ To this end, we next relate $Vol(S)$ to the GUT\ scale
$M_{GUT}$. \ In most of the cases we consider, non-zero flux in the internal
directions of $S$ will partially break the bulk gauge group of the
seven-brane. \ Letting $\sqrt{\left\langle F_{S}\right\rangle }$ denote the
mass scale of the internal flux, we therefore require:%
\begin{equation}
M_{GUT}^{2}\sim\left\langle F_{S}\right\rangle \text{.}%
\end{equation}
Because the flux is measured in units of length$^{-2}$ on the surface $S$,
this implies:%
\begin{equation}
Vol(S)\sim M_{GUT}^{-4}\text{.} \label{VOLSMGUT}%
\end{equation}
Equation (\ref{basicscalerelation}) therefore yields:%
\begin{equation}
Vol(B_{3})\sim\left(  \alpha_{GUT}M_{pl}M_{GUT}^{-4}\right)  ^{2}\text{.}%
\end{equation}
The radii $R_{B}$ and $R_{S}$ are therefore given by:%
\begin{align}
\frac{1}{R_{S}}  &  \sim M_{GUT}=3\times10^{16}\text{ GeV}\\
\frac{1}{R_{B}}  &  \sim M_{GUT}\times\varepsilon^{1/3}\sim10^{16}\text{ GeV}%
\end{align}
where we have introduced the small parameter:%
\begin{equation}
\varepsilon\equiv\frac{M_{GUT}}{\alpha_{GUT}M_{pl}}\sim7.5\times
10^{-2}\text{.}%
\end{equation}
Collecting equations (\ref{basicscalerelation}) and (\ref{VOLSMGUT}), the
parameter $R_{\bot}$ now takes the form:%
\begin{equation}
\frac{1}{R_{\bot}}=M_{GUT}\times\varepsilon^{\gamma}\sim5\times10^{15\pm
0.5}\text{ GeV} \label{RPERP}%
\end{equation}
where $1/3\leq\gamma\leq1$. \ We note that these numerical values for the
radii satisfy the inequality of line (\ref{RADIIINEQU}).

We conclude this Section by discussing the normalization of Yukawa couplings
in models where the superpotential originates from the triple intersection of
matter curves. \ In a holomorphic basis of wave functions, the F- and D-terms
are:%
\begin{align}
L_{F}^{hol}  &  =\underset{p}{\sum}\psi_{i}(p)\psi_{j}(p)\psi_{k}(p)\int
d^{2}\theta\widetilde{\phi}_{i}\widetilde{\phi}_{j}\widetilde{\phi}%
_{k}\label{FTERMS}\\
&  \equiv\lambda_{ijk}^{hol}\int d^{2}\theta\widetilde{\phi}_{i}%
\widetilde{\phi}_{j}\widetilde{\phi}_{k}\\
L_{D}^{hol}  &  =M_{\ast}^{2}\underset{\Sigma}{\int}d^{4}\theta K(\widetilde
{\phi},\widetilde{\phi}^{\dag}) \label{DTERMS}%
\end{align}
where in the above, $\psi_{i}(p)$ denotes the internal value of the wave
function associated with the four-dimensional chiral superfield $\widetilde
{\phi}_{i}$ evaluated at a point $p$ in $S$, and the holomorphic Yukawa
couplings are defined as:
\begin{equation}
\lambda_{ijk}^{hol} = \underset{p}{\sum}\psi_{i}(p)\psi_{j}(p)\psi
_{k}(p)\text{.}%
\end{equation}
The behavior of the wave functions near these points can generate
hierarchically small values near nodal points, and order one values away from
such nodal points.

We eventually wish to extract numerical estimates for the physical Yukawa
couplings, defined in a basis of four-dimensional chiral superfields with
canonically-normalized kinetic terms. However, if we reduce the $D$-term in
\eqref{DTERMS} over $\Sigma$, we find that the kinetic term for $\widetilde
{\phi}$ is multiplied by the $L^{2}$-norm on $\Sigma$ of the corresponding
zero-mode wave function $\psi$.

In general, $\psi$ transforms on $\Sigma$ as a holomorphic section of
${K_{\Sigma}^{1/2} \otimes L}$, where $L$ is a line bundle on $\Sigma$
determined by the gauge field on $S$. Both $K_{\Sigma}^{1/2}$ and $L$ carry
natural hermitian metrics inherited from the bulk metric and gauge field on
$S$. Fixing the holomorphic wave function $\psi$, we are interested in how the
$L^{2}$-norm of $\psi$ scales with the metric on $S$, since the volume of $S$
effectively determines $M_{GUT}$. For concreteness, let us write the metric on
$S$ in local holomorphic coordinates $(z,w)$ as ${ds^{2} = g_{z \overline z}
\, dz d\overline z \,+\, g_{w \overline w} \, dw d\overline w}$, where $z$ is
a local holomorphic coordinate along $\Sigma$ and $w$ is a holomorphic
coordinate normal to $\Sigma$. Under an overall scaling ${g \mapsto\ell g}$,
the hermitian metric on $L$ is unchanged, so the norm of $\psi$ behaves as
\begin{align}
\label{SCALEPSI}\langle\psi|\psi\rangle\,  &  =\, \int_{\Sigma}d^{2} z \, g_{z
\overline z} \big(g^{z \overline z}\big)^{1/2} \psi\overline\psi\,,\cr &
\longmapsto\, \ell^{1/2} \, \langle\psi|\psi\rangle\,.
\end{align}
Since the volume of $\Sigma$ scales with $\ell$, we see from \eqref{SCALEPSI}
that $\langle\psi|\psi\rangle$ scales with $\Vol(\Sigma)^{1/2}$.

At first glance, the dependence of $\langle\psi|\psi\rangle$ on $\ell$ might
appear to be the only source of $\ell$-dependence in the respective $F$- and
$D$-terms in \eqref{FTERMS} and \eqref{DTERMS}, since the $F$-term is
determined by the overlap of fixed holomorphic wavefunctions. However, in
making precise sense of this overlap, an additional $\ell$-dependence also enters.

To explain this $\ell$-dependence, let us consider a slightly simplified
situation, for which the holomorphic curves $\Sigma_{1}$, $\Sigma_{2}$, and
$\Sigma_{3}$ meet transversely at a point inside a Calabi-Yau threefold
$B_{3}$. The role of the line bundle $L$ is inessential, so on each curve we
take the wavefunction $\psi_{i}$ to transform as a holomorphic section of
$K_{\Sigma_{i}}^{1/2}$. In local holomorphic coordinates $(z,w,v)$ around the
point $p$ of intersection, the wavefunction overlap is defined by
\begin{equation}
\label{OVERLAP}\psi_{1}(p) \psi_{2}(p) \psi_{3}(p) \, \frac{\sqrt{dz}
\sqrt{dw} \sqrt{dv}}{\sqrt{\Omega(p)}}\,.
\end{equation}
Here $\Omega$ is a holomorphic three-form on $B_{3}$ which we must introduce
so that the overlap in \eqref{OVERLAP} does not depend on the particular
holomorphic coordinates $(z,w,v)$ chosen at $p$.

Of course, $\Omega$ is unique up to scale --- but it is precisely the scale of
the overlap that we are trying to fix! Given that $B_{3}$ carries a metric, we
fix the norm of $\Omega$ by the requirement that ${-i \, \Omega\^ \bar
\Omega\,=\, \omega\^ \omega\^ \omega}$, where $\omega$ is the K\"ahler form
associated to the metric on $B_{3}$. Once we impose this condition, $\Omega$
scales as ${\Omega\mapsto\ell^{3/2} \, \Omega}$ under an overall scaling of
the metric on $B_{3}$. Hence the wavefunction overlap in \eqref{OVERLAP} and
thus the holomorphic Yukawa coupling $\lambda^{hol}_{i j k}$ actually scales
as $\ell^{-3/4}$.

After canonically normalizing all kinetic terms, the physical Yukawa couplings
are given by
\begin{equation}
\lambda^{phys}_{i j k} \,=\, \frac{\lambda^{hol}_{i j k}}{\sqrt{M_{*}^{2}
\langle\psi_{i}|\psi_{i}\rangle\, M_{*}^{2} \langle\psi_{j}|\psi_{j}\rangle\,
M_{*}^{2} \langle\psi_{k}|\psi_{k}\rangle}}\,.
\end{equation}
By the preceding discussion, under an overall scaling ${g \mapsto\ell\, g}$ of
the metric on $B_{3}$, the physical Yukawa coupling scales as ${\lambda
^{phys}_{i j k} \mapsto\ell^{-3/2} \, \lambda^{phys}_{i j k}}$. Restoring the
dependence on the volumes of each curve, we find the result which one would
naively guess,
\begin{equation}
\label{PHYSYUK}\lambda^{phys}_{i j k} \,=\, \frac{\lambda^{0}_{i j k}}%
{\sqrt{M_{\ast}^{2}\Vol(\Sigma_{i})\,M_{\ast}^{2}\Vol(\Sigma_{j})\,M_{\ast
}^{2}\Vol(\Sigma_{k})}}\,.
\end{equation}
Here $\lambda^{0}_{i j k}$ denotes the fiducial, order one Yukawa coupling
defined by \eqref{OVERLAP} when $B_{3}$ has unit volume.

Although we have phrased the preceding discussion in the very special case
that $\Sigma_{1}$, $\Sigma_{2}$, and $\Sigma_{3}$ are holomorphic curves
intersecting transversely in a Calabi-Yau threefold, the result
\eqref{PHYSYUK} holds quite generally in F-theory. According to the discussion
in \S $5.2$ of \cite{BHV}, when $\Sigma_{1}$, $\Sigma_{2}$, and $\Sigma_{3}$
are matter curves intersecting at a point $p$ inside $S$, one must choose a
trivialization of ${\left(  K^{1/2}_{\Sigma_{1}} \otimes K^{1/2}_{\Sigma_{2}}
\otimes K^{1/2}_{\Sigma_{3}}\right)  \!\Big|_{p}}$ to evaluate the
wavefunction overlap. This choice, analogous to the choice of $\Omega$ in
\eqref{OVERLAP}, introduces the same scaling with $\ell$.

Once we introduce four-dimensional chiral superfields $\{\phi_{i}\}$ with
canonical kinetic terms, the $F$-terms become
\begin{equation}
L_{F}=\lambda_{ijk}^{0} \int d^{2}\theta\frac{\phi_{i}\phi_{j}\phi_{k}}%
{\sqrt{M_{\ast}^{2}\Vol(\Sigma_{i}) \, M_{\ast}^{2}\Vol(\Sigma_{j}) \,
M_{\ast}^{2}\Vol(\Sigma_{k})}}\,. \label{YUKCOUP}%
\end{equation}
We note that when all matter curves have comparable volumes set by the overall
size of $\Vol(S)$, $\Vol(\Sigma)^{2}\sim\Vol(S)$ . \ In this case,
\eqref{GUTCOUP} implies:
\begin{equation}
L_{F}=\alpha_{GUT}^{3/4}\lambda_{ijk}^{0} \int d^{2}\theta\, \phi_{i}\phi
_{j}\phi_{k}\text{.}%
\end{equation}

In rescaling each field by an appropriate power of the volume factor, we shall
typically use the classical value of $\Vol(\Sigma_{i})$. \ Strictly speaking,
this approximation is only valid in the supergravity limit. \ Due to the fact
that in F-theory there is at present no perturbative treatment of quantum
corrections, most of the numerical results obtained throughout this paper can
only be reliably treated as order of magnitude estimates. \

\section{General Overview of the Models\label{GeneralOverview}}

In this Section we provide a guide to the class of models we study. \ The
choice of K\"{a}hler surface $S$ already determines many properties of the low
energy effective theory. \ In keeping with our general philosophy, we require
that the spectrum at low energies must not contain any exotics. \ When
$h^{1,0}(S)\neq0$, we expect the low energy spectrum to contain additional
states obtained by reduction of the bulk supergravity modes of the
compactification. \ For this reason we shall always require that $\pi_{1}(S)$
is a finite group. \ There are two further possible refinements depending on
whether or not the model in question admits a limit in which $M_{GUT}$ remains
finite while $M_{pl}\rightarrow\infty$. \ In order to fully decouple gravity,
the extension of the local metric on $S$ to a local Calabi-Yau fourfold must
possess a limit in which the surface $S$ can shrink to zero size. \ In
particular, this imposes the condition that $K_{S}^{-1}$ must be ample. \ This
is equivalent to the condition that $S$ is a del Pezzo surface, in which case
$h^{2,0}(S)=0$. \ We note that the degree $n\geq2$ Hirzebruch surfaces satisfy
$h^{2,0}(S)=0$ but do not define fully consistent decoupled models.%
\begin{figure}
[ptb]
\begin{center}
\includegraphics[
height=2.4526in,
width=5.7865in
]%
{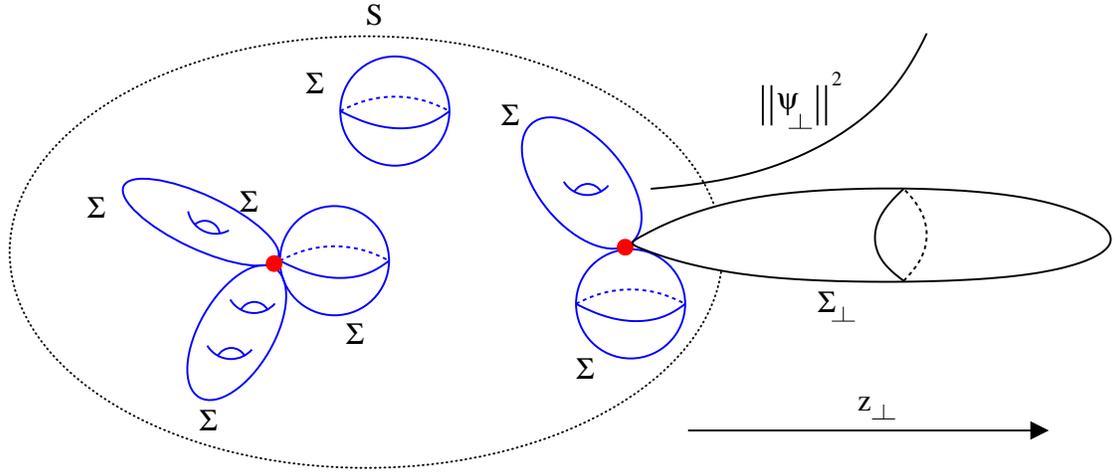}%
\caption{The bulk group on the K\"{a}hler surface $S$ corresponds to a
singularity of type $G_{S}$. \ Over complex codimension one matter curves in
$S$ which we denote by $\Sigma$, this singularity type can further enhance so
that six-dimensional matter fields localize on these curves. \ Over complex
codimension two points in $S$ the singularity type can enhance further. \ On
the left of the figure we depict a triple intersection of matter curves in
$S$. \ It is also possible for one of the matter curves to intersect $S$ at a
point. \ Depending on the background gauge fluxes and local curvatures, wave 
functions localized on curves normal to the GUT brane are either exponentially 
suppressed or of order one near the point of contact with the GUT brane.}%
\label{enhancer}%
\end{center}
\end{figure}

In fact, even the way in which the gauge group of the GUT\ breaks to that of
the MSSM strongly depends on whether or not such a decoupling limit exists.
\ For surfaces with $h^{2,0}(S)\neq0$, the zero mode content will contain
contributions from the bulk holomorphic $(2,0)$ form. \ Because the $(2,0)$
form determines the position of the exceptional brane inside of the threefold
base $B_{3}$, a non-zero vev for the associated zero modes corresponds to the
usual breaking of the GUT\ group via an adjoint-valued chiral
superfield.\footnote{The potential application of this GUT breaking mechanism
was noted in a footnote of \cite{SvrcekKachruMediation} and has also been
discussed in \cite{BHV,WijnholtDonagi}.} \ Along these lines, we present some
examples of four-dimensional GUT\ models which can originate from surfaces of
general type. \ \textit{An important corollary of this condition is that the
usual four-dimensional field theory GUT\ models cannot be fully decoupled from
gravity!} \ We believe this is important because it runs counter to the usual
effective field theory philosophy that issues pertaining to the Planck scale
can always be decoupled. \ This is in accord with the existence of a swampland
of effective field theories which may not admit a consistent UV\ completion
which includes gravity \cite{VafaSwamp}. \ Moreover, as we explain in greater
detail later, it is also possible that a generic surface of general type may
not support sufficiently many matter curves of the type needed to engineer a
fully realistic four-dimensional GUT\ model.

When available, discrete Wilson lines in higher-dimensional theories provide
another way to break the GUT\ group\ to $G_{std}$. \ Indeed, most models based
on compactifications of the heterotic string on Calabi-Yau threefolds require
discrete Wilson lines to break the gauge group and project out exotics from
the low energy spectrum. \ When $\pi_{1}(S)\neq0$, a similar mechanism for
gauge group breaking is available for exceptional seven-brane theories. \ As
an example, we present a toy model where $S$ is an Enriques surface and
$G_{S}=SU(5)$. \ In our specific example, we find that the zero mode content
contains additional vector-like pairs of fields in exotic representations of
$G_{std}$.

We next turn to the primary case of interest for bottom up string
phenomenology where $S$ is a del Pezzo surface.\ \ Because $h^{2,0}(S)=0$ and
$\pi_{1}(S)=0$ for del Pezzo surfaces, the two mechanisms for gauge group
breaking mentioned above are now unavailable. \ In this case, the GUT\ group
breaks to a smaller subgroup due to non-trivial internal fluxes. \ For
example, the group $SU(5)$ can break to $SU(3)\times SU(2)\times U(1)_{Y}$
when the internal flux takes values in the $U(1)_{Y}$ factor. \ In heterotic
compactifications this mechanism is unavailable because a non-zero internal
field strength would generate a string scale mass for the $U(1)$ hypercharge
gauge boson in four dimensions \cite{WittenU(1)}. \ We find that in F-theory
compactifications \textit{without a heterotic dual}, there is a natural
topological condition for the four-dimensional gauge boson to remain massless.
\ Our expectation is that this condition is satisfied for many choices of
compact threefolds $B_{3}$. \ In the remainder of this Section we discuss
further properties of del Pezzo models.

Along these lines, we present models based on $G_{S}=SU(5)$ where the gauge
group of the eight-dimensional theory breaks directly to $G_{std}$ in four
dimensions, as well as a hybrid scenario where $G_{S}=SO(10)$ breaks to
$SU(5)\times U(1)$ in four dimensions and then subsequently descends from a
flipped $SU(5)$ GUT\ model to the MSSM. \ In fact, we also present a general
no go theorem showing that direct breaking of $SO(10)$ to $G_{std}$ via
abelian fluxes always generates extraneous matter in the low energy spectrum.
\ In both the regular $SU(5)$ and flipped $SU(5)$ scenarios, we find that in
order to achieve the exact spectrum of the MSSM, all of the matter fields must
localize on Riemann surfaces. \ In the $G_{S}=SU(5)$ models, the matter fields
organize into the $\overline{5}$ and $10$ of $SU(5)$. \ In the $G_{S}=SO(10)$
models, a complete multiplet in the $16$ of $SO(10)$ localizes on the matter
curves. \ In both cases, all matter localizes on curves so that all of the
tree level superpotential terms descend from the triple intersection of matter
curves. \ When some of the matter localizes on different curves, this leads to
texture zeroes in the Yukawa matrices.

In addition to presenting some examples of minimal del Pezzo models, one of
the primary purposes of this paper is to develop a number of ingredients which
can be of use in further more refined model building efforts. \ A general
overview of these ingredients has already been given in the Introduction, so
rather than repeat this here, we simply summarize the primary themes of the
minimal $SU(5)$ model which recur throughout this paper. \ The most prominent
ingredient is the internal hypercharge flux which facilitates GUT\ breaking.
This hyperflux also provides a natural solution to the doublet-triplet
splitting problem and generates distorted GUT\ mass relations for the lighter
generations. \ More generally, the presence of additional global $U(1)$
symmetries in the low energy theory forbids a number of potentially
problematic interaction terms from appearing in the superpotential.
\ Topologically, the absence of dangerous operators translates into conditions
on how the matter curves intersect inside of $S$. \ For example, proton decay
is automatically suppressed when the Higgs up and down fields localize on
different matter curves. \ When these curves do not intersect, the $\mu$ term
is zero. \ When the Higgs matter curves do intersect, the resulting $\mu$ term
can be naturally suppressed. \ Indeed, an important feature of all the models
we consider is that while expectations from effective field theory would
suggest that vector-like pairs will always develop a suitably large mass, here
we find two distinct possibilities depending on the choice of the sign for the
gauge fluxes: In one case (when the normal wave function is attracted to our
brane) we essentially recover the field theory intuition. On the other hand,
with a different choice of sign (when the normal wave function is repelled
from our brane) we find the opposite situation, where $\mu$ is highly
suppressed. \ The ostensibly large mass term corresponding to the vev of a
gauge singlet is in fact exponentially suppressed since its wave function is
very small near our brane. \ Here, the principle of decoupling is especially
important because the large positive curvature of the del Pezzo surface can
lead to a natural suppression of the normal wave functions. \ This provides an
explanation for why the $\mu$ term is far below the GUT\ scale, as well as why
the neutrino masses are so far below the electroweak scale. \ While we discuss
many of these mechanisms in the specific context of the minimal $SU(5)$ model,
these same features carry over to the flipped $SU(5)$ GUT\ models as well.
\ In such cases, additional well-established field theoretic mechanisms are
also available. \ For example, four-dimensional flipped $SU(5)$ models already
contain an elegant mechanism for doublet-triplet splitting which also
naturally suppresses dangerous dimension five operators responsible for proton
decay. \ In this case, we can also utilize a conventional seesaw mechanism to
generate hierarchically light neutrino masses.

\section{Surfaces of General Type\label{GeneralType}}

In this Section we present some examples of models where Planck scale physics
cannot be decoupled from local GUT\ models. \ Recall that in a traditional
four-dimensional GUT, the GUT\ group breaks to $G_{std}$ when an
adjoint-valued chiral superfield develops a suitable vev. \ In F-theory, this
requires that the seven-brane wraps a surface with $h^{2,0}(S)\neq0$. \ Before
proceeding to a discussion of GUT\ models based on such surfaces, we first
discuss some important constraints on matter curves and supersymmetric gauge
field configurations for such surfaces.

In many cases, some of the chiral fields of the low energy theory will
localize on matter curves in $S$. \ When $h^{2,0}(S)\neq0$, the number of
available matter curves will typically be much smaller than the dimension of
$H_{2}(S,%
\mathbb{Z}
)$ would suggest. \ To see this, suppose that an element of~$H_{2}(S,%
\mathbb{Z}
)$ corresponds to a holomorphic curve $\Sigma$ in $S$. \ We shall also refer
to the class $[\Sigma]$ as an \textquotedblleft effective\textquotedblright%
\ divisor. \ Given a $(2,0)$ form $\Omega$ on $S$, note that:%
\begin{equation}
\underset{\Sigma}{\int}\Omega=\underset{S}{\int}\Omega\wedge PD(\Sigma)=0
\label{holcurveconstraint}%
\end{equation}
where $PD(\Sigma)$ denotes the element of $H^{2}(S,%
\mathbb{Z}
)$ which is Poincar\'{e} dual to $\Sigma$. \ This last equality follows from
the fact that $PD(\Sigma)$ corresponds to the first Chern class of an
appropriate line bundle and therefore is of type $(1,1)$.\footnote{This last
correspondence follows from the link between divisors and line bundles.} \ We
thus see that although the condition $h^{2,0}(S)\neq0$ is satisfied by a large
class of vacua, at generic points in the complex structure moduli space each
element of $H^{2,0}(S,%
\mathbb{C}
)$ imposes an additional constraint of the form given by equation
(\ref{holcurveconstraint}). \ At the level of cohomology, the divisor classes
are parameterized by the Picard lattice of $S$:%
\begin{equation}
Pic(S)=H^{1,1}(S,%
\mathbb{C}
)\cap H^{2}(S,%
\mathbb{Z}
)\text{.} \label{PICARD}%
\end{equation}
For example, we note that for a generic algebraic $K3$ surface, $Pic(S)$ has
rank one. \ Indeed, this lattice is generated by the hyperplane class
inherited from the projective embedding of a general quartic in $\mathbb{P}%
^{3}$. \ It is only at special points in the complex structure moduli space
that additional holomorphic curves are present. An example of a $K3$ surface
of this type occurs when the quartic is of Fermat type. \ In this case, the
rank of $Pic(S)$ is instead $20$. \ Because there is a one to one
correspondence between line bundles and divisors on $S$, we conclude that a
similar condition holds for the available line bundles on a generic surface.

Having stated these caveats on what we expect for generic surfaces of general
type, we now construct an $SO(10)$ GUT\ model with semi-realistic Yukawa
matrices. \ In order to have a sufficient number of matter curves, we consider
a seven-brane with worldvolume gauge group $SO(12)$ wrapping a surface $S$
defined by the blowup at $k$ points of a degree $n\geq5$ hypersurface in
$\mathbb{P}^{3}$ with $n$ odd. \ Some properties of hypersurfaces in
$\mathbb{P}^{3}$ are reviewed in Appendix \ref{HyperReview}. \ We have
introduced these blown up curves in order to simplify several properties of
our example. \ Indeed, as explained around equation (\ref{PICARD}), the Picard
lattice of a surface may have low rank. \ An important point is that some of
the numerical invariants such as $h^{2,0}(S)$ and $\chi(S,\mathcal{O}_{S})$ of
the degree $n$ hypersurface remain invariant under these blowups. \ Thus, for
many purposes we will be able to perform many of our calculations of the zero
mode content as if the surface were a degree $n$ hypersurface in
$\mathbb{P}^{3}$.

For $n\geq5$, we expect to find a large number of additional adjoint-valued
chiral superfields. \ Geometrically, the vevs of these fields correspond to
complex structure moduli in the Calabi-Yau fourfold which can develop a mass
in the presence of a suitable background flux. \ We show that in the present
context, a suitable profile of vevs can simultaneously break the GUT\ group
and lift all excess fields from the low energy spectrum.

As explained in Section \ref{BasicSetup}, in the context of a local model, we
are free to specify the enhancement type along codimension one matter curves
inside of $S$. \ We first introduce four curves $\Sigma_{1},\Sigma_{2}%
,\Sigma_{3},\Sigma_{B}$ where the singularity type enhances to $E_{7}$ so that
a half-hypermultiplet in the $32$ of $SO(12)$ localizes on each curve. \ With
notation as in Appendix \ref{HyperReview}, the homology class of each curve
is:%
\begin{align}
\lbrack\Sigma_{1}]  &  =E_{2}\\
\lbrack\Sigma_{2}]  &  =E_{4}\\
\lbrack\Sigma_{3}]  &  =E_{6}\\
\lbrack\Sigma_{B}]  &  =-a_{1}l_{1}-E_{8}-E_{9}\text{.}%
\end{align}
where we have written $K_{H_{n}}=a_{1}l_{1}+a_{2}l_{2}+...$ for some
generators $l_{i}$ of $H_{2}(H_{n},%
\mathbb{Z}
)$ such that $l_{i}\cdot l_{j}=0$ for $i\neq j$. \ Using the genus formula
$C\cdot(C+K_{S})=2g-2$, we conclude that the genera of $\Sigma_{1},\Sigma
_{2},\Sigma_{3}$ are all zero while $\Sigma_{B}$ has genus one. We note that
in order for $\Sigma_{B}$ to represent a holomorphic curve, it may be
necessary to go to some special points in the moduli space of the surface $S$.
In the presence of a suitable internal flux, a single generation in the $16$
of $SO(10)$ will localize on each of the $\Sigma_{i}$'s. \ The fields
localized on $\Sigma_{B}$ will instead develop a suitable vev to lift
extraneous matter from the low energy spectrum.

We next introduce the curve $\Sigma_{R}$ where the singularity type enhances
to $SO(14)$ so that a hypermultiplet transforming in the $12$ of $SO(12)$
localizes on this curve. \ The homology class of $\Sigma_{R}$ is:%
\begin{equation}
\lbrack\Sigma_{R}]=-a_{2}l_{2}-E_{10}-E_{11}%
\end{equation}
so that $\Sigma_{R}$ has genus one.

A supersymmetric $U(1)$ gauge field configuration can simultaneously break
$SO(12)$ to $SO(10)\times U(1)_{PQ}$ and also induce a net chiral matter
content in the four-dimensional effective theory. \ Representations of
$SO(12)$ decompose under the subgroup $SO(10)\times U(1)_{PQ}$ as:%
\begin{align}
SO(12)  &  \supset SO(10)\times U(1)_{PQ}\\
66  &  \rightarrow45_{0}+1_{0}+10_{2}+10_{-2}\\
32  &  \rightarrow16_{1}+\overline{16}_{-1}\\
12  &  \rightarrow1_{2}+1_{-2}+10_{0}\text{.}%
\end{align}
All candidate Higgs fields in the $10_{-2}$ are equally charged under the
group $U(1)_{PQ}$ and we shall therefore loosely refer to it as a Peccei-Quinn
symmetry. \ We consider configurations such that one generation in the
$16_{1}$ of $SO(10)$ localizes along each $\Sigma_{i}$ for $i=1$,$2,3$. \ In
addition to the matter content of the MSSM, we shall also require that there
is extra vector-like matter in the $16_{1}$ and $\overline{16}_{-1}$ localized
along $\Sigma_{B}$ and a $10_{0}$ and $1_{2}$ localized along $\Sigma_{R}$.
\ When the extra vector-like $16$'s develop a vev at suitably large energy
scales, they will remove an additional $U(1)_{B-L}$ gauge boson from the low
energy spectrum. \ Further, interaction terms between the $10_{0}$ and
$1_{-2}$ can also serve to remove extraneous matter from the spectrum.

The above requirements are satisfied by a large class of supersymmetric line
bundles. \ For concreteness, we consider the line bundle:%
\begin{equation}
L=\mathcal{O}_{S}(E_{1}-E_{2}+E_{3}-E_{4}+E_{5}-E_{6}-E_{10}+E_{12}%
+N(E_{14}-E_{15})) \label{SO10linebundle}%
\end{equation}
where to simplify some cohomology calculations, we shall sometimes take $N$ to
be a large integer. \ By inspection, there exists a parametric family of
K\"{a}hler classes such that the condition:%
\begin{equation}
\omega\wedge c_{1}(L)=0 \label{stabcond}%
\end{equation}
holds. \ In the above, $\omega$ denotes a particular choice of K\"{a}hler form
on $S$.

\subsection{Bulk Matter Content}

While all of the chiral matter of the MSSM\ localizes on the matter curves
$\Sigma_{1}$, $\Sigma_{2}$ and $\Sigma_{3}$, the internal $U(1)$ flux
specified by the line bundle of equation (\ref{SO10linebundle}) will also
induce additional bulk zero modes. \ The bulk matter content all descends from
the adjoint representation of $SO(12)$. \ First consider the number of chiral
superfields transforming in the representation $45_{0}+1_{0}$. \ These fields
are neutral under $U(1)_{PQ}$ so that the total number of chiral superfields
transforming in this representation is $h^{1}(S,\mathcal{O}_{S})+h^{2}%
(S,\mathcal{O}_{S})$. \ In the present case, $h^{1}(S,\mathcal{O}_{S})=0$ so
that it is enough to compute $h^{2}(S,\mathcal{O}_{S})=h^{2,0}(S)$. \ The
Hodge numbers of $S$ are computed in Appendix \ref{HyperReview} with the end
result:%
\begin{equation}
\left(  \frac{1}{6}(n^{3}-6n^{2}+11n)-1\right)  \times\left(  45_{0}%
+1_{0}\right)  \in H_{\overline{\partial}}^{2}(S,\mathcal{O}_{S})\text{.}%
\end{equation}
When these fields develop a suitable vev, the GUT\ group will break to
$G_{std}$.

The chiral superfields transforming in the $10_{\pm2}$ are classified by the
bundle-valued cohomology groups:%
\begin{equation}
10_{\pm2}\in H_{\overline{\partial}}^{0}(S,L^{\mp2})^{\ast}\oplus
H_{\overline{\partial}}^{1}(S,L^{\pm2})\oplus H_{\overline{\partial}}%
^{2}(S,L^{\mp2})^{\ast}\text{.}%
\end{equation}
Now, when the integer $N$ of equation (\ref{SO10linebundle}) is sufficiently
large, both $H_{\overline{\partial}}^{0}(S,L^{\mp2})^{\ast}$ and
$H_{\overline{\partial}}^{2}(S,L^{\mp2})^{\ast}$ will indeed vanish. \ The
resulting dimension of $H_{\overline{\partial}}^{1}(S,L^{\pm2})$ can then be
computed via an index formula:%
\begin{align}
h^{1}(S,L^{\pm2})  &  =-\left(  \chi(\mathcal{O}_{S})+\frac{1}{2}c_{1}(S)\cdot
c_{1}(L^{\pm2})+\frac{1}{2}c_{1}(L^{\pm2})^{2}\right) \\
&  =-\frac{1}{6}(n^{3}-6n^{2}+11n)+(16+4N^{2})
\end{align}
so that there are an equal number of $10_{+2}$ and $10_{-2}$'s. \ Based on
their coupling to the fields localized along the matter curve, we shall
tentatively identify these as Higgs fields.

\subsection{Localized Matter Content}

We now study the chiral matter content localized on matter curves. \ By
construction, $L$ restricts to a degree one line bundle on the genus zero
matter curves $\Sigma_{1},\Sigma_{2},\Sigma_{3}$ so that a single generation
transforming in the $16_{1}$ localizes on each matter curve. \ Further, $L$
restricts to a trivial line bundle on $\Sigma_{B}$ so that a single
vector-like pair of $16_{1}$ and $\overline{16}_{-1}$ localizes along
$\Sigma_{B}$. \ Finally, $L$ restricts to a degree $-1$ bundle, $\mathcal{O}%
_{\Sigma_{R}}(-p)$ on the genus one matter curve $\Sigma_{R}$ where $p$
denotes a degree one divisor of $\Sigma_{R}$. \ In order to achieve one copy
of the $10_{0}$, we also include a contribution to the flux from the other
seven-brane intersecting the GUT\ model seven-brane along $\Sigma_{R}$ so that
$L_{\Sigma_{R}}^{\prime}=$ $\mathcal{O}_{\Sigma_{R}}(p^{\prime})$, where
$p^{\prime}$ is another degree one divisor of $\Sigma_{R}$. \ The total field
content on $\Sigma_{R}$ is therefore given by one $10_{0,1}$, three $1_{-2,1}%
$'s and one $1_{-2,-1}$, where the two subscripts indicate the $U(1)$ charge
with respect to the two $U(1)$ factors.\footnote{As we explain later in
Section \ref{Exotica}, the overall normalization of the $U(1)$ charges is
somewhat inconsequential so long as the fields transform in mathematically
well-defined line bundles.} \ The representation content and type of matter
curve are summarized in the following table:%
\begin{equation}%
\begin{tabular}
[c]{|l|l|l|l|l|l|}\hline
$SO(10)\text{ Model}$ & $\text{Curve}$ & $\text{Class}$ & $g_{\Sigma}$ &
$L_{\Sigma}$ & $L_{\Sigma}^{\prime n}$\\\hline
$1\times16_{1}$ & $\Sigma_{1}$ & $E_{2}$ & $0$ & $\mathcal{O}_{\Sigma_{1}}(1)$
& $\mathcal{O}_{\Sigma_{1}}$\\\hline
$1\times16_{1}$ & $\Sigma_{2}$ & $E_{4}$ & $0$ & $\mathcal{O}_{\Sigma_{2}}(1)$
& $\mathcal{O}_{\Sigma_{2}}$\\\hline
$1\times16_{1}$ & $\Sigma_{3}$ & $E_{6}$ & $0$ & $\mathcal{O}_{\Sigma_{3}}(1)$
& $\mathcal{O}_{\Sigma_{3}}$\\\hline
$1\times\left(  16_{1}+\overline{16}_{-1}\right)  $ & $\Sigma_{B}$ &
$-a_{1}l_{1}-E_{8}-E_{9}$ & $1$ & $\mathcal{O}_{\Sigma_{B}}(0)$ &
$\mathcal{O}_{\Sigma_{B}}$\\\hline
$1\times10_{0,1}+3\times1_{-2,1}+1\times1_{-2,-1}$ & $\Sigma_{R}$ &
$-a_{2}l_{2}-E_{10}-E_{11}$ & $1$ & $\mathcal{O}_{\Sigma_{R}}(-p)$ &
$\mathcal{O}_{\Sigma_{R}}(p^{\prime})$\\\hline
\end{tabular}
\text{.}%
\end{equation}
As will be clear when we discuss the high energy superpotential, although the
$1_{-2,-1}$ couples non-trivially with the $10_{0,1}$ to bulk modes on $S$,
the $1_{-2,1}$'s do not contribute to the cubic superpotential, and we shall
therefore neglect their contribution to the low energy theory. \ To simplify
notation, we shall therefore refer to the $10_{0,1}$ as the $10_{0}$ and the
$1_{-2,-1}$ as the $1_{-2}$.

\subsection{High Energy Superpotential}

In the present model, the Yukawa couplings of the MSSM originate from purely
bulk couplings and couplings between bulk gauge fields and matter fields
localized along matter curves. \ In addition, a background flux configuration
in the Calabi-Yau fourfold will also couple to the complex structure moduli of
the compactification. \ Indeed, as shown in \cite{BHV}, the vevs of the bulk
$(2,0)$ form and fields localized along matter curves all determine complex
deformations of the background compactification. \ In the case of fields
localized along the matter curve, this corresponds to the \textquotedblleft
mesonic\textquotedblright\ branch of moduli space. \ We therefore conclude
that fluxes can induce a non-trivial mass and vev for the corresponding
fields. \ At energy scales close to $M_{GUT}$ but below the energy scale where
the first Kaluza-Klein mode can contribute an appreciable amount, the high
energy superpotential is:%
\begin{equation}
W_{high}=W_{S}+W_{S\Sigma\Sigma}+W_{flux}+W_{np} \label{WHIGH}%
\end{equation}
where:%
\begin{align}
W_{S}  &  =f_{iIJ}10_{+2}^{(I)}\times10_{-2}^{(J)}\times(45_{0}^{(i)}%
+1_{0}^{(i)})\label{WS}\\
W_{S\Sigma\Sigma}  &  =\lambda_{aJ}16_{1}^{(a)}\times16_{1}^{(a)}\times
10_{-2}^{(J)}+\alpha_{a}10_{2}^{(a)}\times10_{0}\times1_{-2}\label{WSIGMA}\\
&  +\left(  \beta_{J}16_{1}\times16_{1}\times10_{-2}^{(J)}+\gamma_{J}%
\overline{16}_{-1}\times\overline{16}_{-1}\times10_{2}^{(J)}\right) \\
W_{flux}  &  =\underset{CY_{4}}{\int}\Omega\wedge G_{4}\label{WFLUX}\\
W_{np}  &  =\mu_{-4}^{(IJ)}10_{+2}^{(I)}\times10_{+2}^{(J)}+\mu_{+4}%
^{(IJ)}10_{-2}^{(I)}\times10_{-2}^{(J)}%
\end{align}
In the above, terms proportional to the coefficients $\lambda_{aJ}$ descend
from the three matter curves $\Sigma_{1},\Sigma_{2},\Sigma_{3}$, while terms
proportional to $\beta_{J}$ and $\gamma_{J}$ descend from the matter curve
$\Sigma_{B}$. \ Here, we have also included the effects of non-perturbatively
generated mass terms for the $10$'s which explicitly violate the $U(1)_{PQ}$
global symmetry. Such terms can originate from exponentially suppressed
higher-dimensional operators which couple the fields of the GUT model to
additional GUT group singlets. \ When these singlets develop a suitable vev,
they can generate terms of the type given by $W_{np}$. \ In this case, the
resulting $\mu$ term will naturally be exponentially suppressed. \ A similar
mechanism has been analyzed in the context of type\ II intersecting D-brane
models as a potential solution to the $\mu$ problem
\cite{CveticBlumenhagenInstantons}.

While stabilizing the moduli in a realistic compactification is certainly a
non-trivial task, in a local model, the vevs of the complex structure moduli
can effectively be tuned to an arbitrary value. \ Letting $\Omega^{(0)}$
denote the value of the holomorphic four form of the Calabi-Yau fourfold with
the desired values of the complex structure moduli, we note that the critical
points of $W_{flux}$ with $G_{4}=\lambda(\Omega^{(0)}+\overline{\Omega}%
^{(0)})$ will indeed yield such a configuration. \ For compact models, this
must be appropriately adjusted because the potential for the overall volume of
the Calabi-Yau fourfold will develop a non-supersymmetric minimum.

\subsection{Low Energy Spectrum}

We now show that an appropriate choice of vevs in $W_{high}$ given by equation
(\ref{WHIGH}) can yield a low energy effective theory with precisely the
matter content of the MSSM\ and a semi-realistic low energy superpotential.
\ We first demonstrate that the above model can indeed remove all excess
matter at sufficiently high energies. \ To this end, first note that when a
$45_{0}^{(i)}$ develops the vev:%
\begin{equation}
\left\langle 45_{0}\right\rangle =i\sigma_{y}\otimes diag(a,a,a,b,b)
\end{equation}
the resulting gauge group will break to $SU(3)_{C}\times SU(2)_{L}\times
U(1)_{Y}\times U(1)_{B-L}$. \ By inspection of equation (\ref{WS}), when
$a\sim M_{GUT}$, this vev will also remove the Higgs triplets of
$10_{-2}^{(J)}$ (and the $10_{+2}^{(I)}$'s) from the low energy spectrum.
\ When the zero mode content contains at least two $45$'s which have distinct
couplings to the product $10_{+2}^{(I)}\times10_{-2}^{(J)}$, a suitable choice
of $b$ for each $45$ can be arranged so that at most one pair of $SU(2)_{L}$
doublets from one linear combination of the $10_{-2}$'s will remain massless.
\ We note that this is simply a variant on the well-known Dimopoulos-Wilczek
mechanism for achieving doublet-triplet splitting in four-dimensional $SO(10)$
GUT\ models \cite{DimopoulosWilczek,BabuBarrImplementation}.

In the absence of other field vevs, the resulting spectrum would contain two
$SU(2)_{L}$ doublets from a bulk $10_{-2}$ as well as its counterpart $10_{2}%
$. \ In fact, we now demonstrate that when the flux induces a suitably large
mass term for the $10_{0}$ as well as a vev for the $1_{-2}$, the resulting
low energy spectrum will not contain any fields transforming in the
representation $10_{2}$. \ With the above choice of fluxes, the mass matrix
for the $10_{0}$ and remaining $10_{2}$ is schematically of the form:%
\begin{equation}
W_{eff}\supset\left[
\begin{array}
[c]{cc}%
10_{2} & 10_{0}%
\end{array}
\right]  \left[
\begin{array}
[c]{cc}%
0 & \left\langle 1_{-2}\right\rangle \\
\left\langle 1_{-2}\right\rangle  & \left\langle M_{flux}\right\rangle
\end{array}
\right]  \left[
\begin{array}
[c]{c}%
10_{2}\\
10_{0}%
\end{array}
\right]
\end{equation}
so that all extraneous $10_{+2}$'s can indeed lift from the low energy spectrum.

The resulting spectrum is almost that of the MSSM at low energies. \ The only
additional matter content is an additional $U(1)_{B-L}$ gauge boson and a
vector-like pair of matter fields $16_{1}$ and $\overline{16}_{-1}$ localized
on $\Sigma_{B}$. \ In fact, when the $16_{1}$ and $\overline{16}_{-1}$ develop
a suitable vev, they will break $U(1)_{B-L}$.

Maximally utilizing conventional four-dimensional field theoretic mechanisms
to achieve the correct matter spectrum, this model yields the spectrum of the
MSSM\ at low energies. \ Moreover, by placing the three generations on three
distinct matter curves, a large hierarchy in scales can be generated by a
suitable choice of K\"{a}hler class.

The effective superpotential is now schematically of the form:
\begin{equation}
W_{eff}=\mu H_{u}H_{d}+\lambda_{ij}^{u}Q^{i}U^{j}H_{u}+\lambda_{ij}^{d}%
Q^{i}D^{j}H_{d}+\lambda_{ij}^{l}L^{i}E^{j}H_{d}+\lambda_{ij}^{\nu}L^{i}N^{j}_{R}H_{u}+...%
\end{equation}
where the $\lambda_{ij}$'s are all diagonal.

While it is of course possible to further refine the above model, we believe
this provides a fruitful starting point for analyzing how traditional
four-dimensional GUT models can embed in F-theory. Again, we emphasize that
strictly speaking, a purely four-dimensional effective field theory approach
breaks down in this case because no decoupling limit between $M_{GUT}$ and
$M_{pl}$ is available.

\section{Surfaces with Discrete Wilson Lines\label{DiscreteWilson}}

In the previous Section we presented an example of a four-dimensional
GUT\ model which breaks to the MSSM\ when a collection of adjoint-valued
chiral superfields develop appropriate vevs. \ This requires that the surface
$S$ wrapped by the seven-brane satisfies $h^{2,0}(S)\neq0$. \ When $\pi
_{1}(S)\neq0$, it is also possible for the GUT\ group to spontaneously break
to the gauge group of the Standard Model via an appropriate choice of Wilson
lines. \ In this Section we describe some features of models based on the case
where $S$ is an Enriques surface. \ After reviewing some basic properties of
such surfaces, we present a toy model with bulk gauge group $G_{S}=SU(5)$.
\ Although the correct matter content of the MSSM can localize on matter
curves, we find that the discrete Wilson lines also generically produce
additional vector-like pairs of zero modes transforming in exotic
representations of $G_{std}$. \ This can be traced back to the fact that the
universal cover of an Enriques surface is a $K3$ surface. \ Although we do not
present a complete model based on an Enriques surface, we discuss how these
problems can be avoided by including further field-theoretic mechanisms to
lift extraneous matter from the low energy spectrum. \ It is also possible
that other surfaces with different fundamental groups may provide additional
possibilities. \ To this end, we conclude by mentioning some other surfaces
which have been studied in the mathematics literature.

We begin by reviewing some relevant features of Enriques surfaces. \ Further
details can be found in \cite{BarthSurfaces}. \ An Enriques surface $S$ is
defined by the conditions:%
\begin{equation}
K_{S}^{2}=\mathcal{O}_{S}\text{ but }K_{S}\neq\mathcal{O}_{S}%
\end{equation}
and that the \textquotedblleft irregularity\textquotedblright\ $h^{1,0}%
(S)=q(S)=0$. \ The non-vanishing Hodge numbers of an Enriques surface are
$h^{1,1}(S)=10$ and $h^{0,0}(S)=h^{2,2}(S)=1$. \ The fundamental group of $S$
is $\pi_{1}(S)=%
\mathbb{Z}
_{2}$. \ Moreover, the universal cover of $S$ is a $K3$ surface. \ Indeed, the
Hodge number $h^{2,0}(K3)=1$ does not survive in the quotient space.
\ Nevertheless, we shall see that in the presence of discrete Wilson lines,
the zero mode content retains some imprint from the underlying $K3$ surface.

Recall that for a $K3$ surface, the intersection form on $H^{2}(K3,%
\mathbb{Z}
)$ is isomorphic to:%
\begin{equation}
H^{2}(K3,%
\mathbb{Z}
)=(-E_{8})\oplus(-E_{8})\oplus U\oplus U\oplus U
\end{equation}
where $-E_{8}$ denotes minus the intersection form for the Lie algebra $E_{8}$
and the \textquotedblleft hyperbolic element\textquotedblright\ $U$ is the
intersection form with entries given by the Pauli matrix $\sigma_{x}$. \ The
intersection form on $S$ is instead given by:%
\begin{equation}
H^{2}(S,%
\mathbb{Z}
)/Tor=(-E_{8})\oplus U
\end{equation}
where in the above we have modded out by possible torsional elements. \ As an
integral lattice, $H^{2}(S,%
\mathbb{Z}
)$ is isomorphic to:%
\begin{equation}
H^{2}(S,%
\mathbb{Z}
)\simeq%
\mathbb{Z}
^{10}\oplus%
\mathbb{Z}
_{2}\text{.}%
\end{equation}
We label the generators of $H^{2}(S,%
\mathbb{Z}
)$ as $\alpha_{1},...,\alpha_{8}$ in correspondence with the roots of $E_{8}$
and $d_{1}$ and $d_{2}$ for the generators associated with $U$ such that
$d_{i}\cdot d_{j}=1-\delta_{ij}$. \ Finally, we label the torsion element as
$t$. \ An important feature of Enriques surfaces is that the Poincar\'{e} dual
homology classes for $d_{1}$ and $d_{2}$ both represent holomorphic elliptic
curves in $S$.

We now present a toy model with $S$ an Enriques surface with bulk gauge group
$G_{S}=SU(5)$ which spontaneously breaks to $G_{std}$ due to a discrete Wilson
line taking values in the $U(1)_{Y}$ factor. \ The example we shall now
present cannot be considered even semi-realistic because in addition to
containing exotic matter, the tree level superpotential contains too many
texture zeroes. \ Nevertheless, it illustrates some of the elements which are
necessary in more realistic constrictions. \ To simplify our discussion, we
shall emphasize elements unique to having non-trivial discrete Wilson models.

Because bulk modes descend from the adjoint representation of $SU(5)$ and all
of the matter of the Standard Model descends from other representations of
$SU(5)$, the chiral superfields of the MSSM\ must localize on matter curves.
\ The generic $G_{S}=SU(5)$ singularity enhances to $SU(6)$ along the Higgs
and $\overline{5}_{M}$ matter curves and enhances to $SO(10)$ along the
$10_{M}$ matter curve. \ The matter curves and choice of line bundle
assignment are given in the following table:
\begin{equation}%
\begin{tabular}
[c]{|l|l|l|l|l|l|l|}\hline
$\text{Enriques Model}$ & $\text{Curve}$ & $K3\text{ curve}$ & $\text{Class}$
& $g_{\Sigma}$ & $L_{\Sigma}$ & $L_{\Sigma}^{\prime n}$\\\hline
$1\times\left(  5_{H}+\overline{5}_{H}\right)  $ & $\Sigma_{H}$ &
$\widetilde{\Sigma}_{H}$ & $d_{1}$ & $1$ & $%
\mathbb{Z}
_{2}\otimes\mathcal{O}_{\Sigma_{H}}$ & $\mathcal{O}_{\Sigma_{H}}$\\\hline
$3\times\overline{5}_{M}$ & $\Sigma_{M}^{(1)}$ & $\widetilde{\Sigma}_{M}%
^{(1)}\amalg\widetilde{\Sigma}_{M}^{\prime(1)}$ & $d_{2}$ & $1$ &
$\mathcal{O}_{\Sigma_{M}^{(1)}}$ & $\mathcal{O}_{\Sigma_{M}^{(1)}}(-3p_{1}%
)$\\\hline
$3\times10_{M}$ & $\Sigma_{M}^{(2)}$ & $\widetilde{\Sigma}_{M}^{(2)}%
\amalg\widetilde{\Sigma}_{M}^{\prime(2)}$ & $d_{1}$ & $1$ & $\mathcal{O}%
_{\Sigma_{M}^{(2)}}$ & $\mathcal{O}_{\Sigma_{M}^{(2)}}(3p_{1})$\\\hline
\end{tabular}
\ \text{.}%
\end{equation}
In the above, we have also indicated how each curve lifts to $K3$. \ In this
case both matter curves $\Sigma_{M}^{(i)}$ lift to the disjoint union of two
curves in $K3$ while the Higgs curve $\Sigma_{H}$ lifts to a curve which is
fixed by the $%
\mathbb{Z}
_{2}$ involution in $K3$. \ As an explicit example, we can consider the case
where the covering space of $S$ is a real $K3$ surface and the $%
\mathbb{Z}
_{2}$ involution corresponds to complex conjugation. \ In this case, the curve
$\Sigma_{M}^{(i)}$ lifts to a generic holomorphic curve and its image under
complex conjugation while $\Sigma_{H}$ lifts to a real algebraic curve in
$K3$. \ We now show that in this case the discrete Wilson line projects out
the Higgs triplet from the low energy spectrum.

Because the Higgs curve is fixed by the $%
\mathbb{Z}
_{2}$ involution, the fields localized on this curve will transform
non-trivially in the presence of a $%
\mathbb{Z}
_{2}$ Wilson line. \ The analysis below equation (\ref{curveIndex}) applies
equally well to fields localized on matter curves. \ Under the breaking
pattern $SU(5)\supset SU(3)\times SU(2)\times U(1)$, the $5$ of $SU(5)$
decomposes to $(1,2)_{3}+(3,1)_{-2}$. \ In this case, the relevant cohomology
group lifts to the $%
\mathbb{Z}
_{2}$ odd eigenspace:%
\begin{equation}
5_{H}\in H_{\overline{\partial}}^{0}(\widetilde{\Sigma}_{H},\mathcal{O}%
_{\widetilde{\Sigma}_{H}})\simeq%
\mathbb{C}
_{(-)}\text{.}%
\end{equation}
Hence, we conclude that the total wave function for the components of the
$5_{H}$ and $\overline{5}_{H}$ take values in the invariant subspaces:%
\begin{align}
(1,\overline{2})_{-3},(1,2)_{3}  &  \in\left[
\mathbb{C}
_{(-)}\otimes H_{\overline{\partial}}^{0}(\widetilde{\Sigma}_{H}%
,\mathcal{O}_{\widetilde{\Sigma}_{H}})\right]  ^{%
\mathbb{Z}
_{2}}\simeq%
\mathbb{C}%
\\
(\overline{3},1)_{-2},(3,1)_{2}  &  \in\left[
\mathbb{C}
_{(+)}\otimes H_{\overline{\partial}}^{0}(\widetilde{\Sigma}_{H}%
,\mathcal{O}_{\widetilde{\Sigma}_{H}})\right]  ^{%
\mathbb{Z}
_{2}}=0\text{.}%
\end{align}
Hence, the Higgs triplet is absent from the low energy spectrum while the
Higgs up and down doublets remain.

The matter content of this example is not fully realistic because it also
contains contributions from the bulk zero modes which appear as vector-like
pairs transforming in exotic representations of $G_{std}$.\ \ To compute the
bulk zero mode content in the presence of the discrete Wilson line, we again
apply the analysis below equation (\ref{curveIndex}) in the special case where
the bundle $\mathcal{T}$ is trivial. \ Decomposing the adjoint representation
of $SU(5)$ to $SU(3)\times SU(2)\times U(1)$, the only irreducible
representations which transform non-trivially under the $U(1)$ factor are the
$(3,\overline{2})_{-5}$ and $(\overline{3},2)_{5}$. \ We now compute the
number of bulk zero modes transforming in the $(3,\overline{2})_{-5}$. \ In
the covering $K3$ space, the contribution to the number of zero modes from the
holomorphic $(2,0)$ is given by the $%
\mathbb{Z}
_{2}$ invariant subspace:%
\begin{equation}
(3,\overline{2})_{-5}\in\left[
\mathbb{C}
_{(-)}\otimes H_{\overline{\partial}}^{2}(K3,\mathcal{O}_{K3})\right]  ^{%
\mathbb{Z}
_{2}}%
\end{equation}
where the $%
\mathbb{C}
_{(-)}$ factors indicates the charge of the representation $(3,\overline
{2})_{-5}$ under the $%
\mathbb{Z}
_{2}$ subgroup of $U(1)_{Y}$. \ Next recall that the $%
\mathbb{Z}
_{2}$ group action on the holomorphic $(2,0)$ form sends $\varphi
\mapsto-\varphi$. \ In particular, this implies that the cohomology group
$H_{\overline{\partial}}^{2}(K3,\mathcal{O}_{K3})\simeq%
\mathbb{C}
_{(-)}$. \ A similar analysis also holds for zero modes transforming in the
representation $(3,\overline{2})_{-5}$. \ Because $%
\mathbb{C}
_{(-)}\otimes%
\mathbb{C}
_{(-)}$ is $%
\mathbb{Z}
_{2}$ invariant, we conclude that the low energy spectrum contains exotic
vector-like pairs.

There are potentially several ways to avoid the presence of these exotics.
\ For example, when $G_{S}=SO(10)$, a combination of $U(1)$ flux breaking and
discrete Wilson line breaking might avoid any contributions from bulk zero
modes. \ Moreover, even if additional exotic particles are present in the low
energy spectrum, it is conceivable that an appropriately engineered
superpotential could cause these exotics to develop a large mass.

It is also possible to consider a more general class of surfaces with
non-trivial discrete Wilson lines. \ In the present context the maximal case
of interest would be surfaces with $h^{1,0}(S)=h^{2,0}(S)=0$ and $\pi_{1}(S)$
a finite group. \ Some examples of surfaces such as the classical Godeaux and
Campadelli surfaces may be found in \cite{BarthSurfaces}. \ As a technical
aside, we note that one particularly interesting class of surfaces can be
obtained by choosing $n$ distinct points of a del Pezzo 9 surface and
performing an order $a_{i}$ logarithmic transformation at the $i^{th}$
point.\footnote{See \cite{BarthSurfaces} for the definition and further
properties of logarithmic transformations of surfaces.} \ The resulting
surface has the same Hodge numbers, Euler character and signature as the del
Pezzo 9 surface and is called a Dolgachev surface, $D(a_{1},...,a_{n})$. \ For
example, when $n=2$ and $a_{1}$ and $a_{2}$ have a common divisor, the
fundamental group is $\pi_{1}(D(a_{1},a_{2}))\simeq%
\mathbb{Z}
_{m}$ where $m=\gcd(a_{1},a_{2})$. \ See
\cite{HambletonKreckOne,HambletonKreckTwo} and references therein for further
discussion of Dolgachev surfaces defined by two logarithmic transformations.
\ We note that the case $a_{1}=2$, $a_{2}=2$ corresponds to the Enriques
surface. \ It is also common in the mathematics literature to treat the more
general case as well. \ When the $a_{i}$ are pairwise co-prime integers, the
resulting fundamental group is \cite{BauerOkonek}:%
\begin{equation}
\pi_{1}(D(a_{1},...,a_{n}))=\left\langle t_{1},...,t_{n}|t_{i}^{a_{i}}%
=1,t_{1}\cdot\cdot\cdot t_{n}=1\right\rangle \text{.}%
\end{equation}
Given the prominent role that the del Pezzo 9 surface has played in recent
heterotic models such as \cite{BouchardDonagi,OvrutHET}, it would be
interesting to study models based on such Dolgachev surfaces.

\section{Geometry of Del Pezzo Surfaces\label{GeomdelPezzReview}}

In the remainder of this paper we focus on the case of primary interest where
$S$ is a del Pezzo surface. \ In this case, it is at least in principle
possible to consistently decouple the Planck scale from the GUT\ scale.
\ Because much of the analysis to follow relies on properties of del Pezzo
surfaces, in this Section we collect various relevant facts about the geometry
of such surfaces. \ After giving the definition of del Pezzo surfaces, we
catalogue the moduli of such surfaces which must be stabilized in a globally
consistent model. \ Next, we review the beautiful connection between the
homology groups of del Pezzo surfaces and the root lattices of exceptional Lie
algebras. \ In particular, we show that the line bundles $L$ on $S$ such that
both $L$ and $L^{-1}$ have trivial cohomology are in one to one correspondence
with the roots of the corresponding exceptional Lie algebra. \ This
classification will prove important when we study vacua with trivial bulk zero
mode content.

The two simplest examples of del Pezzo surfaces are $\mathbb{P}^{1}%
\times\mathbb{P}^{1}$ and $\mathbb{P}^{2}$. \ There are eight additional
surfaces defined as the blowup of $\mathbb{P}^{2}$ at up to eight points in
general position. \ We shall refer to such surfaces as del Pezzo $N$ ($dP_{N}%
$) surfaces for the case of $N$ blown up points.

We now describe the K\"{a}hler and complex structure moduli spaces of these
surfaces. \ First consider the K\"{a}hler moduli of del Pezzo surfaces.
\ $\mathbb{P}^{1}\times\mathbb{P}^{1}$ has two K\"{a}hler moduli corresponding
to the volume of the two $\mathbb{P}^{1}$ factors. \ There is a single
K\"{a}hler modulus which fixes the overall size of $\mathbb{P}^{2}$. \ In
addition to the overall size of the $\mathbb{P}^{2}$, for the del Pezzo $N$
surfaces, there are $N$ further moduli corresponding to the volume of each
blown up cycle. \ Further properties of the K\"{a}hler cone for each del Pezzo
surface are reviewed in Appendix A of \cite{BHV}.

In addition to the K\"{a}hler moduli of each del Pezzo surface, these surfaces
may also possess a moduli space of complex structures. \ For $\mathbb{P}%
^{1}\times\mathbb{P}^{1}$\ and $\mathbb{P}^{2}$ there is a unique choice of
complex structure. \ When $S=dP_{N}$, the overall $PGL(3)$ symmetry of
$\mathbb{P}^{2}$ implies that the number of complex structure moduli is $2N-8$
so that in an isolated local model only surfaces with $5\leq N\leq8$ possess a
moduli space of complex structures. \ In the context of a globally consistent
moduli, this distinction is somewhat artificial because the overall $PGL(3)$
action on $\mathbb{P}^{2}$ may not properly extend to the compact threefold base.

We next describe the homology groups of the del Pezzo surfaces. \ The homology
group $H_{2}(\mathbb{P}^{1}\times\mathbb{P}^{1},%
\mathbb{Z}
)$ is two dimensional and has generators $\sigma_{1}$ and $\sigma_{2}$
corresponding to the two $\mathbb{P}^{1}$ factors. \ These generators have
intersection product:%
\begin{equation}
\sigma_{i}\cdot\sigma_{j}=1-\delta_{ij} \label{p1p1intersection}%
\end{equation}
where $\delta_{ij}$ is the Kronecker delta. \ The canonical class for
$\mathbb{P}^{1}\times\mathbb{P}^{1}$ is:%
\begin{equation}
K_{\mathbb{P}^{1}\times\mathbb{P}^{1}}=-c_{1}(\mathbb{P}^{1}\times
\mathbb{P}^{1})=-2\sigma_{1}-2\sigma_{2}\text{.}%
\end{equation}
In particular, $-K_{\mathbb{P}^{1}\times\mathbb{P}^{1}}$ defines a K\"{a}hler
class on $\mathbb{P}^{1}\times\mathbb{P}^{1}$ where both $\mathbb{P}^{1}$
factors have volume two in an appropriate normalization.

The homology group $H_{2}(dP_{N},%
\mathbb{Z}
)$ is $N+1$ dimensional and has generators $H$, $E_{1},...,E_{N}$ where $H$
denotes the hyperplane class inherited from $\mathbb{P}^{2}$ and the $E_{i}$
denote the exceptional divisors associated with the blowup. \ These generators
have intersection product:%
\begin{equation}
H\cdot H=1\text{, }H\cdot E_{i}=0\text{, }E_{i}\cdot E_{j}=-\delta_{ij}%
\end{equation}
so that the signature of $H_{2}(dP_{N},%
\mathbb{Z}
)$ is $(+,-^{N})$. \ The canonical class for $dP_{N}$ is:%
\begin{equation}
K_{dP_{N}}=-c_{1}(dP_{N})=-3H+E_{1}+...+E_{N}\text{.}%
\end{equation}

There is a beautiful connection between del Pezzo $N\geq2$ surfaces and
exceptional Lie algebras. \ This material is reviewed for example in
\cite{ManinDelPezzo} and has played a role in proposed M-theory dualities
\cite{MysteriousDelPezzos}. \ We now review how the sublattice of
$H_{2}(dP_{N},%
\mathbb{Z}
)$ orthogonal to $K_{dP_{N}}$ is identified with the root space of the
corresponding Lie algebra $E_{N}$. \ Because $dP_{2}$ admits a different
treatment, first consider the $dP_{N}$ surfaces with $N\geq3$. \ The
generators of the lattice $\left\langle K_{dP_{N}}\right\rangle ^{\bot}$ are:
\begin{equation}
\alpha_{1}=E_{1}-E_{2},...,\alpha_{N-1}=E_{N-1}-E_{N},\alpha_{N}=H-E_{1}%
-E_{2}-E_{3}\text{.}%
\end{equation}
The intersection product of the $\alpha_{i}$'s is identical to minus the
Cartan matrix for the dot product of the simple roots for the corresponding
Lie algebra $E_{N}$. \ For $dP_{2}$, the single generator of the lattice
$\left\langle K_{dP_{N}}\right\rangle ^{\bot}$ is given by $E_{1}-E_{2}$,
which we identify as a root of $su(2)$.

This correspondence further extends to include the Weyl group of the
exceptional Lie algebras. \ In the following we shall adopt a
\textquotedblleft geometric\textquotedblright\ convention so that the
signature of the root space is negative definite.\footnote{With this sign
convention, a root $\alpha$ satisfies $\alpha\cdot\alpha=-2$.} \ The Weyl
group for a simply connected Lie algebra with simple roots $\alpha
_{1},...,\alpha_{N}$ is generated by the Weyl reflections $w_{\alpha_{i}}$.
\ Given an element $\alpha$ of the root lattice, the Weyl reflected vector
$w_{\alpha_{i}}(\alpha)$ is:%
\begin{equation}
w_{\alpha_{i}}(\alpha)=\alpha+(\alpha\cdot\alpha_{i})\alpha_{i}\text{.}%
\end{equation}
This is precisely the action of the large group of diffeomorphisms for the del
Pezzo $N$ surfaces on the corresponding generators orthogonal to $K_{dP_{N}}$.
\ Indeed, note that the canonical class is invariant under the action of the
Weyl group.

Anticipating future applications, we now show that when $S$ is a del Pezzo
$N\geq2$ surface, the collection of all line bundles $L$ such that:%
\begin{equation}
H_{\overline{\partial}}^{i}(S,L^{\pm1})=0 \label{vanishcondition}%
\end{equation}
for all $i$ are in one to one correspondence with the roots of the Lie algebra
$E_{N}$.

Because the indices defined by $L$ and $L^{-1}$ must separately vanish, the
difference in the two indices also vanishes:%
\begin{equation}
0=\chi(dP_{N},L)-\chi(dP_{N},L^{-1})=c_{1}(dP_{N})\cdot c_{1}(L)=-K_{dP_{N}%
}\cdot c_{1}(L)\text{.} \label{diffindex}%
\end{equation}
Treating $c_{1}(L)$ as an element of $H_{2}(dP_{N},%
\mathbb{Z}
)$, $c_{1}(L)$ is therefore a vector in the orthogonal complement of the
canonical class. \ Hence, $c_{1}(L)$ corresponds to an element of the root
lattice of $E_{N}$. \ Utilizing equation (\ref{diffindex}), the index
$\chi(dP_{N},L)$ now takes the form:%
\begin{equation}
\chi(dP_{N},L)=1+\frac{1}{2}c_{1}(L)\cdot(c_{1}(L)+c_{1}(dP_{N}))=1+\frac
{1}{2}c_{1}(L)\cdot c_{1}(L)
\end{equation}
which vanishes provided:%
\begin{equation}
c_{1}(L)\cdot c_{1}(L)=-2
\end{equation}
which is the condition for $c_{1}(L)$ to correspond to a root of $E_{N}$.
\ Conversely, we note that given a root $\alpha$ of $\left\langle K_{dP_{N}%
}\right\rangle ^{\bot}$, the line bundle $L=\mathcal{O}_{dP_{N}}(\alpha)$
defines a supersymmetric gauge field configuration. \ The vanishing theorem of
\cite{VafaWitten,BHV} and the vanishing of the corresponding index now imply
that all cohomology groups are trivial.

A similar analysis holds for the remaining del Pezzo surfaces $\mathbb{P}^{2}%
$, $\mathbb{P}^{1}\times\mathbb{P}^{1}$ and $dP_{1}$. \ When $S=\mathbb{P}%
^{2}$, we note that because $H_{2}(\mathbb{P}^{2},%
\mathbb{Z}
)$ has a single generator given by the hyperplane class of $\mathbb{P}^{2}$,
all non-trivial line bundles $L$ have $c_{1}(L)\cdot c_{1}(\mathbb{P}^{2}%
)\neq0$ so that equation (\ref{diffindex}) is never satisfied.

To treat the cases $S=\mathbb{P}^{1}\times\mathbb{P}^{1},dP_{1}$ and in order
to partially widen the scope of our discussion, we note that these del Pezzo
surfaces are also Hirzebruch surfaces. \ More generally, recall that the
middle homology of the degree $n$ Hirzebruch surface $\mathbb{F}_{n}$ has
generators $\sigma$ and $f$ which have intersection pairing:%
\begin{equation}
f\cdot f=0\text{, }f\cdot\sigma=1\text{, }\sigma\cdot\sigma=-n\text{.}%
\end{equation}
The canonical class for $\mathbb{F}_{n}$ is:%
\begin{equation}
K_{\mathbb{F}_{n}}=-c_{1}(\mathbb{F}_{n})=-(n+2)f-2\sigma\text{.}%
\end{equation}
We now show that $\mathbb{F}_{0}$ is the only Hirzebruch surface which admits
line bundles satisfying equation (\ref{vanishcondition}). \ To this end,
consider a line bundle $L=\mathcal{O}_{\mathbb{F}_{n}}(af+b\sigma)$. \ In
order to satisfy equation (\ref{vanishcondition}), we must have:%
\begin{equation}
0=\chi(\mathbb{F}_{n},L)-\chi(\mathbb{F}_{n},L^{-1})=c_{1}(\mathbb{F}%
_{n})\cdot c_{1}(L)=b(n+2)+2a-2bn^{2}\text{.} \label{hirzvanish}%
\end{equation}
When this condition is satisfied, the index $\chi(\mathbb{F}_{n},L)$ vanishes
provided:%
\begin{align}
0  &  =\chi(\mathbb{F}_{n},L)=1+\frac{1}{2}c_{1}(\mathbb{F}_{n})\cdot
c_{1}(L)+\frac{1}{2}c_{1}(L)\cdot c_{1}(L)\\
&  =1+\frac{1}{2}(2ab-b^{2}n^{2})=1+\frac{1}{2}(b^{2}n^{2}-b^{2}(n+2))
\end{align}
or,%
\begin{equation}
-2=b^{2}(n^{2}-(n+2))\text{.}%
\end{equation}
In order for this equation to possess a solution over the integers, $b=\pm1$
and $n^{2}-n=0$ so that $n=0$ or $n=1$. \ First consider the case where $n=1$.
\ Returning to equation (\ref{hirzvanish}), when $n=1$ and $b=\pm1$, we find
that $a=\pm1/2$, which is not an integer. \ We therefore conclude that the
only remaining case is $n=0$. \ For $\mathbb{F}_{0}$, the only line bundles
satisfying equation (\ref{vanishcondition}) are $L=\mathcal{O}_{\mathbb{F}%
_{0}}(\pm f\mp\sigma)=\mathcal{O}_{\mathbb{P}^{1}\times\mathbb{P}^{1}}%
(\pm\sigma_{1}\mp\sigma_{2})$ where in the final equality we have reverted to
the notation of equation (\ref{p1p1intersection}).

\section{GUT\ Breaking via $U(1)$ Fluxes\label{GUTBreakviaFlux}}

When $S$ is a del Pezzo surface, the zero mode content does not contain any
adjoint-valued chiral superfields which could potentially play the role of a
four-dimensional GUT\ Higgs fields.\ \ In this Section we present an
alternative mechanism where the GUT group breaks due to non-trivial
hypercharge flux in the internal directions. \ Experience with other string
compactifications suggests that a non-trivial internal field strength would
cause the photon to develop a string scale mass because this gauge boson
couples non-trivially to the p-form gauge potentials of the closed string
sector. \ In this Section we present a topological criterion for this $U(1)$
gauge boson to remain massless. \ This then provides a novel mechanism for
GUT\ group breaking in F-theory.

To analyze whether the coupling to closed string modes will generate a mass
for the $U(1)$ gauge boson, first recall that the ten-dimensional supergravity
action contains the terms (neglecting the overall normalization of individual
terms by order one constants):\footnote{For D-branes, the relative
normalizations between these terms contains factors of $g_{s}$. \ In the
present class of models, this distinction is ambiguous because these vacua
exist in a regime of strong coupling.}%
\begin{equation}
S^{(10d)}\supset M_{\ast}^{8}\underset{%
\mathbb{R}
^{3,1}\times B_{3}}{\int}C\bigtriangleup_{10}C-M_{\ast}^{4}\underset{%
\mathbb{R}
^{3,1}\times S}{\int}Tr(F\wedge\ast_{8}F)+M_{\ast}^{4}\underset{%
\mathbb{R}
^{3,1}\times S}{\int}C_{(4)}\wedge Tr(F\wedge F)
\end{equation}
where $C_{(4)}$ denotes the RR\ four-form gauge potential and $F$ denotes the
eight-dimensional field strength of the seven-brane. \ Letting $\left\langle
F_{S}\right\rangle $ denote the non-vanishing field strength in the internal
directions, integrating out $C_{(4)}$ yields a term in the effective action of
the form:%
\begin{equation}
S_{eff}^{(10d)}\supset\underset{%
\mathbb{R}
^{3,1}\times B_{3}}{\int}\delta_{%
\mathbb{R}
^{3,1}\times S}\wedge\left\langle F_{S}\right\rangle \wedge F\frac
{1}{\bigtriangleup_{10}}\delta_{%
\mathbb{R}
^{3,1}\times S}\wedge\left\langle F_{S}\right\rangle \wedge F
\end{equation}
where $\delta_{%
\mathbb{R}
^{3,1}\times S}$ denotes the delta function for the seven-brane and we have
dropped the overall trace because our primary interest is in abelian instanton configurations.

Next, expand $\delta_{%
\mathbb{R}
^{3,1}\times S}\wedge\left\langle F_{S}\right\rangle $ in a basis of
eigenmodes so that:%
\begin{equation}
\delta_{%
\mathbb{R}
^{3,1}\times S}\wedge\left\langle F_{S}\right\rangle =\underset{\alpha}{\sum
}f_{\alpha}\psi_{\alpha}%
\end{equation}
where $\bigtriangleup_{6}\psi_{\alpha}=\lambda_{\alpha}\psi_{\alpha}$ denote
eigenmodes of the Laplacian on $B_{3}$ and $f_{\alpha}$ denote the associated
Fourier coefficients. \ We thus arrive at a non-local term in the
four-dimensional effective action:%
\begin{align}
L_{eff}^{(4d)}  &  \supset\underset{\alpha}{\sum}\underset{S}{\int}%
F\wedge\frac{\left\Vert f_{\alpha}\psi_{\alpha}|_{S}\right\Vert ^{2}%
}{\bigtriangleup_{4}+\lambda_{\alpha}}F\\
&  =\underset{\alpha=0}{\sum}\underset{S}{\int}\left\Vert f_{\alpha}%
\psi_{\alpha}|_{S}\right\Vert ^{2}A^{2}+\underset{\alpha\neq0}{\sum}%
\underset{S}{\int}F\wedge\frac{\left\Vert f_{\alpha}\psi_{\alpha}%
|_{S}\right\Vert ^{2}}{\bigtriangleup_{4}+\lambda_{\alpha}}F\text{.}%
\end{align}
so that the contribution from zero modes of $\Delta_{6}$ induces a mass term
for the four-dimensional gauge boson. \ The remaining modes induce a non-local
operator which tends to zero in the decompactification limit.

The zero modes of $\Delta_{6}$ which can potentially couple to the internal
field on $S$ correspond to harmonic representatives of the cohomology group
$H^{2}(B_{3},%
\mathbb{R}
)$ which are Poincar\'{e} dual to elements of $H_{4}(B_{3},%
\mathbb{Z}
)$. \ \ For concreteness, we let $\Gamma$ denote such a four-cycle. \ In the
same spirit as \cite{BuicanVerlinde}, we therefore conclude that the
four-dimensional $U(1)$ gauge boson will remain massless provided the class in
$H^{2}(S,%
\mathbb{Z}
)$ corresponding to $\left\langle F_{S}\right\rangle $ integrates trivially
when wedged with any element of $H^{2}(B_{3},%
\mathbb{Z}
)$. \ In other words, given any four-cycle $\Gamma$ in $B_{3}$, $\Gamma$ must
intersect trivially with the Poincar\'{e} dual of $\left\langle F_{S}%
\right\rangle $ which we denote as $\left[  F_{S}\right]  $ for some element
of $H_{2}(S,%
\mathbb{Z}
)$. \ This implies that the cycle $\left[  F_{S}\right]  $ must be trivial in
$B_{3}$.\footnote{This same observation has been made independently by M.
Wijnholt.} \ We note that just as in \cite{BuicanVerlinde}, this entire
discussion can be phrased in terms of the relative cohomology between $S$ and
$B_{3}$, and we refer the reader there for more details on this type of argument.

Our expectation is that this condition can be met in a large number of cases.
\ Indeed, in backgrounds where the $(2,0)$ form vanishes, a line bundle $L$
corresponds to a supersymmetric gauge field configuration when \cite{BHV}:%
\begin{equation}
\omega\wedge c_{1}(L)=0 \label{stabcond}%
\end{equation}
where $\omega$ denotes the K\"{a}hler form on $S$. \ In particular, if this
$\omega$ descends from the K\"{a}hler form in the threefold base $B_{3}$, this
is a necessary condition for the Poincar\'{e} dual of $\left\langle
F_{S}\right\rangle $ to lift to a trivial class in $H_{2}(B_{3},%
\mathbb{Z}
)$. \ Note that when $\dim H_{2}(B_{3},%
\mathbb{Z}
)=1$, this condition is in fact sufficient.

For illustrative purposes, we now show that there exist compactifications of
F-theory where this condition can be met. \ To this end, we consider an
elliptically fibered Calabi-Yau fourfold with base $B_{3}=\mathbb{P}^{3}$.
\ In this case, the homology ring $H_{\ast}(\mathbb{P}^{3},%
\mathbb{Z}
)$ is generated by the hyperplane class $H_{\mathbb{P}^{3}}$.\ \ Introducing
homogeneous coordinates $x_{0},x_{1},x_{2},x_{3}$, we recall that the
vanishing locus of a generic degree two polynomial in the $x_{i}$ defines a
$\mathbb{P}^{1}\times\mathbb{P}^{1}$ in $B_{3}$, and the vanishing locus of a
generic degree three polynomial defines a del Pezzo $6$ surface in $B_{3}$.
\ As reviewed in Appendix \ref{HyperReview}, a multiple of the generator
$H_{\mathbb{P}^{3}}$ restricts to the anti-canonical class of a degree $n$
hypersurface in $\mathbb{P}^{3}$.

Letting $\sigma_{1}$ and $\sigma_{2}$ denote the generators of $H_{2}%
(\mathbb{P}^{1}\times\mathbb{P}^{1},%
\mathbb{Z}
)$ corresponding to the two $\mathbb{P}^{1}$ factors, the class $\sigma
_{1}-\sigma_{2}$ lifts to a trivial class in $\mathbb{P}^{3}$ due to the fact
that $K_{\mathbb{P}^{1}\times\mathbb{P}^{1}}\cdot(\sigma_{1}-\sigma_{2})=0$.
\ Similar considerations apply for the del Pezzo $6$ surface because all of
the two-cycles corresponding to elements in the root lattice of $E_{6}$ are
orthogonal to $K_{dP_{6}}$.

\subsection{Absence of a Heterotic Analogue}

Given the usual heterotic/F-theory duality, it is natural to ask whether
GUT\ group breaking via internal fluxes can also occur in the heterotic
string. \ A general obstruction to using $U(1)$ fluxes in heterotic models was
already noted in \cite{WittenU(1)}. \ In fact, in all F-theory models which
admit a heterotic dual, the mechanism described above is unavailable! \ To
establish this, first recall that the basic heterotic/F-theory duality relates
compactifications of the heterotic string on an elliptic curve to
compactifications of F-theory on an elliptically fibered $K3$
\cite{VafaFTHEORY}. \ Extending this duality fiberwise, the heterotic string
compactified on an elliptically fibered Calabi-Yau threefold is dual to
F-theory compactified on a $K3$-fibered Calabi-Yau fourfold. \ In this case,
the threefold base of the F-theory compactification is a $\mathbb{P}^{1}$
fibration over a K\"{a}hler surface $S$.

We now establish that in this case, an internal hypercharge flux will always
cause the corresponding four-dimensional gauge boson to lift from the low
energy theory. \ As explained previously, it is enough to determine whether
this internal flux wedges non-trivially with any two forms in $H^{2}(B_{3},%
\mathbb{R}
)$. \ To see why this occurs, first consider the case where the fibration is
trivial so that the threefold base is of the form $S\times\mathbb{P}^{1}%
=B_{3}$. \ In this case, we note that:%
\begin{equation}
H^{2}(B_{3},%
\mathbb{R}
)=H^{2}(S\times\mathbb{P}^{1},%
\mathbb{R}
)\simeq H^{2}(S,%
\mathbb{R}
)\oplus H^{2}(\mathbb{P}^{1},%
\mathbb{R}
)\text{.} \label{cohomproduct}%
\end{equation}
This implies that all non-zero elements of $H^{2}(S,%
\mathbb{R}
)$ wedge non-trivially with some element of $H^{2}(B_{3},%
\mathbb{R}
)$. \ Next consider the case of a non-trivial fibration. \ The only
consequence of the non-trivial fibration structure is that the cohomology
group $H^{2}(B_{3},%
\mathbb{R}
)$ could potentially contain additional contributions on top of those already
present in the product formula of equation (\ref{cohomproduct}).\footnote{At a
more formal level, this is a direct consequence of the Leray-Serre spectral
sequence.} \ In particular, all of the elements of the cohomology group of
$H^{2}(S,%
\mathbb{R}
)$ again wedge non-trivially with some element of $H^{2}(B_{3},%
\mathbb{R}
)$.

\section{Avoiding Exotica\label{Exotica}}

As argued in the previous Section, abelian fluxes provide a potentially
generic mechanism for breaking the GUT\ group to $G_{std}$. \ As shown in
\cite{BHV}, such fluxes also determine the zero mode content of the low energy
effective theory. \ It thus follows that the zero mode content of the theory
may not match to the MSSM. \ In keeping with our general philosophy, we
require that all of the zero modes other than the Higgs fields must organize
into complete GUT\ multiplets. \ Indeed, if these zero modes do not fill out
complete GUT\ multiplets, they can potentially spoil the unification of the
gauge couplings.

It is in principle possible that these restrictions can be relaxed. \ If all
exotics come in vector-like pairs, effective field theory arguments would
appear to suggest that such pairs will develop a large mass and lift from the
low energy spectrum. \ We note that in the present case, all mass terms
descend from cubic or higher order superpotential terms. \ Large mass terms
will only result when a singlet develops a sufficiently large vev. \ As will
be clear in all of the models considered here, such singlets are charged under
additional gauged symmetries. \ In this case, such mass terms may not be
sufficiently large to avoid spoiling gauge coupling unification. \ For these
reasons, we shall always require that the zero mode content of the low energy
theory contains no vector-like pairs of fields in exotic representations of
$G_{std}$.

This constraint imposes important restrictions on admissible gauge bundle
configurations which can break the bulk gauge group $G_{S}$ to $G_{std}$. \ In
particular, when $G_{S}=SU(5)$, we show that the gauge bundle configurations
with no exotica are in one to one correspondence with the roots of an
exceptional Lie algebra corresponding to the del Pezzo surface in question.
\ Moreover, when $G_{S}=SO(10)$, we present a no go theorem which shows that
direct breaking of $G_{S}$ to $G_{std}$ via internal fluxes always produces
exotica in the low energy theory.

\subsection{Fractional Line Bundles \label{FRACBUND}}

In this Section we determine which internal fluxes can break the GUT\ group
and simultaneously do not generate any extraneous zero modes in the low energy
spectrum. \ In fact, a cursory analysis would incorrectly suggest that such
states are unavoidable. \ For example, the decomposition of the adjoint
representation of $SU(5)$ decomposes under $G_{std}$\ as:%
\begin{align}
SU(5)  &  \supset SU(3)\times SU(2)\times U(1)\\
24  &  \rightarrow(1,1)_{0}+(8,1)_{0}+(1,3)_{0}+(3,\overline{2})_{-5}%
+(\overline{3},2)_{5}\text{.}%
\end{align}
We note that no fields of the MSSM\ transform in the representation
$(3,\overline{2})_{-5}$ or $(\overline{3},2)_{5}$. \ Letting $L$ denote the
supersymmetric line bundle associated with this breaking pattern, the bulk
zero mode content therefore descends to:%
\begin{align}
(3,\overline{2})_{-5}  &  \in H_{\overline{\partial}}^{0}(S,L^{5})^{\ast
}\oplus H_{\overline{\partial}}^{1}(S,L^{-5})\oplus H_{\overline{\partial}%
}^{2}(S,L^{5})^{\ast}\label{SU5excessone}\\
(\overline{3},2)_{5}  &  \in H_{\overline{\partial}}^{0}(S,L^{-5})^{\ast
}\oplus H_{\overline{\partial}}^{1}(S,L^{5})\oplus H_{\overline{\partial}}%
^{2}(S,L^{-5})^{\ast}\text{.} \label{SU5excesstwo}%
\end{align}
Mathematically, the collection of admissible line bundles are those which have
vanishing cohomology group. \ As explained in Section \ref{GeomdelPezzReview},
when $S$ is a del Pezzo $N$ surface, such line bundles are in one to one
correspondence with the roots of the Lie algebra $E_{N}$, with a similar
result for $\mathbb{P}^{1}\times\mathbb{P}^{1}$. \ \ By definition, a root
$\alpha$ satisfies the condition that $n\alpha$ is also a root only when
$n=\pm1$. \ It now follows that if $L$ is a line bundle, $L^{5}$ cannot
correspond to a root of the Lie algebra $E_{N}$. \ Said differently, the
integral quantization of fluxes in the bulk theory would appear to present a
general obstruction towards realizing the spectrum of the MSSM\ without any
additional bulk matter with exotic $U(1)_{Y}$ charges.

We now argue that so long as all fields transform in mathematically
well-defined line bundles, fractional powers of line bundles also define
consistent gauge field configurations for the bulk theory. \ To establish
this, first recall that when all fields of a theory with gauge group $SU(N)$
transform in the adjoint representation, all observables are invariant under
$SU(N)$ modulo the center. \ Hence, the actual gauge group of the theory is
$SU(N)/%
\mathbb{Z}
_{N}$ so that the flux quantization condition allows gauge field
configurations with $1/N$ fractional flux units \cite{THooftFracFlux}. \ In
the presence of quark fields charged in the fundamental of $SU(N)$, we note
that the gauge group is indeed $SU(N)$ rather than $SU(N)/%
\mathbb{Z}
_{N}$.

In the present class of models, a similar fractional quantization condition
holds because all of the resulting gauge groups descend from an $E_{8}$ gauge
group. \ Indeed, recall that the $E_{N}$ groups canonically embed in $E_{8}$
as:%
\begin{equation}
\frac{E_{N}\times SU(K)}{%
\mathbb{Z}
_{K}}\subset E_{8} \label{maxE8subgroup}%
\end{equation}
where $N+K=9$. \ This result can be established as follows. \ Decomposing the
adjoint representations of $E_{N}$ and $SU(K)$ to $E_{N-1}\times U(1)$ and
$SU(K-1)\times U(1)$, we find that the resulting representations all have
charge $0$ or $\pm K$. \ As two examples, consider the decomposition of the
adjoint representations of the algebras $E_{6}$ and $E_{5}=SO(10)$:%
\begin{align}
E_{6}  &  \supset SO(10)\times U(1)\\
78  &  \rightarrow1_{0}+45_{0}+16_{-3}+\overline{16}_{3}\\
E_{5}  &  \supset SU(5)\times U(1)\\
45  &  \rightarrow1_{0}+24_{0}+10_{4}+\overline{10}_{-4}\text{.}%
\end{align}
Returning to the weight space decomposition of the charged representations, it
follows that the relative normalization of the matrices which generate the
Cartan subalgebras of $E_{N}$ and $E_{8}$ differ by $1/K$. \ Exponentiating
these matrices, we arrive at the desired condition in the corresponding subgroups.

This fractional quantization condition demonstrates that in the above example,
we may treat $L^{5}$ as a line bundle, with $L$ a \textquotedblleft fractional
power\textquotedblright\ of a line bundle. \ Moreover, fields localized on a
matter curve $\Sigma$ transform as sections of $K_{\Sigma}^{1/2}\otimes
L_{\Sigma}^{a}\otimes L_{\Sigma}^{\prime b}$ for integers $a$ and $b$, where
$L_{\Sigma}$ and $L_{\Sigma}^{\prime}~$respectively denote the restriction of
potentially fractional line bundles on $S$ and $S^{\prime}$. \ Indeed, the
common identification of the centers of the gauge groups in
(\ref{maxE8subgroup}) illustrates that although the individual restrictions of
$L$ and $L^{\prime}$ to $\Sigma$ may correspond to ill-defined line bundles,
their tensor product may still determine a mathematically well-defined line
bundle. \ We therefore conclude that so long as the resulting fields all
transform in well-defined bundles, the corresponding fractional line bundles
are physically well-defined.

\subsection{A No Go Theorem for $G_{S}=SO(10)$\label{NOGO}}

The analysis of the previous subsection establishes that when $G_{S}=SU(5)$,
there are no exotic bulk zero modes if and only if the gauge bundle
corresponds to a fractional line bundle of the form $\mathcal{O}_{S}%
(\alpha)^{1/5}$ where $\alpha$ corresponds to a root associated with an
element of $H_{2}(S,%
\mathbb{Z}
)$. \ In this Section we show that when $G_{S}=SO(10)$, direct breaking to
$G_{std}$ via fluxes always results in exotica in the low energy spectrum.

To establish this result, we note that the classification of Appendix
\ref{BreakClass} shows that the only instanton configurations which break
$SO(10)$ to $G_{std}$ take values in the subgroup $U(1)_{1}\times U(1)_{2}$ so
that the commutant subgroup in $SO(10)$ is $SU(3)\times SU(2)\times
U(1)_{1}\times U(1)_{2}$. \ With respect to this decomposition, the adjoint,
spinor and vector representations of $SO(10)$ decompose as:%
\begin{align}
SO(10)  &  \supset SU(5)\times U(1)_{2}\supset SU(3)\times SU(2)\times
U(1)_{1}\times U(1)_{2}\\
45  &  \rightarrow(1,1)_{0,0}+(1,1)_{0,0}+(8,1)_{0,0}+(1,3)_{0,0}\\
&  +(3,\overline{2})_{-5,0}+(\overline{3},2)_{5,0}+(3,2)_{1,4}+(\overline
{3},\overline{2})_{-1,-4}\\
&  +(\overline{3},1)_{-4,4}+(3,1)_{4,-4}+(1,1)_{6,4}+(1,1)_{-6,-4}\\
16  &  \rightarrow(1,1)_{0,-5}+(\overline{3},1)_{2,3}+(1,\overline{2}%
)_{-3,3}\\
&  +(1,1)_{6,-1}+(3,2)_{1,-1}+(\overline{3},1)_{-4,-1}\\
10  &  \rightarrow(3,1)_{-2,2}+(1,2)_{3,2}+(\overline{3},1)_{2,-2}%
+(1,\overline{2})_{-3,-2}\text{.}%
\end{align}

In the MSSM, fields charged under the subgroup $SU(3)\times SU(2)$ transform
in the representations $(3,2)$, $(1,2)$ and $(\overline{3},1)$. \ Returning to
the decomposition of the $45$, we conclude that the low energy spectrum must
contain no fields transforming in the $(\overline{3},2)_{5,0}$, $(\overline
{3},\overline{2})_{-1,-4}$ or $(3,1)_{4,-4}$.

In F-theory, all of the matter content of the MSSM descend from the $45$,
$16$, $\overline{16}$ or $10$ of $SO(10)$. \ As reviewed in Appendix
\ref{BreakClass}, there are precisely two linear combinations of $U(1)_{1}$
and $U(1)_{2}$ which can correspond to $U(1)_{Y}$ in the Standard Model:%
\begin{align}
U(1)_{Y}  &  =U(1)_{1}\label{firsthyper}\\
U(1)_{Y}  &  =-\frac{1}{5}U(1)_{1}-\frac{6}{5}U(1)_{2}\text{.}
\label{secondhyper}%
\end{align}
While the first case corresponds to embedding hypercharge in the usual way
inside of the $SU(5)$ factor, the second possibility corresponds to a
\textquotedblleft flipped\textquotedblright\ embedding of hypercharge
\cite{BarrFlipped}.

First suppose that $U(1)_{Y}$ is given by equation (\ref{firsthyper}).
\ Letting $A\equiv L_{1}^{5}$ and $B\equiv L_{1}^{-1}\otimes L_{2}^{-4}$, the
condition that the zero mode content must contain no exotic matter requires
that the following cohomology groups must vanish:%
\begin{align}
(\overline{3},2)_{5,0}  &  \in H_{\overline{\partial}}^{1}%
(S,A)=0\label{vanishone}\\
(3,\overline{2})_{-5,0}  &  \in H_{\overline{\partial}}^{1}(S,A^{-1}%
)=0\label{vanishtwo}\\
(\overline{3},\overline{2})_{-1,-4}  &  \in H_{\overline{\partial}}%
^{1}(S,B)=0\label{vanishthree}\\
(3,1)_{4,-4}  &  \in H_{\overline{\partial}}^{1}(S,A\otimes
B)=0\label{vanishfour}\\
(1,1)_{-6,-4}  &  \in H_{\overline{\partial}}^{1}(S,A^{-1}\otimes B)=0\text{.}
\label{vanishfive}%
\end{align}

For a supersymmetric configuration, it follows from the vanishing theorem of
\cite{BHV} that the cohomology groups $H_{\overline{\partial}}^{0}$ and
$H_{\overline{\partial}}^{2}$ vanish for all of the above line bundles. \ The
cohomology group $H_{\overline{\partial}}^{1}$ therefore vanishes when the
index of each line bundle vanishes. \ Equations (\ref{vanishone}) and
(\ref{vanishtwo}) imply:%
\begin{equation}
0=\chi(S,A)+\chi(S,A^{-1})=2+c_{1}(A)\cdot c_{1}(A)\text{.} \label{firstcond}%
\end{equation}
On the other hand, equations (\ref{vanishthree})-(\ref{vanishfive}) imply:%
\begin{equation}
0=\chi(S,A\otimes B)+\chi(S,A^{-1}\otimes B)-2\chi(S,B)=c_{1}(A)\cdot c_{1}(A)
\end{equation}
which contradicts equation (\ref{firstcond}). \ The resulting low energy
spectrum will therefore always contain some exotic matter.

Next consider the flipped embedding of $U(1)_{Y}$ given by equation
(\ref{secondhyper}). \ With notation as above, the condition that the zero
mode content must contain no exotic matter now requires that the following
cohomology groups vanish:%
\begin{align}
(\overline{3},\overline{2})_{-1,-4}  &  \in H_{\overline{\partial}}%
^{1}(S,B)=0\\
(3,2)_{1,4}  &  \in H_{\overline{\partial}}^{1}(S,B^{-1})=0\\
(\overline{3},2)_{5,0}  &  \in H_{\overline{\partial}}^{1}(S,A)=0\\
(3,1)_{4,-4}  &  \in H_{\overline{\partial}}^{1}(S,A\otimes B)=0\\
(1,1)_{6,4}  &  \in H_{\overline{\partial}}^{1}(S,A\otimes B^{-1})=0\text{.}%
\end{align}
These conditions are the same as those of equations (\ref{vanishone}%
-\ref{vanishfive}) with the roles of $A$ and $B$ interchanged. \ We therefore
conclude that in all cases, the resulting spectrum will contain exotic matter.

More generally, we note that the classification of possible breaking patterns
provided in Appendix \ref{BreakClass} requires at least one $U(1)$ factor.
\ When $G_{S}$ has rank five or more, direct breaking to $G_{std}$ therefore
requires the instanton configuration to take values in a subgroup of $G_{S}$
with rank at least two. \ We note that while only abelian instanton
configurations are available for rank four and five bulk gauge groups, it is
in principle possible that an $SU(2)$ valued instanton could partially break
the bulk gauge group when $G_{S}=E_{6}$. \ However, decomposing the adjoint
representation to $G_{std}$, the number of different exotic representations
appears to always be greater than the rank of the subgroup in which the
instanton takes values. \ The requirement that so many different cohomology
groups must simultaneously vanish is then an over-constrained problem so that
in such cases exotics are unavoidable.

\subsection{MSSM\ Spectrum\label{MSSMSpectrum}}

In this Section we explain how to obtain the exact spectrum of the MSSM\ when
$S$ is a del Pezzo surface. \ As explained in subsection \ref{NOGO}, direct
breaking via internal fluxes will generate exotics when the bulk gauge group
is not $SU(5)$. \ Restricting to the case $G_{S}=SU(5)$, the only candidate
bundles which will not generate exotic bulk zero modes are in one to
correspondence with the roots of an exceptional Lie algebra. \ In this case,
all of the matter content of the MSSM must localize on matter curves.

\begin{figure}
[ptb]
\begin{center}
\includegraphics[
height=1.5783in,
width=2.5901in
]%
{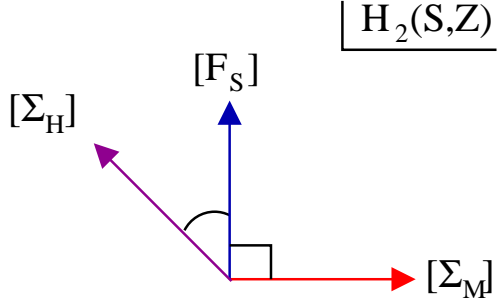}%
\caption{Letting $[F_{S}]$ denote the two-cycle in $H_{2}(S,\mathbb{Z} )$
which is Poincar\'{e} dual to the background hypercharge flux $\left\langle
F_{S}\right\rangle $, there is a natural distinction between the class of the
Higgs curve $[\Sigma_{H}]$ and the class of the chiral matter curves
$[\Sigma_{M}]$. \ Indeed, while the net flux on $\Sigma_{M}$ must vanish to
preserve a full GUT\ multiplet, the gauge field configuration must restrict
non-trivially on the Higgs curves in order to solve the doublet triplet
splitting problem. \ When the net flux on the Higgs curve is not zero, this
corresponds to the condition that $[\Sigma_{M}]$ and $\left[  F_{S}\right]  $
are orthogonal while $[\Sigma_{H}]$ and $\left[  F_{S}\right]  $ are not.}%
\label{parorthog}%
\end{center}
\end{figure}

Individual components of a GUT\ multiplet will interact differently with the
internal hypercharge flux. \ In keeping with our general philosophy, we
require that a complete GUT\ multiplet must localize on a given matter curve
so that on such curves, the net hypercharge flux must vanish. \ Otherwise, a
different index will determine the number of zero modes coming from each
component of a complete GUT\ multiplet. \ On the other hand, the gauge field
must restrict non-trivially on the Higgs curves in order to solve the
doublet-triplet splitting problem. \ See figure \ref{parorthog} for a
depiction of how the corresponding elements in $H_{2}(S,%
\mathbb{Z}
)$ intersect.%

In order to achieve a chiral matter spectrum in four dimensions, the net flux
on the matter curve cannot vanish. \ As an example, consider a six-dimensional
hypermultiplet in the $5_{1}$ of $G_{S}\times G_{S^{\prime}}=SU(5)\times U(1)$
which localizes on an exceptional curve $\Sigma$ with homology class $E_{1}$.
\ The overall normalization of the $U(1)$ charge is not particularly important
because we shall consider vacua with fractional line bundles. \ When
$L=\mathcal{O}_{S}(E_{2}-E_{3})^{1/5}$ the restriction of $L$ to $\Sigma$ is
trivial. $\ \ $Letting $L^{^{\prime}}$ denote the supersymmetric line bundle
on the seven-brane which intersects the GUT model seven-brane along $\Sigma$,
the restriction of $L^{\prime}$ to $\Sigma$ must be non-trivial in order to
achieve a chiral matter spectrum. \ For example, when $L_{\Sigma}^{\prime
}=\mathcal{O}_{\Sigma}(-3)$, the zero mode content is:%
\begin{align}
0\times5  &  \in H_{\overline{\partial}}^{0}(\Sigma,K_{\Sigma}^{1/2}%
\otimes\mathcal{O}_{\Sigma}(-3))=0\\
3\times\overline{5}  &  \in H_{\overline{\partial}}^{0}(\Sigma,K_{\Sigma
}^{1/2}\otimes\mathcal{O}_{\Sigma}(3))=H_{\overline{\partial}}^{0}%
(\Sigma,\mathcal{O}_{\Sigma}(2))
\end{align}
where we have also indicated the multiplicity of the zero modes. \ Similar
considerations apply for other GUT\ multiplets.

On the other hand, the Higgs fields of the MSSM\ do not fill out complete
GUT\ multiplets at low energies. \ In this case, the net hypercharge flux
piercing this matter curve must be non-zero. \ More precisely, recall that for
minimal supersymmetric $SU(5)$ GUT models, the Higgs up and down fields
respectively descend from the $5$ and $\overline{5}$ of $SU(5)$, where the $5$
decomposes to $(3,1)_{-2}+(1,2)_{3}$. \ Letting $L_{\Sigma}$ denote the
restriction of the bulk gauge bundle $L$ to the matter curve $\Sigma$ with
similar notation for $L_{\Sigma}^{\prime}$, we note that the zero mode content
is determined by the cohomology groups:%
\begin{align}
(1,2)_{3}  &  \in H_{\overline{\partial}}^{0}(\Sigma,K_{\Sigma}^{1/2}\otimes
L_{\Sigma}^{3}\otimes L_{\Sigma}^{\prime n})\\
(3,1)_{-2}  &  \in H_{\overline{\partial}}^{0}(\Sigma,K_{\Sigma}^{1/2}\otimes
L_{\Sigma}^{-2}\otimes L_{\Sigma}^{\prime n})
\end{align}
where $n$ is an integer associated with the $U(1)$ charge associated with the
brane wrapping $S^{\prime}$. \ Mathematically, we wish to find line bundles
such that $K_{\Sigma}^{1/2}\otimes L_{\Sigma}^{3}\otimes L_{\Sigma}^{\prime
n}$ has non-vanishing cohomology whereas $K_{\Sigma}^{1/2}\otimes L_{\Sigma
}^{-2}\otimes L_{\Sigma}^{\prime n}$ has trivial cohomology. \ A necessary
condition for $K_{\Sigma}^{1/2}\otimes L_{\Sigma}^{-2}\otimes L_{\Sigma
}^{\prime n}$ to have trivial cohomology is that the degree of the line bundle
$L_{\Sigma}^{-2}\otimes L_{\Sigma}^{\prime n}$ must vanish. As a brief aside,
we recall the well-known fact that degree zero line bundles are in one to one
correspondence with points on the Jacobian of the curve.

As an example, consider a genus one matter curve where the line bundles
$L_{\Sigma}$ and $L_{\Sigma}^{\prime}$ are given by:%
\begin{align}
L_{\Sigma}  &  =\mathcal{O}_{\Sigma}(-np_{1}+np_{2})\\
L_{\Sigma}^{\prime}  &  =\mathcal{O}_{\Sigma}(3p_{1}-3p_{2})
\end{align}
where $p_{1}$ and $p_{2}$ denote distinct degree one divisors on $\Sigma$
which are not linearly equivalent. \ Because these divisors are not linearly
equivalent, the divisor $p_{1}-p_{2}$ is not effective.\footnote{More
generally, recall that on a general genus $g$ Riemann surface, a divisor $D$
with degree $\geq g$ is linearly equivalent to an effective divisor
\cite{Griffiths}. \ This imposes a non-trivial constraint on the ways in which
doublet-triplet splitting can arise for a general matter curve.} \ Assuming
that $K_{\Sigma}^{1/2}$ is trivial, we have:%
\begin{align}
(1,2)_{3}  &  \in H_{\overline{\partial}}^{0}(\Sigma,\mathcal{O}_{\Sigma
}(0))\simeq%
\mathbb{C}%
\\
(3,1)_{-2}  &  \in H_{\overline{\partial}}^{0}(\Sigma,\mathcal{O}_{\Sigma
}(5n(p_{1}-p_{2}))=0
\end{align}
since the divisor $p_{1}-p_{2}$ is not effective. \ In this case, we achieve a
vector-like pair of Higgs up/down fields on the curve $\Sigma$.

Now, there is no reason that the Higgs up and down fields must localize on the
same matter curve. \ In a certain sense, the above implementation of doublet
triplet splitting is somewhat artificial precisely because the distinguishing
feature of the Higgs curve is that a non-trivial flux is present. With this in
mind, it seems far more natural to consider line bundles which have
non-trivial degree on the Higgs curves. \ In this case, a given Higgs curve
will automatically contain more Higgs up than Higgs down fields.

To give an explicit example of this type, consider a six-dimensional
hypermultiplet in the $5$ of $SU(5)$ localized on a genus zero curve $\Sigma$.
\ In this case, the zero mode content is determined by the cohomology groups:%
\begin{align}
(1,2)_{3}  &  \in H_{\overline{\partial}}^{0}(\Sigma,K_{\Sigma}^{1/2}\otimes
L_{\Sigma}^{3}\otimes L_{\Sigma}^{\prime})\\
(1,\overline{2})_{-3}  &  \in H_{\overline{\partial}}^{0}(\Sigma,K_{\Sigma
}^{1/2}\otimes L_{\Sigma}^{-3}\otimes L_{\Sigma}^{\prime-1})\\
(3,1)_{-2}  &  \in H_{\overline{\partial}}^{0}(\Sigma,K_{\Sigma}^{1/2}\otimes
L_{\Sigma}^{-2}\otimes L_{\Sigma}^{\prime})\\
(\overline{3},1)_{2}  &  \in H_{\overline{\partial}}^{0}(\Sigma,K_{\Sigma
}^{1/2}\otimes L_{\Sigma}^{2}\otimes L_{\Sigma}^{\prime-1})\text{.}%
\end{align}
The zero mode content on $\Sigma$ yields precisely one Higgs up field for
fractional line bundle assignments:%
\begin{equation}
L_{\Sigma}=\mathcal{O}_{\Sigma}(1)^{1/5}\text{ and }L_{\Sigma}^{\prime
}=\mathcal{O}_{\Sigma}(1)^{2/5}\text{.}%
\end{equation}
Similarly, a single Higgs down field can also localize on another matter curve.

It is also possible to localize a single Higgs up field on a higher genus
matter curve. \ For example, with notation as above, when $\Sigma$ is a genus
one curve, the fractional line bundle assignments:%
\begin{equation}
L_{\Sigma}=\mathcal{O}_{\Sigma}(p_{1})^{1/5}\otimes\mathcal{O}_{\Sigma}%
(p_{1}-p_{2})^{1/5}\text{ and }L_{\Sigma}^{\prime}=\mathcal{O}_{\Sigma}%
(p_{1})^{2/5}\otimes\mathcal{O}_{\Sigma}(p_{1}-p_{2})^{-3/5}%
\end{equation}
will again yield a single $H_{u}$ field localized on $\Sigma$.

In fact, in Section \ref{DTS} we will show that in order to remain in accord
with current bounds on the lifetime of the proton, the Higgs fields
\textit{must} localize on different matter curves. \ These matter curves may
or may not intersect inside of $S$. \ When these curves do not intersect,
these fields do not couple in the superpotential and the $\mu$ term is
automatically zero. \ Moreover, when these curves do intersect, they must
interact with a third gauge singlet which localizes on a curve that only
intersects $S$ at a point. \ In Section \ref{Suppression} we estimate the
behavior of this gauge singlet wave function near the surface $S$ and show
that this naturally yields an exponentially suppressed $\mu$ term.

\subsection{Candidates For Dark Matter}

In the MSSM with R-parity, the lightest supersymmetric partner (LSP) could be
a viable dark matter candidate. \ In fact, in the context of a local model, it
is natural to expect a large number of additional gauge degrees of freedom
which only interact gravitationally with the MSSM. \ This appears to be an
automatic feature of many consistent string compactifications which will
typically contain several hidden sectors. \ For example, in the perturbative
heterotic string, this role can be played by the hidden $E_{8}$ factor. \ A
rough comparison of the two $E_{8}$ factors would then suggest that half of
the matter content in such a model could be visible, and the other half could
be dark matter. \ In F-theory, the analogue of the hidden $E_{8}$ factor could
be the additional seven-branes which are required for the compactification to
be globally consistent. \ For example, F-theory compactified on $K3$
corresponds to a configuration of $24$ seven-branes. \ More generally, it
would be of interest to estimate the number of seven-branes which only
interact gravitationally with the MSSM. \ In this case, the total class of the
seven-branes in a threefold base $B_{3}$ is given by $12c_{1}(B_{3})$.
\ Integrating this Chern class over an appropriate two-cycle would then yield
a rough estimate on the amount of dark matter from seven-branes. \ It is also
in principle possible that the total number of three-branes in the
compactification could also contribute to the dark matter content of the
model. \ In the absence of fluxes, the total number of three-branes is given
by $\chi(CY_{4})/24$. \ We note that in order for the Calabi-Yau fourfold to
be elliptically fibered, the threefold base $B_{3}$ must be a Fano variety.
\ For example, $B_{3}=\mathbb{P}^{3}$, gives $48$ seven-branes. Note that the
GUT group involves a bound state of $O(10)$ such seven-branes. We find it
quite amusing that this is in rough agreement with the observed ratio between
visible and dark matter in our Universe! Of course, this depends on the
relative masses for the various visible and hidden fields. There is a finite
list of such manifolds \cite{ShafFANO}, and it would therefore be of interest
to compare the relative number of three-branes and seven-branes in such compactifications.

\section{Geometry and Matter Parity\label{R-parity}}

From a phenomenological viewpoint, matter parity provides a simple way to
forbid renormalizable terms in the four-dimensional superpotential which can
potentially induce proton decay. \ It also naturally leads to an LSP\ which
could potentially be a dark matter candidate. \ In a Lorentz invariant theory,
this is equivalent to assigning an appropriate R-parity to the individual
components of a superfield. \ Indeed, the essential point is that this
discrete symmetry distinguishes the Higgs superfields from all of the other
chiral superfields of the MSSM. \ In this Section we argue that the presence
of such a $%
\mathbb{Z}
_{2}$ symmetry is quite natural from the perspective of F-theory. \

As explained in subsection \ref{MSSMSpectrum}, the Higgs fields localize on
matter curves pierced by a net amount of internal hypercharge flux while the
chiral matter localizes on curves where the net hypercharge flux is trivial.
\ This is a discrete choice which naturally distinguishes the\ Higgs
superfields from the rest of the chiral superfields of the MSSM.

From a more global perspective, these fluxes correspond to the localization of
four-form $G$-flux in the compactification. \ If the Calabi-Yau fourfold
admits a geometric $%
\mathbb{Z}
_{2}$ symmetry, then these fluxes will decompose into even and odd elements of
$H^{4}(CY_{4},%
\mathbb{Z}
)$ which we denote by $H^{4}(CY_{4},%
\mathbb{Z}
)_{+}$ and $H^{4}(CY_{4},%
\mathbb{Z}
)_{-}$. \ If this symmetry is well-defined, it follows that on a given
seven-brane, the corresponding line bundles must have a definite parity under
this choice of sign. \ For example, the parity of the line bundle on the $S$
brane can be even while the parity of the line bundles on the other branes may
have other parities.

It now follows that the net flux on a matter curve can only be non-zero when
the flux and matter curve have the same parity. \ Indeed, letting $F^{\pm}$
denote a flux with parity $\pm1$ with similar notation for matter curves
$\Sigma^{\pm}$, the unbroken $%
\mathbb{Z}
_{2}$ symmetry implies:%
\begin{align}
\underset{\Sigma^{-}}{\int}F_{\Sigma^{-}}^{+}  &  =\underset{S}{\int}%
F^{+}\wedge PD(\Sigma^{-})=-\underset{S}{\int}F^{+}\wedge PD(\Sigma^{-})=0\\
\underset{\Sigma^{+}}{\int}F_{\Sigma^{+}}^{-}  &  =\underset{S}{\int}%
F^{-}\wedge PD(\Sigma^{+})=-\underset{S}{\int}F^{-}\wedge PD(\Sigma^{+})=0
\end{align}
where $PD(\Sigma^{\pm})\in H^{2}(S,%
\mathbb{Z}
)$ denotes the Poincar\'{e} dual element of $\left[  \Sigma^{\pm}\right]  \in
H_{2}(S,%
\mathbb{Z}
)$. In other words, when the integral of the flux over a curve does not
vanish, the flux and curve have the $\mathit{same}$ parity.

In order for this group action to remain well-defined on the matter curves,
the internal wave functions which are sections of appropriate bundles must
also have a definite sign under the group action. \ First consider the parity
of the Higgs fields. \ These wave functions are defined as sections of line
bundles which depend non-trivially on the restriction of line bundles from
both $S$ as well as other transversely intersecting seven-branes. \ We
therefore conclude that both fluxes must have the same parity. \ In
particular, we conclude if the parity of the bulk gauge field is even, then
the Higgs fields will also have even parity.

Next consider the parity of the remaining matter fields. \ Here it is
essential that the net flux contribution from $S$ is trivial on all such
matter curves. \ In particular, if the gauge bundle from the transversely
intersecting seven-brane is odd under matter parity, then the corresponding
sections on each matter curve will also be odd under the $%
\mathbb{Z}
_{2}$ action on the Calabi-Yau fourfold. \ Hence, we obtain on rather general
grounds a geometric version of matter parity.

\section{Proton Decay and Doublet-Triplet Splitting\label{DTS}}

As argued in subsection \ref{MSSMSpectrum}, there exist vacua which yield the
exact spectrum of the MSSM\ for an appropriate choice of flux in a local
intersecting seven-brane configuration. \ In particular, we found that the
Higgs triplets can typically be removed from the low energy spectrum. \ While
this mechanism provides a natural way to achieve the correct zero mode
spectrum in the Higgs sector, when the Higgs up and down fields localize on
the same matter curve, the higher Kaluza-Klein modes of the corresponding
six-dimensional fields will generate higher order superpotential terms of the
form $QQQL/M_{KK}$ with order one coefficients. \ While here we have presented
the operator in terms of the Kaluza-Klein mass scale $M_{KK}$, for minimal
$SU(5)$ GUT\ models, we can reliably approximate $M_{KK}$ by $M_{GUT}%
$.\footnote{When we present some examples of four-dimensional flipped $SU(5)$
models which descend from an eight-dimensional $SO(10)$ model, there can be a
small discrepancy between the four-dimensional GUT\ scale $M_{GUT}$ and
$M_{KK}$.}\ If present, such operators can significantly shorten the lifetime
of the proton.

\begin{figure}
[ptb]
\begin{center}
\includegraphics[
height=3.5094in,
width=4.9623in
]%
{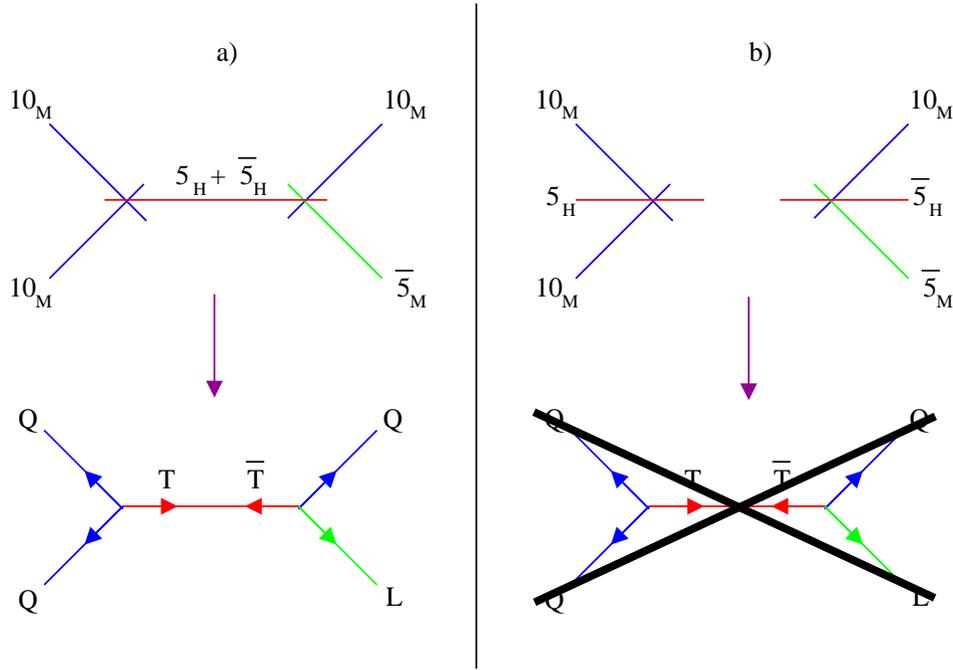}%
\caption{Depiction of how the geometry of matter curves directly translates
into amplitudes in the low energy theory. \ In case a), the Higgs up and down
fields localize on the same matter curve. \ The resulting field theory diagram
which generates the operator $QQQL$ is given by interpreting each matter curve
as the leg of a Feynman diagram. \ In case b), the Higgs up and down fields
localize on distinct matter curves. \ In this case, the Feynman diagram
involving the exchange of massive Higgs triplets is unavailable.}%
\label{qqql}%
\end{center}
\end{figure}

We now explain how such terms could potentially be generated in our class of
models. \ When all Yukawa couplings to the Higgs triplets are order one
parameters, the superpotential terms:%
\begin{equation}
W_{GUT}=QQT_{u}+QLT_{d}+M_{KK}T_{u}T_{d} \label{GUTTripletproblem}%
\end{equation}
will give a large mass to the Higgs triplets $T_{u}T_{d}$ of order $M_{KK}$.
\ Integrating out $T_{u}$ and $T_{d}$, the coefficient of the operator
$QQQL/M_{KK}$ would then be too large to satisfy present constraints. \ In
fact, the geometry of the matter curves indicates precisely when we can expect
such terms to be generated. \ The tree level diagram which generates the
offending operator is given by drawing the intersection locus of the matter
curves and interpreting each matter curve as a leg of the corresponding
Feynman diagram. \ See figure \ref{qqql} for a depiction of how the geometry
of the matter curves quite literally translates into a statement about
diagrams in the low energy theory.%

While it is in principle possible to suppress the value of this coefficient by
incorporating flavor symmetries, in the context of four dimensional
supersymmetric GUT\ models, this problem can be avoided by having $T_{u}$ and
$T_{d}$ develop masses by pairing with additional heavy triplet states
$T_{u}^{\prime}$ and $T_{d}^{\prime}$ so that the superpotential instead takes
the form:\footnote{We thank S. Raby for emphasizing this point to us.}%
\begin{equation}
W_{GUT}=QQT_{u}+QLT_{d}+MT_{u}T_{d}^{\prime}+MT_{d}T_{u}^{\prime}%
\end{equation}
which does not generate the offending dimension five operator from integrating
out massive fields at tree level. \ Note that this occurs automatically when
the Higgs up and down fields localize on distinct matter curves.

In compactifications of the heterotic string on Calabi-Yau threefolds, the
Higgs triplet is typically projected out of the low energy spectrum by
discrete Wilson lines. \ In general, it is not clear to us whether this
sufficiently suppresses proton decay. \ Indeed, while the Higgs triplet zero
mode may be absent from the spectrum, there is an entire tower of
Kaluza\ Klein modes which must also be considered. \ If any of these modes
contribute an interaction term of the form given by equation
(\ref{GUTTripletproblem}), the coefficient of the offending dimension five
operator may still be too large to remain in accord with observation.

To summarize, we have seen that the proton decays too rapidly when the Higgs
up and down fields localize on the same matter curve. \ As a necessary first
step, we have shown that when these Higgs fields localize on distinct matter
curves, integrating out the higher\ Kaluza-Klein modes for the Higgs fields
does not generate the offending baryon number violating term $QQQL$. \ Even
so, it is still in principle possible that some exotic process could generate
the operator $QQQL$. \ In fact, placing the\ Higgs fields on different matter
curves automatically equips them with additional global symmetries in the low
energy effective theory. \ As we now explain, these symmetries significantly
extend the lifetime of the proton.

\section{Extra $U(1)$'s and Higher Dimension Operators\label{ExtraU1}}

In Section \ref{DTS} we have shown that the dimension five operators
responsible for proton decay are naturally suppressed when the Higgs up and
Higgs down fields localize on different matter curves. \ In this Section we
explain from a different perspective why this suppression occurs and also
discuss on more general grounds when we expect other higher dimensional
operators to suffer a similar fate.

Imposing additional global symmetries provides one common way to suppress
undesirable interaction terms in field theory. \ Indeed, so long as the global
symmetry remains unbroken, all of the higher order terms of the effective
superpotential will also respect this symmetry. \ In F-theory, these $U(1)$
factors occur automatically because the breaking direction in the Cartan
subalgebra of a given singularity determines the location of the matter curves
in the geometry. \ Matter localizes on the curve precisely when it is charged
under the appropriate subgroup. \ While this generically allows local triple
intersections of matter curves to take place, all of the fields of the MSSM
will therefore be charged under additional $U(1)$ factors. \ These extra
$U(1)$'s can therefore naturally suppress higher dimension operators. \ When
two curves do not intersect inside of $S$, fields localized on each curve will
be charged under distinct $U(1)$ groups. \ This can forbid cubic interaction
terms as well as many higher order contributions to the effective
superpotential. \ It would be very interesting to determine the precise
mapping between topological properties of intersecting curves and the
associated $U(1)$ fields.

From a bottom up perspective, the fields of the MSSM\ contain various
accidental symmetries. \ Assuming generic values of the Yukawa couplings and
that the $\mu$ term originates from the vev of a gauge singlet, the classical
action is invariant under four $U(1)$ symmetries. \ These can be identified
with $U(1)_{Y}$ hypercharge, $U(1)_{B}$ baryon number, $U(1)_{B-L}$ baryon
minus lepton number and a $U(1)_{PQ}$ Peccei-Quinn symmetry. \ Of these four
possibilities, only $U(1)_{Y}$ and $U(1)_{B-L}$ are potentially non-anomalous.

In a quantum theory of gravity, any global symmetry must be promoted to a
gauge symmetry. \ One potential worry is that because the fields of the
MSSM\ are naturally charged under these $U(1)$'s, the presence of these gauge
bosons could lead to conflict with experiment. \ While these $U(1)$'s will
typically be anomalous and therefore lift from the low energy spectrum, it is
interesting to ask whether a massless $U(1)$ of this type is already ruled out
by experiment. \ This is not very promising because current constraints from
fifth force experiments have set a strong limit on the gauge coupling of extra
massless $U(1)$ gauge bosons:%
\begin{equation}
g_{extra}\lesssim\frac{m_{n}}{M_{pl}}\sim10^{-19} \label{extrau1bound}%
\end{equation}
where $m_{n}$ denotes the mass of the neutron. \ In the absence of a natural
explanation for why the gauge coupling would be so weak for such couplings,
this appears quite fine-tuned. \ In fact, such a small value is already in
conflict with the conjecture that gravity is the weakest force
\cite{WeakGravConjecture}. \ See \cite{FayetExtraU(1),DobrescuU(1)} for
further discussion on extra massless $U(1)$ gauge bosons.

The analogue of equation (\ref{GUTCOUP}) for $\alpha_{extra}=g_{extra}%
^{2}/4\pi$ is of the form:%
\begin{equation}
\alpha_{extra}^{-1}=M_{\ast}^{4}Vol(S_{extra})\sim M_{\ast}^{4}R_{\bot}%
^{2}R_{S}^{2}%
\end{equation}
where as before, $R_{\bot}$ denotes the length scale associated with the
direction normal to the surface $S$. \ In tandem with equation (\ref{GUTCOUP})
this implies:
\begin{equation}
\alpha_{extra}\sim\alpha_{GUT}\frac{R_{S}^{2}}{R_{\bot}^{2}}=\alpha
_{GUT}\times\varepsilon^{\gamma}\sim7\times10^{-3\pm0.5}\text{.}%
\end{equation}

Based on the above estimate, we conclude that all additional $U(1)$ gauge
bosons must develop a sufficiently large mass in order to lift from the low
energy spectrum. \ In fact, our expectation is that this only imposes a mild
constraint on the compactification. \ When the $U(1)$ symmetry is anomalous,
the Green-Schwarz mechanism will generate a string scale mass for the gauge
boson. \ Even when the $U(1)$ symmetry is non-anomalous, the gauge boson can
still develop a large mass. \ Indeed, although the analysis of Section
\ref{GUTBreakviaFlux} shows that four-dimensional $U(1)$ gauge bosons can
remain massless in the presence of internal fluxes, it also establishes
sufficient conditions for such bosons to develop a large mass on the order of
$R_{\bot}^{-1}$. \ In either case, we therefore expect that it is always
possible for all extraneous $U(1)$ gauge bosons to develop a suitably large
mass. \ In the low energy effective theory, some imprint of the gauge symmetry
will remain as an approximate global symmetry in the low energy effective
theory. \ These global symmetries can be violated by non-perturbative
contributions to the superpotential from Euclidean branes wrapping the various
K\"{a}hler surfaces of the compactification. \ Such contributions are
naturally suppressed by an exponential factor of the form $\exp(-c/\alpha
_{extra})$ where $c$ is an order one positive number. \ Similar instanton
effects have been proposed as a possible solution to the cosmological constant
problem \cite{SvrcekCC}. \ Such exponentials could also provide a novel method
of generating contributions to the flavor sector of the theory. \ We present
one brief speculation along these lines in Section \ref{TowardsYukawas}. \ As
a brief aside, recall that in Section \ref{GeneralType} we presented an
example of a four-dimensional GUT model where an appropriate operator
generated by non-perturbative contributions could produce an effective $\mu$
term. \ Indeed, when a strict decoupling limit does not exist, it is likely
that non-perturbative contributions to the superpotential could play a more
prominent role in the effective theory.

\section{Towards Realistic Yukawa Couplings\label{TowardsYukawas}}

Finding vacua with the correct matter spectrum of the MSSM is only the first
step in constructing a semi-realistic model. \ In models where all chiral
matter localizes on matter curves, the leading order contribution to the
four-dimensional effective superpotential originates from the triple
intersection of matter curves. \ After presenting a general analysis of how
matter curves can form triple intersections in $S$, we show that in order to
achieve one generation with mass which is hierarchically larger than the two
lighter generations, some of the matter curves must self-intersect or
\textquotedblleft pinch\textquotedblright\ inside of $S$. \ See figure
\ref{selfintersect} for a depiction of a pinched curve. \ While a complete
theory of flavor is beyond the scope of this paper, we can nevertheless
provide a qualitative explanation for why the heaviest generation obeys an
approximate GUT mass relation which is violated by the lighter generations.
\ In fact, the effect we discover is \textit{generically realized} in vacua
with non-zero internal hypercharge flux because the Aharanov-Bohm effect
distorts the wave functions of individual components of a GUT\ multiplet by
different amounts. \ Moreover, this distortion becomes more pronounced as the
mass of the generation decreases. \ We conclude by presenting some
speculations on how more detailed properties of flavor physics could originate
from a local del Pezzo model.

\subsection{Criteria For Triple Intersections \label{TripCurveCrit}}

As reviewed in Section \ref{BasicSetup}, cubic contributions to the
superpotential of an exceptional seven-brane can originate from three sources.
\ These correspond to interactions amongst three bulk zero modes, interactions
between a single bulk zero mode and two zero modes localized on a matter
curve, and interaction terms between three zero modes on matter curves. \ As
explained in subsection \ref{MSSMSpectrum}, in a minimal $SU(5)$ GUT all of
the field content of the MSSM localizes on curves. \ Thus, the leading order
contribution to the effective superpotential comes from the triple
intersection of matter curves.

Locally, the triple intersection of matter curves in $S$ occurs when the bulk
singularity type $G_{S}$ undergoes an at least twofold enhancement to a
singularity of type $G_{p}\supset G_{S}\times U(1)_{1}\times U(1)_{2}$.
\ Following the general philosophy of \cite{KatzVafa}, we note that matter
localized along curves in $S$ is charged under the corresponding
$U(1)_{1}\times U(1)_{2}$ subgroup. \ Indeed, letting $t_{1}$ and $t_{2}$
denote the local deformation parameters associated with the two $U(1)$
factors, this curve is locally described by an equation of the form:%
\begin{equation}
at_{1}+bt_{2}=0\text{.}%
\end{equation}
In the above, the constants $a$ and $b$ are determined by the decomposition of
the adjoint representation of $G_{p}$ to $G_{S}\times U(1)_{1}\times U(1)_{2}$
so that the appropriate irreducible representation of $G_{S}$ has
$U(1)_{1}\times U(1)_{2}$ charge $(a,b)$. \ This is simply the statement that
because the Cartan subgroup is visible to the geometry, this local enhancement
in singularity type has been Higgsed in the bulk to $G_{S}$.

The triple intersection of three curves $\Sigma_{1}$, $\Sigma_{2}$ and
$\Sigma_{3}$ requires that the intersection product of the corresponding
homology classes satisfies $[\Sigma_{i}]\cdot\lbrack\Sigma_{j}]>0$ for $i\neq
j$. \ Even so, generic curves representing each class which all intersect
pairwise will not form a triple intersection in $S$. \ However, in certain
cases there exist representative holomorphic curves of each homology class
which can form a triple intersection inside of $S$. \ For this to occur, it
must be possible to deform the point of intersection of one pair of curves to
coincide with the point of intersection of another pair. \ In other words, the
normal bundle $N_{\Sigma/S}$ of one of the curves must possess at least one
global section. \ Although from the perspective of the surface $S$ this may
appear to be a somewhat non-generic situation, we note that in F-theory such
points of triple intersection occur automatically. \ Indeed, as explained in
\cite{BHV}, this follows from the fact that in F-theory, rank two enhancements
in the singularity type will generically occur at points in $S$. \ The claim
now follows from group theoretic considerations.

At a pragmatic level, given curves $\Sigma_{i}=(f_{i}=0)$, it is possible to
engineer a triple intersection by requiring that one of the $f_{i}$ is a
linear combination of the other two $f_{i}$'s in the ring of sections on $S$.
\ Assuming without loss of generality that $f_{3}$ is given by a linear
combination of $f_{1}$ and $f_{2}$, this can be written as:%
\begin{equation}
f_{3}=\alpha_{1}f_{1}+\alpha_{2}f_{2} \label{triplecondition}%
\end{equation}
where the $\alpha_{i}$ correspond to holomorphic sections of some line bundles
on $S$. \ For example, this condition is satisfied when both $[\Sigma
_{3}]-[\Sigma_{1}]$ and $[\Sigma_{3}]-[\Sigma_{2}]$ are \textquotedblleft
effective\textquotedblright\ divisors, namely divisors which correspond to
holomorphic curves.

This geometric condition can be used to narrow the search for vacua which are
phenomenologically viable. \ For example, to forbid cubic matter parity
violating contributions to the superpotential, it is enough to require that
the curves supporting the chiral matter of the MSSM\ must not form a triple
intersection. \ On the other hand, in order to have at non-trivial interaction
terms, some of the matter content of the MSSM must localize on a curve which
\textit{is not exceptional}. \ Indeed, three exceptional curves cannot triple
intersect in $S$. \ This follows from the fact that the normal bundle of a
curve in $S$ has degree $[\Sigma]\cdot\lbrack\Sigma]$ which equals $-1$ for an
exceptional curve. \ Because $H_{\overline{\partial}}^{0}(\mathbb{P}%
^{1},\mathcal{O}(-1))=0$, none of the pairwise intersection points in such a
configuration can be deformed to a point of triple intersection.

\subsection{Textures\label{TextureConsequence}}

At zeroth order, it is most important to obtain a naturally heavy third
generation in the quark sector. \ Indeed, the mass of the top quark is roughly
$170$ GeV, which is significantly higher than the next heaviest up type quark.
\ This requires that the corresponding Yukawa coupling must be sufficiently
large. \ In a suitable basis, we therefore require that the up-type Yukawa
couplings are of the form:%
\begin{equation}
\lambda^{u}\sim\left[
\begin{array}
[c]{ccc}%
\varepsilon_{11} & \varepsilon_{12} & \varepsilon_{13}\\
\varepsilon_{21} & \varepsilon_{22} & \varepsilon_{23}\\
\varepsilon_{31} & \varepsilon_{32} & 1
\end{array}
\right]  \label{YukawaHierarchy}%
\end{equation}
where the $\varepsilon$'s are all parametrically smaller than $1$.

When all of the cubic terms of the superpotential originate from the triple
intersection of matter curves in $S$, there is additional structure in the
form of the Yukawa couplings. \ First consider the Yukawa couplings for fields
charged in the $10$ of $SU(5)$. \ In this case, the interaction terms:%
\begin{equation}
W\supset\lambda_{ij}^{u}5_{H}\times10_{M}^{(i)}\times10_{M}^{(j)}%
\end{equation}
are non-zero whenever the curves defined by $\Sigma_{H}$, $\Sigma_{i}$ and
$\Sigma_{j}$ form a triple intersection. \ When none of the $\Sigma_{i}$
self-intersect, or \textquotedblleft pinch\textquotedblright, it follows that
the general form of $\lambda_{ij}^{u}$ is:%
\begin{equation}
\lambda_{ij}^{u}=\left[
\begin{array}
[c]{ccc}%
0 & A & B\\
A & 0 & C\\
B & C & 0
\end{array}
\right]  \label{upyukawa}%
\end{equation}
where $A$, $B$ and $C$ are constants given by evaluating wave function
overlaps. \ We now argue that this matrix cannot yield one generation which is
hierarchically heavier than the first two generations. \ In order for such a
hierarchy to exist, we require that there exists a limit in the parameters
$A$, $B$ and $C$ where two of the masses determined by $\lambda_{ij}^{u}$ tend
to zero while the third mass remains large.

In the limit in which one of the generations has zero mass, the determinant of
the matrix $\lambda_{ij}^{u}$ vanishes:%
\begin{equation}
2ABC=0
\end{equation}
so that without loss of generality, we may assume that the strictly massless
limit corresponds to $A=0$. \ Since the trace of $\lambda^{u}$ is zero, we
conclude that when $A=0$, two of the eigenvalues of $\lambda^{u}$ are equal in
magnitude and have opposite sign. \ This implies that there does not exist a
limit in which two of the generations are parametrically lighter than the
third. \ On the contrary, this would suggest that two of the generations are
significantly heavier than the lightest generation. \ We emphasize that this
result holds independent of how the kinetic terms are normalized. \ This is
because it is always possible to switch to a basis of fields where the kinetic
terms are canonically normalized. \ This alters the form of $\lambda^{u}$ by a
similarity transformation and an overall rescaling. In this new basis, the
determinant and trace will still vanish so that the above argument proceeds as before.

\subsubsection{Self-Intersecting or Pinched Curves\label{SelfIntersect}}

\begin{figure}
[ptb]
\begin{center}
\includegraphics[
height=1.4875in,
width=4.2341in
]%
{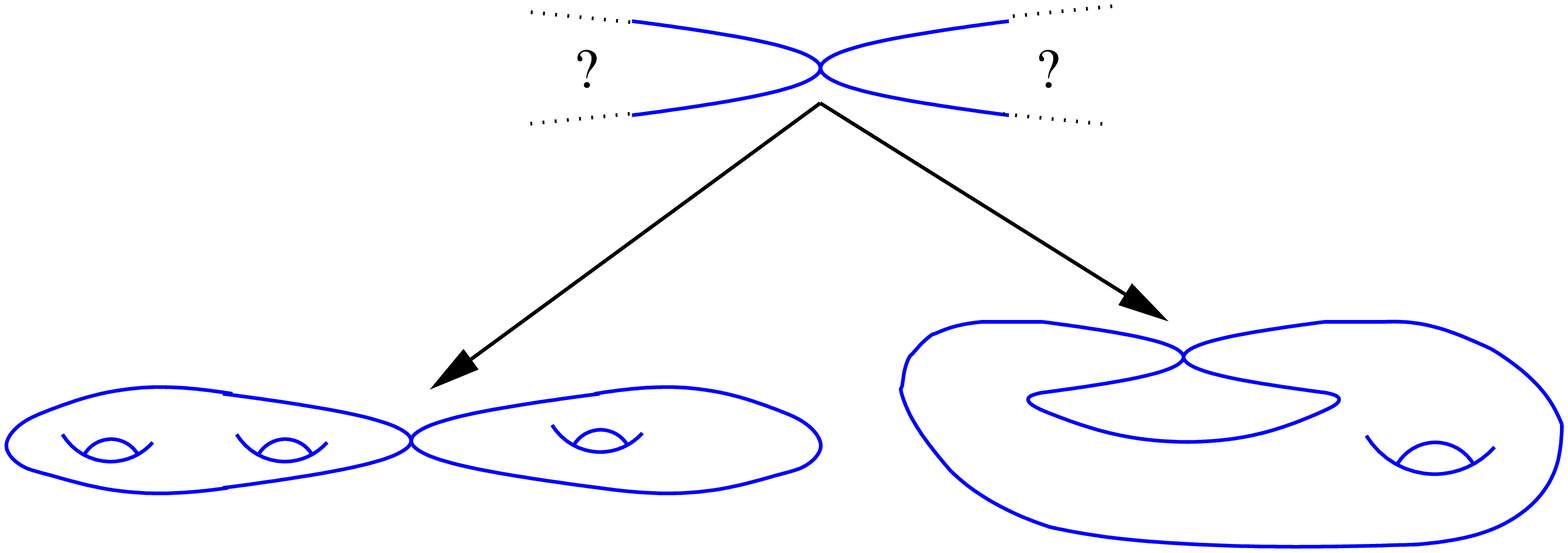}%
\caption{Depiction of how a local enhancement in the singularity type can
enhance to the intersection of two distinct curves (left), or a single curve
which self-intersects (right).}%
\label{selfintersect}%
\end{center}
\end{figure}

Rather than appeal to non-perturbative effects, we note that such a hierarchy
can easily be achieved provided the Yukawa matrix possesses at least one
non-zero diagonal element. \ Geometrically, this requires that one of the
matter curves must pinch off so that globally, the curve intersects itself
inside of $S$. \ We caution that this notion of self-intersection is somewhat
stronger than what is usual meant by self-intersection at the level of
homology. \ At the level of homology, a class is typically said to self
intersect when two distinct representatives of a given homology class
intersect inside of $S$. \ See figure \ref{selfintersect} for a depiction of
how a curve can self-intersect by pinching off inside of $S$.%

We now extend the analysis of \cite{BHV} for smooth matter curves to the
present case of interest where the curve may pinch off, or self-intersect.
Before describing the case of self intersection, let us recall what happens
when two distinct curves intersect. In this case near the generic intersection
point the two curves can be modeled by the equation:%
\begin{equation}
z_{1}z_{2}=0
\end{equation}
where $z_{1}=0$ describes one curve and $z_{2}=0$ denotes the other so that
the intersection point is located at $z_{1}=z_{2}=0$. \ By group theory
considerations explained in \cite{BHV}, it is clear that a third matter curve
will also pass through this point, with a local defining equation $z_{3}%
=z_{1}+z_{2}=0$. This gives rise to a Yukawa interaction of the form:%
\begin{equation}
W\supset\phi_{1}\phi_{2}\phi_{3}%
\end{equation}
where $\phi_{i}$ denotes a field associated with the local vanishing locus
$z_{i}=0$.

From a global perspective, this description does not specify whether $\phi
_{1}$ and $\phi_{2}$ localize on distinct matter curves or whether they
localize on the same curve. \ In the case where these fields localize on the
same curve, the locus $z_{1}=0$ curve must connect to the $z_{2}=0$ curve in a
more global description inside of $S$. \ In other words, these two loci must
form a single Riemann surface. Hence, a self-intersecting curve corresponds to
a genus $g+1$ curve which pinches to a genus $g$ curve in such a way that this
pinching process does not lead to two disconnected surfaces. \ Conversely,
when this pinching process produces two disconnected curves, this describes
the case where the matter curves are distinct.

To analyze the matter content localized on a self-intersecting curve, we note
that the overlap of wave functions at the pinching point determines a single
linear relation amongst the various zero modes of the form:%
\begin{equation}
\alpha_{i}\phi^{(i)}=0
\end{equation}
where the $\phi^{(i)}$ label the zero modes of the genus $g$ curve obtained by
pinching the associated genus $g+1$ curve. \ This identification reduces the
value of the associated index by one.

The number of self-intersection points as well as their proximity will clearly
have an impact on the properties of the Yukawa couplings in the low energy
theory. \ To illustrate this point, it is enough to consider the up type
Yukawa couplings of the minimal GUT\ model which descend from the cubic
interaction term:%
\begin{equation}
W\supset\lambda_{ij}^{u}5_{H}\times10_{M}^{i}\times10_{M}^{j}\text{.}%
\end{equation}
Suppose that three generations in the $10$ of $SU(5)$ all localize on the same
self-intersecting matter curve. \ If there is only one point of
self-intersection which we denote by $0$, the Yukawa matrix is given by the
outer product of the wave function for the three generations:%
\begin{equation}
\lambda_{ij}^{u}=\psi_{H}(0)\psi_{i}(0)\psi_{j}(0)
\end{equation}
so that it automatically has rank one. \ By a suitable change of basis, the
leading order behavior of the up-type Yukawa couplings is given by equation
(\ref{YukawaHierarchy}) as required in a semi-realistic model. \ Additional
points of self-intersection will increase the rank of the up-type Yukawa
coupling matrix. \ In this case, the relative proximity between these points
of intersection as well the analogous expressions for the down-type Yukawa
couplings will control the masses and mixing angles in the quark sector. \ It
would be interesting to determine whether a hierarchical pattern of masses and
mixing angles could emerge from such a treatment.

\subsection{GUT\ Mass Relations\label{GUTMASS}}

In this subsection we show that the usual GUT\ mass relations present in the
simplest four-dimensional GUT\ models can be significantly distorted in the
presence of an internal hypercharge flux. \ In the simplest four-dimensional
GUT models, the masses of the up and down type quarks are determined by the
superpotential terms:%
\begin{equation}
W\supset\lambda_{ij}^{u}5_{H}\times10_{M}^{i}\times10_{M}^{j}+\lambda_{ij}%
^{d}\overline{5}_{H}\times\overline{5}_{M}^{i}\times10_{M}^{j}\text{.}
\label{WSUP}%
\end{equation}
Assuming that the individual components of a GUT multiplet have the same wave
function normalization, this would imply that $m_{q}=m_{l}$ for the quarks and
leptons which unify in a $\overline{5}_{M}$ of $SU(5)$. \ Evolving the values
of the masses observed at low energies up to the GUT scale, it is well-known
that only the third generation obeys a relation of the form $m_{b}\sim
m_{\tau}$. \ At the level of precision we can perform here, the original
analysis of mass relations in the non-supersymmetric $SU(5)$ GUT\ analyzed in
\cite{GeorgiJarlskog} is certainly sufficient. \ \ In this case, the actual
mass relations at the GUT scale are:
\begin{equation}
m_{b}\sim m_{\tau}\text{, }m_{s}\sim m_{\mu}/3\text{, }m_{d}\sim3m_{e}\text{.}
\label{approxrelation}%
\end{equation}
See \cite{RossUpdateMassRelations} for an updated analysis of the various mass
relations obtained by extrapolating the observed values of the masses to the
GUT\ scale. \ This problem is even more pronounced for the simplest $SO(10)$
GUTs where all interaction terms descend from the coupling $16_{M}\times
16_{M}\times10_{H}$. \ Letting $i=1,2,3$, we can parameterize the violation of
the expected mass relation for each generation:%
\begin{equation}
\delta_{i}=\frac{m_{q}^{(i)}-m_{l}^{(i)}}{m_{q}^{(i)}+m_{l}^{(i)}}\text{.}%
\end{equation}
Returning to equation (\ref{approxrelation}), the violation of the simplest
mass relation for each generation is:%
\begin{equation}
\delta_{3}=0\text{, }\delta_{2}\sim50\%\text{, }\delta_{1}\sim-50\%\text{.}%
\end{equation}

In purely four-dimensional GUT\ models, one popular way to rectify the above
problems requires introducing higher-dimensional representations which couple
differently to the individual components of a full GUT\ multiplet. \ It is
also common to introduce adjoint-valued chiral superfields which can couple to
the chiral matter of the MSSM\ through higher dimension operators. \ In both
approaches, the field content necessary to avoid some of the problematic mass
relations of the simplest GUT\ models is unavailable in a del Pezzo model!

In higher dimensional theories, additional mechanisms are potentially
available. \ In compactifications of the heterotic string on Calabi-Yau
manifolds, the particle content can organize into GUT multiplets while the
wave functions corresponding to a given generation may not admit such a simple
interpretation. \ For example, in the presence of a discrete Wilson line which
breaks the GUT\ group to the Standard Model gauge group, individual components
of a GUT\ multiplet may be projected out. \ In this way, some of the usual
mass relations could become ambiguous \cite{WittenSymmBreak}. \ Further,
additional mixing terms between vector-like pairs of massive Kaluza-Klein
modes can also obscure the meaning of simple GUT\ mass relations. \ An example
of this type is discussed in \cite{WittenDecon}. \ Similar ideas are also
quite common in orbifold GUT\ models. \ In a minimal $SU(5)$ GUT model of the
type treated here, one extreme solution would be to invoke the mechanism of
doublet triplet splitting via fluxes described in Section \ref{DTS} so that
individual components of a full GUT\ multiplet could localize on distinct
matter curves.

While this provides one possible way to avoid incorrect mass relations amongst
members of the lighter generations, we find it somewhat anti-thetical to the
whole idea of grand unification that the matter content of the Standard Model
neatly fits into GUT multiplets.\ \ Indeed, it would seem unfortunate to
sacrifice such an aesthetic motivation for grand unification. \ Moreover, the
usual GUT mass relation does work relatively well for the third generation.
\ We now argue that even when a complete GUT multiplet localizes on a matter
curve, the relative normalization of the kinetic terms between different
components of the GUT multiplet will in general be different. Moreover, we
give a qualitative explanation for why the mass relations become increasingly
distorted for the lighter generations.

Recall that in the minimal $SU(5)$ GUT, the net hypercharge flux vanishes on
curves supporting complete GUT multiplets. \ Indeed, the converse of this
condition for the Higgs curves provides a qualitative explanation for why
these fields do not fill out full GUT\ multiplets. \ Although the
\textit{average} hypercharge flux vanishes on chiral matter curves, the field
strength will in general \textit{not vanish pointwise}. \ Because the
individual components of a GUT\ multiplet have different hypercharge, the
corresponding wave functions will couple differently to this background flux
leading to distinct zero mode wave functions. The fact that the zero mode wave
functions are not the same, and may in particular have different magnitudes,
can be interpreted as Aharanov-Bohm interferences in a varying B-field background.

In a minimal $SU(5)$ GUT, all of the interaction terms originate from
evaluating the wave functions at points of triple intersection and now there
is no reason why the magnitude of different matter fields within a GUT
multiplet are the same. This leads to different Yukawa couplings and thus to
different mass relations. In particular, assuming for simplicity no mixing
between generations, we have modified mass relations of the form:%
\begin{equation}
m_{q}=m_{l}\left\vert \frac{\psi_{q}(0)}{\psi_{l}(0)}\right\vert .
\end{equation}
It would be interesting to examine whether modified GUT mass relations for the
lighter generations of the general type proposed in \cite{GeorgiJarlskog}
admit a geometric interpretation.

We now estimate the expected distortion in the usual GUT mass relations due to
the Aharanov-Bohm effect with a varying B-field. \ To this end, let
$F_{\Sigma}$ denote the internal $U(1)$ hypercharge field strength on the
matter curve $\Sigma$. \ The overall scaling dependence of the mass relation
violation $\delta$ can be determined by rescaling the overall volume of
$\Sigma$ by $\varepsilon$. \ Because the reduction of the instanton to
$\Sigma$ scales as $|F_{\Sigma}|^{2}/\varepsilon$, it follows that $F_{\Sigma
}$ rescales by a factor of $\sqrt{\varepsilon}$. \ This reduction is explained
in further detail in \cite{BershadskyVafa2d}.\ \ It now follows that the
violation of the mass relation will be proportional to:%
\begin{equation}
\delta\sim\sqrt{\varepsilon}\text{.}%
\end{equation}
Note that as the volume of $\Sigma$ tends to zero, the amount of violation in
the mass relation also vanishes. \ Equation (\ref{PHYSYUK}) implies that the
masses of fields localized on $\Sigma$ scale as:%
\begin{equation}
M\sim1/Vol(\Sigma)\sim1/\varepsilon\text{,}%
\end{equation}
because in a canonical normalization of all fields, each wave function
contributes a factor of $\psi(0)/\sqrt{M_{\ast}^{2}Vol(\Sigma)}$ to the Yukawa
couplings. \ Hence, the violation of the mass relation obeys the scaling law:%
\begin{equation}
\delta\sim1/\sqrt{M}\text{.}%
\end{equation}
While a mass relation will still hold for each generation, the particular
numerical coefficient relating the masses will depend on the generation in question.

To conclude this Section, we note that a common theme running throughout much
of this paper is the central role of the internal hypercharge flux. \ Indeed,
an intra-generational distortion in the usual GUT\ mass relations
\textit{requires} the presence of an internal hypercharge flux. \ In a sense,
we can view the violation of the GUT mass relation as the first experimental
evidence for the existence of extra dimensions!

\subsection{Generating Semi-Realistic Hierarchies and Mixing
Angles\label{FroggattNielsen}}

In this subsection we speculate on one possible way to achieve semi-realistic
mass hierarchies and mixing angles in the context of our compactification.
\ To frame the discussion to follow, we first review the field theory
Froggatt-Nielsen Mechanism for generating a hierarchical structure in both the
masses and mixing angles of the quark sector. \ As observed in
\cite{FroggattNielsen}, this naturally occurs when the up and down Yukawa
couplings assume the form:%
\begin{equation}
\lambda_{ij}^{u}=g_{ij}^{u}\varepsilon^{a_{i}+b_{j}}\text{, }\lambda_{ij}%
^{d}=g_{ij}^{d}\varepsilon^{a_{i}+c_{j}}\text{,} \label{FNYukawas}%
\end{equation}
where the $g$'s are order one $3\times3$ matrices and $\varepsilon$ is a small
parameter which is related to the Cabbibo angle $\theta_{c}\sim0.2$. \ With
this ansatz, the quark sector exhibits hierarchical masses and mixing angles
determined by appropriate powers of $\varepsilon$ \cite{FroggattNielsen}.

From a field theory perspective, this type of power law suppression naturally
occurs in theories with additional global $U(1)$ symmetries. \ For example, if
the superfields $Q^{i}$, $U^{i}$ and $D^{i}$ have charges $a_{i}$, $b_{i}$ and
$c_{i}$ under a global $U(1)$ symmetry, then the corresponding fields interact
by coupling to an appropriate power of a gauge singlet charged under this
global $U(1)$. \ For example, letting $\phi$ denote a gauge singlet superfield
with charge $+1$ under this global symmetry, the lowest order coupling in the
superpotential is given by:%
\begin{equation}
W\supset g_{ij}^{u}\left(  \frac{\phi}{M_{pl}}\right)  ^{-a_{i}-b_{j}}%
Q^{i}U^{j}H_{u}+g_{ij}^{d}\left(  \frac{\phi}{M_{pl}}\right)  ^{-a_{i}-c_{j}%
}Q^{i}D^{j}H_{d}%
\end{equation}
where for the purposes of this discussion we assume that $H_{u}$ and $H_{d}$
are neutral under the global $U(1)$ symmetry. \ When $\phi$ develops a vev
less than $M_{pl}$, we obtain the expected hierarchy in the Yukawa couplings
of equation (\ref{FNYukawas}).

We now speculate as to how such a hierarchy could potentially occur in
compactifications of F-theory. \ Given a sufficiently generic configuration of
matter curves which form triple intersections, in a holomorphic basis of wave
functions the resulting holomorphic Yukawa couplings introduced in Section
\ref{MASSSCALES} will be given by order one complex numbers. \ To extract the
values of the physical up and down type Yukawa couplings, all of these fields
must be rescaled to a canonical normalization of all kinetic terms. \ In the
large volume limit, this simply rescales each wave function by an appropriate
power of the overall volume factor so that the up and down type Yukawa
couplings are:%
\begin{equation}
\lambda_{ij}^{u}=g_{ij}^{u}Z_{i}^{(10)}Z_{j}^{(10)}Z_{H_{u}}\text{, }%
\lambda_{ij}^{d}=g_{ij}^{d}Z_{i}^{(10)}Z_{j}^{(\overline{5})}Z_{H_{d}}
\label{YukawaZs}%
\end{equation}
where we have introduced the notation $Z=\left(  M_{\ast}^{2}Vol(\Sigma
)\right)  ^{-1/2}$. \ In the above, the superscript on each $Z$ denotes the
representation and as usual, the indices $i$ and $j$ label the generations.
\ In the extreme case where the volumes of the matter curves are hierarchical,
this would provide a crude analogue of the Froggatt-Nielsen mechanism. \ It is
not clear to us, however, that such a hierarchy is always available for
self-intersecting curves. \ Indeed, it is likely that the $Z$'s differ by
order one factors. \ While this is typically enough to sufficiently distort
the usual GUT\ mass relations, it may prove insufficient to produce the large
hierarchy in mass scales between the top quark and the charm quark, for example.

Implicit in the above discussion is the assumption that the $Z$'s of equation
(\ref{YukawaZs}) only depend on the classical volumes of the matter curves.
\ Indeed, as explained in Section \ref{MASSSCALES}, the overall normalization
of each wave function will receive quantum corrections away from the large
volume limit. \ While we do not have a systematic method for computing these
corrections, experience in perturbative string theory strongly suggests that
these corrections are exponentially suppressed as functions of the K\"{a}hler
moduli. \ Moreover, these corrections may induce small off-diagonal terms in
the K\"{a}hler metric for the fields of the required type to generate a
hierarchical structure in the physical Yukawa couplings.

In a similar vein, it is also tempting to speculate that non-perturbative
contributions to the superpotential from Euclidean 3-branes wrapping divisors
in the Calabi-Yau fourfold base could also contribute to a viable model of
flavor physics. \ Indeed, because such corrections will typically violate
global $U(1)$ symmetries present in the low energy effective theory, the
corresponding exponential factor can in principle have a form compatible with
the Froggatt-Nielsen mechanism. \ While these remarks are admittedly
speculative, it would be interesting to see whether there exist calculable
examples of the desired type.

\subsection{Textures From Discrete Symmetries and Large
Diffeomorphisms\label{DISCRETESYMMS}}

Discrete symmetries provide another possible way to achieve semi-realistic
Yukawa couplings and interaction terms because such models can mimic the
primary features of the Froggatt-Nielsen mechanism, but with the global
continuous symmetry replaced by a discrete symmetry. \ In this approach, it is
common to search for finite groups which admit two- and three-dimensional
irreducible representations. \ For example, the two lightest generations could
transform in a two-dimensional representation while the heaviest generation
could transform as a singlet. \ As one application, these symmetries are
typically enough to alleviate potential problems with FCNCs in gravity
mediation scenarios.\footnote{We thank K.S. Babu for emphasizing this point to
us.} \ A list of candidate discrete flavor groups with order at most thirty
one which are of phenomenological interest has been tabulated in
\cite{FramptonKephartScan}. \ Some common choices in the model building
literature are the symmetric group on three or four letters denoted by $S_{3}$
and $S_{4}$ as well as $A_{4}$, the alternating subgroup of $S_{4}$. \ See
\cite{Ma2004Review,Ma2007Review} for a recent review of some possibilities
along these lines. \ In the present context, the group of large
diffeomorphisms of a del Pezzo surface provide a potentially attractive
starting point for a theory of flavor based on discrete symmetries. \ We note
that some version of this gauged symmetry will survive even away from the
large volume regime. \ It is therefore possible that such symmetries could
undergird a theory of flavor.

The group of large diffeomorphisms for the del Pezzo surfaces has a natural
action on the matter curves of the del Pezzo which automatically lifts to a
group action on the matter fields of the MSSM. \ For example, the del Pezzo 3
surface corresponds to the exceptional group $E_{3}=SU(3)\times SU(2)$ which
has Weyl group $S_{3}\times S_{2}$. \ The $S_{3}$ factor could potentially
play the role of the desired flavor group.

One potential caveat to the above proposal is that the action of the Weyl
group on the matter curves corresponds to an integral representation. \ In
other words, the corresponding characters take values in the integers. \ This
follows from the fact that the Weyl group naturally permutes the exceptional
curves of the del Pezzo surface. \ In particular, because the entries in the
character tables for the phenomenologically most interesting representations
of $A_{4}$ and $S_{4}$ are given by various powers of a third root of unity,
this direct application of discrete symmetries may be too trivial.

We note that no similar obstruction is present in the case of the discrete
group $S_{3}$. \ Indeed, consider as a toy model the case where the three
generations have localized on the exceptional curves $E_{1}$, $E_{2}$ and
$E_{3}$ of the del Pezzo 3 surface. \ In this case, the $S_{3}$ Weyl group
permutes the exceptional curves. \ The three dimensional representation
spanned by the three curves also determines how $S_{3}$ acts on the three
generations. \ This three dimensional representation decomposes to the sum of
a two dimensional representation and singlet which are respectively spanned
by:%
\begin{equation}
\left\langle E_{1},E_{2},E_{3}\right\rangle \simeq\left\langle E_{1}%
-E_{2},E_{2}-E_{3}\right\rangle _{\text{doublet}}\oplus\left\langle
E_{1}+E_{2}+E_{3}\right\rangle _{\text{singlet}}\text{.}%
\end{equation}
This suggests that the wave function for the heavy generation transforms as
the singlet, while the two light generations transform as the doublet. \ It
would be interesting to develop such a theory of flavor in more detail.

\section{Suppression Factors From Singlet Wave Functions\label{Suppression}}

So far we have only considered contributions to the superpotential from matter
fields which all transform as non-trivial representations of $G_{std}$. \ A
fully realistic model will most likely contain contributions to the effective
superpotential from chiral superfields which transform as gauge singlets under
$G_{std}$. \ For example, the $\mu$ term could originate from a cubic
interaction term between the Higgs fields and a gauge singlet. \ The vev of
this singlet would then set the size of $\mu$. \ As another example, we note
that because neutrino oscillations are now well-established, the
superpotential must contain terms of the form $LN_{R}H_{u}$ where $N_{R}$
denotes the right-handed neutrino superfields which transform as gauge singlets.

Generating appropriately small neutrino masses as well as a value for the
$\mu$ term near the scale of electroweak symmetry breaking has historically
been a challenge in string-based models. Some discussion on neutrino masses in
string theory may be found for example in \cite{GiedtLangackerNeutrinos}. \ In
type II D-brane constructions, contributions to the superpotential from
wrapped Euclidean branes can produce an appropriately large Majorana mass term
for right-handed neutrinos \cite{CveticBlumenhagenInstantons,IbanezUrangaMajorana}.
Similar effects may also generate exponentially suppressed $\mu$ terms
\cite{CveticBlumenhagenInstantons}. More recently, it has also been shown that
D-brane instantons can also potentially generate suppressed Dirac neutrino
masses \cite{CveticDiracNeutrinos}. In this Section, we show that the Yukawa
couplings which involve a singlet of $G_{S}$ can in suitable circumstances be
exponentially suppressed relative to the Yukawa couplings which only involve
fields charged under $G_{S}$.

The rest of this Section is organized as follows. \ In subsection
\ref{ATTREPEL}, we study the behavior of gauge singlet wave functions which
contribute to the low energy superpotential. \ After performing this analysis,
in subsection \ref{YUKESTIMATE}, we estimate the overall normalization of the
Yukawa couplings for such gauge singlet wave functions. \ For interaction
terms involving three singlets, there is a natural volume suppression effect.
\ For gauge singlets which are attracted to the GUT\ model seven-brane, the
wave function behaves as if it had localized on a matter curve inside of $S$.
\ For gauge singlets which are repelled away from the GUT model seven-brane,
we find that the Yukawa couplings are naturally suppressed. \ In the remaining
subsections we show that these effects can naturally generate both
hierarchically small $\mu$ terms and neutrino masses. \ In both cases, we find
that order one parameters in the high energy theory naturally can yield values
which are in rough agreement with observation.

\subsection{Wave Function Attraction and Repulsion\label{ATTREPEL}}

To setup notation, we consider three seven-branes which wrap surfaces $S$,
$S^{\prime}$, and $S^{\prime\prime}$ inside the compactification threefold
$B_{3}$ and which carry respective gauge groups $G_{S}$, $G_{S^{\prime}}$, and
$G_{S^{\prime\prime}}$. By assumption, $S$, $S^{\prime}$, and $S^{\prime
\prime}$ intersect transversely along smooth curves
\begin{equation}
\Sigma_{X}\,=\,S\cap S^{\prime}\,,\qquad\qquad\Sigma_{Y}=\,S\cap
S^{\prime\prime}\,,\qquad\qquad\Sigma_{\perp}\,=\,S^{\prime}\cap
S^{\prime\prime}\,, \label{MCURVES}%
\end{equation}
which give rise to corresponding chiral superfields $X$, $Y$, and $\Phi$ in
four dimensions. Each superfield transforms as a bifundamental\footnote{See
\S $4.2$ of \cite{BHV} for a description of the generalized notion of
\textquotedblleft bifundamental\textquotedblright\ relevant for intersecting
seven-branes in F-theory.} under the respective products ${G_{S}\times
G_{S^{\prime}}}$, ${G_{S}\times G_{S^{\prime\prime}}}$, and ${G_{S^{\prime}%
}\times G_{S^{\prime\prime}}}$. Finally, if the curves $\Sigma_{X}$,
$\Sigma_{Y}$, and $\Sigma_{\perp}$ themselves intersect transversely at a
single point, the low-energy effective superpotential contains a cubic
coupling of the form
\begin{equation}
W_{\perp}\,=\,\lambda\,\Phi XY\,, \label{WPERP}%
\end{equation}
invariant under ${G_{S}\times G_{S^{\prime}}\times G_{S^{\prime\prime}}}$.

By assumption, the kinetic terms for $X$, $Y$, and $\Phi$ have the canonical
normalization in four dimensions, so the dimensionless coupling $\lambda$ in
$W_{\perp}$ depends upon the $L^{2}$-norms of the associated zero-mode
wavefunctions on the curves in \eqref{MCURVES}. Since both $\Sigma_{X}$ and
$\Sigma_{Y}$ are compact curves inside $S$, the norms of wavefunctions for $X$
and $Y$ merely scale with the volumes of the curves in $S$. However, unlike
$\Sigma_{X}$ and $\Sigma_{Y}$, the curve $\Sigma_{\perp}$ is not embedded in
$S$ but rather intersects $S$ transversely at a point in $B_{3}$. From the
perspective of the four-dimensional effective theory, this distinction in
geometry is reflected by the fact that $\Phi$ transforms as a singlet under
$G_{S}$, whereas $X$ and $Y$ form a vector-like pair. We are interested in the
limit that $S$ contracts inside $B_{3}$, or equivalently, in the limit that
the volume of $\Sigma_{\perp}$ goes to infinity. In the limit that
$\Sigma_{\perp}$ becomes non-compact, we clearly need to be careful in our
estimate for the norm of the wavefunction $\psi$ associated to the singlet
$\Phi$.

We are ultimately interested in the behavior of $\psi$ near the point where
$\Sigma_{\perp}$ intersects $S$, so let us introduce local holomorphic and
anti-holomorphic coordinates $(z,\overline z)$ on $\Sigma_{\perp}$ such that
${z=0}$ is the location of the intersection with $S$. As we reviewed in
Section \ref{BasicSetup}, $\psi$ generally transforms on $\Sigma_{\perp}$ as a
holomorphic section of the bundle ${K_{\Sigma_{\perp}}^{1/2} \otimes L}$,
\begin{equation}
\psi\,\in\,H^{0}_{\overline\partial}\big(\Sigma_{\perp}, K_{\Sigma_{\perp}%
}^{1/2} \otimes L\big)\,,\qquad\qquad L \,=\, L^{\prime}\big|_{\Sigma_{\perp}%
}\otimes L^{\prime\prime}\big|_{\Sigma_{\perp}}\,,
\end{equation}
where $L^{\prime}$ and $L^{\prime\prime}$ are line bundles on $S^{\prime}$ and
$S^{\prime\prime}$. Because $\psi$ is holomorphic, $\psi$ satisfies
\begin{equation}
\label{HOLPSI}\overline\partial^{\dagger}\overline\partial\psi\,=\, 0\,,
\end{equation}
where $\overline\partial$ is the Dolbeault operator acting on ${K_{\Sigma
_{\perp}}^{1/2} \otimes L}$, and $\overline\partial^{\dagger}$ is the adjoint
operator defined with respect to the induced metric on $\Sigma_{\perp}$ and
the hermitian metric on $L$ inherited from $L^{\prime}$ and $L^{\prime\prime}$.

Besides the Dolbeault operator $\overline\partial$, the bundle ${K_{\Sigma
_{\perp}}^{1/2} \otimes L}$ also carries a unitary connection which defines a
covariant derivative $\nabla$ and an associated Laplacian ${\triangle=
\nabla^{\dagger}\nabla}$. By a standard Hodge identity reviewed in Appendix E
of \cite{BHV}, the Laplacian $\triangle$ is related to the operator
$\overline\partial^{\dagger}\overline\partial$ via
\begin{equation}
\label{WEITZENBOCK}\triangle\,=\, 2 \, \overline\partial^{\dagger}%
\overline\partial\,-\, {\frac{1}{2}} \mathcal{R} \,+\, \mathcal{F}\,.
\end{equation}
Here $\mathcal{R}$ is the scalar curvature of the metric on $\Sigma_{\perp}$,
and $\mathcal{F}$ is the scalar curvature of the unitary connection on $L$.

The positive constants in \eqref{WEITZENBOCK} will not be important for the
following analysis, but the signs will be essential. First, the relative sign
between $\mathcal{R}$ and $\mathcal{F}$ in \eqref{WEITZENBOCK} arises because
$\mathcal{R}$ is the scalar curvature of the induced metric on $\Sigma_{\perp
}$ and hence is the curvature of a connection on the holomorphic tangent
bundle ${T\Sigma_{\perp}\cong K_{\Sigma_{\perp}}^{-1}}$, as opposed to a
connection on the spin\footnote{The factor `$1/2$' multiplying $\mathcal{R}$
in \eqref{WEITZENBOCK} arises from the square-root in the spin bundle
$K_{\Sigma_{\perp}}^{1/2}$.} bundle $K_{\Sigma_{\perp}}^{1/2}$. To fix the
overall sign multiplying $\mathcal{R}$, we note that the Laplacian $\triangle$
is a positive-definite hermitian operator. On the other hand, because $\psi$
is holomorphic,
\begin{equation}
\triangle\psi=\left(  -{\frac{1}{2}}\mathcal{R}\,+\,\mathcal{F}\right)
\!\psi\,. \label{VAN}%
\end{equation}
According to \eqref{VAN}, if ${\mathcal{F}=0}$ and ${\mathcal{R}>0}$ is
strictly positive, then $\psi$ must vanish. Such a vanishing is consistent
with the fact that ${K_{\Sigma_{\perp}}^{1/2}=\mathcal{O}(-1)}$ admits no
holomorphic sections on ${\Sigma_{\perp}={\mathbb{P}}^{1}}$, and this
observation fixes the sign of ${\mathcal{R}}$ in the Hodge identity \eqref{WEITZENBOCK}.

In a local unitary frame, the Laplacian $\triangle$ takes the standard
Euclidean form ${\triangle= -4 \, \partial^{2} / \partial z \, \partial
\overline z}$, and \eqref{VAN} reduces to the wave equation
\begin{equation}
4 \, \frac{\partial^{2}\psi}{\partial z \partial\overline z} \,+\, \left(
\mathcal{F} \,-\, {\frac{1}{2}} \mathcal{R}\right)  \psi\,=\, 0\,.
\end{equation}
Thus if $\psi$ is normalized so that ${\psi(0)=1}$, then $\psi$ behaves near
${z=0}$ as
\begin{align}
\label{LOCPSI}\psi(z,\overline z) \,  &  =\, \exp{\!\left(  -\frac{1}{4} \,
m^{2}_{0} \, |z|^{2}\right)  } \,+\, \cdots\,,\cr m^{2}_{0} \, & =\, \left[
\mathcal{F} \,-\, {\frac{1}{2}} \mathcal{R}\right]  _{z=0}\,,
\end{align}
where the `$\cdots$' indicate terms in $\psi$ that vanish at ${z=0}$, and the
curvatures which define $m^{2}_{0}$ are evaluated at that point. In general,
$m^{2}_{0}$ can be either\footnote{Holomorphy of $\psi$ implies that the total
curvature satisfies ${\int_{\Sigma_{\perp}}\!\!\! \star(\mathcal{F} \,-\,
{\frac{1}{2}} \mathcal{R}) \ge0}$, but the sign of $\mathcal{F} \,-\,
{\frac{1}{2}} \mathcal{R}$ may vary from point to point on $\Sigma_{\perp}$.}
negative or positive, and the sign of $m^{2}_{0}$ determines whether $\psi$
exponentially grows or decays away from the origin.

At first glance, one might be perplexed as to how such exponential behavior in
$\psi$ can arise, since nothing so far really distinguishes the point ${z=0}$.
In fact, given that $\psi$ is written in a unitary frame, the behavior in
\eqref{LOCPSI} merely reflects the behavior of the metric on ${K_{\Sigma
_{\perp}}^{1/2} \otimes L}$.

As a very concrete example, let us take $\Sigma_{\perp}$ to be ${\mathbb{P}%
}^{1}$, with a metric which we parameterize in Liouville form as
\begin{equation}
ds^{2}\,=\,\mathrm{e}^{2\,\phi(z,\overline{z})}\,dz\,d\overline{z}\,.
\label{LIOUVILLE}%
\end{equation}
For instance, if the metric on ${\mathbb{P}}^{1}$ is round with constant
curvature $\Lambda^{2}$, then
\begin{equation}
\phi(z,\overline{z})\,=\,-\ln{\!\left(  1\,+\,\frac{1}{4}\,\Lambda^{2}%
|z|^{2}\right)  }\,. \label{FUBINI}%
\end{equation}
The role of the particular line bundle ${K_{\Sigma_{\perp}}^{1/2}\otimes L}$
is inessential, so for simplicity we just take $\psi$ to transform in the
holomorphic tangent bundle $T{\mathbb{P}}^{1}$. As is well-known, holomorphic
tangent vectors on ${\mathbb{P}}^{1}$ take the global form
\begin{equation}
u(z)\,\frac{\partial}{\partial z}\,,\qquad\qquad u(z)\,=\,a_{0}\,+\,a_{1}%
z\,+\,a_{2}z^{2}\,,
\end{equation}
where $(a_{0},a_{1},a_{2})$ are complex parameters. However, if $\phi
(z,\overline{z})$ in \eqref{LIOUVILLE} varies non-trivially over ${\mathbb{P}%
}^{1}$, the holomorphic vector ${\partial/\partial z}$ does not have constant
length. To describe $\psi$ in a unitary frame, we instead introduce a new
basis vector $\hat{e}$ for $T{\mathbb{P}}^{1}$,
\begin{equation}
\hat{e}\,=\,{\frac{1}{2}}\,\mathrm{e}^{-\phi(z,\overline{z})}\,\frac{\partial
}{\partial z}\,.
\end{equation}
Though $\hat{e}$ is not holomorphic, $\hat{e}$ does have constant, unit length
in the metric \eqref{LIOUVILLE}. In the frame described by $\hat{e}$, a
holomorphic tangent vector $\psi$ therefore takes the form
\begin{equation}
\psi\,=\,\mathrm{e}^{\phi(z,\overline{z})}\,u(z)\,\hat{e}. \label{UNTPSI}%
\end{equation}
Because the scalar curvature of the metric in \eqref{LIOUVILLE} is given in
terms of $\phi$ as
\begin{equation}
\mathcal{R}\,=\,-4\,\mathrm{e}^{-2\,\phi}\,\frac{\partial^{2}\phi}{\partial
z\,\partial\overline{z}}\,, \label{SCALARR}%
\end{equation}
the behavior near ${z=0}$ of $\psi$ in \eqref{UNTPSI} is controlled by the
local curvature.\footnote{Because of the conventions adopted, $\mathcal{R}$ in
\eqref{SCALARR} plays the role of $\mathcal{F}$ in \eqref{LOCPSI}.}

To make use of \eqref{LOCPSI}, we must still estimate $m^{2}_{0}$ at the point
where $\Sigma_{\perp}$ intersects the surface $S$. Since $m^{2}_{0}$ receives
contributions from both $\mathcal{R}$ and $\mathcal{F}$, we consider each
contribution in turn.

To estimate $\mathcal{R}$, we recall that $S$ is a del Pezzo surface shrinking
to zero size inside the elliptic Calabi-Yau fourfold $\mathcal{X}$. As a
result, the scalar curvature on $S$ is large and positive, of order
$M_{GUT}^{2}$. On the other hand, since $\mathcal{X}$ is Calabi-Yau, the total
scalar curvature on $\mathcal{X}$ vanishes. Because the elliptic fiber of
$\mathcal{X}$ is generically non-degenerate, with negligible curvature, the
large positive curvature of $S$ near its point of intersection with
$\Sigma_{\perp}$ must be locally cancelled by a corresponding negative
curvature on $\Sigma_{\perp}$ itself. The scalar curvature $\mathcal{R}$ on
$\Sigma_{\perp}$ near ${z=0}$ is thus negative and of order
\begin{equation}
\mathcal{R}\,\sim\,-M_{GUT}^{2}\,.
\end{equation}
We note that if $\Sigma_{\perp}$ has genus zero or one, then $\mathcal{R}$
must become positive elsewhere on $\Sigma_{\perp}$ as dictated by the Euler characteristic.

We apply a similar argument to estimate the curvature $\mathcal{F}$ on $L$
near ${z=0}$. By definition, the line bundle $L$ is a tensor product
${L^{\prime}\big|_{\Sigma_{\perp}} \otimes L^{\prime\prime}\big|_{\Sigma
_{\perp}}}$ of line bundles $L^{\prime}$ and $L^{\prime\prime}$ on respective
surfaces $S^{\prime}$ and $S^{\prime\prime}$, and both $L^{\prime}$ and
$L^{\prime\prime}$ carry anti-self-dual connections. The following
observations are symmetric between $L^{\prime}$ and $L^{\prime\prime}$, but
for concreteness let us focus on the bundle $L^{\prime}$ over $S^{\prime}$.

The surface $S^{\prime}$ contains two curves ${\Sigma_{X}= S \cap S^{\prime}}$
and ${\Sigma_{\perp}= S^{\prime\prime}\cap S^{\prime}}$ which intersect
transversely at the point ${z=0}$ on $\Sigma_{\perp}$. Since $S$ is shrinking
inside $\mathcal{X}$, the curve $\Sigma_{X}$ is similarly shrinking inside the
surface $S^{\prime}$. In this situation, an anti-self-dual connection on
$L^{\prime}$ over $S^{\prime}$ must restrict to a solution of the
two-dimensional Yang-Mills equations on the shrinking curve $\Sigma_{X}$.
Hence the curvature of $L^{\prime}$ on $\Sigma_{X}$ must be constant and
uniform, of order ${d \Vol(\Sigma_{X})^{-1} \sim d \, M_{GUT}^{2}}$, where $d$
is the degree of $L^{\prime}$ on $\Sigma_{X}$.

Without loss, we assume that the metric on $S^{\prime}$ at the intersection of
$\Sigma_{X}$ and $\Sigma_{\perp}$ takes the diagonal form ${ds^{2} = dz
d\overline z \,+\, dw d\overline w}$, where $w$ is a local holomorphic
coordinate on $\Sigma_{X}$ and $z$ is a local holomorphic coordinate on
$\Sigma_{\perp}$. Because the curvature of the connection on $L^{\prime}$ is
anti-self-dual, the curvature at ${z=0}$ along $\Sigma_{\perp}$ must be
opposite to the curvature along $\Sigma_{X}$. Hence the curvature of
$L^{\prime}$ on $\Sigma_{\perp}$ is of order ${-d \, M_{GUT}^{2}}$.

Including a similar contribution from $L^{\prime\prime}$, we find
\begin{equation}
\mathcal{F} \,\sim\, -\left[  \frac{\deg(L^{\prime}\big|_{\Sigma_{X}}%
)}{\Vol(\Sigma_{X})} \,+\, \frac{\deg(L^{\prime\prime}\big|_{\Sigma_{Y}}%
)}{\Vol(\Sigma_{Y})}\right]  \,\sim\, \pm M_{GUT}^{2}\,.
\end{equation}
Both $\mathcal{R}$ and $\mathcal{F}$ are of roughly the same magnitude, but
whereas the sign of $\mathcal{R}$ is fixed, the sign of $\mathcal{F}$
generally depends upon the degrees of $L^{\prime}$ and $L^{\prime\prime}$ as
well as the relative volumes of the matter curves $\Sigma_{X}$ and $\Sigma
_{Y}$ in $S$. We see no particular reason why the contributions to
$\mathcal{F}$ from $\Sigma_{X}$ and $\Sigma_{Y}$ should be correlated in
either sign or absolute value. So depending upon the choices for $L^{\prime}$
and $L^{\prime\prime}$, the parameter ${m_{0}^{2}=\mathcal{F}\,-\,{\frac{1}%
{2}}\mathcal{R}}$ can be either positive or negative, of order $M_{GUT}^{2}$.

We are left to estimate the norm of the singlet wavefunction $\psi$. Now, the
great virtue of writing $\psi$ in a unitary frame is that the $L^{2}$-norm of
$\psi$ is given directly by
\begin{equation}
\label{LTWON}||\psi||^{2} \,=\, M^{2}_{*} \, \int_{\Sigma_{\perp}}
\!\!\!\omega\,|\psi|^{2}\,,\qquad\omega\,=\, \frac{i}{2} \, \mathrm{e}^{2
\phi(z,\overline z)} \, dz\wedge d\overline z\,,\qquad|\psi|^{2}
\equiv\overline\psi\psi\,.
\end{equation}
Here $\omega$ is the K\"ahler form for the induced metric on $\Sigma_{\perp}$,
which for concreteness we parameterize in the Liouville form
\eqref{LIOUVILLE}. According to \eqref{LOCPSI} and \eqref{SCALARR}, the
integrand of \eqref{LTWON} then behaves to leading order near ${z=0}$ as
\begin{align}
\label{LTWONII}\mathrm{e}^{2 \phi(z, \overline z)} \, |\psi|^{2} \,  &
\approx\, \exp{\!\left[  -{\frac{1}{2}} \left(  m_{0} \,+\, \mathcal{R}%
\right)  |z|^{2}\right]  }\,,\cr & =\, \exp{\!\left[  -{\frac{1}{2}} \left(
\mathcal{F} \,+\, {\frac{1}{2}} \mathcal{R}\right)  |z|^{2}\right]  }\,.
\end{align}

If the combination ${\mathcal{F}\,+\, {\frac{1}{2}} \mathcal{R}}$ is positive
at ${z=0}$, the integral over $\Sigma_{\perp}$ in \eqref{LTWON} has rapid
Gaussian decay at the scale $M_{GUT}$, so immediately
\begin{equation}
\label{NOCUTNORM}||\psi||^{2} \,\sim\, \frac{M_{*}^{2}}{M_{GUT}^{2}}%
\,,\qquad\qquad\qquad\left[  \mathcal{F} \,+\, {\frac{1}{2}} \mathcal{R}%
\right]  _{z=0}\!\!> 0\,.
\end{equation}
In this case the normal wave function is \textit{attracted} to our brane.

Conversely, if ${\mathcal{F}\,+\,{\frac{1}{2}}\mathcal{R}}$ is negative at
${z=0}$, the expression in \eqref{LTWONII} rapidly blows up away from the
origin. In this case the normal wave function is \textit{repelled} from our
brane. To make sense of $||\psi||^{2}$, we impose a cutoff in the integral
over $\Sigma_{\perp}$ at a scale ${|z|\sim R_{\perp}}$. As we discuss briefly
below, we expect the Gaussian approximation in \eqref{LTWONII} to be valid up
to the cutoff, so we estimate $||\psi||^{2}$ as
\begin{equation}
\left\langle \psi|\psi\right\rangle =||\psi||^{2}\,\sim\,\frac{M_{\ast}^{2}%
}{M_{GUT}^{2}}\,\exp{\!\left(  c\,M_{GUT}^{2}R_{\perp}^{2}\right)  }%
\,,\qquad\qquad\left[  \mathcal{F}\,+\,{\frac{1}{2}}\mathcal{R}\right]
_{z=0}\!\!<0\,. \label{CUTNORM}%
\end{equation}
In this estimate, ${c>0}$ is an order one constant which our analysis does not
fix, though the expression in \eqref{CUTNORM} depends sensitively upon its
value. Similarly, the estimate depends upon our choice of $R_{\perp}$, which
roughly encodes the behavior of the metric on $B_{3}$ away from $S$. We recall
that $R_{\perp}$ is parameterized as
\begin{equation}
R_{\perp}=M_{GUT}^{-1}\,\varepsilon^{-\gamma}\,,\qquad\qquad\varepsilon
\,=\,\frac{M_{GUT}}{\alpha_{GUT}M_{pl}}\,,
\end{equation}
where $\gamma$ typically lies in the range ${1/3<\gamma<1}$.

In making the estimate \eqref{CUTNORM} for $||\psi||^{2}$, we assume that the
curvature of the Calabi-Yau metric on $\mathcal{X}$ (and similarly the
connection on $L$) is slowly varying and of order $M_{GUT}^{2}$ in a region of
size $R_{\perp}$ away from $S$. This behavior of the Calabi-Yau metric on
$\mathcal{X}$ is suggested by similar behavior of the local Calabi-Yau metric
on the cotangent bundle $T^{*}{\mathbb{C}} {\mathbb{P}}^{1}$, as exhibited for
instance in \S $3$ of \cite{OoguriVF}. In the case of $T^{*}{\mathbb{C}%
}{\mathbb{P}}^{1}$, the scalar curvature $\mathcal{R}$ along the cotangent
fiber experiences only a slow, power-law decay away from ${\mathbb{C}%
}{\mathbb{P}}^{1}$, and we roughly expect the same behavior normal to $S$ in
$\mathcal{X}$. However, a more precise estimate of $||\psi||^{2}$ clearly
demands a more detailed analysis of the local Calabi-Yau metric on
$\mathcal{X}$.

\subsection{Estimating Yukawa Couplings\label{YUKESTIMATE}}

Having estimated the local behavior of gauge singlet wave functions near the
del Pezzo surface, we now determine the corresponding values of the Yukawa
couplings in the low energy theory. \ With notation as above, to estimate the
size of the Yukawa coupling in equation (\ref{WPERP}), we introduce the wave
function $x$ (resp. $y$) for the chiral superfield $X$ (resp. $Y$) which
localizes on the matter curve $\Sigma_{X}$ (resp. $\Sigma_{Y}$) in $S$. \ The
superpotential term of equation (\ref{WPERP}) due to a triple overlap between
$\Sigma_{X}$, $\Sigma_{X}$, $\Sigma_{\bot}$ at a point $p$ is:
\begin{align}
W_{\bot}  &  =\lambda\Phi XY\\
&  =\frac{x(p)}{\sqrt{M_{\ast}^{2}Vol(\Sigma_{X})}}\frac{y(p)}{\sqrt{M_{\ast
}^{2}Vol(\Sigma_{Y})}}\frac{\psi(p)}{\sqrt{\left\langle \psi|\psi\right\rangle
}}\Phi XY
\end{align}
where in the above, we have adopted the physical normalization of Yukawa
couplings detailed in Section \ref{MASSSCALES}. \ The value of the Yukawa
coupling strongly depends on whether the del Pezzo surface attracts or repels
the gauge singlet wave function from the point of intersection. \ By contrast,
we note that because $X$ and $Y$ localize on matter curves inside of $S$, the
values of $x(p)$ and $y(p)$ are order one numbers. \ Making the rough
approximation $M_{\ast}^{2}Vol(\Sigma)\sim\alpha_{GUT}^{-1/2}$, the resulting
Yukawa coupling is:%
\begin{equation}
\lambda=\alpha_{GUT}^{1/2}\frac{\psi(p)}{\sqrt{\left\langle \psi
|\psi\right\rangle }}\text{.} \label{Yukawaestimate}%
\end{equation}

We now estimate the value of the Yukawa coupling depending on whether the
GUT\ model seven-brane attracts or repels the gauge singlet wave function.
\ To this end, we shall frequently refer back to the estimates of the various
length scales obtained in Section \ref{MASSSCALES}. \ In the repulsive case,
equation (\ref{CUTNORM}) now implies:%
\begin{align}
\lambda_{\text{repel}}  &  \sim\alpha_{GUT}^{1/2}\times\left(  \alpha
_{GUT}^{1/4}\frac{R_{S}}{R_{\bot}}\exp\left(  -\frac{c}{\varepsilon^{2\gamma}%
}\right)  \right) \label{repel}\\
&  =\alpha_{GUT}^{3/4}\times\varepsilon^{\gamma}\exp\left(  -\frac
{c}{\varepsilon^{2\gamma}}\right)
\end{align}
where the second equality follows from equation (\ref{RPERP}) and as in the
previous subsection, $c$ is a positive order one number.

By contrast, in the undamped case described by equation (\ref{NOCUTNORM}), the
associated Yukawa coupling is:%
\begin{equation}
\lambda_{\text{attract}}\sim\alpha_{GUT}^{1/2}\frac{M_{GUT}}{M_{\ast}}%
\sim\alpha_{GUT}^{3/4}\text{.} \label{attract}%
\end{equation}
Physically, the value of $\lambda_{\text{attract}}$ agrees with the intuition
that in the attractive case, all details of the compactification decouple
because the gauge singlet behaves as though it localizes on a curve in $S$.
\ In general, we see that:%
\begin{equation}
\left\vert \lambda_{\text{attract}}\right\vert \gg\left\vert \lambda
_{\text{repel}}\right\vert \text{.}%
\end{equation}

In addition to interaction terms between matter fields inside of $S$ and a
single gauge singlet, it is also possible for three gauge singlet wave
functions to interact outside of $S$. \ When one such gauge singlet develops a
non-zero vev, the resulting interaction term will determine the mass of the
remaining gauge singlets. \ Letting $\psi_{i}$ denote gauge singlet wave
functions for $i=1,2,3$, the value of the physical Yukawa coupling from wave
function overlap at a point $b$ outside of $S$ is now given by:%
\begin{align}
\lambda_{\text{singlet}}  &  \sim\frac{\psi_{1}(b)}{\sqrt{M_{\ast}%
^{2}Vol(\Sigma_{1})}}\frac{\psi_{2}(b)}{\sqrt{M_{\ast}^{2}Vol(\Sigma_{2})}%
}\frac{\psi_{3}(b)}{\sqrt{M_{\ast}^{2}Vol(\Sigma_{3})}}\sim\frac{1}{\left(
M_{\ast}R_{\bot}\right)  ^{3}}\label{singlet}\\
&  \sim\alpha_{GUT}^{3/4}\left(  \frac{R_{S}}{R_{\bot}}\right)  ^{3}%
=\alpha_{GUT}^{3/4}\times\varepsilon^{3\gamma}\text{.} \label{secsinglet}%
\end{align}
We note that in comparison to Yukawa couplings on $S$ which are on the order
of $\alpha_{GUT}^{3/4}$, this naturally yields an overall suppression factor
by a non-trivial power of $\varepsilon$.

\subsection{$\mu$ Term}

We now discuss a natural mechanism for obtaining small supersymmetric $\mu$
terms. \ For concreteness, suppose that the bulk gauge group $G_{S}=SU(5)$ and
that the $H_{u}$ and $H_{d}$ fields localize on distinct matter curves where
the singularity type enhances to $SU(6)$. \ In the case where these curves do
not intersect, the $\mu$ term is automatically zero. \ In the case where they
do intersect, the matter fields will interact with a gauge singlet which
localizes on a curve normal to $S$. \ Letting $\Phi$ denote the chiral
superfield for this gauge singlet, the superpotential now contains the
interaction term:%
\begin{equation}
W_{\mu}\supset\lambda\Phi H_{u}H_{d}\sim\alpha_{GUT}^{1/2}\frac{\psi(p)}%
{\sqrt{\left\langle \psi|\psi\right\rangle }}\Phi H_{u}H_{d}%
\end{equation}
with notation as in equation (\ref{Yukawaestimate}). \ When $\Phi$ develops a
vev, the superpotential will contain a $\mu$ term for the Higgs up and Higgs
down fields. \ The value of this vev is controlled by the dynamics orthogonal
to $S$ and therefore scales as:%
\begin{equation}
\left\langle \Phi\right\rangle \sim\frac{1}{R_{\bot}}\sim M_{GUT}%
\times\varepsilon^{\gamma}\text{.}%
\end{equation}
Returning to equations (\ref{repel}) and (\ref{attract}), it thus follows that
in the attractive case, the resulting value of $\mu$ is far above the
electroweak scale, and would lift the Higgs doublets from the low energy
spectrum. \ On the other hand, in the exponentially damped case, the value of
the $\mu$ term is:%
\begin{equation}
\mu=\lambda_{\text{repel}}\left\langle \Phi\right\rangle \sim\alpha
_{GUT}^{3/4}\times\varepsilon^{2\gamma}\exp\left(  -\frac{c}{\varepsilon
^{2\gamma}}\right)  \text{.} \label{muterm}%
\end{equation}
This leads to a large hierarchy between the $\mu$ term and the GUT\ scale.
\ For example, with $\gamma=1$ and $c=1/7$ we find $\mu\sim140$ GeV. \ In
Section \ref{NUMEROLOGY} we present some additional estimates of $\mu$.

\subsection{Neutrino Masses}

At a conceptual level, the $\mu$ term and Dirac mass terms for the neutrinos
both originate from interactions between two fields on curves in $S$ and a
third field which localizes on a curve normal to $S$. \ Indeed, in the
previous subsection we found that when the gauge singlet wave function is
exponentially suppressed near $S$, the $\mu$ term is hierarchically suppressed
below the GUT scale. \ We now estimate the values of the light neutrino masses
of the MSSM depending on the profile of the right-handed neutrino wave
function near the surface $S$. \ When the gauge singlet is attracted to $S$, a
variant on the usual seesaw mechanism yields neutrino masses which are
approximately correct. \ On the other hand, when the gauge singlet is repelled
away from $S$, the value of the Dirac masses is already quite low, and the
seesaw mechanism would yield unviable neutrino masses. \ In fact, the Dirac
mass terms are already in a viable range so that in this case the neutrinos
are purely of Dirac type.

For simplicity, we perform our estimates for a single neutrino species,
because as explained in Section \ref{TowardsYukawas}, a detailed model of
flavor is currently beyond our reach. \ In this case, the neutrino sector of
the superpotential is:%

\begin{equation}
W_{\nu}=\lambda_{D}LN_{R}H_{u}+\lambda_{\text{singlet}}\Theta N_{R}N_{R}
\label{neutrinointeraction}%
\end{equation}
where $N_{R}$ denotes the right-handed neutrino chiral superfield, and
$\Theta$ is another gauge singlet. \ In certain cases, the second interaction
term may not be present. \ In the following we analyze the interplay between
the behavior of the right-handed neutrino wave functions near $S$ and this
second interaction term.

\subsubsection{Majorana Masses and a Seesaw}

We now consider the case where the second interaction term $\Theta N_{R}N_{R}$
does not vanish and show that a phenomenologically viable scenario requires
that the right-handed neutrino wave function is attracted to $S$. \ When
$\Theta$ develops a vev, it induces a Majorana mass term for the right-handed
neutrinos. \ Using the value of $\lambda_{\text{singlet}}$ given by equation
(\ref{secsinglet}), this yields the Majorana mass:%
\begin{equation}
m_{M}\equiv\lambda_{\text{singlet}}\left\langle \Theta\right\rangle
=\frac{\lambda_{\text{singlet}}}{R_{\bot}}=\alpha_{GUT}^{3/4}M_{GUT}%
\times\varepsilon^{4\gamma}\sim3\times10^{12\pm1.5}\text{ GeV.}%
\end{equation}

The value of the Dirac masses strongly depends on the profile of the gauge
singlet wave function near $S$. \ By inspection of equations (\ref{repel}) and
(\ref{attract}), the value of $\lambda_{\text{attract}}$ will induce a Dirac
mass term for neutrinos which is around the electroweak scale, while the value
of $\lambda_{\text{repel}}$ will induce a far smaller Dirac mass term. \ The
mass matrix for the neutrinos is:%
\begin{equation}
M_{\nu}=\left[
\begin{array}
[c]{cc}%
0 & \frac{1}{2}m_{D}\\
\frac{1}{2}m_{D} & m_{M}%
\end{array}
\right]  \sim\alpha_{GUT}^{3/4}\left[
\begin{array}
[c]{cc}%
0 & \left\langle H_{u}\right\rangle \\
\left\langle H_{u}\right\rangle  & M_{GUT}\times\varepsilon^{4\gamma}%
\end{array}
\right]  \text{.}%
\end{equation}

Because the Majorana mass term is non-zero, it is much larger than the Dirac
mass terms so that the smaller eigenvalue of $M_{\nu}$ is given by the usual
seesaw mechanism:%
\begin{equation}
m_{\text{light}}\sim\frac{m_{D}^{2}}{m_{M}}\text{.}%
\end{equation}
Due to the fact that the Majorana mass term is in the usual range expected for
a seesaw mechanism, $m_{D}$ must be on the order of the electroweak scale in
order to yield a viable light neutrino mass. \ Restricting to this case,
$m_{\text{light}}$ is now given by:%
\begin{equation}
m_{\text{light}}\sim\left(  \alpha_{GUT}^{3/4}\times\varepsilon^{-4\gamma
}\right)  \times\frac{\left\langle H_{u}\right\rangle ^{2}}{M_{GUT}}%
\sim2\times10^{-1\pm1.5}\text{ eV.} \label{light}%
\end{equation}
We note that in this case, we automatically find an enhancement over the naive
seesaw value $\left\langle H_{u}\right\rangle ^{2}/M_{GUT}$! \ Indeed, in the
GUT\ literature it is often necessary to lower the Majorana mass term below
$M_{GUT}$ to obtain more realistic neutrino masses.

\subsubsection{Suppressed Dirac Masses}

Next consider the possibility that the interaction term between $\Theta$ and
$N_{R}$ in equation (\ref{neutrinointeraction}) does not exist so that the
neutrinos are purely of Dirac type. \ In the previous subsection we found that
a variant of the standard seesaw mechanism requires that the right-handed
neutrino wave function is attracted towards $S$. \ Indeed, the Dirac mass
terms for the undamped wave functions were automatically on the order of the
electroweak scale. \ In the absence of a seesaw mechanism, this profile for
the wave functions would yield an unacceptably large value for the neutrino
masses. \ On the other hand, the wave functions which are repelled away from
$S$ will naturally generate much smaller Dirac neutrino mass terms.

Restricting to the repulsive case, the Dirac mass term is:%
\begin{equation}
m_{\text{Dirac}}=\lambda_{\text{repel}}\left\langle H_{u}\right\rangle
\sim\left\langle H_{u}\right\rangle \times\left[  \alpha_{GUT}^{3/4}%
\times\varepsilon^{\gamma}\exp\left(  -\frac{c}{\varepsilon^{2\gamma}}\right)
\right]  \text{.} \label{DiracMass}%
\end{equation}
The essential point of the above formula is that the Dirac mass can be quite
light, and for an appropriate order one value of $c$, yields a
phenomenologically viable mass for the light neutrinos. \ For example, setting
$c=5$ and $\gamma=1/3$ yields $m_{\text{Dirac}}\sim6\times10^{-3}$ eV. Before
closing this subsection, we note that while large Majorana mass terms which
violate lepton number are typically invoked as a primary cause of leptogenesis
in early universe cosmology, there do exist viable alternative scenarios which
only require Dirac neutrino masses. See \cite{DiracLeptoReview} and references
therein for a recent account of Dirac leptogenesis.

\subsection{Relating $\mu$ and $\nu$\label{MODINDEPPREDICTION}}

In the previous subsection we presented a general formula which naturally
generates an exponentially suppressed value for the masses of purely Dirac
type neutrinos. \ Indeed, the exponential damping terms for both the $\mu$
term of equation (\ref{muterm}) and the Dirac mass term of equation
(\ref{DiracMass}) are both sensitive to an order one parameter which we denote
by $c$. \ We now present a relation between $\mu$ and $m_{\text{Dirac}}$ in
which the overall dependence on this exponential factor cancels out. \ This
expression is model independent in the sense that it does not depend as
strongly on the details of the exponential suppression factor.

The exponential suppression factors of the $\mu$ term and the purely Dirac
mass term both originate from a gauge singlet wave function which is repelled
away from the surface $S$ so that:%
\begin{align}
m_{\text{Dirac}}  &  =\lambda_{\text{repel}}(c)\left\langle H_{u}\right\rangle
\\
\mu &  =\lambda_{\text{repel}}(c^{\prime})\left\langle \Phi\right\rangle
\end{align}
where $\left\langle \Phi\right\rangle $ denotes the vev of a gauge singlet
which localizes on a matter curve normal to $S$. \ In the above, we have
allowed two potentially different suppression factors such that $c$ and
$c^{\prime}$ may differ by some small amount.

Making the simplifying assumption $c=c^{\prime}$, all exponential effects
cancel, and we obtain the rough estimate:%
\begin{equation}
m_{\text{Dirac}}=\mu\frac{\left\langle H_{u}\right\rangle }{\left\langle
\Phi\right\rangle }=\frac{\mu\varepsilon^{-\gamma}}{\langle H_{u}\rangle
}\times\frac{\langle H_{u}\rangle^{2}}{M_{GUT}}\sim5\times10^{-3\pm0.5}\text{
eV} \label{MODELINDEP}%
\end{equation}
for $\mu\sim100$ GeV. \ Of course, for small mismatches between the parameters
$c$ and $c^{\prime}$, slightly higher (or lower) values are also in principle possible.

\section{Supersymmetry Breaking\label{SUSYBREAK}}

Up to now, our analysis has assumed that the four-dimensional effective theory
preserves $\mathcal{N}=1$ supersymmetry. \ See
\cite{IbanezSUSYFTHEORY,BuchbinderSUSY} for recent discussions of
supersymmetry breaking in F-theory and \cite{FloratosKokorelis} for an
explicit realization of gauge mediated supersymmetry breaking in an
intersecting D-brane model. \ In this Section we briefly sketch how
supersymmetry breaking can be communicated to the MSSM in a gauge mediation
scenario. \ Further details will appear in \cite{SUSYBREAK}. \ A more general
framework which interpolates between gauge mediation and gravity mediation is
given in \cite{SvrcekKachruMediation}. \ In that context, supersymmetry
breaking takes place on a seven-brane distinct from a GUT\ model seven-brane.
\ When these branes intersect, supersymmetry breaking is communicated via
gauge mediation. \ As the separation between the seven-branes increases, this
interpolates to a gravity mediation scenario. \ In the present case, most of
our seven-branes form non-trivial topological intersections which cannot
disappear. \ While we shall present some brief speculations on generating
hierarchically small values for the scale of supersymmetry breaking, a
complete analysis would entail a broader discussion which is beyond the scope
of this paper.

To frame the discussion to follow, we now briefly sketch the basic features of
gauge mediated supersymmetry breaking. See \cite{GiudiceSUSYReview} for a
review of gauge mediation. \ In general, most mediation mechanisms consist of
three sectors. \ These are given by the sector of the theory which breaks
supersymmetry, the sector of communication, and the MSSM\ itself. \ Although
we do not specify how supersymmetry can be broken, we can still parameterize
this breaking in terms of at least one chiral superfield $X$ which develops a
supersymmetry breaking vev:%
\begin{equation}
\left\langle X\right\rangle =x+\theta^{2}F\text{.} \label{SUSYVEV}%
\end{equation}

To specify the messenger sector, we introduce vector-like pairs of
GUT\ multiplets which will communicate supersymmetry breaking to the MSSM.
\ As an explicit example, we take $Y$ to transform in the fundamental of
$SU(5)$ and $Y^{\prime}$ in the anti-fundamental. \ These fields can then
localize on matter curves inside of $S$. The messengers couple to $X$ via an
interaction term of the form:%
\begin{equation}
W_{4d}\supset W_{mess}=\lambda XYY^{^{\prime}}\text{.} \label{messenger}%
\end{equation}
Once $X$ develops a vev of the type given by equation (\ref{SUSYVEV}), the
messengers will get a mass:%
\begin{equation}
M_{mess}=\lambda x\text{.}%
\end{equation}
Supersymmetry breaking then communicates to the MSSM\ because the messenger
fields interact with the gauge bosons of the MSSM. \ In this setup, the soft
masses for the gauginos are generated at one loop order while the soft scalar
masses are generated at two loop order. \ One attractive feature of the gauge
mediation scenario is that FCNCs are automatically suppressed.

Although precise numerical estimates are beyond the scope of the present
paper, to simply get a sense of the mass scales involved, recall that in gauge
mediation, the masses of the gauginos are:%
\begin{equation}
m_{i}\sim\frac{\alpha_{i}(M_{weak})}{4\pi}\frac{F}{x}\text{.}%
\end{equation}
We note that this estimate does not require any knowledge of the overall
normalization factors appearing in equation (\ref{messenger}). \ The lightest
gaugino in this case is the Bino which in viable models has a mass of
$\sim100$ GeV. \ Plugging in the properly normalized value of the hypercharge
coupling at the weak scale given by $\alpha_{1}(M_{weak})\sim(5/3)\times
(1/128)\sim10^{-2}$, we see that the scale of supersymmetry breaking $\sqrt
{F}$ and the messenger scale $x$ are related via:%
\begin{equation}
\sqrt{F}\sim300\text{ GeV}^{1/2}\sqrt{x}\text{.}%
\end{equation}

Depending on the origin of the $X$ field in the F-theory GUT\ model, the
resulting messenger mass scales can potentially be quite different. \ In the
following subsections we discuss three natural candidates for $X$ in the
present class of compactifications. \ The field $X$ can correspond to a bulk
gauge boson on a transversely intersecting seven-brane, or a field which
localizes on a matter curve orthogonal to $S$. \ In the latter case, there are
two further refinements depending on whether the GUT model seven-brane
attracts or repels the corresponding gauge singlet wave function.

\subsection{Bulk Gauge Boson Breaking}

When the matter fields $Y$ and $Y^{^{\prime}}$ localize on the same curve,
these fields will automatically couple to the bulk gauge fields of a
seven-brane which transversely intersects the GUT model seven-brane. \ In this
case, we can interpret $x$ as the supersymmetric vev of the bulk gauge field.
\ The value of $x$ depends on the volume of the matter curve containing the
messenger fields as well as the remaining bulk worldvolume of the other
seven-brane. \ Using the basic scaling relations obtained in Section
\ref{MASSSCALES}, we estimate $\langle X\rangle\sim1/R_{\bot}$ so that the
resulting messenger mass is:
\begin{align}
M_{mess}  &  =\alpha_{GUT}M_{GUT}\varepsilon^{2\gamma}\\
&  \sim1\times10^{15\pm0.5}\text{ GeV.}%
\end{align}

\subsection{Gauge Singlet Breaking}

It is also possible that $X$ could correspond to a gauge singlet which
localizes on a matter curve which intersects $S$ at a point. \ In this case,
much of the analysis performed in Section \ref{Suppression} carries over.
\ For example when the gauge singlet wave function for $X$ is attracted
towards the seven-brane, it couples to the messenger fields with the same
strength as a field inside of $S$. \ In this case, the messenger mass is on
the order of:%
\begin{equation}
M_{mess}=\frac{\lambda_{\text{attract}}}{R_{\bot}}=\alpha_{GUT}^{3/4}%
M_{GUT}\times\varepsilon^{\gamma}\sim5\times10^{14\pm0.5}\text{ GeV.}%
\end{equation}

On the other hand, the seven-brane can also repel the gauge singlet wave
function. \ In this case, the messenger mass scale can be hierarchically much
lighter than the GUT\ scale due to the exponential suppression factor present
at the point of intersection with the seven-brane. \ In this case, the
resulting messenger mass is given by a similar expression to that derived for
the $\mu$ term in equation (\ref{muterm}):%
\begin{equation}
M_{mess}=\frac{\lambda_{\text{repel}}}{R_{\bot}}\sim M_{GUT}\times\alpha
_{GUT}^{3/4}\varepsilon^{2\gamma}\exp\left(  -\frac{c}{\varepsilon^{2\gamma}%
}\right)  \text{.}%
\end{equation}
In this case, the messenger mass scale can potentially range over many
candidate values. \ For example, we obtain a value of $\sim10^{12}$ GeV when
$c=1$ and $\gamma=1/3$, and a value of $\sim300$ TeV when $c=1/10$ and
$\gamma=1$.

\subsection{Soft Breaking Boundary Conditions}

A well-known difficulty with the gauge mediation scenario is that it is
typically difficult to simultaneously generate the correct values for the
$\mu$ and $B\mu$ terms. \ In the present context, we note that the $\mu$ term
is naturally light and on the order of the electroweak scale. \ Indeed, this
setup decouples the issue of supersymmetry breaking from the $\mu$ problem.
\ In fact, at the GUT\ scale, the $B\mu$ term is zero at high energies, and is
instead radiatively generated. \ Phenomenological fits to this range of
parameter space favor larger values for $\tan\beta= \langle H_{u} \rangle/
\langle H_{d} \rangle$ \cite{RattazziSaridLargeTanBeta}.

We also expect that higher order terms in the superpotential of the form:
\begin{equation}
W_{quart}=\frac{c_{ijk}}{M_{KK}}X\Lambda_{i}\Lambda_{j}\Lambda_{k}%
\end{equation}
where $\Lambda_{i}$ denotes a generic field of the MSSM cannot be generated by
integrating out Kaluza-Klein modes. \ As explained in Section \ref{ExtraU1},
this is due to the fact that such terms will typically violate a global
$U(1)~$symmetry in the low energy theory. \ Indeed, matter fields in F-theory
are always charged under additional $U(1)$ factors of precisely this type.
Letting $\sigma_{i}$ denote the bosonic component of the chiral superfield
$\Lambda_{i}$, this suggests that the values of the soft breaking $A$-terms in
the effective potential:%
\begin{equation}
V_{eff}=A_{ijk}\sigma_{i}\sigma_{j}\sigma_{k}%
\end{equation}
will automatically vanish at the scale set by $x$. \ Because both the $B\mu$
and $A$ terms vanish, there is a common rephasing symmetry of the fields which
naturally avoids additional CP\ violating phases.

\subsection{Speculations on Supersymmetry Breaking}

To conclude this Section, we now briefly speculate on ways in which
supersymmetry breaking can take place in the various scenarios outlined above.
\ First consider the case where $X$ is identified with a bulk gauge field on a
seven-brane which intersects the GUT seven-brane. \ Returning to the equations
of motion for fields on the transversely intersecting seven-brane $S^{\prime}$
derived in \cite{BHV}, the value of $F$ is:%
\begin{equation}
F^{\ast}=\overline{\partial}^{\prime}\phi^{\prime}+\delta_{\Sigma}\left\langle
\left\langle Y^{c},Y\right\rangle \right\rangle _{ad(P)}+...
\end{equation}
where $\phi^{\prime}$ denotes the holomorphic $(2,0)$ form for this brane and
the $...$ denotes contributions to the F-term localized on other matter curves
in the surface $S^{\prime}$. \ When the righthand side of the above equation
is non-zero, this will break supersymmetry. \ This can easily occur when the
background value of the G-flux in the Calabi-Yau fourfold is incompatible with
the complex structure on $S^{\prime}$. \ Because this difference can be quite
small in principle, we can obtain small values for $F$ in this case.

Next consider scenarios where the $X$ field corresponds to a gauge singlet
localized on a matter curve intersecting $S$ at a point. \ While we have
primarily focussed on the behavior of this wave function in supersymmetric
backgrounds, presumably a similar analysis will also carry through in a
non-supersymmetric background. \ In this vein, it may be possible to extend
the discussion of Section \ref{Suppression} to this more general case. \ It
would be interesting to see whether a suitable hierarchy in the scale of
supersymmetry could be arranged in this way.

\section{$SU(5)$ Model\label{SU5MODEL}}

Having presented a number of potential model building ingredients in the
previous Sections, we now proceed to some semi-realistic examples of models
based on a del Pezzo 8 surface which incorporates at least some of these
ideas. \ Our expectation is that significant refinements are possible in the
actual examples we present. \ As explained in previous sections, the
GUT\ group directly breaks to $G_{std}$ via an internal hypercharge flux.
\ Moreover, to avoid exotic matter representations, the available internal
fluxes are in one to one correspondence with the roots of an exceptional Lie
algebra. \ In this case, all of the matter content of the MSSM\ must localize
on curves in $S$. \ The fields in the $5$ or $\overline{5}$ localize on curves
where the bulk $SU(5)$ singularity enhances to $SU(6)$, while fields in the
$10$ and $\overline{10}$ localize on curves where $SU(5)$ enhances to $SO(10)$.

As explained in \cite{BHV}, the interaction terms $\overline{5}_{H}%
\times\overline{5}_{M}\times10_{M}$ originate from points where the bulk
singularity $G_{S}=SU(5)$ undergoes a twofold enhancement in rank to an
$SO(12)$ singularity. \ Similarly, the interaction terms $5_{M}\times
10_{M}\times10_{M}$ originate from a twofold enhancement in rank to an $E_{6}$
singularity. \ As in \cite{BHV}, we deduce the local behavior of the matter
curves near such points by decomposing the adjoint representations of $SO(12)$
and $E_{6}$ to the product $SU(5)\times U(1)\times U(1)$:%
\begin{align}
SO(12)  &  \supset SU(5)\times U(1)_{1}\times U(1)_{2}\\
66  &  \rightarrow1_{0,0}+1_{0,0}+24_{0,0}\\
&  +5_{2,2}+5_{-2,2}+\overline{5}_{-2,2}+\overline{5}_{-2,-2}+10_{4,0}%
+\overline{10}_{-4,0}\\
E_{6}  &  \supset SU(5)\times U(1)_{a}\times U(1)_{b}\\
78  &  \rightarrow1_{0,0}+1_{0,0}+1_{-5,-3}+1_{5,3}+24_{0,0}\\
&  +5_{-3,3}+\overline{5}_{3,-3}+10_{-1,-3}+\overline{10}_{1,3}+10_{4,0}%
+\overline{10}_{-4,0}\text{.}%
\end{align}
Consider first the fields associated with the Cartan of $SO(12)$. \ Labeling
the local Cartan generators as $t_{1}$, $t_{2}$, we conclude that a
six-dimensional field in the $5$ localizes on the matter curve $(t_{1}%
+t_{2}=0)$ and another field in the $5$ localizes along $(t_{1}-t_{2}=0)$,
while a $10$ localizes on the matter curve $(t_{1}=0)$. \ Similar
considerations apply for $E_{6}$, from which we conclude that a
six-dimensional field in the $5$ localizes on the matter curve $(-t_{a}%
+t_{b}=0)$, while distinct six-dimensional $10$'s localize on the matter
curves $(t_{a}+3t_{b}=0)$ and $(t_{a}=0)$. \ The gauge singlets of $SU(5)$
localize on curves which only intersect $S$ at a discrete set of points.\ To
generate naturally suppressed $\mu$ terms and light Dirac masses for the
neutrinos, we also consider local enhancements to $SU(7)$.

For illustrative purposes, we first present an example which we shall refer to
as \textquotedblleft Model I\textquotedblright\ which exhibits the correct
matter spectrum of the MSSM\ at low energies, but which also contains
unrealistic interaction terms. \ Indeed, in this model the third generation is
not hierarchically heavier than the two lighter generations. \ Moreover, the
neutrinos of the Standard Model are exactly massless. \ Finally, the model
contains superpotential terms which lead to rapid proton decay. \ After
explaining the primary features of this model, we next present a more refined
example of admissible matter curves which rectifies all of the above issues.

As a first example, consider a model with fractional line bundle
$L=\mathcal{O}_{S}(E_{3}-E_{4})^{1/5}$ and matter content localized on the
following choice of matter curves:%
\begin{equation}%
\begin{tabular}
[c]{|l|l|l|l|l|l|}\hline
$\text{Model I}$ & $\text{Curve}$ & $\text{Class}$ & $g_{\Sigma}$ &
$L_{\Sigma}$ & $L_{\Sigma}^{\prime n}$\\\hline
$1\times\left(  5_{H}+\overline{5}_{H}\right)  $ & $\Sigma_{H}$ & $-K_{S}$ &
$1$ & $\mathcal{O}_{\Sigma_{H}}(p_{1}-p_{2})^{1/5}$ & $\mathcal{O}_{\Sigma
_{H}}(p_{1}-p_{2})^{-3/5}$\\\hline
$3\times\overline{5}_{M}$ & $\Sigma_{M}^{(1)}$ & $E_{1}$ & $0$ &
$\mathcal{O}_{\Sigma_{M}^{(1)}}$ & $\mathcal{O}_{\Sigma_{M}^{(1)}}%
(-3)$\\\hline
$2\times10_{M}$ & $\Sigma_{M}^{(2)}$ & $H-E_{1}-E_{2}$ & $0$ & $\mathcal{O}%
_{\Sigma_{M}^{(2)}}$ & $\mathcal{O}_{\Sigma_{M}^{(2)}}(2)$\\\hline
$1\times10_{M}$ & $\Sigma_{M}^{(3)}$ & $E_{2}$ & $0$ & $\mathcal{O}%
_{\Sigma_{M}^{(3)}}$ & $\mathcal{O}_{\Sigma_{M}^{(3)}}(1)$\\\hline
\end{tabular}
\label{SU5Example}%
\end{equation}
where $p_{1}$ and $p_{2}$ denote two divisors on $\Sigma_{H}$ which are not
linearly equivalent and we have indicated how $L$ restricts on each matter
curve as well as the gauge bundle content of each GUT\ multiplet due to the
restriction of the line bundle $L^{\prime}$ on $S^{\prime}$ to the various
matter curves. \ By construction, we find that a vector-like pair of Higgs
doublets localizes on $\Sigma_{H}$. \ The degree of the line bundles on each
of the chiral matter curves has been chosen to reproduce the correct
multiplicity in the MSSM.

In terms of $SU(5)$ GUT\ multiplets, the schematic form of the superpotential
is:%
\begin{equation}
W_{SU(5)}=\lambda_{ij}^{d} \cdot \overline{5}_{H}\times\overline{5}_{M}^{(i)}%
\times10_{M}^{(j)}+\lambda_{j}^{u} \cdot 5_{H}\times10_{M}\times10_{M}^{(j)}%
\end{equation}
where $i=1,2,3$ labels the three generations of $\overline{5}_{M}$ all
localized on a single matter curve and $j=1,2$ labels the two generations of
$10_{M}$ localized on the matter curve $\Sigma_{M}^{(2)}$. More generally, the superpotential may also contain interactions which involve gauge singlets which take the schematic form $1 \times 5 \times \overline{5}$. Such interactions can then lead to a $\mu$ term for the Higgs and a Dirac mass term for the neutrinos.

As the above example demonstrates, there are potentially many admissible local
models of this type which can all yield the matter content of the MSSM.
\ Although this model possesses non-trivial interaction terms, it is unclear
whether these terms are consistent with constraints from low energy physics.
\ As argued in subsection \ref{TextureConsequence}, when no curves
self-intersect or pinch inside of $S$, the corresponding Yukawa couplings do
not produce the correct hierarchy in quark masses. \ Moreover, as explained in
Section \ref{DTS}, because $H_{u}$ and $H_{d}$ localize on the same matter
curve, lifting the Higgs triplets via fluxes can still induce quartic terms in
the superpotential of the form $QQQL/M_{KK}$ with order one coefficients.
\ Finally, in addition to an incorrect hierarchy for the quarks, the neutrinos
are exactly massless in this model.

\begin{figure}
[ptb]
\begin{center}
\includegraphics[
height=3.1004in,
width=4.4036in
]%
{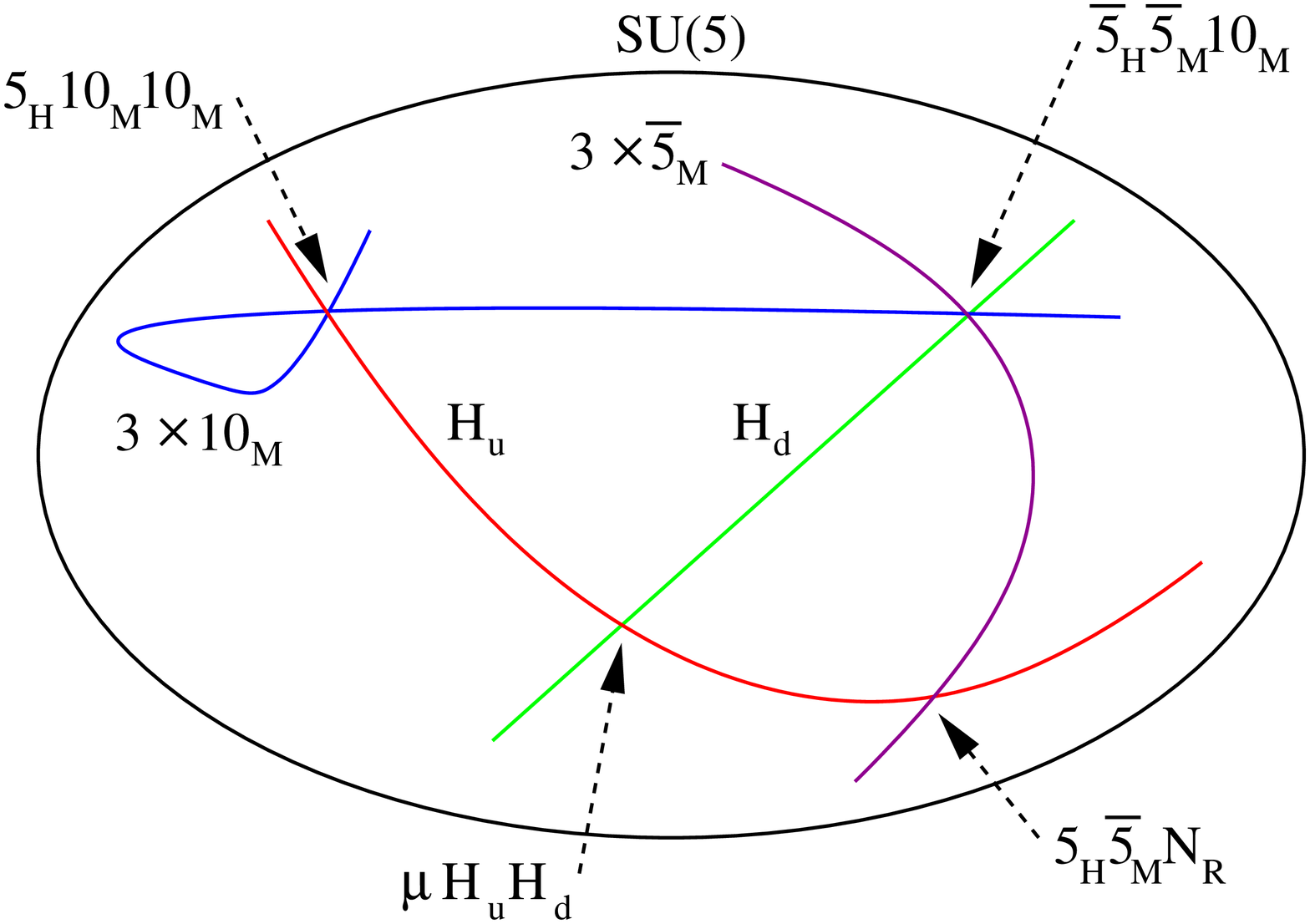}%
\caption{Depiction of the various matter curves in the $SU(5)$ model referred
to as \textquotedblleft Model II\textquotedblright. \ In this case, all three
generations in the $10$ of $SU(5)$ localize on one curve and three generations
in the $\overline{5}$ localize on another curve. \ The Higgs up and down
curves localize on distinct matter curves and intersect at a point in $S$.
\ The contributions to the superpotential from the intersection of various
matter curves is also indicated.}%
\label{su5curves}%
\end{center}
\end{figure}

We now present a different configuration of matter curves which resolves all
of the problems mentioned above. \ To this end, we require that at least one
generation localize on a self-intersecting curve. \ For simplicity, we place
all three generations of $10_{M}$'s on a self-intersecting $\mathbb{P}^{1}$
and all three generations of $\overline{5}_{M}$'s on a smooth $\mathbb{P}^{1}$ which does not
self-intersect. \ With the same choice of $L=\mathcal{O}_{S}(E_{3}%
-E_{4})^{1/5}$ as in the previous example, the matter content, line bundle
assignments and effective class of each matter curve are:%
\begin{equation}%
\begin{tabular}
[c]{|l|l|l|l|l|l|}\hline
$\text{Model II}$ & $\text{Curve}$ & $\text{Class}$ & $g_{\Sigma}$ &
$L_{\Sigma}$ & $L_{\Sigma}^{\prime n}$\\\hline
$1\times5_{H}$ & $\Sigma_{H}^{(u)}$ & $H-E_{1}-E_{3}$ & $0$ & $\mathcal{O}%
_{\Sigma_{H}^{(u)}}(1)^{1/5}$ & $\mathcal{O}_{\Sigma_{H}^{(u)}}(1)^{2/5}%
$\\\hline
$1\times\overline{5}_{H}$ & $\Sigma_{H}^{(d)}$ & $H-E_{2}-E_{4}$ & $0$ &
$\mathcal{O}_{\Sigma_{H}^{(d)}}(-1)^{1/5}$ & $\mathcal{O}_{\Sigma_{H}^{(d)}%
}(-1)^{2/5}$\\\hline
$3\times10_{M}$ & $\Sigma_{M}^{(1)}\text{ (pinched)}$ & $2H-E_{1}-E_{5}$ & $0$
& $\mathcal{O}_{\Sigma_{M}^{(1)}}$ & $\mathcal{O}_{\Sigma_{M}^{(1)}}%
(3)$\\\hline
$3\times\overline{5}_{M}$ & $\Sigma_{M}^{(2)}$ & $H$ & $0$ & $\mathcal{O}%
_{\Sigma_{M}^{(2)}}$ & $\mathcal{O}_{\Sigma_{M}^{(2)}}(3)$\\\hline
\end{tabular}
\ \text{.}%
\end{equation}
See figure \ref{su5curves} for a depiction of the various matter curves in
this model. \ In computing the multiplicities on the self-intersecting curve
we have neglected all subtleties which could occur based on viewing this curve
as a pinched genus one curve because the flux data from the non-compact brane
is a free discrete parameter which we can always tune to give the correct
number of generations.%
The superpotential now takes the form:%
\begin{align}
W_{SU(5)}  &  =\lambda_{ij}^{d} \cdot \overline{5}_{H}\times\overline{5}_{M}^{(i)}
\times10_{M}^{(j)}+\lambda_{ij}^{u} \cdot 5_{H}\times10_{M}^{(i)}\times10_{M}
^{(j)}\\
&  +\rho_{\text{repel}}^{ia} \cdot 5_{H}\times\overline{5}_{M}^{(i)}\times
N_{R}^{(a)}+\lambda_{\text{repel}} \cdot \Phi\times5_{H}\times\overline{5}_{H}%
\end{align}

where in the above, the intersection between $\Sigma_{H}^{(u)}$ and
$\Sigma_{M}^{(2)}$ leads to a two-fold enhancement in rank to an $SU(7)$
singularity so that the singlet $N_{R}^{(a)}$ may be identified with the
right-handed neutrinos and the vev of $\Phi$ determines the supersymmetric
$\mu$ term. \ In this model, the neutrino masses are purely of Dirac type.
\ As explained in Section \ref{Suppression}, these gauge singlet wave
functions can generate an exponential suppression of the expected type.
\ Finally, as explained in greater detail in Section \ref{DTS}, because the
$H_{u}$ and $H_{d}$ fields localize on distinct matter curves, the operator
$QQQL$ is automatically suppressed by a phenomenologically acceptable amount.

\section{Evading the No Go Theorem and Flipped Models\label{FLIPPED}}

In the previous sections we have presented many potential ingredients for
building models based on $G_{S}=SU(5)$. \ This is partially due to the
analysis of subsection \ref{NOGO} which shows that for $G_{S}=SO(10)$, direct
breaking to $G_{std}$ via internal fluxes will always as a byproduct generate
exotic matter fields. \ For surfaces of general type, a partial breaking to
$SU(5)\times U(1)$ would not present a serious obstruction because after
breaking to a four-dimensional GUT group, the remaining breaking can proceed
when an adjoint-valued field develops a suitable vev. \ For del Pezzo models,
a similar mechanism exists for flipped GUT models.

We now recall the primary features of four-dimensional flipped $SU(5)$
GUT\ models \cite{BarrFlipped,DerendingerFlipped,HagelinEllisFlipped}. \ The
gauge group of flipped $SU(5)$ is $SU(5)\times U(1)$, which naturally embeds
in $SO(10)$. \ Indeed, the chiral matter content of the Standard Model is
given by the flipped $SU(5)$ multiplets:%
\begin{align}
\text{Matter}  &  :3\times(1_{-5}+\overline{5}_{3}+10_{-1})\\
\text{MSSM Higgs}  &  :1\times(5_{2}+\overline{5}_{-2})\\
\text{GUT Higgs}  &  :1\times(10_{-1}+\overline{10}_{1})
\end{align}
where $U(1)_{Y}$ of the MSSM\ corresponds to a linear combination of the
$U(1)$ generator in $SU(5)$ and the overall $U(1)$ factor. \ Due to the fact
that the $U(1)$ hypercharge is given by a flipped embedding, the $5_{2}$ contains
the Higgs down of the MSSM, while the $\overline{5}_{-2}$ contains the Higgs
up. \ In addition to interaction terms which descend from the $16_{M}%
\times16_{M}\times10_{H}$ in an $SO(10)$ GUT, a flipped $SU(5)$ model includes
the interaction terms $5_{2}\times10_{-1}\times10_{-1}$ and $\overline{5}%
_{-2}\times\overline{10}_{1}\times\overline{10}_{1}$ between the MSSM\ and
GUT\ Higgs fields. \ These interaction terms descend from $16_{h}\times
16_{h}\times10_{H}$ and $\overline{16}_{h}\times\overline{16}_{h}\times10_{H}$
in an $SO(10)$ GUT. \ As explained in \cite{HagelinEllisFlipped} there is a
unique F- and D-flat direction along which the GUT\ Higgs $10_{-1}$ and
$\overline{10}_{1}$ develop a vev. \ This vev simultaneously breaks
$SU(5)\times U(1)$ to $SU(3)\times SU(2)\times U(1)$ while also giving a large
mass to the Higgs triplets of the $5_{2}$ and $\overline{5}_{-2}$. \ In order
to emphasize the embedding in $SO(10)$, we shall organize all of the matter
content in terms of representations of $SO(10)$. \ Explicitly, we have:%
\begin{align}
SO(10)  &  \supset SU(5)\times U(1)\\
16_{M}  &  =1_{-5}+\overline{5}_{3}+10_{-1}\\
10_{H}  &  =5_{2}+\overline{5}_{-2}%
\end{align}
Because the GUT Higgs fields $10_{-1}$ and $\overline{10}_{1}$ do not fill out
a complete $SO(10)$ multiplet, we shall refer to these fields as $\Pi$ and
$\overline{\Pi}$, respectively.

We now explain how in F-theory a higher dimensional $SO(10)$ GUT\ can
naturally break to a four-dimensional flipped $SU(5)$ GUT. \ For concreteness,
we consider models based on the del Pezzo 8 surface. \ The adjoint
representation of $SO(10)$ decomposes into representations of $SU(5)\times
U(1)$ as:%
\begin{align}
SO(10)  &  \supset SU(5)\times U(1)\\
45  &  \rightarrow1_{0}+24_{0}+10_{4}+\overline{10}_{-4}\text{.}%
\end{align}
By inspection, the $U(1)$ charge assignment of the $10_{4}$ does not
correspond to the representation content of any field in a flipped $SU(5)$
model. \ We therefore require that the zero mode content of the theory must
not contain any $10_{4}$'s or $\overline{10}_{-4}$'s. \ In this case, the only
gauge bundle configurations which do not contain any such exotics are all of
the form $\mathcal{O}_{S}(\alpha)^{1/4}$ where $\alpha$ corresponds to a
simple root of $H_{2}(S,%
\mathbb{Z}
)$.

So long as the instanton configuration breaks $G_{S}$ to a four-dimensional
flipped GUT\ group with all matter fields in well-defined flipped
GUT\ multiplets, we can avoid additional exotica in the low energy spectrum.
\ For example, in breaking $E_{6}$ to $SO(10)\times U(1)$, the adjoint
decomposes as:%
\begin{align}
E_{6}  &  \supset SO(10)\times U(1)\\
78  &  \rightarrow1_{0}+45_{0}+16_{-3}+\overline{16}_{3}\text{.}
\label{E6adjointdecomp}%
\end{align}
Further breaking $SO(10)$ to $SU(5)\times U(1)$, if we again require that no
zero modes descend from the $45_{0}$ of $SO(10)\times U(1)$, we will
generically produce zero modes which descend from the $16_{-3}$ and
$\overline{16}_{3}$. \ We note that in this case, the zero modes can still
organize into complete flipped multiplets.

\subsection{Flipped $SU(5)$ Model}

We now present a hybrid model which partially unifies to a flipped
$SU(5)$ GUT\ as a four-dimensional model and then further unifies to a higher
dimensional $SO(10)$ GUT model. \ Because none of the matter fields of the
flipped model descend from the adjoint representation of $SO(10)$, all of the
chiral matter content of the flipped $SU(5)$ model must localize on matter
curves. \ Hence, the $SO(10)$ interaction term $16_{M}\times16_{M}\times
10_{H}$ must originate from the triple intersection of matter curves. \ To
this end, we consider a geometry where the generic $SO(10)$ singularity
undergoes a twofold enhancement in rank to $E_{7}$ and $SO(14)$ singularities.

Decomposing the adjoint representation of $E_{7}$
with respect to the subgroup $SO(10)\times U(1)\times U(1)$ yields:%
\begin{align}
E_{7}  &  \supset SO(10)\times U(1)_{1}\times U(1)_{2}\label{E7decompone}\\
133  &  \rightarrow1_{0,0}+1_{0,2}+1_{0,-2}+1_{0,0}+45_{0,0}\\
&  +10_{2,0}+10_{-2,0}+16_{-1,1}+16_{-1,-1}+\overline{16}_{1,1}+\overline
{16}_{1,-1} \label{E7decompthree}%
\end{align}
so that six-dimensional hypermultiplets in the $16$ localize on the two matter
curves $(-t_{1}+t_{2}=0)$ and $(-t_{1}-t_{2}=0)$ and a six-dimensional
hypermultiplet in the $10$ localizes on the matter curve $(t_{1}=0)$.  By
inspection, we see that a local enhancement to $E_{7}$ can accommodate interaction terms of the form
$16 \times 16 \times 10$ and $\overline{16} \times \overline{16} \times 10$.  A similar analysis establishes
that a local enhancement to $SO(14)$ can accommodate an interaction term of the form $1 \times 10 \times 10$.\footnote{In fact, in a previous version
of this paper, these local $U(1)$ charge assignments for the explicit flipped
models considered were not properly taken into account. \ We thank J. Marsano,
N. Saulina and S. Sch\"{a}fer-Nameki for bringing this error to our attention.}

We now present a toy hybrid scenario which we refer to as the \textquotedblleft
Hybrid I\textquotedblright\ model. \ Some deficiencies with this example will
be rectified in the \textquotedblleft Hybrid II\textquotedblright\ model.
\ The $SO(10)$ GUT\ group breaks to $SU(5)\times U(1)$ with no bulk exotics
when the gauge bundle configuration corresponds to the fractional line bundle
$L=\mathcal{O}_{S}(E_{1}-E_{2})^{1/4}$. \ In the Hybrid I model, the matter
curves and gauge bundle assignments for each curve are:%
\begin{equation}%
\begin{tabular}
[c]{|l|l|l|l|l|l|}\hline
$\text{Hybrid I}$ & $\text{Curve}$ & $\text{Class}$ & $g_{\Sigma}$ &
$L_{\Sigma}$ & $L_{\Sigma}^{\prime n}$\\\hline
$1\times16_{M}$ & $\Sigma_{M}^{(1)}$ & $E_{3}$ & $0$ & $\mathcal{O}%
_{\Sigma_{M}^{(1)}}$ & $\mathcal{O}_{\Sigma_{M}^{(1)}}(1)$\\\hline
$2\times16_{M}$ & $\Sigma_{M}^{(2)}$ & $H-E_{3}-E_{4}$ & $0$ & $\mathcal{O}%
_{\Sigma_{M}^{(2)}}$ & $\mathcal{O}_{\Sigma_{M}^{(2)}}(2)$\\\hline
$1\times10_{H}^{(d)}$ & $\Sigma_{H}^{(d)}$ & $2H-E_{1}-E_{3}$ & $0$ &
$\mathcal{O}_{\Sigma_{H}^{(d)}}(1)^{1/4}$ & $\mathcal{O}_{\Sigma_{H}^{(d)}}%
(1)^{1/2}$\\\hline
$1\times10_{H}^{(u)}$ & $\Sigma_{H}^{(u)}$ & $2H-E_{2}-E_{3}$ & $0$ &
$\mathcal{O}_{\Sigma_{H}^{(u)}}(-1)^{1/4}$ & $\mathcal{O}_{\Sigma_{H}^{(u)}%
}(-1)^{1/2}$\\\hline
$1\times(\Pi+\overline{\Pi})$ & $\Sigma_{h}\text{ (pinched)}$ & $3H-E_{1}%
-E_{2}$ & $1$ & $\mathcal{O}_{\Sigma_{h}}(p_{1}-p_{2})^{1/4}$ & $\mathcal{O}%
_{\Sigma_{h}}(p_{1}-p_{2})^{1/4}$\\\hline
\end{tabular}
\end{equation}
with notation as in (\ref{SU5Example}).

By construction, we find one chiral generation of the MSSM\ localized on
$\Sigma_{M}^{(1)}$ with two generations localized on $\Sigma_{M}^{(2)}$. \ The
matter curve $\Sigma_{H}^{(d)}$ supports a zero mode transforming in the
representation $5_{2}^{(d)}$ which contains the Higgs down of a flipped GUT
model, and $\Sigma_{H}^{(u)}$ supports a single zero mode in the $\overline
{5}_{-2}^{(u)}$. \ Finally, in addition to the matter content of the MSSM, we
have also included a single vector-like pair of GUT Higgs fields $\Pi$ and $\overline{\Pi}$.

Including terms up to quartic order, the resulting superpotential of the
four-dimensional flipped $SU(5)$ model is therefore:%
\begin{equation}
W_{SU(5)\times U(1)}=W_{Matter}+W_{Higgs}+W_{Quartic} \label{flippedSU5W}%
\end{equation}
where the interaction terms for the chiral matter are:%
\begin{align}
W_{Matter}  &  =\lambda^{u}_{i}(\overline{5}_{-2}^{(u)}\times\overline{5}_{3}^{(i)}\times
10_{-1}^{(3)}+\overline{5}_{-2}^{(u)}\times10_{-1}^{(i)}\times\overline{5}%
_{3}^{(i)})\\
&  +\lambda^{d}_{i}(5_{2}^{(d)}\times1_{-5}^{(i)}\times\overline
{5}_{3}^{(3)}+5_{2}^{(d)}\times10_{-1}^{(i)}\times10_{-1}^{(3)})
\end{align}
and $i=1,2$ runs over two of the generations of the MSSM. \ The interaction
terms in the Higgs sector are:%
\begin{equation}
W_{Higgs}= \lambda_{\text{repel}}\cdot \Phi\times5_{2}^{(d)}\times\overline{5}_{-2}%
^{(u)} + \lambda_{\Pi}\cdot 5_{2}^{(d)}\times\Pi\times\Pi+\lambda_{\overline{\Pi}}\cdot \overline{5}_{-2}%
^{(u)}\times\overline{\Pi}\times\overline{\Pi}  \text{.}%
\end{equation}
The final term $W_{Quartic}$ originates from integrating out the heavy
Kaluza-Klein modes associated with the Higgs fields:%
\begin{equation}
W_{Quartic}=\frac{c_{i}}{M_{KK}}10_{-1}^{(i)}\times\overline{\Pi}%
\times10_{-1}^{(3)}\times\overline{\Pi}\text{.}%
\end{equation}
In the above, the mass scale $M_{KK}$ is the overall Kaluza-Klein mass scale.
\ In general, this can be slightly higher than the vev of the GUT\ Higgs
fields. \ We note that when $\overline{\Pi}$ develops a vev which also lifts
the Higgs triplets from the low energy spectrum, it also generates a large
Majorana mass term for the right-handed neutrinos.

Because the matter curves $\Sigma_{M}^{(1)}$ and $\Sigma_{M}^{(2)}$ do not
self-intersect, the resulting model has two heavy generations. \ In contrast
to the minimal $SU(5)$ models considered previously, the field-theoretic
missing partner mechanism already lifts the Higgs triplets and prevents the
higher dimension $QQQL$ operator from being generated. \ Moreover, the model
already incorporates a natural seesaw mechanism.

Before proceeding to a slightly more realistic model, we note that although it would at first appear
to be more economical to place the Higgs up and Higgs down on the same matter
curve, this leads to certain undesirable consequences in the low energy
theory. \ The reason is that the Higgs up and down fields would then be
equally or oppositely charged under a common $U(1)$ symmetry. \ This would
either forbid the coupling $16\times16\times10$ or $\overline{16}%
\times\overline{16}\times10$ in the low energy theory. \ The former
interaction is necessary for generating semi-realistic Yukawa couplings, while
the latter is necessary for implementing doublet-triplet splitting using the
missing partner mechanism. \ In order to achieve both couplings, it appears
necessary to localize these fields on different matter curves.

\begin{figure}
[ptb]
\begin{center}
\includegraphics[
height=3.6443in,
width=4.7625in
]%
{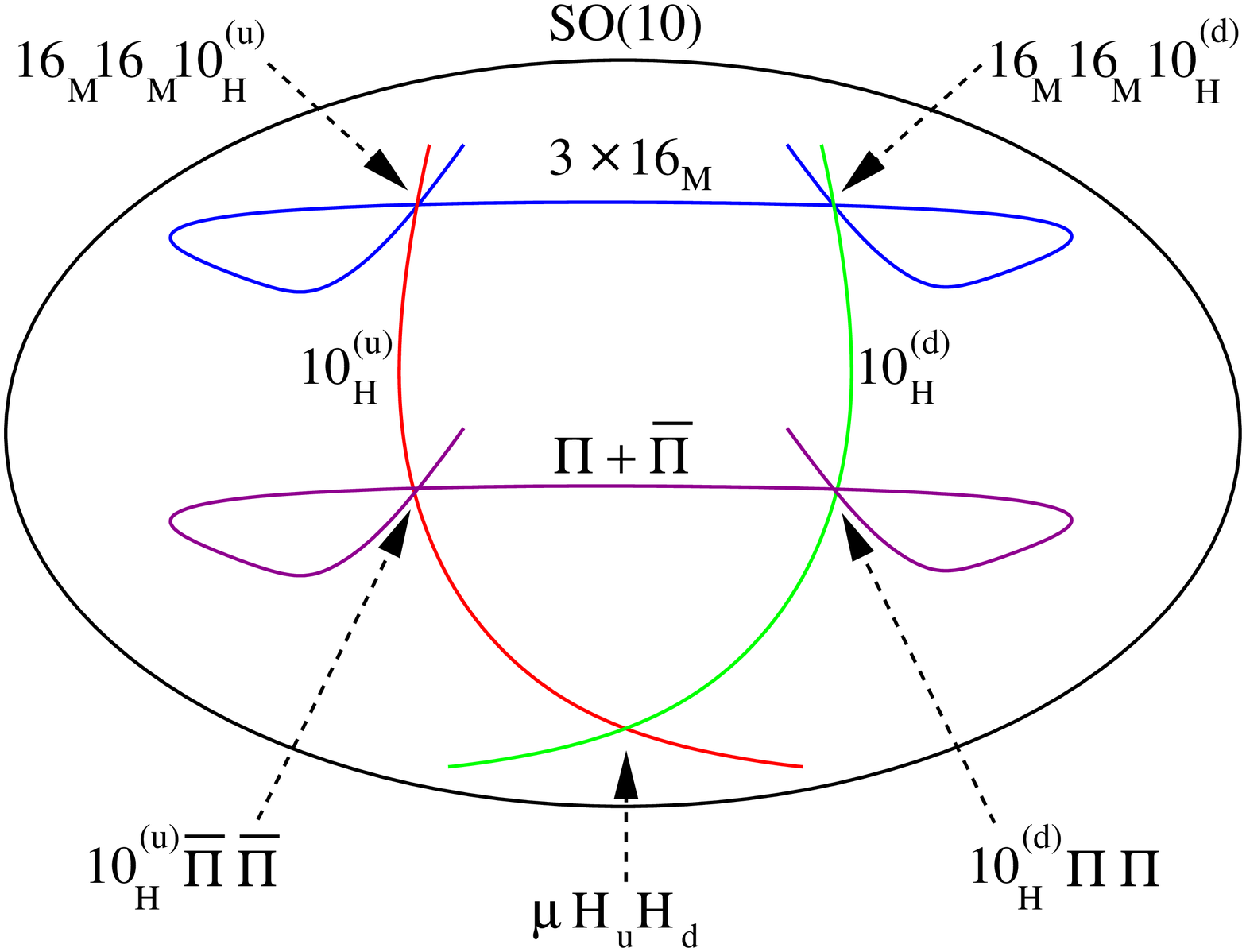}%
\caption{Depiction of the various matter curves in the flipped $SU(5)$ model
referred to as \textquotedblleft Hybrid II\textquotedblright\ in the text.
\ The background instanton configuration breaks the bulk gauge group $SO(10)$
to $SU(5)\times U(1)$. \ In this case, all three generations transform in the
$16$ of $SO(10)$ and localize on a single self-intersecting matter curve.
\ The MSSM Higgs fields descend from two different $10$'s of $SO(10)$. \ The
model also contains a single vector-like pair transforming in the $10_{-1}$
and $\overline{10}_{+1}$ of $SU(5)\times U(1)$ which facilitates GUT group
breaking and doublet-triplet splitting. \ These GUT\ Higgs fields descend from
a six-dimensional hypermultiplet transforming in the $16$ of $SO(10)$.}%
\label{flippedsu5curves}%
\end{center}
\end{figure}

A more realistic hierarchy in quark masses can be achieved when the chiral
matter curves self-intersect. \ As a small refinement on the above model, we
take $L=\mathcal{O}_{S}(E_{1}-E_{2})^{1/4}$ as before, while the matter curves
and gauge bundle assignments for each curve are now:%
\begin{equation}%
\begin{tabular}
[c]{|l|l|l|l|l|l|}\hline
$\text{Hybrid II}$ & $\text{Curve}$ & $\text{Class}$ & $g_{\Sigma}$ &
$L_{\Sigma}$ & $L_{\Sigma}^{\prime n}$\\\hline
$1\times10_{H}^{(d)}$ & $\Sigma_{H}^{(d)}$ & $2H-E_{1}-E_{3}$ & $0$ &
$\mathcal{O}_{\Sigma_{H}^{(d)}}(1)^{1/4}$ & $\mathcal{O}_{\Sigma_{H}^{(d)}}%
(1)^{1/2}$\\\hline
$1\times10_{H}^{(u)}$ & $\Sigma_{H}^{(u)}$ & $2H-E_{2}-E_{3}$ & $0$ &
$\mathcal{O}_{\Sigma_{H}^{(u)}}(-1)^{1/4}$ & $\mathcal{O}_{\Sigma_{H}^{(u)}%
}(-1)^{1/2}$\\\hline
$3\times16_{M}$ & $\Sigma_{M}\text{ (pinched)}$ & $3H$ & $1$ & $\mathcal{O}%
_{\Sigma_{M}}$ & $\mathcal{O}_{\Sigma_{M}}(3p^{\prime})$\\\hline
$1\times\left(  \Pi+\overline{\Pi}\right)  $ & $\Sigma_{h}\text{ (pinched)}$ &
$3H-E_{1}-E_{2}$ & $1$ & $\mathcal{O}_{\Sigma_{h}}(p_{1}-p_{2})^{1/4}$ &
$\mathcal{O}_{\Sigma_{h}}(p_{1}-p_{2})^{1/4}$\\\hline
\end{tabular}
\end{equation}
so that all three generations localize on the matter curve $\Sigma_{M}$. \ See
figure \ref{flippedsu5curves} for a depiction of the Hybrid II model.%
While the zero mode content of this case is the same as the Hybrid I model,
the self-intersection of the matter curves allows the model to have one
generation which is hierarchically heavier than the lighter two generations,
much as in the second minimal $SU(5)$ example of Section \ref{SU5MODEL}.
\ Aside from this difference, the structure of the superpotential is quite
similar to that given by equation (\ref{flippedSU5W}). \ Indeed, just as in the Hybrid I model,
there exist higher dimension operators which can generate large Majorana mass
terms for the right-handed neutrinos.

\section{Numerology\label{NUMEROLOGY}}

Throughout this paper we have given numerical estimates of various quantities
which appear to be in rough agreement with observation. \ In this Section we
demonstrate that for an appropriate choice of order one constants, many of the
relations obtained throughout are in agreement with experimental observation.
\ Our point here is not so much to show that we can match to the precise
numerical values, but rather that the numbers we have obtained are not wildly
different from the expected ranges. \ Indeed, although we shall typically
evaluate all quantities at the GUT scale, in a more accurate analysis these
quantities would of course have to be evolved under renormalization group flow
to low energies. \ In this regard, our order of magnitude estimates will be
somewhat naive, although we believe it still gives a reliable guide for the
ranges of energy scales involved in our models. \ Moreover, for concreteness,
in this Section we focus on the case of the minimal $SU(5)$ model.

At the level of precision with which we can reliably estimate parameters, all
of our estimates depend on order one coefficients, the Planck mass $M_{pl}$,
the GUT scale $M_{GUT}$, the Higgs up vev, and the value of the gauge coupling
constants at the GUT scale, $\alpha_{GUT}$. \ Throughout, we use the following
approximate values:%
\begin{align}
M_{pl}  &  \sim1\times10^{19}\text{ GeV}\label{MPLVAL}\\
M_{GUT}  &  \sim3\times10^{16}\text{ GeV}\\
\left\langle H_{u}\right\rangle  &  \sim246\text{ GeV}\label{AGUTVAL}\\
\alpha_{GUT}  &  =\frac{g_{YM}^{2}(M_{GUT})}{4\pi}\sim\frac{1}{25}\text{.}%
\end{align}
In general, factors of $2$ and $\pi$ are typically beyond the level of
precision which we can reliably estimate.

The above parameters appear geometrically as the length scale $R_{S}$
associated with the size of the del Pezzo, $R_{B}$ which is associated with
the size of the threefold base, and $R_{\bot}$ which may be viewed as a local
cutoff on the behavior of wave functions in the model. \ These length scales
are related by appropriate powers of the small parameter:%
\begin{equation}
\varepsilon=\frac{M_{GUT}}{\alpha_{GUT}M_{pl}}\sim7.5\times10^{-2}\text{.}%
\end{equation}
The various length scales are then given by:%
\begin{align}
\frac{1}{R_{S}}  &  =M_{GUT}\sim3\times10^{16}\text{ GeV}\\
\frac{1}{R_{B}}  &  =M_{GUT}\times\varepsilon^{1/3}\sim1\times10^{16}\text{
GeV}\\
\frac{1}{R_{\bot}}  &  =M_{GUT}\times\varepsilon^{\gamma}\sim5\times
10^{15\pm0.5}\text{ GeV}%
\end{align}
where the parameter $1/3\lesssim\gamma\lesssim1$ ranges from $1/3$ when
$B_{3}$ is homogeneous, to $1$ when $B_{3}$ is given by a tubular geometry.

We now collect and slightly expand on the estimates obtained throughout this
paper. \ We begin by discussing the mass scales associated with quarks. \ In
this case, the masses of the quarks at the GUT\ scale are very roughly given
by:%
\begin{equation}
m_{q}\sim\alpha_{GUT}^{3/4}\left\langle H_{u}\right\rangle \sim20\text{ GeV.}%
\end{equation}
Note that the top quark mass is about a factor of 3 higher than this (taking
into account the RG flow), which suggests that perhaps the corresponding
curves are smaller by that factor to give the correct wave function normalization.

We have also seen that matter fields which localize on curves normal to $S$ in
the threefold base $B_{3}$ can provide a natural mechanism for generating
light neutrino masses as well an exponentially suppressed $\mu$ term. \ As an
intermediate case, we have shown that right-handed neutrino wave functions
which are attracted to the seven-brane can potentially realize a viable seesaw
mechanism. \ Reproducing equation (\ref{light}) for the convenience of the
reader, the light neutrino mass in the seesaw scenario is:%
\begin{equation}
m_{\text{light}}\sim\alpha_{GUT}^{3/4}\frac{\left\langle H_{u}\right\rangle
^{2}}{M_{GUT}}\varepsilon^{-4\gamma}\sim2\times10^{-1\pm1.5}\text{ eV.}%
\end{equation}

Gauge singlet wave functions can also exhibit more extreme behavior. \ Indeed,
when the Higgs up and down fields localized on different matter curves which
intersect, they interact with a gauge singlet wave function outside of $S$.
\ When this wave function is exponentially suppressed, the induced $\mu$ term
is given by equation (\ref{muterm}):%
\begin{equation}
\mu(c,\gamma)\sim M_{GUT}\times\alpha_{GUT}^{3/4}\varepsilon^{2\gamma}%
\exp\left(  -\frac{c}{\varepsilon^{2\gamma}}\right)  \text{.}%
\end{equation}
We find that when $c$ and $\gamma$ are order one numbers, this value can
naturally fall near the electroweak scale. \ For example, we have:%
\begin{align}
\mu(c  &  =1/7,\gamma=1)\sim140\text{ GeV}\\
\mu(c  &  =1,\gamma=0.64)\sim107\text{ GeV.}%
\end{align}

In a scenario where the neutrinos are purely of Dirac type, an exponentially
small value can also be achieved when the gauge singlet wave function is
exponentially damped near $S$. \ The Dirac mass is given by equation
(\ref{DiracMass}):%
\begin{equation}
m_{\text{Dirac}}(c,\gamma)\sim\left\langle H_{u}\right\rangle \times
\alpha_{GUT}^{3/4}\varepsilon^{\gamma}\exp\left(  -\frac{c}{\varepsilon
^{2\gamma}}\right)  \text{.}%
\end{equation}
As for the $\mu$ term, order one values of $c$ and $\gamma$ yield reasonable
values for the masses. $\ $Indeed, as explained in subsection
\ref{MODINDEPPREDICTION}, when the exponential suppression factors are
identical for the Dirac neutrino mass and $\mu$ term, we obtain the estimate:%
\begin{equation}
m_{\text{Dirac}}(c,\gamma)\sim\mu(c,\gamma)\frac{\left\langle H_{u}%
\right\rangle }{M_{GUT}}\times\varepsilon^{-\gamma}\sim0.5\times10^{-2\pm
0.5}\text{ eV}%
\end{equation}
when $\mu(c,\gamma)\sim100$ GeV. \ We have also observed that a similar
analysis of Yukawa couplings also applies in estimates of the messenger mass
scales for gauge mediated supersymmetry breaking scenarios.

\section{Conclusions\label{CONCLUSIONS}}

F-theory provides a natural setup for studying GUT models in string theory.
\ In this paper we have adopted a bottom up approach to string phenomenology
and have found that it provides a surprisingly powerful constraint on low
energy physics. \ One's natural expectation is that there should be a great
deal of flexibility in local models where issues pertaining to a globally
consistent compactification can always be deferred to a later stage of
analysis. \ This is indeed the case in models where a sufficiently loose
definition of \textquotedblleft local data\textquotedblright\ is adopted so
that gravity need not decouple, and\ we have given some examples along these
lines. \ Strictly speaking, though, a local model is well-defined by local
data when the model admits a limit where it is in principle possible to
decouple the GUT scale from the Planck scale. \ Perhaps surprisingly, this
qualitative condition endows these GUT models with considerable predictive power.

The main lesson we have learned is that the mere \textit{existence} of a
decoupling limit constrains both the local geometry of the compactification as
well as the type of seven-brane which can wrap a compact surface in the local
model. \ To realize a GUT model with no low energy exotics, the bulk gauge
group of the seven-brane can only have rank four, five or six, and in order
for a decoupling limit to even exist in principle, the seven-brane must wrap a
del Pezzo surface. Moreover, all of the vacua which descend at low energies to
the MSSM in four dimensions all possess an internal $U(1)$ hypercharge flux on
the del Pezzo which at least partially breaks the GUT\ group. \ For
concreteness, in this paper we have primarily focused on the cases where the
bulk gauge group in eight dimensions is $SU(5)$ or $SO(10)$.

In the minimal $SU(5)$ model, all of the matter content at low energies
derives from the intersection of the GUT\ model seven-brane with additional
non-compact seven-branes. \ We have explained how the fields which localize at
such intersections can only transform in the $5$, $10$ or complex conjugate
representations. \ Moreover, the interaction terms are all cubic in the matter
fields because the superpotential derives from the triple intersection of
matter curves. \ Matter fields which are neutral under the GUT group localize
on matter curves which are orthogonal to the brane. \ When the gauge singlet
is attracted to the seven-brane, the corresponding Yukawa couplings behave as
though the gauge singlet had localized inside of $S$. \ On the other hand,
\ when the gauge singlet wave function is repelled away from the seven-brane,
this can yield a significant exponential suppression in the value of the
Yukawa couplings on the order of $\exp(-c/\varepsilon^{2\gamma})$ where $c$
and $\gamma$ are order one positive numbers and $\varepsilon\sim\alpha
_{GUT}^{-1}M_{GUT}/M_{pl}$. \ In particular, vector-like pairs in such
compactifications do not always develop masses on the order of $M_{GUT}$.
\ This runs counter to a coarse effective field theory analysis which would
otherwise suggest that such pairs should always develop large masses. \ In
fact, we have seen that this is consistent with a more refined effective field
theory analysis because there are typically additional global symmetries
present in the low energy theory.

The exponential suppression of such Yukawa couplings naturally solves the
$\mu$ problem and also provides a natural mechanism for generating acceptably
light neutrino masses. \ The wave function for the right-handed neutrino is
either attracted or repelled away from the del Pezzo surface. \ In the
repulsive case, the neutrino mass term is purely of Dirac type and is on the
order of $\ 0.5\times10^{-2\pm0.5}$ eV. \ In the attractive case, we find a
natural implementation of a modified seesaw mechanism so that the light
neutrinos masses are $2\times10^{-1\pm1.5}$ eV and the Majorana mass is
$\sim3\times10^{12\pm1.5}$ GeV, which is naturally smaller than the simplest
GUT seesaw models.

The combination of non-trivial hypercharge flux in the internal dimensions and
the existence of additional fluxes derived from the transversally intersecting
seven-branes alleviates a number of problems which plague four-dimensional
supersymmetric GUT models. The doublet-triplet splitting problem reduces to
the condition that the hypercharge flux and flux from the other seven-branes
both pierce the Higgs matter curves, while the net hypercharge flux vanishes
on curves which support full GUT\ multiplets.

The internal $U(1)$ hypercharge flux also provides a qualitative explanation
for why the $b-\tau$ GUT\ mass relation approximately matches with observation
while the lighter two generations at best obey distorted versions of this
relation. \ This is in a sense the remnant of the mechanism that solves the
doublet-triplet splitting problem. \ Even though the net hypercharge flux
vanishes on a matter curve which supports a complete GUT\ multiplet, the field
strength is not identically zero. \ In this way, the GUT\ multiplet wave
functions experience an Aharanov-Bohm effect which increasingly distorts the
GUT\ mass relations as the mass of the GUT multiplet decreases. \ In fact,
this mechanism \textit{requires} that the internal hypercharge gauge field be non-trivial.

This flux will also typically generate a threshold correction to the
unification of the gauge couplings. While there are potentially many other
such threshold corrections due to Kaluza-Klein modes, it would clearly be of
interest to see whether at least some of these corrections can be reliably
estimated in our setup.

The geometry of the compactification can also prevent the proton from decaying
too rapidly. \ Cubic terms in the superpotential are typically excluded in a
bottom up approach by requiring that the theory is invariant under R-parity.
\ We have found two ways that the geometry can forbid the same interaction
terms which R-parity removes. \ In one case, R-parity corresponds to a
suitable $\mathbb{Z}_{2}$ symmetry in the geometry of the Calabi-Yau fourfold.
\ At a topological level, the absence of R-parity violating cubic interaction
terms corresponds to a technically natural restriction on which matter curves
intersect. \ In the scenario where R-parity descends from a $%
\mathbb{Z}
_{2}$ group action on the Calabi-Yau, the hypercharge flux and the Higgs
matter curves are invariant under this group action while the matter curves
are odd. \ Due to the $%
\mathbb{Z}
_{2}$ symmetry, the net hypercharge flux must vanish on matter curves which
are odd under this group action. \ Hence, this \textit{automatically}
\textit{forces} the localized matter to organize in complete GUT\ multiplets.
\ Note that this symmetry also permits a non-vanishing hypercharge flux on the
Higgs curves, which is consistent with our solution to the doublet-triplet
splitting problem. \ At higher order in the effective superpotential, the
topological condition determining which curves intersect also forbids
potentially dangerous baryon number violating quartic operators in the
superpotential. \ Indeed, placing the Higgs up and down fields on distinct
matter curves equips the matter fields with additional global symmetries which
can forbid such operators.

We have also shown how the geometry of the matter curves translates in the low
energy effective theory into non-trivial structure in the Yukawa couplings.
The coarsest features of textures follow from the discrete data determining
how matter curves intersect inside the seven-brane so that texture zeroes are
generically present. \ We have also presented some speculations on potential
ways that additional structure in the Yukawa couplings could arise from a
geometrical realization of the Froggatt-Nielsen mechanism, or through an
interpretation of the discrete automorphism group of a del Pezzo surface as a
flavor group symmetry.

Communicating supersymmetry breaking is also straightforward in this setup.
\ Indeed, we have shown that the geometry of del Pezzo surfaces can easily
accommodate vector-like pairs of GUT\ multiplets localized on isolated matter
curves. \ These vector-like pairs can then serve as the messenger fields in
gauge mediated supersymmetry breaking. \ We have presented different scenarios
showing the flexibility of this approach. \ Depending on the case at hand, the
messenger masses can range from near the GUT\ scale, to energy scales which
are significantly lower. \ Moreover, because we have an independent mechanism
for naturally suppressing the $\mu$ term, this class of models preserves the
best features of gauge mediation models while avoiding the notoriously
difficult issue of generating $\mu$ and $B\mu$ at around the electroweak scale.

It is perhaps surprising that a few key ideas seem to resolve many problems
simultaneously. \ Indeed, the overall economy in these ingredients lends
substantial credence to the basic framework. \ On the other hand, it is also
clear that we have by no means exhausted the potential avenues of
investigation. \ A more systematic study of textures and choices of matter
curves, as well as the geometric underpinning of the corresponding Calabi-Yau
fourfold are all issues which deserve further attention. In addition, the
communication of supersymmetry breaking is simple enough in our setup that it
could potentially lead to observable predictions at the LHC. \ It would
clearly be of interest to study such a scenario further.

\section*{Acknowledgements}

We thank K.S. Babu, V. Bouchard, F. Denef, A.L. Fitzpatrick, J. Marsano, S.
Raby, N. Saulina, S. Sch\"{a}fer-Nameki, P. Svr\v{c}ek, A. Tomasiello, M. Wijnholt, E.
Witten and S.-T. Yau for helpful discussions. \ The work of the authors is
supported in part by NSF grants PHY-0244821 and DMS-0244464. \ The research of
JJH is also supported by an NSF Graduate Fellowship.

\appendix

\section*{Appendices}

\section{Gauge Theory Anomalies and Seven-Branes}

In this Appendix we further elaborate on the geometric condition for the low
energy spectrum to be free of gauge theory anomalies. \ First recall the
well-studied case of perturbatively realized gauge theories obtained as the
low energy limit of D-brane probes of non-compact Calabi-Yau singularities.
\ The condition that all gauge theory anomalies must cancel is equivalent to
the requirement that in a consistent bound state of D3-, D5- and D7-branes,
the total RR\ flux measured over a compact cycle must vanish
\cite{CachazoVafaGeomUnif}. \ Even in a non-compact Calabi-Yau threefold given
by the total space $\mathcal{O}(K_{S})\rightarrow S$ with $S$ a K\"{a}hler
surface, the theory of a stack of D7-branes wrapping $S$ is inconsistent
because the self-intersection of the divisor $S$ in the Calabi-Yau threefold
is a compact Riemann surface. \ In a globally consistent model, additional
O7-planes must be introduced to cancel the corresponding RR\ tadpole.
\ Indeed, a consistent compactification of F-theory on an elliptically fibered
Calabi-Yau fourfold will automatically contain similar contributions so that
the net monodromy around all seven-branes is trivial.

Next consider the potential contribution from D5-branes to a candidate bound
state. \ Letting $[\Sigma_{D5}]$ denote the total homology class of D5-branes
wrapping compact two-cycles in $H_{2}(S,%
\mathbb{Z}
)$, the resulting theory is consistent provided:%
\begin{equation}
\lbrack\Sigma_{D5}]\cdot K_{S}=0\text{.} \label{pertflux}%
\end{equation}
There is no analogous condition for D3-branes in a non-compact model because
the flux lines can escape to infinity in the non-compact model.

In this Appendix we consider more general intersecting seven-brane
configurations with chiral matter induced from a non-trivial field strength.
\ Using the fact that a low energy theory must be free of non-abelian gauge
anomalies, we determine the geometric analogue of equation (\ref{pertflux})
for intersecting $A\times A$ and $D\times A$ brane configurations in a broader
class of F-theory compactifications. \ We also present an example of anomaly
cancelation for an $E_{7}$ exceptional brane theory.

\subsection{$A\times A$ Anomalies\label{AXAANOM}}

We now consider seven-branes wrapping two K\"{a}hler surfaces $S$ and
$S^{\prime}$ such that the gauge group of the respective seven-branes is
$G_{S}=SU(N)$ and $G_{S^{\prime}}=SU(N^{\prime})$\ with a six-dimensional
bifundamental localized along a matter curve $\Sigma=S\cap S^{\prime}$.
\ Because only instanton configurations with an overall $U(1)$ factor can
induce chirality in the bulk and on matter curves, it is enough to consider
instanton configurations in $S$ and $S^{\prime}$ taking values in $U(1)^{n}$
and $U(1)^{n^{\prime}}$ for some $n\leq N-1$ and $n^{\prime}\leq N^{\prime}-1$.

We consider a breaking pattern such that $SU(N)$ decomposes into non-abelian
subgroup factors $SU(N_{1}),...,SU(N_{n})$. \ Similar conventions will also
hold for the decomposition of the gauge group $SU(N^{\prime})$. \ Letting
$\overrightarrow{q}$ denote the charge of a representation under the
$U(1)^{n-1}$ subgroup, the fundamental and adjoint representation decompose
as:%
\begin{align}
SU(N_{1}+...+N_{n})  &  \supset SU(N_{1})\times...\times SU(N_{n})\times
U(1)^{n-1}\\
N  &  \rightarrow(N_{1})_{\overrightarrow{q}_{1}}\oplus....\oplus
(N_{n})_{\overrightarrow{q}_{n}}\\
A_{N}  &  \rightarrow\left(  A_{N_{1}}\right)  _{2\overrightarrow{q}_{1}%
}\oplus....\oplus(A_{N_{n}})_{2\overrightarrow{q}_{n}}\\
&  \oplus\left[  \underset{i<j}{\oplus}(N_{i}\times N_{j})_{\overrightarrow
{q}_{i}+\overrightarrow{q}_{j}}\right] \\
ad(SU(N))  &  \rightarrow\underset{i=1}{\overset{n}{\oplus}}ad(SU(N_{i}%
))_{0}\oplus\left[  \underset{i\neq j}{\oplus}(N_{i}\times\overline{N}%
_{j})_{\overrightarrow{q}_{i}-\overrightarrow{q}_{j}}\right]
\end{align}
where for future use we have also indicated how the two index anti-symmetric
representation $A_{N}$ decomposes. \ In the above, the charge assignments
$\overrightarrow{q}_{i}$ satisfy the tracelessness condition:%
\begin{equation}
\underset{i=1}{\overset{n}{\sum}}N_{i}\overrightarrow{q}_{i}=0
\label{traceless}%
\end{equation}
Letting $L_{1},...,L_{n-1}$ denote the line bundles which determine the
$U(1)^{n-1}$ gauge field configuration with similar conventions for
$L_{i}^{\prime}$, the chiral matter content transforming in the fundamental
representation $N_{i}$ of $SU(N_{i})$ in $S$ and $\Sigma$ are given by the
indices derived in \cite{BHV}:%
\begin{align}
\#(N_{i}\times\overline{N}_{j})_{\overrightarrow{q}_{i}-\overrightarrow{q}%
_{j}}  &  =-c_{1}(S)\cdot c_{1}\left(  L_{1}^{\pi_{1}(\overrightarrow{q}%
_{i}-\overrightarrow{q}_{j})}\right)  +...\\
&  +-c_{1}(S)\cdot c_{1}\left(  L_{n-1}^{\pi_{n-1}(\overrightarrow{q}%
_{i}-\overrightarrow{q}_{j})}\right) \\
\#(N_{i})_{\overrightarrow{q}_{i}}\times(N_{i^{\prime}}^{\prime}%
)_{\overrightarrow{q}_{i^{\prime}}^{\prime}}  &  =\deg L_{1|\Sigma}^{\pi
_{1}(\overrightarrow{q}_{i})}+...+\deg L_{n|\Sigma}^{\pi_{n}(\overrightarrow
{q}_{i})}\\
&  +\deg L_{1|\Sigma}^{\prime\pi_{1}^{\prime}(\overrightarrow{q}_{i^{\prime}%
}^{\prime})}+...+\deg L_{n^{\prime}|\Sigma}^{\prime\pi_{n}^{\prime
}(\overrightarrow{q}_{i^{\prime}}^{\prime})}%
\end{align}
where $\pi_{i}$ denotes the projection onto the $i^{th}$ component of a given
charge vector, and a negative number indicates the net chiral matter content
transforms in the complex conjugate representation.

The net anomaly coefficient $a_{i}$ of the $SU(N_{i})$ factor is given by
summing over all contributions to the fundamental representation of
$SU(N_{i})$. \ Letting $d_{i}=c_{1}(S)\cdot c_{1}(L_{i})$ and $d_{i|\Sigma
}=\deg L_{i|\Sigma}$, we find:%
\begin{align}
a_{i}  &  =-\underset{j=1}{\overset{n}{\sum}}N_{j}\underset{k=1}{\overset
{n-1}{\sum}}\pi_{k}(\overrightarrow{q}_{i}-\overrightarrow{q}_{j})d_{k}\\
&  +\underset{i^{\prime}=1}{\overset{n^{\prime}}{\sum}}N_{i^{\prime}}^{\prime
}\left(  \underset{k=1}{\overset{n-1}{\sum}}\pi_{k}(\overrightarrow{q}%
_{i})d_{k|\Sigma}+\underset{k^{\prime}=1}{\overset{n^{\prime}-1}{\sum}}%
\pi_{k^{\prime}}^{\prime}(\overrightarrow{q}_{i^{\prime}}^{\prime
})d_{k^{\prime}|\Sigma}^{\prime}\right)  \text{.}%
\end{align}
Simplifying this expression using the tracelessness condition of equation
(\ref{traceless}) and the analogous condition for the $N_{i^{\prime}}$ now
implies:%
\begin{align}
a_{i}  &  =-N\underset{k=1}{\overset{n-1}{\sum}}\pi_{k}(\overrightarrow{q}%
_{i})d_{k}+N^{\prime}\underset{k=1}{\overset{n-1}{\sum}}\pi_{k}%
(\overrightarrow{q}_{i})d_{k|\Sigma}\\
&  =\underset{S}{\int}c_{1}(L_{1}^{\pi_{1}(\overrightarrow{q}_{i})}%
\otimes...\otimes L_{n-1}^{\pi_{n-1}(\overrightarrow{q}_{i})})c_{1}\left(
\mathcal{O}_{S}(K_{S})^{N}\otimes\mathcal{O}_{S}(\Sigma)^{N^{\prime}}\right)
\text{.} \label{AXAanomfin}%
\end{align}
The condition for $a_{i}$ to vanish is the direct analogue of the
perturbatively realized condition in equation (\ref{pertflux}).

\subsection{$A\times D$ Anomalies\label{AXDANOM}}

We now consider seven-branes wrapping two K\"{a}hler surfaces $S$ and
$S^{\prime}$ such that the gauge group of the respective seven-branes is
$G_{S}=SU(N)$ and $G_{S^{\prime}}=SO(2R+2M)$\ with six-dimensional matter
fields localized along the curve $\Sigma=S\cap S^{\prime}$. \ Decomposing
$SO(2N+2R+2M)\supset SU(N)\times SO(2R+2M)\times U(1)$, the six-dimensional
fields localized on $\Sigma$ now transform in the representation
$(A_{N},1)\oplus(N,2R)$ of $SU(N)\times SO(2R+2M)$. \ As before, it is enough
to treat instanton configurations taking values in the subgroups
$U(1)^{n}\subset SU(N)$ and $U(1)^{t}\subset SO(2R)$. \ In order to simplify
the combinatorics associated with breaking patterns of the $SO$ gauge group
factor, we confine our analysis to the breaking pattern $SO(2R+2M)\supset
SO(2R)\times SU(M)\times U(1)$. \ The fundamental and adjoint representations
of $SO(2R+2M)$ decompose into the commutant subgroup of $U(1)$ as:%
\begin{align}
SO(2R+2M)  &  \supset SO(2R)\times SU(M)\times U(1)\\
2R  &  \rightarrow(2R)_{0}\oplus\left(  (M)_{p}\oplus(\overline{M}_{i}%
)_{-p}\right) \\
ad(SO(2R))  &  \rightarrow1_{0}\oplus ad(SO(2R_{i}))_{0}\oplus ad(SU(M_{i}%
))_{0}\\
&  \oplus(A_{M})_{2p}\oplus(\overline{A_{M}})_{-2p}\oplus(2R,M)_{p}%
\oplus(2R,\overline{M})_{-p}\text{.}%
\end{align}

Consider first non-abelian anomalies associated to the gauge group factor
$SU(N_{i})$. \ In this case, we recall that in a normalization of group
generators where the fundamental has anomaly coefficient $+1$, the two index
anti-symmetric representation has anomaly coefficient $N_{i}-4$. \ Repeating a
similar analysis to that given in the previous Section, the total anomaly
coefficient for the non-abelian group $SU(N_{i})$ is:%
\begin{align}
a_{i}  &  =-N\underset{k=1}{\overset{n-1}{\sum}}\pi_{k}(\overrightarrow{q}%
_{i})d_{k}+(2R+2M)\underset{k=1}{\overset{n-1}{\sum}}\pi_{k}(\overrightarrow
{q}_{i})d_{k|\Sigma}\\
&  +(N_{i}-4)\underset{k=1}{\overset{n-1}{\sum}}\pi_{k}(2\overrightarrow
{q}_{i})d_{k|\Sigma}+\underset{j\neq i}{\sum}N_{j}\left(  \underset
{k=1}{\overset{n-1}{\sum}}\pi_{k}(\overrightarrow{q}_{i}+\overrightarrow
{q}_{j})d_{k|\Sigma}\right) \\
&  =-2N\underset{k=1}{\overset{n-1}{\sum}}\pi_{k}(\overrightarrow{q}_{i}%
)d_{k}+(2R+2M)\underset{k=1}{\overset{n-1}{\sum}}\pi_{k}(\overrightarrow
{q}_{i})d_{k|\Sigma}-8\underset{k=1}{\overset{n-1}{\sum}}\pi_{k}%
(\overrightarrow{q}_{i})d_{k|\Sigma}\\
&  =2\underset{S}{\int}c_{1}(L_{1}^{\pi_{1}(\overrightarrow{q}_{i})}%
\otimes...\otimes L_{n-1}^{\pi_{n-1}(\overrightarrow{q}_{i})})c_{1}\left(
\mathcal{O}_{S}(K_{S})^{N}\otimes\mathcal{O}_{S}(\Sigma)^{R+M-4}\right)
\text{.} \label{AXDanomfinfirst}%
\end{align}
Comparing equations (\ref{AXAanomfin}) and (\ref{AXDanomfinfirst}), the shift
$R+M\rightarrow R+M-4$ indicates the presence of an $O7$-plane.

Next consider the anomaly coefficient of the $SU(M)$ factor. \ In this case,
the total anomaly coefficient for the non-abelian group $SU(M)$ is:%
\begin{align}
b_{i}  &  =-2p(M-4)d^{\prime}-2Rpd^{\prime}+2pNd_{\Sigma}^{\prime}\\
&  =2\underset{S^{\prime}}{\int}c_{1}(L^{\prime p})c_{1}\left(  \mathcal{O}%
_{S^{\prime}}(\Sigma)^{N}\otimes\mathcal{O}_{S^{\prime}}(K_{S^{\prime}%
})^{R+M-4}\right)  \text{.}%
\end{align}
Proceeding by induction, it now follows that a similar result also holds for
the more general breaking pattern where each $SU(M)$ and $SO(2R)$ factor
decomposes further.

\subsection{$E_{7}$ Anomalies}

The analysis of the previous subsections demonstrates that for $A$- and $D$-
type seven-branes, the geometric condition for anomaly cancelation in the
four-dimensional effective theory relates the total matter content in the bulk
with that localized on matter curves. \ We now determine the analogous
condition for a seven-brane with gauge group $G_{S}=E_{7}$ and $M$ copies of
the $56$ localized on a curve $\Sigma$. \ We consider a $U(1)$ gauge field
configuration which breaks $E_{7}$ to $SU(7)\times U(1)$. \ The representation
content of $E_{7}$ decomposes as:%
\begin{align}
E_{7}  &  \supset SU(8)\supset SU(7)\times U(1)\\
56  &  \rightarrow7_{-6}+\overline{7}_{6}+21_{2}+\overline{21}_{-2}\\
133  &  \rightarrow1_{0}+7_{8}+\overline{7}_{-8}+48_{0}+\overline{35}%
_{4}+35_{-4}%
\end{align}
where the $21$, $35$ and $\overline{35}$ denote the two, three and four index
anti-symmetric representations of $SU(7)$. \ It now follows that the chiral
matter content derived from $S$ and $\Sigma$ is:%
\begin{align}
\#7_{8}  &  =-c_{1}(S)\cdot c_{1}(L^{8})\label{SE7}\\
\#35  &  =-c_{1}(S)\cdot c_{1}(L^{-4})\\
\#7_{-6}  &  =M\deg L_{|\Sigma}^{-6}\label{SIE7}\\
\#21  &  =M\deg L_{|\Sigma}^{2}\text{.} \label{SE7FIN}%
\end{align}

To compute the anomaly of the $SU(7)$ theory, we first recall that the anomaly
coefficient for the $i$-index anti-symmetric representation $A_{k}^{(i)}$ of
$SU(n)$ in $2(k-1)$ dimensions is \cite{BradenSiegwartSUN}:%
\begin{align}
A_{k}^{(2)}  &  =n-2^{k-1}\\
A_{k}^{(3)}  &  =\frac{1}{2}n^{2}-\frac{1}{2}n(2^{k}+1)+3^{k-1}\\
A_{k}^{(4)}  &  =\frac{1}{12}(2n^{3}-3n^{2}(2^{k}+2)+n(4\times3^{k}%
+3\times2^{k}+4)-3\times4^{k})
\end{align}
so that in four dimensions, the anomaly coefficients of the $SU(7)$ theory are
$A_{k}^{(2)}=3,$ $A_{k}^{(3)}=2$, $A_{k}^{(4)}=-2$. \ Returning to equations
(\ref{SE7})-(\ref{SE7FIN}), we note that the contribution to the total anomaly
from $S$ and $\Sigma$ separately cancel in this particular case so that we do
not deduce an analogue of equation (\ref{pertflux}).

\section{Hypersurfaces in $\mathbb{P}^{3}$\label{HyperReview}}

In this Section we review some properties of degree $n$ hypersurfaces $H_{n}$
in $\mathbb{P}^{3}$. \ Further details can be found for example in
\cite{DonaldsonKronheimer}. \ Letting $H$ denote the hyperplane class of
$\mathbb{P}^{3}$, the total Chern class of $H_{n}$ is given by the adjunction
formula:%
\begin{equation}
c(H_{n})=\frac{c(\mathbb{P}^{3})}{c(N_{H_{n}/\mathbb{P}^{3}})}%
=1+(4-n)H+(6-4n+n^{2})H^{2}\text{.} \label{chernsum}%
\end{equation}
It thus follows that the Euler character $e(H_{n})$, holomorphic Euler
characteristic $\chi(\mathcal{O}_{H_{n}})$ and signature $\tau\left(
H_{n}\right)  $ are:%
\begin{align}
e(H_{n})  &  =\underset{H_{n}}{\int}c_{2}(H_{n})=\underset{\mathbb{P}^{3}%
}{\int}n(6-4n+n^{2})H^{3}=n^{3}-4n^{2}+6n\label{Eulerchar}\\
\chi(\mathcal{O}_{H_{n}})  &  =\underset{H_{n}}{\int}\frac{c_{1}(H_{n}%
)^{2}+c_{2}(H_{n})}{12}=\frac{1}{6}(n^{3}-6n^{2}+11n)\\
\tau(H_{n})  &  =\underset{H_{n}}{\int}\frac{c_{1}(H_{n})^{2}-2c_{2}(H_{n}%
)}{3}=-\frac{1}{3}(n^{3}-4n)\text{.} \label{signature}%
\end{align}
We next determine the Hodge numbers of $H_{n}$. \ Using the Lefschetz
hyperplane theorem, $h^{1,0}(\mathbb{P}^{3})=0$ implies $h^{1,0}(H_{n})=0$.
\ Moreover, because $e(H_{n})=2+2h^{2,0}+h^{1,1}$ and $\chi(\mathcal{O}%
_{H_{n}})=1-h^{0,1}+h^{0,2}=1+h^{0,2}$, equations (\ref{Eulerchar}%
)-(\ref{signature}) imply:%
\begin{align}
h^{1,1}(H_{n})  &  =\frac{1}{3}(2n^{3}-6n^{2}+7n)\\
h^{2,0}(H_{n})  &  =\frac{1}{6}(n^{3}-6n^{2}+11n)-1\\
b_{2}(H_{n})  &  =n^{3}-4n^{2}+6n-2\text{.}%
\end{align}
The last expression determines the dimension of $H_{2}(H_{n},%
\mathbb{Z}
)$ as a lattice over the integers. \ It follows from Poincar\'{e} duality that
when equipped with the intersection pairing of the geometry, this lattice is
self-dual. \ Moreover, returning to equation (\ref{chernsum}), reduction of
$c_{1}(H_{n})$ mod $2$ implies that $H_{n}$ is spin when $n$ is
even.\footnote{This follows from Wu's theorem and the fact that $H_{n}$ is
simply connected.} \ This in turn implies that the lattice $H_{2}(H_{n},%
\mathbb{Z}
)$ is even (resp. odd) for $n$ even (resp. odd). \ Because the signature and
dimension uniquely determine a lattice with indefinite signature, we conclude
that the lattice is of the general form:%
\begin{align}
H_{2}(H_{n},%
\mathbb{Z}
)  &  \simeq(+1)^{\oplus(b_{2}+\tau)/2}\oplus(-1)^{\oplus(b_{2}-\tau)/2}\text{
\ \ }(n\text{ odd})\\
H_{2}(H_{n},%
\mathbb{Z}
)  &  \simeq(-E_{8})^{\tau/8}\oplus U^{\oplus(b_{2}-\tau)/2}\text{
\ \ \ \ \ \ \ \ \ \ \ \ }(n\text{ even})
\end{align}
where $-E_{8}$ is minus the Cartan matrix for $E_{8}$ and $U$ is the
\textquotedblleft hyperbolic element\textquotedblright\ with entries specified
by the Pauli matrix $\sigma_{x}$. The canonical class has self intersection number:%
\begin{equation}
K_{H_{n}}\cdot K_{H_{n}}=\underset{H_{n}}{\int}c_{1}(H_{n})^{2}=n(n-4)^{2}%
\text{.}%
\end{equation}

For many purposes, it is of practical use to have a large number of
contractible rational curves inside of a given surface which can serve as
matter curves for a given model. \ We note, however, that general results from
the mathematics literature \cite{XUHyperONE,XUHyperTWO} demonstrate that for a
generic hypersurface of degree at least five, the minimal genus of a curve is
at least two. \ Indeed, typically a given homology class only corresponds to a
holomorphic curve for a specific choice of complex structure. \ To avoid such
subtleties, we consider the blowup of a degree $n$ hypersurface at $k$ points,
$B_{k}H_{n}$. \ While the value of $h^{2,0}$ remains invariant under this
process, the canonical class of the resulting space is now given by:%
\begin{equation}
K_{B_{k}H_{n}}=K_{H_{n}}+E_{1}+...+E_{k}\text{.}%
\end{equation}
where the $E_{i}$ denote the effective classes associated with blown up rational curves.

\section{Classification of Breaking Patterns\label{BreakClass}}

In this Appendix we classify all possible breaking patterns via instantons for
a theory defined by a seven-brane filling $%
\mathbb{R}
^{3,1}\times S$ with bulk gauge group $G_{S}$ such that the resulting spectrum
can in principle contain the matter content of the Standard Model. \ While
breaking patterns for GUT groups is certainly a well-studied topic in the
phenomenology literature, as far as we are aware, this question has not been
studied from the perspective of F-theory. \ Indeed, although much of our
analysis in this paper has focussed on the cases where the bulk gauge group is
$SU(5)$ or $SO(10)$, it seems of use for future potential efforts in this
direction to catalogue a broader class of candidate breaking patterns which
could in principle arise from compactifications of F-theory. \ We note that by
appealing to gauge invariance and certain basic phenomenological requirements,
a partial classification of candidate breaking patterns which can appear in
string theory has been given in \cite{TatarClassify}.

Throughout our analysis, we shall assume that our model is generic in the
sense that along complex codimension one and two subspaces, the rank of the
singularity type can enhance by one or two. \ While in this paper we have
focussed on a minimal class of models where the bulk gauge group is
$G_{S}=SU(5)$ or $SO(10)$, there are additional possibilities at higher rank.
For example, in higher rank cases it may be possible to allow some of the
matter fields of the MSSM to originate from bulk zero modes. \ We now proceed
to an analysis of all possible breaking patterns via instantons which can
accommodate the matter content of the MSSM. \ The relevant group theory
material on the decomposition of various irreducible representations may be
found in \cite{Slansky,McKayPatera}.

In keeping with our general philosophy, we shall also assume that the group
corresponding to the rank two enhancement in singularity type is a subgroup of
$E_{8}$. \ For this reason, the rank of the singularity type can be at most
six. \ Moreover, because the Standard Model gauge group has rank four, it is
enough to classify breaking patterns associated with singularities of rank
four, five and six. \ The relevant $ADE$-type of the singularities are
therefore:%
\begin{align}
\text{Rank 4}  &  \text{: \ }A_{4}\text{, }D_{4}\\
\text{Rank 5}  &  \text{: \ }A_{5}\text{, }D_{5}\\
\text{Rank 6}  &  \text{: \ }A_{6}\text{, }D_{6}\text{, }E_{6}\text{.}%
\end{align}
The singularity type does not fully determine the gauge group $G_{S}$. \ When
the collapsed cycles of the singularity type are permuted under a monodromy in
the fiber direction, the resulting gauge group is given by the quotient of the
original simply laced group by an outer automorphism. \ In this way, we can
also obtain all non-simply laced groups such as $SO(2n+1)$, $USp(2n)$, $F_{4}$
and $G_{2}$. \ In what follows we adopt the convention $USp(2)\simeq SU(2)$.
\ It therefore follows that we must analyze the breaking patterns for the
following possibilities:%
\begin{align}
\text{Rank 4}  &  \text{: \ }SU(5)\text{, }SO(8)\text{, }SO(9)\text{, }%
F_{4}\label{rankfour}\\
\text{Rank 5}  &  \text{: \ }SU(6)\text{, }SO(10)\text{, }%
SO(11)\label{rankfive}\\
\text{Rank 6}  &  \text{: \ }SU(7)\text{, }SO(12)\text{, }E_{6}\text{.}
\label{ranksix}%
\end{align}
Note in particular that the bulk gauge group is never of $USp$ type. \ There
are in general many possible ways in which the Standard Model gauge group can
embed in the above gauge groups. \ To classify admissible breaking patterns to
the Standard Model gauge group, we shall require that all of the matter
content of the Standard Model must be present. \ While much of our analysis
will hold for non-supersymmetric theories as well, we shall typically focus on
the field content and interactions of the MSSM. \ In terms of the gauge group
$SU(3)\times SU(2)\times U(1)$, the representation content of the fields of
the MSSM are:%
\begin{equation}%
\begin{tabular}
[c]{|l|l|l|l|l|l|}\hline
$Q$ & $U$ & $D$ & $H_{d},L$ & $E$ & $H_{u}$\\\hline
$(3,2)_{1}$ & $(\overline{3},1)_{-4}$ & $(\overline{3},1)_{2}$ & $(1,2)_{-3}$
& $(1,1)_{6}$ & $(1,2)_{3}$\\\hline
\end{tabular}
\text{.}%
\end{equation}
\ In addition, any realistic model must allow the three superpotential terms:%
\begin{equation}
W\supset QUH_{u}+QDH_{d}+ELH_{d}\text{.}%
\end{equation}

Starting from representations which descend from the decomposition of the
adjoint representation of $E_{8}$, our strategy will be to rule out as many
possible breaking patterns as possible because the representation content is
incorrect, or because gauge invariance in the parent theory forbids a required
superpotential term. \ For $SO$ gauge groups, we assume the matter organizes
into the fundamental, spinor or adjoint representations. \ For $SU$ gauge
groups, we assume that in addition to the adjoint representation, the matter
organizes into one, two or three index anti-symmetric representations.

To classify the possible breaking patterns of a given bulk gauge group $G_{S}%
$, we first list all maximal subgroups. \ Next, we determine all maximal
subgroups of each such subgroup and proceed iteratively until we arrive at the
Standard Model gauge group. We note that even for a unique nested sequence of
subgroups, there may be several distinct subgroups whose commutant contains
the Standard Model gauge group. \ The classification of these possible
subgroups is aided by the fact that the gauge group of the Standard Model has rank 
four so that the corresponding instanton configuration can only take values in
a rank one or two subgroup of a given bulk gauge group $G_{S}$.

Although they cannot serve as a bulk gauge group, it is also convenient to
list all maximal subgroups of some common lower rank groups which appear
frequently. \ The maximal subgroups of $SO(7)$, $SU(4)$, $USp(6)$, $USp(4)$
and $G_{2}$ are:%
\begin{align}
SO(7)  &  \supset SU(4)\label{SO7top}\\
SO(7)  &  \supset SU(2)\times SU(2)\times SU(2)\\
SO(7)  &  \supset USp(4)\times U(1)\\
SO(7)  &  \supset G_{2}\\
SU(4)  &  \supset SU(3)\times U(1)\\
SU(4)  &  \supset SU(2)\times SU(2)\times U(1)\\
SU(4)  &  \supset USp(4)\\
SU(4)  &  \supset SU(2)\times SU(2)\\
USp(6)  &  \supset SU(3)\times U(1)\label{USpsixone}\\
USp(6)  &  \supset SU(2)\times USp(4)\label{USpsixtwo}\\
USp(6)  &  \supset SU(2)\label{USpsixthree}\\
USp(6)  &  \supset SU(2)\times SU(2)\label{USpsixfour}\\
USp(4)  &  \supset SU(2)\times SU(2)\label{USp4one}\\
USp(4)  &  \supset SU(2)\times U(1)\label{USp4two}\\
USp(4)  &  \supset SU(2)\label{USp4three}\\
G_{2}  &  \supset SU(3)\label{G2one}\\
G_{2}  &  \supset SU(2)\times SU(2)\label{G2two}\\
G_{2}  &  \supset SU(2)\text{.} \label{G2bot}%
\end{align}

In the remainder of this Appendix, we classify possible breaking patterns via
instantons of the bulk gauge group. To further specify the order of breaking
in a nested sequence of subgroups, we shall sometimes enclose separate
subgroup factors in square brackets.

\subsection{Rank Four}

We now classify all breaking patterns of rank four groups. \ Although $SU(5)$
is the only group of line (\ref{rankfour}) which contains complex
representations, for our higher dimensional theories, it is a priori possible
that a suitable $U(1)$ field strength in either the compact or non-compact
directions of an intersecting seven-brane theory can induce a net chirality in
the resulting gauge group. \ In the rank four case we list all maximal
subgroups even if they do not contain the Standard Model gauge group. \ This
is done because for the higher rank cases, such breaking patterns may become
available. \ In the rank four case, we find that only $G_{S}=SU(5)$ is a
viable possibility.

\subsubsection{$SU(5)$}

There is a single maximal subgroup of $SU(5)$ which contains the Standard
Model gauge group. \ Indeed, the representation content is given by the
Georgi-Glashow model:%
\begin{align}
SU(5)  &  \supset SU(3)_{C}\times SU(2)_{L}\times U(1)_{Y}\equiv G_{std}\\
5  &  \rightarrow(1,2)_{3}+(3,1)_{-2}\\
10  &  \rightarrow(1,1)_{6}+(\overline{3},1)_{-4}+(3,2)_{1}\\
24  &  \rightarrow(1,1)_{0}+(1,3)_{0}+(3,2)_{-5}+(\overline{3},2)_{5}%
+(8,1)_{0}\text{.}%
\end{align}
By turning on an instanton in $U(1)_{Y}$, we break to the desired gauge group.

\subsubsection{$SO(8)$}

We now proceed to the case of $SO(8)$. \ The maximal subgroups of $SO(8)$ are
\cite{Slansky}:%
\begin{align}
SO(8)  &  \supset SU(2)\times SU(2)\times SU(2)\times SU(2)\\
SO(8)  &  \supset SU(4)\times U(1)\\
SO(8)  &  \supset SU(3)\\
SO(8)  &  \supset SO(7)\\
SO(8)  &  \supset SU(2)\times USp(4)\text{.}%
\end{align}
Returning to lines (\ref{SO7top}-\ref{G2bot}), it follows that there does not
exist a breaking pattern which yields $G_{std}$.

\subsubsection{$SO(9)$\label{SectionSO9}}

The maximal subgroups of $SO(9)$ are:%
\begin{align}
SO(9)  &  \supset SO(8)\label{firstSO9}\\
SO(9)  &  \supset SU(2)\times SU(2)\times USp(4)\\
SO(9)  &  \supset SU(2)\times SU(4)\label{thirdSO9}\\
SO(9)  &  \supset SU(2)\\
SO(9)  &  \supset SU(2)\times SU(2)\text{.} \label{lastSO9}%
\end{align}
Of the above possibilities, only line (\ref{thirdSO9}) contains $G_{std}$.
\ Breaking to $G_{std}$ via a $U(1)$ instanton yields:%
\begin{align}
SO(9)  &  \supset SU(2)\times SU(4)\supset SU(2)\times\left[  SU(3)\times
\left[  U(1)\right]  \right] \\
9  &  \rightarrow(3,1)_{0}+(1,3)_{2}+(1,\overline{3})_{-2}\\
16  &  \rightarrow(2,1)_{3}+(2,3)_{-1}+(2,1)_{-3}+(2,\overline{3})_{+1}\\
36  &  \rightarrow(3,1)_{0}+(1,1)_{0}+(1,3)_{-4}+(1,\overline{3}%
)_{4}+(1,8)_{0}+(3,3)_{2}+(3,\overline{3})_{-2}\text{.}%
\end{align}
By inspection, all singlets of $SU(2)\times SU(3)$ are also neutral under the
$U(1)$ factor. \ It thus follows that $SO(9)$ is ruled out as a candidate.

\subsubsection{$F_{4}$}

The maximal subgroups of $F_{4}$ are:%
\begin{align}
F_{4}  &  \supset SO(9)\\
F_{4}  &  \supset SU(3)\times SU(3)\label{F4two}\\
F_{4}  &  \supset SU(2)\times USp(6)\label{F4three}\\
F_{4}  &  \supset SU(2)\\
F_{4}  &  \supset SU(2)\times G_{2}%
\end{align}
the first case is excluded by the previous analysis of $SO(9)$, leaving only
lines (\ref{F4two}) and (\ref{F4three}).

First consider the breaking pattern of (\ref{F4two}):%
\begin{align}
F_{4}  &  \supset SU(3)_{1}\times SU(3)_{2}\\
26  &  \rightarrow(8,1)+(3,3)+(\overline{3},\overline{3})\\
52  &  \rightarrow(8,1)+(1,8)+(6,\overline{3})+(\overline{6},3)\text{.}%
\end{align}
Breaking either factor of $SU(3)\supset SU(2)\times U(1)$ via a $U(1)$
instanton, we note that all resulting $SU(3)\times SU(2)$ singlets are also
neutral under $U(1)$. \ We therefore conclude that the breaking pattern of
line (\ref{F4two}) is also excluded.

Next consider the remaining breaking pattern of (\ref{F4three}) which can
descend to the Standard Model gauge group:%
\begin{align}
F_{4}  &  \supset SU(2)\times USp(6)\supset SU(2)\times\lbrack SU(3)\times
U(1)]\\
26  &  \rightarrow(2,3)_{1}+(2,\overline{3})_{-1}+(2,3)_{-2}+(2,\overline
{3})_{2}+(2,8)_{0}\\
52  &  \rightarrow(3,1)_{0}+(1,1)_{0}+(1,6)_{2}+(1,\overline{6})_{-2}%
+(1,8)_{0}\\
&  +(2,1)_{3}+(2,1)_{-3}+(2,6)_{-1}+(2,\overline{6})_{1}\text{.}%
\end{align}
As before, the resulting singlets of the non-abelian factor are also neutral
under the $U(1)$ factor. \ Summarizing, we find that the only available rank
four bulk gauge group which can contain the Standard Model is $SU(5)$.

\subsection{Rank Five}

We now proceed to rank five bulk gauge groups. \ While it is in principle
possible that an $SU(2)$ instanton configuration could produce a consistent
breaking pattern to the particle content of the Standard Model, we find that
in all cases, the relevant breaking pattern is again always an instanton
configuration with structure group $U(1)$ or $U(1)\times U(1)$.

\subsubsection{$SU(6)$}

We assume that the matter content organizes into the representations $6$,
$15$, $20$ and $35$ of $SU(6)$, as well as their dual representations. \ The
maximal subgroups of $SU(6)$ are:%
\begin{align}
SU(6)  &  \supset SU(5)\times U(1)\label{SU6one}\\
SU(6)  &  \supset SU(2)\times SU(4)\times U(1)\label{SU6two}\\
SU(6)  &  \supset SU(3)\times SU(3)\times U(1)\label{SU6three}\\
SU(6)  &  \supset SU(3)\\
SU(6)  &  \supset SU(4)\\
SU(6)  &  \supset USp(6)\\
SU(6)  &  \supset SU(2)\times SU(3) \label{SU6last}%
\end{align}
of which only the first three contain $G_{std}$. \ By inspection, it now
follows that for $n\geq2$, an $SU(n)$ instanton will break too much of the
gauge group to preserve $G_{std}$. \ Moreover, it follows from lines
(\ref{SU6one})-(\ref{SU6three}) that up to linear combinations of the $U(1)$
charge for the other breaking patterns, it is enough to analyze the $U(1)^{2}$
instanton configuration which breaks $SU(6)$ via the nested sequence
$SU(6)\supset SU(5)\times U(1)\supset G_{std}\times U(1)$. \ Restricting to
$U(1)^{2}$ valued instanton configurations, the decomposition of the one two
and three index anti-symmetric and adjoint representations of $SU(6)$ are:%
\begin{align}
SU(6)  &  \supset SU(5)\times\lbrack U(1)]\supset\lbrack SU(3)\times
SU(2)\times\lbrack U(1)]]\times\lbrack U(1)]\label{SU6decompone}\\
6  &  \rightarrow(1,1)_{0,5}+(3,1)_{-2,-1}+(1,2)_{3,-1}\\
15  &  \rightarrow(1,2)_{3,-4}+(3,1)_{-2,-4}+(1,1)_{6,2}\\
&  +(\overline{3},1)_{-4,-3}+(3,2)_{1,2}\\
20  &  \rightarrow(1,1)_{6,-3}+(\overline{3},1)_{-4,-3}+(3,2)_{1,-3}\\
&  +(1,1)_{-6,3}+(3,1)_{4,3}+(\overline{3},2)_{-1,3}\\
35  &  \rightarrow(1,1)_{0}+(1,2)_{3,6}+(3,1)_{-2,6}\\
&  +(1,2)_{-3,-6}+(\overline{3},1)_{2,-6}+(1,1)_{0,0}+(1,3)_{0,0}\\
&  +(3,2)_{-5,0}+(\overline{3},2)_{5,0}+(8,1)_{0,0}\text{.}%
\end{align}
The above decomposition illustrates the fact that there are a priori different
ways in which the representation content of the MSSM\ can be packaged into
higher dimensional representations.

We now determine all possible choices consistent with obtaining the correct
spectrum and interaction terms. \ We first require that at least one linear
combination of the $U(1)$ charges may be identified with $U(1)_{Y}$ of the
Standard Model. \ Labeling the $U(1)$ charges as $a$ and $b$, this implies
that the charges of the MSSM fields must satisfy the relations:%
\begin{align}
E  &  :5b=\pm6\text{ or }6a+2b=\pm6\text{ or }6a-3b=\pm6\\
Q  &  :a+2b=\pm1\text{ or }a-3b=\pm1\text{ or }-5a=\pm1\text{ }\\
U  &  :2a+b=-4\text{ or }2a+4b=-4\text{ or }-4a-3b=-4\text{ or }2a-6b=-4\\
D  &  :2a+b=2\text{ or }2a+4b=2\text{ or }-4a-3b=2\text{ or }2a-6b=2\\
H_{d},L  &  :3a-b=\pm3\text{ or }3a+6b=\pm3\\
H_{u}  &  :3a-b=\pm3\text{ or }3a+6b=\pm3\text{.}
\end{align}

First suppose that the $E$-relation $5b=\pm6$ holds. \ In this case, the
remaining candidate solutions for $a$ are:%
\begin{align}
Q  &  \Longrightarrow\pm a=\frac{17}{5}\text{ or }\frac{7}{5}\text{ or }%
\frac{23}{5}\text{ or }\frac{13}{5}\text{ or }\frac{1}{5}\\
L  &  \Longrightarrow\pm a=\frac{7}{5}\text{ or }\frac{3}{5}\text{ or }%
\frac{17}{5}\\
D  &  \Longrightarrow a=\frac{8}{5}\text{ or }\frac{2}{5}\text{ or }-\frac
{7}{5}\text{ or }\frac{17}{5}\text{ or }-\frac{7}{5}\text{ or }\frac{4}%
{5}\text{ or }\frac{23\text{ }}{5}\text{ or }-\frac{13}{5}\\
U  &  \Longrightarrow a=-\frac{13}{5}\text{ or }-\frac{7}{5}\text{ or}%
-\frac{22}{5}\text{ or }\frac{2}{5}\text{ or }\frac{1}{10}\text{ or }\frac
{19}{10}\text{ or }\frac{8}{5}\text{ or}-\frac{28}{5}
\end{align}
so that the only common solution to all of the above conditions requires
$a=-7/5$. \ Note, however, that this is an inconsistent assignment because
whereas the $U$ condition requires $a=-7/5$ and $b=-6/5$, the $Q$ condition
requires $a=-7/5$ and $b=+6/5$.

Next suppose that the $E$-relation $6a+2b=\pm6$ holds. \ In this case, $b$ is
now determined by the relations:%
\begin{align}
Q  &  \Longrightarrow b=\pm\frac{6}{5}\text{ or }\pm\frac{3}{5}\text{ or }%
\pm\frac{18}{5}\text{ or }\pm\frac{12}{5}\text{ or }0\\
U  &  \Longrightarrow b=-18\text{ or }-6\text{ or }-\frac{9}{5}\text{ or
}-\frac{3}{5}\text{ or }\frac{24}{5}\text{ or }\frac{9}{10}\text{ or }\frac
{3}{10}\text{ or }0\text{.}%
\end{align}
It thus follows that in this case that the only consistent choice of
$U(1)_{Y}$ requires $b=0$. \ Note that in this case the $U(1)$ charge
assignments match to those of the $SU(5)$ GUT.

Finally, suppose that the $E$-relation $6a-3b=\pm6$ holds. \ In this case, $b$
is now determined by the possible $Q$-relations to be:%
\begin{align}
Q  &  \Longrightarrow b=\pm\frac{4}{5}\text{ or }\pm\frac{12}{5}\text{ or }%
\pm\frac{8}{5}\text{ or }0\\
U  &  \Longrightarrow b=-3\text{ or }-1\text{ or }\pm\frac{6}{5}\text{ or }%
\pm\frac{2}{5}\text{ or }\frac{8}{5}\text{ or }0\\
D  &  \Longrightarrow b=2\text{ or }\pm\frac{4}{5}\text{ or }-\frac{6}%
{5}\text{ or }0
\end{align}
so that the only consistent solution requires $b=0$, as before.

\subsubsection{$SO(10)$}

We assume that the matter content organizes into the representations $10$,
$16$, $\overline{16}$ and $45$ of $SO(10)$. \ The maximal subgroups of
$SO(10)$ which contain $G_{std}$ are:%
\begin{align}
SO(10)  &  \supset SU(5)\times U(1)\label{SO10one}\\
SO(10)  &  \supset SU(2)\times SU(2)\times SU(4)\label{SO10two}\\
SO(10)  &  \supset SO(9)\label{SO10three}\\
SO(10)  &  \supset SU(2)\times SO(7)\label{SO10four}\\
SO(10)  &  \supset SO(8)\times U(1)\\
SO(10)  &  \supset USp(4)\\
SO(10)  &  \supset USp(4)\times USp(4)\text{.} \label{SO10last}%
\end{align}
Of the above maximal subgroups, only the first four contain $SU(3)\times
SU(2)$ as a subgroup. \ Whereas lines (\ref{SO10one}) and (\ref{SO10two}) lead
to well-known GUTs, the maximal subgroups of lines (\ref{SO10three}) and
(\ref{SO10four}) are typically not treated in the GUT\ literature.

We now demonstrate that no breaking pattern of the latter two cases can yield
the MSSM\ spectrum. \ In the case $SO(10)\supset SO(9)$, the $10$, $16$,
$\overline{16}$ and $45$ of $SO(10)$ descend to the $9$, $16$, and $36$ of
$SO(9)$. \ It now follows from the analysis of subsection \ref{SectionSO9}
that no breaking pattern will yield the matter content of the Standard Model.

Next consider the maximal subgroup $SU(2)\times SO(7)$. \ Because there is
only one maximal subgroup of $SO(7)$ which contains $SU(3)$, the unique
candidate breaking pattern in this case is:%
\begin{align}
SO(10)  &  \supset SU(2)\times SO(7)\supset SU(2)\times SU(4)\supset
SU(2)\times\lbrack SU(3)\times\lbrack U(1)]]\\
10  &  \rightarrow(3,1)_{0}+(1,1)_{0}+(1,3)_{2}+(1,\overline{3})_{-2}\\
16  &  \rightarrow(2,1)_{3}+(2,3)_{-1}+(2,1)_{-3}+(2,\overline{3})_{1}\\
45  &  \rightarrow(3,1)_{0}+(1,3)_{2}+(1,\overline{3})_{-2}+(1,1)_{0}%
+(1,3)_{-4}+(1,\overline{3})_{4}\text{.}
\end{align}
By inspection, we note that all singlets of $SU(3)\times SU(2)$ are also
neutral under the $U(1)$ factor. \ We therefore conclude that such a breaking
pattern cannot include $E$-fields.

We now analyze breaking patterns of the two remaining cases of lines
(\ref{SO10one}) and (\ref{SO10two}) which are both well-known in the
GUT\ literature. \ In the present context, we wish to determine whether a
non-standard embedding of the fields in an $SO(10)$ representation could also
be consistent with the field content of the MSSM.

\paragraph{$SO(10)\supset SU(5)\times U(1)$}

Consider first the maximal subgroup $SU(5)\times U(1)$. \ In this case, the
unique nested sequence of maximal subgroups which contains the gauge group
$G_{std}$ is:%
\begin{align}
SO(10)  &  \supset SU(5)\times\lbrack U(1)]\supset SU(3)\times SU(2)\times
\lbrack U(1)_{a}]\times\lbrack U(1)_{b}]\\
10  &  \rightarrow(1,2)_{3,2}+(3,1)_{-2,2}+(1,2)_{-3,-2}+(\overline
{3},1)_{2,-2}\\
16  &  \rightarrow(1,1)_{0,-5}+(1,2)_{-3,3}+(\overline{3},1)_{2,3}%
+(1,1)_{6,-1}\\
&  +(\overline{3},1)_{-4,-1}+(3,2)_{1,-1}\\
45  &  \rightarrow(1,1)_{0}+(1,1)_{6,4}+(\overline{3},1)_{-4,4}+(3,2)_{1,4}\\
&  +(1,1)_{-6,-4}+(3,1)_{4,-4}+(\overline{3},2)_{-1,-4}+(1,1)_{0,0}\\
&  +(1,3)_{0,0}+(8,1)_{0,0}+(3,2)_{-5,0}+(\overline{3},2)_{5,0}\text{.}%
\end{align}
As usual, we require that at least one linear combination of the $U(1)$
charges may be identified with $U(1)_{Y}$ of the Standard Model and that all
of the necessary interaction terms of the MSSM\ are present. \ We begin by
classifying all possible combinations of $Q$-, $U$- and $D$-fields which can
yield the gauge invariant combination $QUH_{u}$:%
\begin{equation}%
\begin{tabular}
[c]{|l|l|l|l|l|}\hline
& $Q$ & $U$ & $H_{u}$ & $(a,b)$\\\hline
$1$ & $(3,2)_{1,-1}$ & $(\overline{3},1)_{2,-2}$ & $(2,1)_{-3,3}$ &
$OUT$\\\hline
$2$ & $(3,2)_{1,-1}$ & $(\overline{3},1)_{2,3}$ & $(2,1)_{-3,-2}$ &
$(-1/5,-6/5)$\\\hline
$3$ & $(3,2)_{1,-1}$ & $(\overline{3},1)_{-4,-1}$ & $(2,1)_{3,2}$ &
$(1,0)$\\\hline
$4$ & $(3,2)_{1,-1}$ & $(\overline{3},1)_{-4,4}$ & $(2,1)_{3,-3}$ &
$(1,0)$\\\hline
$5$ & $(3,2)_{1,4}$ & $(\overline{3},1)_{2,-2}$ & $(2,1)_{-3,-2}$ &
$(-7/5,3/5)$\\\hline
$6$ & $(3,2)_{1,4}$ & $(\overline{3},1)_{2,3}$ & $OUT$ & $OUT$\\\hline
$7$ & $(3,2)_{1,4}$ & $(\overline{3},1)_{-4,-1}$ & $(2,1)_{3,-3}$ &
$(1,0)$\\\hline
$8$ & $(3,2)_{1,4}$ & $(\overline{3},1)_{-4,4}$ & $OUT$ & $OUT$\\\hline
$9$ & $(3,2)_{-5,0}$ & $(\overline{3},1)_{2,-2}$ & $(2,1)_{3,2}$ &
$(-1/5,9/5)$\\\hline
$10$ & $(3,2)_{-5,0}$ & $(\overline{3},1)_{2,3}$ & $(2,1)_{3,-3}$ &
$(-1/5,-6/5)$\\\hline
$11$ & $(3,2)_{-5,0}$ & $(\overline{3},1)_{-4,-1}$ & $OUT$ & $OUT$\\\hline
$12$ & $(3,2)_{-5,0}$ & $(\overline{3},1)_{-4,4}$ & $OUT$ & $OUT$\\\hline
\end{tabular}
\end{equation}
In the above list, entries in the $H_{u}$ column listed by \textquotedblleft%
$OUT$\textquotedblright\ indicate that of the available representations, no
choice yields a gauge invariant quantity in the parent theory. \ Similarly, an
\textquotedblleft$OUT$\textquotedblright\ entry in the $(a,b)$ column
indicates that no consistent solution of $U(1)_{Y}$ exists in this case. \ We
next require that a consistent choice of representation for $D$ and $H_{d}$ to
admit the interaction $QDH_{d}$ also exists amongst the remaining
possibilities:%
\begin{equation}%
\begin{tabular}
[c]{|l|l|l|l|l|}\hline
& $Q$ & $D$ & $H_{d}$ & $(a,b)$\\\hline
$2$ & $(3,2)_{1,-1}$ & $(\overline{3},1)_{-4,-1}$ & $(2,1)_{3,2}$ &
$(-1/5,-6/5)$\\\hline
$3$ & $(3,2)_{1,-1}$ & $(\overline{3},1)_{2,3}$ & $(2,1)_{-3,-2}$ &
$(1,0)$\\\hline
$4$ & $(3,2)_{1,-1}$ & $(\overline{3},1)_{2,-2}$ & $(2,1)_{-3,3}$ &
$(1,0)$\\\hline
$5$ & $(3,2)_{1,4}$ & $OUT$ & $(2,1)_{3,2}$ & $(-7/5,3/5)$\\\hline
$7$ & $(3,2)_{1,4}$ & $OUT$ & $(2,1)_{-3,3}$ & $(1,0)$\\\hline
$9$ & $(3,2)_{-5,0}$ & $OUT$ & $(2,1)_{-3,-2}$ & $(-1/5,9/5)$\\\hline
$10$ & $(3,2)_{-5,0}$ & $OUT$ & $(2,1)_{-3,3}$ & $(-1/5,-6/5)$\\\hline
\end{tabular}
\ \text{.}%
\end{equation}
Of the three remaining possibilities, we next require that the interaction
term $ELH_{d}$ be present:%
\begin{equation}%
\begin{tabular}
[c]{|l|l|l|l|l|}\hline
& $E$ & $L$ & $H_{d}$ & $(a,b)$\\\hline
$2a$ & $(1,1)_{0,-5}$ & $(1,2)_{-3,3}$ & $(2,1)_{3,2}$ & $(-1/5,-6/5)$\\\hline
$2b$ & $(1,1)_{-6,-4}$ & $(1,2)_{3,2}$ & $(2,1)_{3,2}$ & $(-1/5,-6/5)$\\\hline
$3a$ & $(1,1)_{6,-1}$ & $(1,2)_{-3,-2}$ & $(2,1)_{-3,-2}$ & $(1,0)$\\\hline
$3b$ & $(1,1)_{6,4}$ & $(1,2)_{-3,3}$ & $(2,1)_{-3,-2}$ & $(1,0)$\\\hline
$4a$ & $(1,1)_{6,-1}$ & $(1,2)_{-3,-2}$ & $(2,1)_{-3,3}$ & $(1,0)$\\\hline
$4b$ & $(1,1)_{6,4}$ & $OUT$ & $(2,1)_{-3,3}$ & $(1,0)$\\\hline
\end{tabular}
\ \text{.}%
\end{equation}
We therefore conclude that there are in fact five distinct ways in which the
field content of the MSSM\ can be packaged in representations of $SO(10)$.
\ We note in particular that in some cases, the chiral matter of the
MSSM\ does not descend from either of the spinor representations of $SO(10)$.
\ The above classification can also be obtained without imposing the condition
that non-trivial interaction terms be present in the superpotential. \ Indeed,
by listing all possible consistent choices of $U(1)$ charge assignments, we
arrive at the same list of admissible configurations. \ Finally, we note that
the choice $b=0$ corresponds to the breaking pattern where $U(1)_{Y}$ embeds
in $SU(5)$ and the other consistent choice corresponds to the flipped
embedding of hypercharge \cite{BarrFlipped}.

\paragraph{$SO(10)\supset SU(2)\times SU(2)\times SU(4)$}

We next analyze the other nested sequence of maximal subgroups given by
decomposing $SO(10)$ as:%
\begin{align}
SO(10)  &  \supset SU(2)\times SU(2)\times SU(4)\supset SU(2)\times
SU(2)\times\lbrack SU(3)\times U(1)]\\
10  &  \rightarrow(2,2,1)_{0}+(1,1,3)_{2}+(1,1,\overline{3})_{-2}\\
16  &  \rightarrow(2,1,1)_{3}+(2,1,3)_{-1}+(2,1,1)_{-3}+(2,1,\overline{3}%
)_{1}\\
45  &  \rightarrow(3,1,1)_{0}+(1,3,1)_{0}+(1,1,1)_{0}+(1,1,3)_{-4}\\
&  \text{ \ \ \ }+(1,1,\overline{3})_{4}+(1,1,8)_{0}+(2,2,3)_{2}%
+(2,2,\overline{3})_{-2}\text{.}%
\end{align}
While an $SU(2)$ instanton configuration can indeed yield the gauge group
$G_{std}$, we note that the putative $U(1)_{Y}$ would then be incorrect. \ It
thus follows that it is enough to consider $U(1)\times U(1)$ instanton
configurations. \ Because the representation content of this decomposition is
identical to that of the previous case, we conclude that there are again two
possible ways to package the MSSM\ fields into $SO(10)$ representations.

\subsubsection{$SO(11)$}

We assume that the matter content organizes into the representations $11$,
$32$ and $55$ of $SO(11)$. \ The maximal subgroups of $SO(11)$ are:%
\begin{align}
SO(11)  &  \supset SO(10)\\
SO(11)  &  \supset SU(2)\times SO(8)\\
SO(11)  &  \supset USp(4)\times SU(4)\\
SO(11)  &  \supset SU(2)\times SU(2)\times SO(7)\\
SO(11)  &  \supset SO(9)\times U(1)\\
SO(11)  &  \supset SU(2)
\end{align}
so that only the first five maximal subgroups contain $G_{std}$.

\paragraph{$SO(11)\supset SO(10)$}

In the case $SO(11)\supset SO(10)$, the representations of $SO(11)$ decompose
as:%
\begin{align}
SO(11)  &  \supset SO(10)\\
11  &  \rightarrow1+10\\
32  &  \rightarrow16+\overline{16}\\
55  &  \rightarrow10+45
\end{align}
so that all of the analysis of breaking patterns performed for $SO(10)$
carries over to this case as well. \ In this case, it less clear whether the
resulting matter spectrum can be chiral, but all matter fields of the
MSSM\ can indeed be present.

\paragraph{$SO(11)\supset SU(2)\times SO(8)$}

In the case $SO(11)\supset SU(2)\times SO(8)$, the representation content of
$SO(11)$ decomposes as:%
\begin{align}
SO(11)  &  \supset SU(2)\times SO(8)\\
11  &  \rightarrow(3,1)+(1,8^{v})\\
32  &  \rightarrow(2,8^{s})+(2,8^{c})\\
55  &  \rightarrow(3,1)+(1,28)+(3,8^{v})\text{.}%
\end{align}
The two maximal subgroups of $SO(8)$ which contain an $SU(3)$ factor are
$SU(4)\times U(1)$ and $SO(7)\supset SU(4)$.

\subparagraph{$SO(11)\supset SU(2)\times SO(8)\supset SU(2)\times\lbrack
SU(4)\times\lbrack U(1)]]$}

The decomposition to $SU(2)\times\lbrack SU(4)\times\lbrack U(1)]]$ is:%
\begin{align}
SO(11)  &  \supset SU(2)\times SO(8)\supset SU(2)\times\lbrack SU(4)\times
\lbrack U(1)]]\\
11  &  \rightarrow(3,1)_{0}+(1,1)_{2}+(1,1)_{-2}+(1,6)_{0}\\
32  &  \rightarrow(2,8^{s})+(2,8^{c})\rightarrow(2,4)_{1}+(2,\overline
{4})_{-1}+(2,4)_{-1}\\
&  +(2,\overline{4})_{1}\\
55  &  \rightarrow(3,1)_{0}+(1,1)_{0}+(1,6)_{2}+(1,6)_{-2}+(1,15)_{0}\\
&  +(3,1)_{2}+(3,1)_{-2}+(3,6)_{0}%
\end{align}
so that the decomposition to $G_{std}\times U(1)$ along this path is:%
\begin{align}
SO(11)  &  \supset SU(2)\times\lbrack SU(3)\times\lbrack U(1)]_{a}%
\times\lbrack U(1)]_{b}]\\
11  &  \rightarrow(3,1)_{0,0}+(1,1)_{0,2}+(1,1)_{0,-2}+(1,3)_{2,0}%
+(1,\overline{3})_{-2,0}\\
32  &  \rightarrow(2,1)_{3,1}+(2,3)_{-1,1}+(2,1)_{-3,-1}+(2,\overline
{3})_{1,-1}\\
&  +(2,1)_{3,-1}+(2,3)_{-1,-1}+(2,1)_{-3,1}+(2,\overline{3})_{1,1}\\
55  &  \rightarrow(3,1)_{0,0}+(1,1)_{0,0}+(1,3)_{2,2}+(1,\overline{3}%
)_{-2,2}\\
&  +(1,3)_{2,-2}+(1,\overline{3})_{-2,-2}+(1,1)_{0,0}+(1,3)_{-4,0}\\
&  +(1,\overline{3})_{4,0}+(1,8)_{0,0}+(3,1)_{0,2}+(3,1)_{0,-2}\\
&  +(3,3)_{2,0}+(3,\overline{3})_{-2,0}\text{.}%
\end{align}
In order to achieve the correct $U(1)_{Y}$ charge assignment for the
$E$-fields and $Q$-fields, we require:%
\begin{align}
2b  &  =\pm6\\
-a\pm b  &  =1
\end{align}
so that:%
\begin{align}
b  &  =\pm3\\
a  &  =-4\text{ or }2\text{.}%
\end{align}
In order to achieve the correct $U(1)_{Y}$ charge assignment for the
$L$-fields, we must also require:%
\begin{equation}
\pm3a\pm b=\pm3
\end{equation}
so that $a=2$ and without loss of generality, we may choose a sign convention
for $b$ so that $b=3$. \ In this case, the candidate representations for $Q$,
$D$ and $H_{d}$ are:%
\begin{equation}%
\begin{tabular}
[c]{|l|l|l|}\hline
$Q$ & $D$ & $H_{d}$\\\hline
$(2,3)_{-1,1}$ & $(1,\overline{3})_{-2,2}$ & $(2,1)_{-3,-1}$\\\hline
\end{tabular}
\end{equation}
so that the product $QDH_{d}$ is not neutral under $U(1)_{a}$. \ We therefore
conclude that this breaking pattern cannot yield the spectrum of the Standard Model.

\subparagraph{$SO(11)\supset SU(2)\times SO(7)\supset SU(2)\times SU(4)\supset
SU(2)\times SU(3)\times U(1)$}

In this case, breaking to $G_{std}$ proceeds via the nested sequence:%
\begin{align}
SO(11)  &  \supset SU(2)\times SO(8)\supset SU(2)\times SO(7)\\
&  \supset SU(2)\times SU(4)\supset SU(2)\times SU(3)\times U(1)\\
11  &  \rightarrow(3,1)_{0}+(1,1)_{0}+(1,1)_{0}+(1,3)_{2}+(1,\overline
{3})_{-2}\\
32  &  \rightarrow(2,1)_{3}+(2,3)_{-1}+(2,1)_{-3}+(2,\overline{3}%
)_{1}+(2,1)_{3}\\
&  +(2,3)_{-1}+(2,1)_{-3}+(2,\overline{3})_{1}\\
55  &  \rightarrow(3,1)_{0}+(1,1)_{0}+(1,3)_{2}+(1,\overline{3})_{-2}%
+(1,3)_{2}\\
&  +(1,\overline{3})_{-2}+(1,1)_{0}+(1,3)_{-4}+(1,\overline{3})_{4}%
+(1,8)_{0}\\
&  +(3,1)_{0}+(3,1)_{0}+(3,3)_{2}+(3,\overline{3})_{-2}\text{.}
\end{align}
By inspection, the above decomposition does not contain any $E$-fields. \ We
therefore conclude that in all cases, breaking patterns of $SO(11)$ with
maximal subgroup $SU(2)\times SO(8)$ cannot contain $G_{std}$.

\paragraph{$SO(11)\supset USp(4)\times SU(4)$}

Because $USp(4)$ does not contain $SU(3)$ as a subgroup, it follows that in
this case, $SU(4)$ must decompose to $SU(3)\times U(1)$. \ The decomposition
must therefore proceed via the path:%
\begin{align}
SO(11)  &  \supset USp(4)\times SU(4)\supset USp(4)\times SU(3)\times U(1)\\
11  &  \rightarrow(5,1)+(1,6)\rightarrow(5,1)_{0}+(1,3)_{2}+(1,\overline
{3})_{-2}\\
32  &  \rightarrow(4,4)+(4,\overline{4})\rightarrow(4,1)_{3}+(4,3)_{-1}%
+(4,1)_{-3}\\
55  &  \rightarrow(10,1)+(1,15)+(5,6)\rightarrow(10,1)_{0}+(1,1)_{0}\\
&  +(1,3)_{-4}+(1,\overline{3})_{4}+(1,8)_{0}\text{.}%
\end{align}
To proceed further, we specify a maximal subgroup of $USp(4)$ among the ones
listed in lines (\ref{USp4one})-(\ref{USp4three}). \ Because a given instanton
configuration must preserve the non-abelian factor $SU(3)\times SU(2)$ of the
$G_{std}$, we conclude that only the first two are viable breaking patterns.

\subparagraph{$SO(11)\supset USp(4)\times SU(4)\supset USp(4)\times
SU(3)\times U(1)\supset\lbrack SU(2)\times SU(2)]\times\lbrack SU(3)\times
U(1)]$}

In this case, the decomposition of the matter content contains the
representation content of the breaking pattern $SO(10)\supset SU(2)\times
SU(2)\times SU(4)$. \ Explicitly:%
\begin{align}
SO(11)  &  \supset USp(4)\times SU(4)\supset\lbrack SU(2)\times SU(2)]\times
\lbrack SU(3)\times\lbrack U(1)]]\\
11  &  \rightarrow(1,1,1)_{0}+(2,2,1)_{0}+(1,1,3)_{2}+(1,1,\overline{3}%
)_{-2}\\
32  &  \rightarrow(2,1,1)_{3}+(2,1,3)_{-1}+(1,2,1)_{3}+(2,1,3)_{-1}\\
&  +(2,1,1)_{-3}+(2,1,\overline{3})_{1}+(1,2,1)_{-3}+(2,1,\overline{3})_{1}\\
55  &  \rightarrow(3,1,1)_{0}+(1,1,1)_{0}+(1,1,3)_{-4}+(1,1,\overline{3}%
)_{4}\\
&  +(1,1,8)_{0}+(1,1,3)_{2}+(1,1,\overline{3})_{-2}+(2,2,3)_{2}\\
&  +(2,2,\overline{3})_{-2}+(1,3,1)_{0}+(2,2,1)_{0}\text{.}%
\end{align}
It follows that the analysis of breaking patterns for $SO(10)$ directly
carries over to this case as well.

$SO(11)\supset USp(4)\times SU(3)\times U(1)\supset\lbrack SU(2)\times
U(1)]\times\lbrack SU(3)\times U(1)]$

While this is seemingly quite similar to the breaking pattern described
previously, we now show that the embedding of the $U(1)$ factor in $USp(4)$
does not admit an embedding of the matter content of the Standard Model. \ To
this end, we first decompose $SO(11)$ via:%
\begin{align}
SO(11)  &  \supset USp(4)\times SU(4)\supset\lbrack SU(2)\times\lbrack
U(1)]_{a}]\times\lbrack SU(3)\times\lbrack U(1)]_{b}]\\
11  &  \rightarrow(1,1)_{2,0}+(1,1)_{-2,0}+(3,1)_{0,0}+(1,3)_{0,2}%
+(1,\overline{3})_{0,-2}\\
32  &  \rightarrow(2,1)_{1,3}+(2,3)_{1,-1}+(2,1)_{-1,3}+(2,3)_{-1,-1}%
+(2,1)_{-1,-3}\\
&  +(2,\overline{3})_{-1,1}+(2,1)_{1,-3}+(2,\overline{3})_{1,1}\\
55  &  \rightarrow(1,1)_{0,0}+(3,1)_{0,0}+(3,1)_{2,0}+(3,1)_{-2,0}%
+(1,1)_{0,0}\\
&  +(1,3)_{0,-4}+(1,\overline{3})_{0,4}+(1,8)_{0,0}+(1,3)_{2,2}+(1,\overline
{3})_{2,-2}\\
&  +(1,\overline{3})_{-2,-2}+(1,3)_{-2,2}+(3,3)_{0,2}+(3,\overline{3}%
)_{0,-2}\text{.}%
\end{align}
It follows from the above decomposition that the $E$-fields correspond to the
representation $(1,1)_{\pm2,0}$ of the above decomposition. \ It thus follows
that $a=\pm3$. \ Because the $Q$-fields correspond to the representation
$(2,3)_{\pm1,-1}$ and the $L$ fields correspond to the representation
$(2,1)_{\pm1,\pm3}$, we conclude that without loss of generality, fixing the
sign of $a$ to be positive so that $a=+3$, there is a unique linear
combination of $U(1)$ charges so that $a=3$ and $b=2$. \ The field content of
the MSSM\ thus descends from the above representations as:%
\begin{equation}%
\begin{tabular}
[c]{|l|l|l|l|l|l|l|}\hline
$E$ & $Q$ & $U$ & $D$ & $L$ & $H_{u}$ & $H_{d}$\\\hline
$(1,1)_{2,0}$ & $(2,3)_{1,-1}$ & $(1,\overline{3})_{0,-2}$ & $(1,\overline
{3})_{2,-2}$ & $(2,1)_{1,-3}$ & $(2,1)_{-1,3}$ & $(2,1)_{1,-3}$\\\hline
\end{tabular}
\end{equation}
By inspection, we note that whereas the product $QUH_{u}$ is indeed invariant
under all gauge group factors, $QDH_{d}$ violates $U(1)_{b}$. \ We therefore
conclude that the above breaking pattern cannot yield the MSSM.

\paragraph{$SO(11)\supset SU(2)\times SU(2)\times SO(7)$}

Because there is a single maximal subgroup of $SO(7)$ which contains $SU(3)$,
we find that the unique breaking pattern which can reproduce $G_{std}$
proceeds as:%
\begin{align}
SO(11)  &  \supset SU(2)\times SU(2)\times SO(7)\supset SU(2)\times
SU(2)\times SU(4)\\
11  &  \rightarrow(2,2,1)+(1,1,1)+(1,1,6)\\
32  &  \rightarrow(1,2,4)+(1,2,\overline{4})+(2,1,4)+(2,1,\overline{4})\\
55  &  \rightarrow(3,1,1)+(1,3,1)+(2,2,1)+(2,2,6)+(1,1,6)+(1,1,15)
\end{align}
By inspection, this decomposition again contains all of the matter content of
the $SO(10)$ breaking pattern which proceeds via $SO(10)\supset SU(2)\times
SU(2)\times SU(4)$. \ We therefore conclude that the analysis of the breaking
patterns via instantons is identical to this case.

\paragraph{$SO(11)\supset SO(9)\times U(1)$}

The final maximal subgroup which contains $G_{std}$ is given by $SO(9)\times
U(1)$. \ In this case, $SU(2)\times SU(4)$ is the only maximal subgroup of
$SO(9)$ which contains the product $SU(3)\times SU(2)$. \ Decomposing with
respect to this path yields:%
\begin{align}
SO(11)  &  \supset SO(9)\times\lbrack U(1)]_{b}\supset\lbrack SU(2)\times
SU(4)]\times\lbrack U(1)]_{b}\\
&  \supset\lbrack SU(2)\times SU(3)\times\lbrack U(1)]_{a}]\times\lbrack
U(1)]_{b}\\
11  &  \rightarrow(1,1)_{0,-2}+(1,1)_{0,2}+(3,1)_{0,0}+(1,3)_{2,0}%
+(1,\overline{3})_{-2,0}\\
32  &  \rightarrow(2,1)_{3,1}+(2,3)_{-1,1}+(2,1)_{-3,1}+(2,\overline{3}%
)_{1,1}\\
&  +(2,1)_{-3,-1}+(2,\overline{3})_{1,-1}+(2,1)_{3,-1}+(2,3)_{-1,-1}\\
55  &  \rightarrow(1,1)_{0,0}+(3,1)_{0,2}+(1,3)_{2,2}+(1,\overline{3}%
)_{-2,2}+(3,1)_{0,-2}\\
&  +(1,3)_{2,-2}+(1,\overline{3})_{-2,-2}+(3,1)_{0,0}+(3,3)_{2,0}%
+(3,\overline{3})_{-2,0}\\
&  +(1,1)_{0,0}+(1,3)_{-4,0}+(1,\overline{3})_{4,0}+(1,8)_{0,0}\text{.}%
\end{align}
In this case, the $E$-fields must correspond to the representation
$(1,1)_{0,\pm2}$. \ This implies the relation $b=\pm3$. \ Moreover, because
the $Q$ and $L$-fields respectively correspond to the representations
$(2,3)_{-1,\pm1}$ and $(2,1)_{\pm3,\pm1}$, it follows that without loss of
generality $a=2$ and $b=+3$ is the unique choice of $U(1)$ charges which can
yield the correct value of $U(1)_{Y}$ for all fields. \ In this case, the
representation content of the $Q$, $D$ and $H_{d}$ fields is uniquely
determined to be:%
\begin{equation}%
\begin{tabular}
[c]{|l|l|l|}\hline
$Q$ & $D$ & $H_{d}$\\\hline
$(2,3)_{-1,1}$ & $(1,\overline{3})_{-2,2}$ & $(2,1)_{-3,-1}$\\\hline
\end{tabular}
\text{.}%
\end{equation}
Because the product $QDH_{d}$ violates $U(1)_{b}$, we conclude that the
corresponding breaking pattern cannot lead to the MSSM.

\subsection{Rank Six}

We now proceed to the classification of all breaking patterns of rank six
groups. \ Because it is the case of primary phenomenological interest in many
cases, we begin our analysis with breaking patterns of $E_{6}$. \ We next
determine all possible breaking patterns of $SU(7)$ and conclude with an
analysis of breaking patterns of $SO(12)$.

\subsubsection{$E_{6}$}

The non-trivial representations of $E_{6}$ which can descend from the adjoint
representation of $E_{8}$ are the $27$, $\overline{27}$ and $78$ of $E_{6}$.
\ The maximal subgroups of $E_{6}$ are:%
\begin{align}
E_{6}  &  \supset SO(10)\times U(1)\label{Esixone}\\
E_{6}  &  \supset SU(2)\times SU(6)\\
E_{6}  &  \supset SU(3)\times SU(3)\times SU(3)\\
E_{6}  &  \supset USp(8)\\
E_{6}  &  \supset F_{4}\\
E_{6}  &  \supset SU(3)\times G_{2}\label{Esixsix}\\
E_{6}  &  \supset G_{2}\\
E_{6}  &  \supset SU(3)\text{.}%
\end{align}
Of the above configurations, only the maximal subgroups of lines
(\ref{Esixone})-(\ref{Esixsix}) contain $G_{std}$. \ In particular, the first
three breaking patterns can descend to more conventional GUT\ theories. \ We
begin our analysis by demonstrating that none of the remaining possibilities
can produce a consistent embedding of the MSSM.

\paragraph{$E_{6}\supset USp(8)$}

The maximal subgroups of $USp(8)$ are:%
\begin{align}
USp(8)  &  \supset SU(4)\times U(1)\\
USp(8)  &  \supset SU(2)\times USp(6)\label{USp8two}\\
USp(8)  &  \supset USp(4)\times USp(4)\\
USp(8)  &  \supset SU(2)\\
USp(8)  &  \supset SU(2)\times SU(2)\times SU(2)\text{.}%
\end{align}
Of these possibilities, only line (\ref{USp8two}) contains $SU(3)\times
SU(2)$. \ Further, by inspection of lines (\ref{USpsixone})-(\ref{USpsixfour}%
), the only maximal subgroup of $USp(6)$ which contains $SU(3)$ is:%
\begin{equation}
USp(6)\supset SU(3)\times U(1)\text{.}%
\end{equation}
In this case, the unique candidate breaking pattern is:%
\begin{equation}
E_{6}\supset USp(8)\supset SU(2)\times USp(6)\supset SU(2)\times\lbrack
SU(3)\times\lbrack U(1)]]
\end{equation}
which is obtained by a non-trivial $U(1)$ instanton in the $USp(6)$ factor.
\ In this case, the representations of $E_{6}$ decompose as:%
\begin{align}
E_{6}  &  \supset USp(8)\supset SU(2)\times USp(6)\supset SU(2)\times\lbrack
SU(3)\times\lbrack U(1)]]\\
27  &  \rightarrow(2,3)_{1}+(2,\overline{3})_{-1}+(1,3)_{-2}+(1,\overline
{3})_{2}+(1,8)_{0}+(1,1)_{0}\\
78  &  \rightarrow(3,1)_{0}+(1,1)_{0}+(1,6)_{2}+(1,\overline{6})_{-2}%
+(1,8)_{0}+(2,3)_{1}\\
&  +(2,\overline{3})_{-1}+(1,3)_{-2}+(1,\overline{3})_{2}+(1,8)_{0}%
+(2,1)_{3}+(2,1)_{-3}\\
&  +(2,6)_{-1}+(2,\overline{6})_{1}\text{.}%
\end{align}
By inspection, all singlets of $SU(3)\times SU(2)$ are neutral under the only
$U(1)$ factor so that the resulting model cannot contain any $E$-fields.

\paragraph{$E_{6}\supset F_{4}$}

The representation content of $E_{6}$ decomposes under $F_{4}$ as:%
\begin{align}
E_{6}  &  \supset F_{4}\\
27  &  \rightarrow26+1\\
78  &  \rightarrow26+52\text{.}%
\end{align}
Returning to our previous analysis of breaking patterns for $F_{4}$, we
therefore conclude that this breaking pattern cannot produce the correct
matter content of the MSSM.

\paragraph{$E_{6}\supset SU(3)\times G_{2}$}

Although $G_{2}$ contains $SU(3)$ as a maximal subgroup, it is not possible to
arrange for an instanton configuration to break $G_{2}$ to $SU(3)$. \ For this
reason, we conclude that the $SU(3)$ factor of $G_{std}$ must be identified
with the $SU(3)$ factor of the maximal subgroup $SU(3)\times G_{2}$ of $E_{6}%
$. \ In this case, it now follows that the factor $SU(2)\times U(1)$ must
descend from $G_{2}$. \ Returning to lines (\ref{G2one})-(\ref{G2bot}), it
follows that the maximal subgroups $SU(3)$ and $SU(2)\times SU(2)$ contain
$SU(2)\times U(1)$.

First consider the decomposition of representations of $E_{6}$ via the nested
sequence of maximal subgroups:%
\begin{align}
E_{6}  &  \supset SU(3)\times G_{2}\supset SU(3)\times\lbrack SU(3)]\supset
SU(3)\times\lbrack SU(2)\times\lbrack U(1)]]\\
27  &  \rightarrow(\overline{6},1)_{0}+(3,1)_{0}+(3,2)_{1}+(3,1)_{-2}%
+(3,2)_{-1}+(3,1)_{2}\\
78  &  \rightarrow(8,1)_{0}+(1,1)_{-2}+(1,2)_{1}+(1,1)_{2}+(1,2)_{-1}\\
&  +(1,1)_{0}+(1,2)_{3}+(1,2)_{-3}+(1,3)_{0}\text{.}%
\end{align}
Because the ratio of the $U(1)$ charge for the candidate $E$- and $Q$-fields
does not equal six, we conclude that this is not a viable breaking pattern.

Next consider the decomposition associated with the nested sequence of maximal
subgroups:%
\begin{align}
E_{6}  &  \supset SU(3)\times G_{2}\supset SU(3)\times\lbrack SU(2)\times
SU(2)]\\
27  &  \rightarrow(\overline{6},1,1)+(3,1,3)+(3,2,2)\\
78  &  \rightarrow(8,1,1)+(1,1,3)+(1,3,1)+(1,2,4)+(8,1,3)+(8,2,2)\text{.}%
\end{align}
Decomposing the above representations with respect to a $U(1)$ subgroup of
either $SU(2)$ factor, we find that the ratio of $U(1)$ charges for the
candidate $E$- and $Q$-fields again does not equal six. \ Hence, neither
nested sequence of maximal subgroups yields the correct spectrum of the MSSM.

\paragraph{$E_{6}\supset SU(3)\times SU(3)\times SU(3)$}

In order to make the $%
\mathbb{Z}
_{3}$ outer automorphism of $E_{6}$ more manifest, we assume that the
decomposition of $E_{6}$ to the maximal subgroup $SU(3)\times SU(3)\times
SU(3)$ is given by:%
\begin{align}
E_{6}  &  \supset SU(3)_{1}\times SU(3)_{2}\times SU(3)_{3}\label{E6trini}\\
27  &  \rightarrow(\overline{3},3,1)+(3,1,\overline{3})+(1,\overline{3},3)\\
\overline{27}  &  \rightarrow(3,\overline{3},1)+(\overline{3}%
,1,3)+(1,3,\overline{3})\\
78  &  \rightarrow(8,1,1)+(1,8,1)+(1,1,8)+(3,3,3)+(\overline{3},\overline
{3},\overline{3})\text{.}%
\end{align}
While it is also common to conjugate the representation content of the third
$SU(3)$ factor, this is a choice of convention. \ Indeed, because of the $%
\mathbb{Z}
_{3}$ outer automorphism, without loss of generality we require that the first
$SU(3)$ factor is common to $G_{std}$ as well. \ First note that while an
$SU(3)\times U(1)$ instanton can break $E_{6}$ to $G_{std}$, we note that the
resulting $U(1)$ factor of $G_{std}$ must descend from one of the remaining
$SU(3)$ factors. \ By inspection of the above decomposition of line
(\ref{E6trini}), the purported $U(1)_{Y}$ is incorrect.

To proceed further, we next consider the maximal subgroups of the last two
$SU(3)$ factors. \ The maximal subgroups of $SU(3)$ are:%
\begin{align}
a)  &  :SU(3)\supset SU(2)\times U(1)\label{aSU3}\\
b)  &  :SU(3)\supset SU(2)\text{.} \label{bSU3}%
\end{align}
We therefore conclude that there are four distinct maximal subgroups of
$SU(3)\times SU(3)\times SU(3)$ which can potentially yield $G_{std}$.
\ Moreover, in order to achieve the subgroup $SU(2)\times U(1)$ of $G_{std}$,
we must assume that at least one $SU(3)$ factor descends to a maximal subgroup
via line (\ref{aSU3}).

$E_{6}\supset SU(3)\times SU(3)\times SU(3)\supset SU(3)\times\lbrack
SU(2)]\times\lbrack SU(2)\times U(1)]$

We first treat the nested sequence of maximal subgroups where the second
$SU(3)$ factor descends to $SU(2)$ as in line (\ref{bSU3}) while the third
descends to $SU(2)\times U(1)$ as in line (\ref{aSU3}). \ Because
interchanging the last two $SU(3)$ factors of $E_{6}\supset SU(3)\times
SU(3)\times SU(3)$ complex conjugates all representations, a similar analysis
will hold in that case as well. \ The representation content of $E_{6}$
decomposes as:%
\begin{align}
E_{6}  &  \supset SU(3)_{1}\times SU(3)_{2}\times SU(3)_{3}\supset
SU(3)_{1}\times\lbrack SU(2)]_{2}\times\lbrack SU(2)\times U(1)]_{3}\\
27  &  \rightarrow(\overline{3},3,1)+(3,1,1_{2})+(3,1,2_{-1})+(1,3,1_{-2}%
)+(1,3,2_{1})\\
78  &  \rightarrow(8,1,1)+(1,3,1)+(1,5,1)+(1,1,1_{0})+(1,1,2_{3}%
)+(1,1,2_{-3})\\
&  +(1,1,3_{0})+(3,3,1_{2})+(3,3,2_{-1})+(\overline{3},3,1_{-2})+(\overline
{3},3,2_{1})\text{.}%
\end{align}
There are several ways in which an instanton configuration can yield the gauge
group $G_{std}$. \ First consider configurations obtained via a non-trivial
$SU(2)$ instanton configuration. \ Because the $SU(2)$ factor of $SU(3)_{2}$
either breaks completely or to a $U(1)$ subgroup of $SU(2)$, we conclude that
only $SU(2)$ instantons with values in the factor $SU(3)_{2}$ of line
(\ref{E6trini}) can preserve the gauge group $G_{std}$. \ In this case, the
$U(1)$ charge assignments for the $Q$- and $E$-fields are incompatible with
the $U(1)_{Y}$ assignments of the Standard Model.

Next consider abelian instanton configurations which break one of the $SU(2)$
factors. \ Decomposing the factor $SU(2)_{2}$ with respect to a $U(1)$
subgroup, the resulting representation content is:%
\begin{align}
E_{6}  &  \supset SU(3)_{1}\times SU(3)_{2}\times SU(3)_{3}\supset
SU(3)_{1}\times\lbrack SU(2)]_{2}\times\lbrack SU(2)\times U(1)]_{3}\\
&  \supset SU(3)_{1}\times\lbrack U(1)_{a}]_{2}\times\lbrack SU(2)\times
U(1)_{b}]_{3}\\
27  &  \rightarrow(\overline{3},1_{2},1_{0})+(\overline{3},1_{-2}%
,1_{0})+(\overline{3},1_{0},1_{0})+(3,1_{0},1_{2})+(3,1_{0},2_{-1})\\
&  +(1,1_{2},1_{-2})+(1,1_{-2},1_{-2})+(1,1_{0},1_{-2})+(1,1_{2}%
,2_{1})+(1,1_{-2},2_{1})\\
&  +(1,1_{0},2_{1})\\
78  &  \rightarrow(8,1_{0},1_{0})+(1,1_{2},1_{0})+(1,1_{-2},1_{0}%
)+(1,1_{0},1_{0})+(1,1_{4},1_{0})\\
&  +(1,1_{2},1_{0})+(1,1_{0},1_{0})+(1,1_{-2},1_{0})+(1,1_{-4},1_{0}%
)+(1,1_{0},1_{0})\\
&  +(1,1_{0},2_{3})+(1,1_{0},2_{-3})+(1,1_{0},3_{0})+(3,1_{2},1_{2}%
)+(3,1_{-2},1_{2})\\
&  +(3,1_{0},1_{2})+(3,1_{2},2_{-1})+(3,1_{-2},2_{-1})+(3,1_{0},2_{-1}%
)+(\overline{3},1_{-2},1_{-2})\\
&  +(\overline{3},1_{2},1_{-2})+(\overline{3},1_{0},1_{-2})+(\overline
{3},1_{-2},2_{1})+(\overline{3},1_{2},2_{1})+(\overline{3},1_{0}%
,2_{1})\text{.}%
\end{align}
The representation content of each MSSM\ field therefore descends from the
following representations:%
\begin{align}
E  &  :(1,1_{\pm2},1_{\pm2})\label{Efield}\\
Q  &  :(3,1_{0},2_{-1})\text{ or }(3,1_{\pm2},2_{-1})\\
H_{d},L  &  :(1,1_{\pm2},2_{\pm1})\text{ or }(1,1_{0},2_{\pm3})\\
U  &  :(\overline{3},1_{\pm2},1_{0})\text{ or }(\overline{3},1_{0}%
,1_{-2})\text{ or }(\overline{3},1_{\pm2},1_{-2})\\
D  &  :(\overline{3},1_{\pm2},1_{0})\text{ or }(\overline{3},1_{0}%
,1_{-2})\text{ or }(\overline{3},1_{\pm2},1_{-2})\\
H_{u}  &  :(1,1_{\pm2},2_{\pm1})\text{ or }(1,1_{0},2_{\pm3})\text{.}%
\end{align}
There are four possible assignments for the $Q,U,H_{u}$ fields which can yield
a non-trivial $QUH_{u}$ term:%
\begin{equation}%
\begin{tabular}
[c]{|l|l|l|}\hline
$Q$ & $U$ & $H_{u}$\\\hline
$(3,1_{0},2_{-1})$ & $(\overline{3},1_{\pm2},1_{0})$ & $(1,1_{\mp2},2_{+1}%
)$\\\hline
$(3,1_{0},2_{-1})$ & $(\overline{3},1_{0},1_{-2})$ & $(1,1_{0},2_{+3}%
)$\\\hline
$(3,1_{\pm2},2_{-1})$ & $(\overline{3},1_{\mp2},1_{-2})$ & $(1,1_{0},2_{+3}%
)$\\\hline
\end{tabular}
\label{QUHU}%
\end{equation}
so that in the first three cases, the $U(1)_{Y}$ charge of $Q$ requires $b=-1$
while in the final case the $U(1)_{Y}$ charge of $H_{u}$ requires $b=+1$. \ In
particular, this implies that the second choice of charge assignments in line
(\ref{QUHU}) is inconsistent. \ Next consider the first choice of charge
assignments. \ In order to obtain the correct $U(1)_{Y}$ charge assignment for
the $U$-field, we must therefore require $a=\mp2$. \ Finally, the fourth
choice of charge assignments requires $a=\pm1$. \ Of these possible charge
assignments, only the first yields a choice consistent with the $U(1)_{Y}$
charge of the $E$-field in line (\ref{Efield}). \ We therefore find that
$a=-2$ and $b=-1$ where without loss of generality we have chosen a sign for
$a$. \ It now follows that the only candidate charge assignments for the
fields are:%
\begin{equation}%
\begin{tabular}
[c]{|l|l|l|l|l|l|l|}\hline
$E_{27}$ & $Q_{27,78}$ & $U_{27}$ & $D_{\overline{27},78}$ & $L_{\overline
{27}}$ & $H_{u27}$ & $H_{d78}$\\\hline
$(1,1_{-2},1_{-2})$ & $(3,1_{0},2_{-1})$ & $(\overline{3},1_{2},1_{0})$ &
$(\overline{3},1_{0},1_{-2})$ & $(1,1_{2},2_{-1})$ & $(1,1_{-2},2_{+1})$ &
$(1,1_{0},2_{3})$\\\hline
\end{tabular}
\end{equation}
where we have also indicated the $E_{6}$ representation content. \ The
interaction term $QUH_{u}$ therefore descends from a $27^{3}$ term so that in
particular, $Q$ descends from the $27$ of $E_{6}$. \ In order to obtain a
non-trivial $QDH_{d}$ term, this in turn requires $D$ to descend from the $78$
of $E_{6}$ so that we finally obtain the representation content:%
\begin{equation}%
\begin{tabular}
[c]{|l|l|l|l|l|l|l|}\hline
$E_{27}$ & $Q_{27}$ & $U_{27}$ & $D_{\overline{27}}$ & $L_{\overline{27}}$ &
$H_{u27}$ & $H_{d78}$\\\hline
$(1,1_{-2},1_{-2})$ & $(3,1_{0},2_{-1})$ & $(\overline{3},1_{2},1_{0})$ &
$(\overline{3},1_{0},1_{-2})$ & $(1,1_{2},2_{-1})$ & $(1,1_{-2},2_{+1})$ &
$(1,1_{0},2_{3})$\\\hline
\end{tabular}
\end{equation}
we therefore conclude that a $U(1)_{2}\times U(1)_{3}$ of the above type can
indeed yield a spectrum consistent with the MSSM.

$E_{6}\supset SU(3)\times SU(3)\times SU(3)\supset SU(3)\times\lbrack
SU(2)\times U(1)]\times\lbrack SU(2)\times U(1)]$

We next treat the nested sequence of maximal subgroups where the second and
third $SU(3)$ factors of the decomposition $E_{6}\supset SU(3)\times
SU(3)\times SU(3)$ descend to $SU(2)\times U(1)$ as in line (\ref{aSU3}).
\ Under this decomposition, the resulting representation content is:%
\begin{align}
E_{6}  &  =SU(3)_{1}\times SU(3)_{2}\times SU(3)_{3}\label{trinifirst}\\
&  \supset SU(3)_{1}\times\lbrack SU(2)\times U(1)_{a}]_{2}\times\lbrack
SU(2)\times U(1)_{b}]_{3}\\
27  &  \rightarrow(\overline{3},1_{-2},1_{0})+(\overline{3},2_{1}%
,1_{0})+(3,1_{0},1_{2})+(3,1_{0},2_{-1})+(1,1_{2},1_{-2})\\
&  +(1,2_{-1},1_{-2})+(1,1_{2},2_{1})+(1,2_{-1},2_{1})\\
78  &  \rightarrow(8,1_{0},1_{0})+(1,1_{0},1_{0})+(1,2_{3},1_{0}%
)+(1,2_{-3},1_{0})+(1,3_{0},1_{0})\\
&  +(1,1_{0},1_{0})+(1,1_{0},2_{3})+(1,1_{0},2_{-3})+(1,1_{0},3_{0}%
)+(3,1_{-2},1_{-2})\\
&  +(3,1_{-2},2_{1})+(3,2_{1},1_{-2})+(3,2_{1},2_{1})+(\overline{3}%
,1_{2},1_{2})+(\overline{3},1_{2},2_{-1})\\
&  +(\overline{3},2_{-1},1_{2})+(\overline{3},2_{-1},2_{-1})\text{.}
\label{trinilast}%
\end{align}
As opposed to previous examples, we now show that a non-abelian instanton can
indeed yield the spectrum of the MSSM. \ To this end, we first show that the
representation content under the subgroup $SU(3)_{1}\times\lbrack
U(1)_{a}]_{2}\times\lbrack SU(2)\times U(1)_{b}]_{3}$ can yield the desired
spectrum. \ We note that this will then establish the same result for a $U(1)$
instanton which breaks this $SU(2)$ factor to $U(1)$.

The representation content of the candidate fields is given by ignoring the
first $SU(2)$ factor:%
\begin{align}
E  &  :(1,1_{2\varepsilon},1_{-2\varepsilon})\text{ or }(1,2_{\varepsilon
},1_{2\varepsilon})\text{ or }(1,2_{3\varepsilon},1_{0})\\
Q  &  :(3,1_{0},2_{-1})\text{ or }(3,1_{-2},2_{1})\text{ or }(3,2_{1},2_{1})\\
U  &  :(\overline{3},1_{-2},1_{0})\text{ or }(\overline{3},2_{1},1_{0})\text{
or }(\overline{3},1_{0},1_{-2})\\
&  \text{ or }(\overline{3},1_{2},1_{2})\text{ or }(\overline{3},2_{-1}%
,1_{2})\\
D  &  :(\overline{3},1_{-2},1_{0})\text{ or }(\overline{3},2_{1},1_{0})\text{
or }(\overline{3},1_{0},1_{-2})\\
&  \text{ or }(\overline{3},1_{2},1_{2})\text{ or }(\overline{3},2_{-1}%
,1_{2})\\
H_{d},H_{u},L  &  :(1,1_{2\varepsilon},2_{\varepsilon})\text{ or
}(1,2_{-\varepsilon},2_{\varepsilon})\text{ or }(1,1_{0},2_{3\varepsilon})
\end{align}
where $\varepsilon=\pm1$. \ We begin by listing all possible distinct
combinations of fields which can potentially descend to the MSSM\ interaction
term $QUH_{u}$:%
\begin{equation}%
\begin{tabular}
[c]{|l|l|l|l|l|}\hline
& $Q$ & $U$ & $H_{u}$ & $(a,b)$\\\hline
$1$ & $(3,1_{0},2_{-1})$ & $(\overline{3},1_{-2},1_{0})$ & $(1,1_{2},2_{1})$ &
$(2,-1)$\\\hline
$2$ & $(3,1_{0},2_{-1})$ & $(\overline{3},2_{1},1_{0})$ & $(1,2_{-1},2_{1})$ &
$(-4,-1)$\\\hline
$3$ & $(3,1_{0},2_{-1})$ & $(\overline{3},1_{0},1_{-2})$ & $(1,1_{0},2_{3})$ &
$OUT$\\\hline
$4$ & $(3,1_{0},2_{-1})$ & $(\overline{3},1_{2},1_{2})$ & $(1,1_{-2},2_{-1})$
& $(-1,-1)$\\\hline
$5$ & $(3,1_{0},2_{-1})$ & $(\overline{3},2_{-1},1_{2})$ & $(1,2_{1},2_{-1})$
& $(2,-1)$\\\hline
$6$ & $(3,1_{-2},2_{1})$ & $(\overline{3},1_{-2},1_{0})$ & $OUT$ &
$OUT$\\\hline
$7$ & $(3,1_{-2},2_{1})$ & $(\overline{3},2_{1},1_{0})$ & $(1,2_{1},2_{-1})$ &
$(-4,-7)$\\\hline
$8$ & $(3,1_{-2},2_{1})$ & $(\overline{3},1_{0},1_{-2})$ & $(1,1_{2},2_{1})$ &
$OUT$\\\hline
$9$ & $(3,1_{-2},2_{1})$ & $(\overline{3},1_{2},1_{2})$ & $(1,1_{0},2_{-3})$ &
$(-1,-1)$\\\hline
$10$ & $(3,1_{-2},2_{1})$ & $(\overline{3},2_{-1},1_{2})$ & $OUT$ &
$OUT$\\\hline
$11$ & $(3,2_{1},2_{1})$ & $(\overline{3},1_{-2},1_{0})$ & $(1,2_{1},2_{-1})$
& $(2,-1)$\\\hline
$12$ & $(3,2_{1},2_{1})$ & $(\overline{3},2_{1},1_{0})$ & $(1,1_{-2},2_{-1})$
& $(-4,5)$\\\hline
$13$ & $(3,2_{1},2_{1})$ & $(\overline{3},1_{0},1_{-2})$ & $(1,2_{-1},2_{1})$
& $(-1,2)$\\\hline
$14$ & $(3,2_{1},2_{1})$ & $(\overline{3},1_{2},1_{2})$ & $OUT$ &
$OUT$\\\hline
$15$ & $(3,2_{1},2_{1})$ & $(\overline{3},2_{-1},1_{2})$ & $(1,1_{0},2_{-3})$
& $(2,-1)$\\\hline
\end{tabular}
\end{equation}
where we have also solved for the linear combination of $U(1)_{a}$ and
$U(1)_{b}$ consistent with $U(1)_{Y}$ charge assignments in the MSSM. \ Next,
we list all possible combinations of fields consistent with the above
classification which also allow the interaction term $QDH_{d}$.%
\begin{align}
&
\begin{tabular}
[c]{|l|l|l|l|}\hline
& $Q$ & $U$ & $D$\\\hline
$1$ & $(3,1_{0},2_{-1})$ & $(\overline{3},1_{-2},1_{0})$ & $(\overline
{3},2_{1},1_{0})\text{ or }(\overline{3},1_{0},1_{-2})\text{ or }(\overline
{3},1_{2},1_{2})$\\\hline
$2$ & $(3,1_{0},2_{-1})$ & $(\overline{3},2_{1},1_{0})$ & $(\overline{3}%
,1_{0},1_{-2})\text{ or }(\overline{3},2_{-1},1_{2})$\\\hline
$4$ & $(3,1_{0},2_{-1})$ & $(\overline{3},1_{2},1_{2})$ & $(\overline
{3},1_{-2},1_{0})\text{ or }(\overline{3},1_{0},1_{-2})\text{ }$\\\hline
$5$ & $(3,1_{0},2_{-1})$ & $(\overline{3},2_{-1},1_{2})$ & $(\overline
{3},2_{1},1_{0})\text{ or }(\overline{3},1_{0},1_{-2})\text{ or }(\overline
{3},1_{2},1_{2})$\\\hline
$7$ & $(3,1_{-2},2_{1})$ & $(\overline{3},2_{1},1_{0})$ & $OUT$\\\hline
$9$ & $(3,1_{-2},2_{1})$ & $(\overline{3},1_{2},1_{2})$ & $(\overline{3}%
,1_{0},1_{-2})\text{ }$\\\hline
$11$ & $(3,2_{1},2_{1})$ & $(\overline{3},1_{-2},1_{0})$ & $(\overline
{3},2_{1},1_{0})\text{ or }(\overline{3},1_{0},1_{-2})$\\\hline
$12$ & $(3,2_{1},2_{1})$ & $(\overline{3},2_{1},1_{0})$ & $(\overline{3}%
,1_{2},1_{2})$\\\hline
$13$ & $(3,2_{1},2_{1})$ & $(\overline{3},1_{0},1_{-2})$ & $(\overline
{3},1_{-2},1_{0})$\\\hline
$15$ & $(3,2_{1},2_{1})$ & $(\overline{3},2_{-1},1_{2})$ & $(\overline
{3},2_{1},1_{0})\text{ or }(\overline{3},1_{0},1_{-2})$\\\hline
\end{tabular}
\\
&
\begin{tabular}
[c]{|l|l|l|l|}\hline
& $H_{u}$ & $H_{d}$ & $(a,b)$\\\hline
$1$ & $(1,1_{2},2_{1})$ & $(1,2_{-1},2_{1})\text{ or }(1,1_{0},2_{3})\text{ or
}(1,1_{-2},2_{-1})$ & $(2,-1)$\\\hline
$2$ & $(1,2_{-1},2_{1})$ & $(1,1_{0},2_{3})\text{ or }(1,2_{1},2_{-1})$ &
$(-4,-1)$\\\hline
$4$ & $(1,1_{-2},2_{-1})$ & $(1,1_{2},2_{1})\text{ or }(1,1_{0},2_{3})$ &
$(-1,-1)$\\\hline
$5$ & $(1,2_{1},2_{-1})$ & $(1,2_{-1},2_{1})\text{ or }(1,1_{0},2_{3})\text{
or }(1,1_{-2},2_{-1})$ & $(2,-1)$\\\hline
$7$ & $(1,2_{1},2_{-1})$ & $OUT$ & $(-4,-7)$\\\hline
$9$ & $(1,1_{0},2_{-3})$ & $(1,1_{2},2_{1})$ & $(-1,-1)$\\\hline
$11$ & $(1,2_{1},2_{-1})$ & $(1,1_{-2},2_{-1})\text{ or }(1,2_{-1},2_{1})$ &
$(2,-1)$\\\hline
$12$ & $(1,1_{-2},2_{-1})$ & $OUT$ & $(-4,5)$\\\hline
$13$ & $(1,2_{-1},2_{1})$ & $(1,2_{1},2_{-1})$ & $(-1,2)$\\\hline
$15$ & $(1,1_{0},2_{-3})$ & $(1,1_{-2},2_{-1})\text{ or }(1,2_{-1},2_{1})$ &
$(2,-1)$\\\hline
\end{tabular}
\ \text{.}%
\end{align}
To further narrow the possible combinations of fields, we next require that
the interactions in question properly descend from $E_{6}$ invariant terms of
the full theory. \ We find that there many ways to package the field content
of the MSSM\ into representations of $E_{6}$. \ The complete list of
possibilities is:%
\begin{equation}%
\begin{tabular}
[c]{|l|l|l|l|l|}\hline
& $Q$ & $U$ & $D$ & $L$\\\hline
$1a$ & $(3,1_{0},2_{-1})\in27$ & $(\overline{3},1_{-2},1_{0})\in27$ &
$(\overline{3},2_{1},1_{0})\in27$ & $(1,2_{-1},2_{1})\in27$\\\hline
$1b$ & $(3,1_{0},2_{-1})\in27$ & $(\overline{3},1_{-2},1_{0})\in27$ &
$(\overline{3},1_{0},1_{-2})\in\overline{27}$ & $(1,1_{-2},2_{-1})\in
\overline{27}$\\\hline
$1c$ & $(3,1_{0},2_{-1})\in27$ & $(\overline{3},1_{-2},1_{0})\in27$ &
$(\overline{3},1_{0},1_{-2})\in\overline{27}$ & $OUT$\\\hline
$1.5$ & $(3,1_{0},2_{-1})\in27$ & $(\overline{3},1_{-2},1_{0})\in27$ &
$(\overline{3},1_{2},1_{2})\in78$ & $(1,2_{-1},2_{1})\in27$\\\hline
$2a$ & $(3,1_{0},2_{-1})\in27$ & $(\overline{3},2_{1},1_{0})\in27$ &
$(\overline{3},1_{0},1_{-2})\in\overline{27}$ & $OUT$\\\hline
$2b$ & $(3,1_{0},2_{-1})\in27$ & $(\overline{3},2_{1},1_{0})\in27$ &
$(\overline{3},1_{0},1_{-2})\in\overline{27}$ & $(1,2_{1},2_{-1})\in
\overline{27}$\\\hline
$2.5a$ & $(3,1_{0},2_{-1})\in27$ & $(\overline{3},2_{1},1_{0})\in27$ &
$(\overline{3},2_{-1},1_{2})\in78$ & $(1,2_{1},2_{-1})\in\overline{27}%
$\\\hline
$2.5b$ & $(3,1_{0},2_{-1})\in27$ & $(\overline{3},2_{1},1_{0})\in27$ &
$(\overline{3},2_{-1},1_{2})\in78$ & $(1,1_{0},2_{3})\in78$\\\hline
$4$ & $(3,1_{0},2_{-1})\in27$ & $(\overline{3},1_{2},1_{2})\in78$ &
$(\overline{3},1_{-2},1_{0})\in27$ & $OUT$\\\hline
$4.5$ & $(3,1_{0},2_{-1})\in27$ & $(\overline{3},1_{2},1_{2})\in78$ &
$(\overline{3},1_{0},1_{-2})\in\overline{27}$ & $OUT$\\\hline
$5a$ & $(3,1_{0},2_{-1})\in27$ & $(\overline{3},2_{-1},1_{2})\in78$ &
$(\overline{3},2_{1},1_{0})\in27$ & $OUT$\\\hline
$5b$ & $(3,1_{0},2_{-1})\in27$ & $(\overline{3},2_{-1},1_{2})\in78$ &
$(\overline{3},2_{1},1_{0})\in27$ & $(1,1_{-2},2_{-1})\in\overline{27}%
$\\\hline
$5.3a$ & $(3,1_{0},2_{-1})\in27$ & $(\overline{3},2_{-1},1_{2})\in78$ &
$(\overline{3},1_{0},1_{-2})\in\overline{27}$ & $OUT$\\\hline
$5.3b$ & $(3,1_{0},2_{-1})\in27$ & $(\overline{3},2_{-1},1_{2})\in78$ &
$(\overline{3},1_{0},1_{-2})\in\overline{27}$ & $OUT$\\\hline
$5.6a$ & $(3,1_{0},2_{-1})\in27$ & $(\overline{3},2_{-1},1_{2})\in78$ &
$(\overline{3},1_{2},1_{2})\in78$ & $OUT$\\\hline
$5.6b$ & $(3,1_{0},2_{-1})\in27$ & $(\overline{3},2_{-1},1_{2})\in78$ &
$(\overline{3},1_{2},1_{2})\in78$ & $(1,2_{-1},2_{1})\in27$\\\hline
$9$ & $(3,1_{-2},2_{1})\in78$ & $(\overline{3},1_{2},1_{2})\in78$ &
$(\overline{3},1_{0},1_{-2})\in\overline{27}$ & $OUT$\\\hline
$11a$ & $(3,2_{1},2_{1})\in78$ & $(\overline{3},1_{-2},1_{0})\in27$ &
$(\overline{3},2_{1},1_{0})\in27$ & $OUT$\\\hline
$11b$ & $(3,2_{1},2_{1})\in78$ & $(\overline{3},1_{-2},1_{0})\in27$ &
$(\overline{3},2_{1},1_{0})\in27$ & $(1,2_{-1},2_{1})\in27$\\\hline
$11.5a$ & $(3,2_{1},2_{1})\in78$ & $(\overline{3},1_{-2},1_{0})\in27$ &
$(\overline{3},1_{0},1_{-2})\in\overline{27}$ & $OUT$\\\hline
$11.5b$ & $(3,2_{1},2_{1})\in78$ & $(\overline{3},1_{-2},1_{0})\in27$ &
$(\overline{3},1_{0},1_{-2})\in\overline{27}$ & $(1,1_{-2},2_{1})\in
\overline{27}$\\\hline
$13$ & $(3,2_{1},2_{1})\in78$ & $(\overline{3},1_{0},1_{-2})\in\overline{27}$
& $(\overline{3},1_{-2},1_{0})\in27$ & $(1,2_{1},2_{-1})\in\overline{27}%
$\\\hline
$15a$ & $(3,2_{1},2_{1})\in78$ & $(\overline{3},2_{-1},1_{2})\in78$ &
$(\overline{3},2_{1},1_{0})\in27$ & $(1,1_{0},2_{3})\in78$\\\hline
$15b$ & $(3,2_{1},2_{1})\in78$ & $(\overline{3},2_{-1},1_{2})\in78$ &
$(\overline{3},2_{1},1_{0})\in27$ & $(1,2_{-1},2_{1})\in27$\\\hline
$15.5a$ & $(3,2_{1},2_{1})\in78$ & $(\overline{3},2_{-1},1_{2})\in78$ &
$(\overline{3},1_{0},1_{-2})\in\overline{27}$ & $(1,2_{-1},2_{1})\in
27$\\\hline
$15.5b$ & $(3,2_{1},2_{1})\in78$ & $(\overline{3},2_{-1},1_{2})\in78$ &
$(\overline{3},1_{0},1_{-2})\in\overline{27}$ & $(1,1_{-2},2_{-1})\in
\overline{27}$\\\hline
\end{tabular}
\ \text{.}%
\end{equation}%
\begin{equation}%
\begin{tabular}
[c]{|l|l|l|l|l|}\hline
& $E$ & $H_{u}$ & $H_{d}$ & $(a,b)$\\\hline
$1a$ & $(1,1_{2},1_{-2})\in27$ & $(1,1_{2},2_{1})\in27$ & $(1,2_{-1},2_{1}%
)\in27$ & $(2,-1)$\\\hline
$1b$ & $(1,1_{2},1_{-2})\in27$ & $(1,1_{2},2_{1})\in27$ & $(1,1_{0},2_{3}%
)\in78\text{ }$ & $(2,-1)$\\\hline
$1c$ & $(1,2_{3},1_{0})\in78$ & $(1,1_{2},2_{1})\in27$ & $(1,1_{0},2_{3}%
)\in78$ & $(2,-1)$\\\hline
$1.5$ & $(1,2_{3},1_{0})\in78$ & $(1,1_{2},2_{1})\in27$ & $(1,1_{-2}%
,2_{-1})\in\overline{27}$ & $(2,-1)$\\\hline
$2a$ & $(1,1_{-2},1_{2})\in\overline{27}$ & $(1,2_{-1},2_{1})\in27$ &
$(1,1_{0},2_{3})\in78$ & $(-4,-1)$\\\hline
$2b$ & $(1,2_{-1},1_{-2})\in27$ & $(1,2_{-1},2_{1})\in27$ & $(1,1_{0}%
,2_{3})\in78$ & $(-4,-1)$\\\hline
$2.5a$ & $(1,1_{-2},1_{2})\in\overline{27}$ & $(1,2_{-1},2_{1})\in27$ &
$(1,2_{1},2_{-1})\in\overline{27}$ & $(-4,-1)$\\\hline
$2.5b$ & $(1,2_{-1},1_{-2})\in27$ & $(1,2_{-1},2_{1})\in27$ & $(1,2_{1}%
,2_{-1})\in\overline{27}$ & $(-4,-1)$\\\hline
$4$ & $OUT$ & $(1,1_{-2},2_{-1})\in\overline{27}$ & $(1,1_{2},2_{1})\in27$ &
$(-1,-1)$\\\hline
$4.5$ & $OUT$ & $(1,1_{-2},2_{-1})\in\overline{27}$ & $(1,1_{0},2_{3})\in78$ &
$(-1,-1)$\\\hline
$5a$ & $(1,1_{2},1_{-2})\in\overline{27}$ & $(1,2_{1},2_{-1})\in\overline{27}$
& $(1,2_{-1},2_{1})\in27$ & $(2,-1)$\\\hline
$5b$ & $(1,2_{3},1_{0})\in78$ & $(1,2_{1},2_{-1})\in\overline{27}$ &
$(1,2_{-1},2_{1})\in27$ & $(2,-1)$\\\hline
$5.3a$ & $(1,1_{2},1_{-2})\in\overline{27}$ & $(1,2_{1},2_{-1})\in
\overline{27}$ & $(1,1_{0},2_{3})\in78$ & $(2,-1)$\\\hline
$5.3b$ & $(1,2_{3},1_{0})\in78$ & $(1,2_{1},2_{-1})\in\overline{27}$ &
$(1,1_{0},2_{3})\in78$ & $(2,-1)$\\\hline
$5.6a$ & $(1,1_{2},1_{-2})\in\overline{27}$ & $(1,2_{1},2_{-1})\in
\overline{27}$ & $(1,1_{-2},2_{-1})\in\overline{27}$ & $(2,-1)$\\\hline
$5.6b$ & $(1,2_{3},1_{0})\in78$ & $(1,2_{1},2_{-1})\in\overline{27}$ &
$(1,1_{-2},2_{-1})\in\overline{27}$ & $(2,-1)$\\\hline
$9$ & $OUT$ & $(1,1_{0},2_{-3})\in78$ & $(1,1_{2},2_{1})\in27$ &
$(-1,-1)$\\\hline
$11a$ & $(1,1_{2},1_{-2})\in\overline{27}$ & $(1,2_{1},2_{-1})\in\overline
{27}$ & $(1,1_{-2},2_{-1})\in\overline{27}$ & $(2,-1)$\\\hline
$11b$ & $(1,2_{3},1_{0})\in78$ & $(1,2_{1},2_{-1})\in\overline{27}$ &
$(1,1_{-2},2_{-1})\in\overline{27}$ & $(2,-1)$\\\hline
$11.5a$ & $(1,1_{2},1_{-2})\in\overline{27}$ & $(1,2_{1},2_{-1})\in
\overline{27}$ & $(1,2_{-1},2_{1})\in27$ & $(2,-1)$\\\hline
$11.5b$ & $(1,2_{3},1_{0})\in78$ & $(1,2_{1},2_{-1})\in\overline{27}$ &
$(1,2_{-1},2_{1})\in27$ & $(2,-1)$\\\hline
$13$ & $(1,1_{-2},1_{2})\in\overline{27}$ & $(1,2_{-1},2_{1})\in27$ &
$(1,2_{1},2_{-1})\in\overline{27}$ & $(-1,2)$\\\hline
$15a$ & $(1,1_{2},1_{-2})\in\overline{27}$ & $(1,1_{0},2_{-3})\in78$ &
$(1,1_{-2},2_{-1})\in\overline{27}$ & $(2,-1)$\\\hline
$15b$ & $(1,2_{3},1_{0})\in78$ & $(1,1_{0},2_{-3})\in78$ & $(1,1_{-2}%
,2_{-1})\in\overline{27}$ & $(2,-1)$\\\hline
$15.5a$ & $(1,1_{2},1_{-2})\in\overline{27}$ & $(1,1_{0},2_{-3})\in78$ &
$(1,2_{-1},2_{1})\in27$ & $(2,-1)$\\\hline
$15.5b$ & $(1,2_{3},1_{0})\in78$ & $(1,1_{0},2_{-3})\in78$ & $(1,2_{-1}%
,2_{1})\in27$ & $(2,-1)$\\\hline
\end{tabular}
\end{equation}
where the numbering convention has been chosen in order to trace the origin of
each possible permutation, and as before, $OUT$ denotes an entry which has
been ruled out because it cannot yield the correct $U(1)_{Y}$ charge
assignment or interaction term.

\paragraph{$E_{6}\supset SU(10)\times U(1)$}

We now analyze breaking patterns of $E_{6}$ which descend from the maximal
subgroup $SO(10)\times U(1)$ such that:%
\begin{align}
E_{6}  &  \supset SO(10)\times\lbrack U(1)]\\
27  &  \rightarrow1_{4}+10_{-2}+16_{1}\\
78  &  \rightarrow1_{0}+16_{-3}+\overline{16}_{3}+45_{0}\text{.}%
\end{align}
Of the maximal subgroups of $SO\left(  10\right)  $ listed in lines
(\ref{SO10one})-(\ref{SO10last}), only the first four contain the non-abelian
group $SU(3)\times SU(2)$ so that the unique nested sequence of maximal
subgroups of $E_{6}$ is uniquely determined by the paths:%
\begin{align}
E_{6}  &  \supset SO(10)\times\lbrack U(1)]\supset\lbrack SU(5)\times
U(1)]\times U(1)\\
&  \supset\lbrack SU(3)\times SU(2)\times U(1)]\times U(1)]\times U(1)\\
E_{6}  &  \supset SO(10)\times\lbrack U(1)]\supset SU(2)\times SU(2)\times
SU(4)\times\lbrack U(1)]\\
&  \supset SU(2)\times SU(2)\times\lbrack SU(3)\times U(1)]\times U(1)\\
E_{6}  &  \supset SO(10)\times\lbrack U(1)]\supset SO(9)\times\lbrack
U(1)]\supset\lbrack SU(2)\times SU(4)]\times\lbrack U(1)]\\
&  \supset\lbrack SU(2)\times\lbrack SU(3)\times U(1)]]\times\lbrack U(1)]\\
E_{6}  &  \supset SO(10)\times\lbrack U(1)]\supset SU(2)\times SO(7)\supset
\lbrack SU(2)\times SU(4)]\times\lbrack U(1)]\\
&  \supset\lbrack SU(2)\times\lbrack SU(3)\times U(1)]]\times\lbrack
U(1)]\text{.}%
\end{align}
Because the previous analysis of abelian instanton configurations of $SO(10)$
which can yield the MSSM\ spectrum carry over to this case as well, we focus
on breaking patterns which do not embed purely in $SO(10)$. \ While it is in
principle possible to package the field content of the MSSM\ fields into
representations of $E_{6}$ in more exotic ways using the additional $U(1)$
charge, all of these configurations still correspond to generic abelian
instanton configurations.

$E_{6}\supset SO(10)\times\lbrack U(1)]\supset SU(2)\times SO(7)\times\lbrack
U(1)]\supset SU(2)\times SU(4)\times\lbrack U(1)]\supset SU(2)\times\lbrack
SU(3)\times\lbrack U(1)]]\times\lbrack U(1)]$

Decomposing the $27$ and $78$ with respect to this nested sequence of maximal
subgroups, we find:%
\begin{align}
E_{6}  &  \supset...\supset SU(2)\times\lbrack SU(3)\times\lbrack
U(1)_{a}]]\times\lbrack U(1)_{b}]\\
27  &  \rightarrow(1,1)_{0,4}+(3,1)_{0,-2}+(1,1)_{0,-2}+(1,3)_{2,-2}\\
&  +(1,\overline{3})_{-2,-2}+(2,1)_{3,1}+(2,3)_{-1,1}+(2,1)_{-3,1}%
+(2,\overline{3})_{1,1}\\
78  &  \rightarrow(1,1)_{0,0}+(2,1)_{3,-3}+(2,3)_{-1,-3}+(2,1)_{-3,-3}\\
&  +(2,\overline{3})_{1,-3}+(2,1)_{-3,3}+(2,\overline{3})_{1,3}+(2,1)_{3,3}%
+(2,3)_{-1,3}\\
&  +(3,1)_{0,0}+(1,3)_{2,0}+(1,\overline{3})_{-2,0}+(3,3)_{2,0}+(3,\overline
{3})_{-2,0}\\
&  +(3,1)_{0,0}+(1,1)_{0,0}+(1,3)_{-4,0}+(1,\overline{3})_{4,0}+(1,8)_{0,0}%
\text{.}%
\end{align}
We begin by classifying all combinations of representations which can yield
the non-trivial interaction term $QUH_{u}$:%
\begin{align}%
\begin{tabular}
[c]{|l|l|l|l|l|}\hline
& $Q$ & $U$ & $D$ & $L$\\\hline
$1$ & $(2,3)_{-1,1}\in27$ & $(1,\overline{3})_{-2,-2}\in27$ & $(1,\overline
{3})_{-2,2}\in\overline{27}$ & $(2,1)_{-3,-1}\in\overline{27}$\\\hline
$2$ & $(2,3)_{-1,1}\in27$ & $(1,\overline{3})_{-2,-2}\in27$ & $(1,\overline
{3})_{4,0}\in78$ & $(2,1)_{3,-3}\in78$\\\hline
$3$ & $(2,3)_{-1,-1}\in\overline{27}$ & $(1,\overline{3})_{-2,2}\in
\overline{27}$ & $(1,\overline{3})_{-2,-2}\in27$ & $(2,1)_{-3,1}\in27$\\\hline
$4$ & $(2,3)_{-1,-1}\in\overline{27}$ & $(1,\overline{3})_{-2,2}\in
\overline{27}$ & $(1,\overline{3})_{4,0}\in78$ & $(2,1)_{3,3}\in78$\\\hline
\end{tabular}
& \\%
\begin{tabular}
[c]{|l|l|l|l|l|}\hline
& $E$ & $H_{u}$ & $H_{d}$ & $(a,b)$\\\hline
$1$ & $(1,1)_{0,4}\in27$ & $(2,1)_{3,1}\in27$ & $(2,1)_{3,-3}\in78$ &
$(1/2,3/2)$\\\hline
$2$ & $(1,1)_{0,4}\in27$ & $(2,1)_{3,1}\in27$ & $(2,1)_{-3,-1}\in\overline
{27}$ & $(1/2,3/2)$\\\hline
$3$ & $(1,1)_{0,-4}\in\overline{27}$ & $(2,1)_{3,-1}\in\overline{27}$ &
$(2,1)_{3,3}\in78$ & $(1/2,-3/2)$\\\hline
$4$ & $(1,1)_{0,-4}\in\overline{27}$ & $(2,1)_{3,-1}\in\overline{27}$ &
$(2,1)_{-3,1}\in27$ & $(1/2,-3/2)$\\\hline
\end{tabular}
&
\end{align}
so that in this case a non-standard embedding of a $U(1)\times U(1)$ instanton
can indeed yield the spectrum of the MSSM.

$E_{6}\supset SO(10)\times\lbrack U(1)]\supset SO(9)\times\lbrack U(1)]\supset
SU(2)\times SU(4)\times\lbrack U(1)]\supset SU(2)\times\lbrack SU(3)\times
\lbrack U(1)]]\times\lbrack U(1)]$

Decomposing the $27$ and $78$ with respect to this nested sequence of maximal
subgroups, we find:%
\begin{align}
E_{6}  &  \supset...\supset SU(2)\times\lbrack SU(3)\times\lbrack
U(1)_{a}]]\times\lbrack U(1)_{b}]\\
27  &  \rightarrow(1,1)_{0,4}+(1,1)_{0,-2}+(3,1)_{0,-2}+(1,3)_{2,-2}%
+(1,\overline{3})_{-2,-2}\\
&  +(2,1)_{3,1}+(2,3)_{-1,1}+(2,1)_{-3,1}+(2,\overline{3})_{1,1}\\
78  &  \rightarrow(1,1)_{0,0}+(2,1)_{3,-3}+(2,3)_{-1,-3}+(2,1)_{-3,-3}%
+(2,\overline{3})_{1,-3}\\
&  +(2,1)_{3,3}+(2,3)_{-1,3}+(2,1)_{-3,3}+(2,\overline{3})_{1,3}\\
&  +(3,1)_{0,0}+(1,1)_{0,0}+(1,3)_{-4,0}+(1,\overline{3})_{4,0}+(1,8)_{0,0}%
+(3,3)_{2,0}\\
&  +(3,\overline{3})_{-2,0}+(3,1)_{0,0}+(1,3)_{2,0}+(1,\overline{3}%
)_{-2,0}\text{.}%
\end{align}
By inspection, this is precisely the same matter content as in the previous
example. \ We therefore conclude that the abelian instanton configurations
analyzed previously produce an identical MSSM\ spectrum.

$E_{6}\supset SO(10)\times\lbrack U(1)]\supset SU(2)\times SU(2)\times
SU(4)\times\lbrack U(1)]\supset SU(2)\times SU(2)\times\lbrack SU(3)\times
U(1)]\times U(1)$

The decomposition of the $27$ and $78$ of $E_{6}$ in this case yields:%
\begin{align}
E_{6}  &  \supset SO(10)\times\lbrack U(1)]\supset SU(2)\times SU(2)\times
SU(4)\times\lbrack U(1)]\\
&  \supset SU(2)_{1}\times SU(2)_{2}\times\lbrack SU(3)\times U(1)_{a}]\times
U(1)_{b}\\
27  &  \rightarrow(1,1,1)_{0,4}+(2,2,1)_{0,-2}+(1,1,3)_{2,-2}+(1,1,\overline
{3})_{-2,-2}\\
&  +(2,1,1)_{3,1}+(2,1,3)_{-1,1}+(1,2,1)_{-3,1}+(1,2,\overline{3})_{1,1}\\
78  &  \rightarrow(1,1,1)_{0,0}+(2,1,1)_{3,-3}+(2,1,3)_{-1,-3}+(1,2,1)_{-3,-3}%
\\
&  +(1,2,\overline{3})_{1,-3}+(2,1,1)_{-3,3}+(2,1,\overline{3})_{1,3}%
+(1,2,1)_{3,3}\\
&  +(1,2,3)_{-1,3}+(3,1,1)_{0,0}+(1,3,1)_{0,0}+(1,1,1)_{0,0}\\
&  +(1,1,3)_{-4,0}+(1,1,\overline{3})_{4,0}+(1,1,8)_{0,0}+(2,2,3)_{2,0}\\
&  +(2,2,\overline{3})_{-2,0}\text{.}%
\end{align}
In fact, the representation content of this decomposition is identical to that
obtained via the previously treated nested sequence of maximal subgroups given
by lines (\ref{trinifirst})-(\ref{trinilast}):%
\begin{equation}
E_{6}\supset SU(3)\times SU(3)\times SU(3)\supset SU(3)\times\lbrack
SU(2)\times U(1)_{c}]\times\lbrack SU(2)\times U(1)_{d}]
\end{equation}
under the linear change in $U(1)$ charges:%
\begin{align}
U(1)_{a}  &  =\frac{1}{2}U(1)_{c}+\frac{1}{2}U(1)_{d}\\
U(1)_{b}  &  =\frac{1}{2}U(1)_{c}-\frac{1}{2}U(1)_{d}\text{.}%
\end{align}

\paragraph{$E_{6}\supset SU(2)\times SU(6)$}

Decomposing the $27$ and $78$ of $E_{6}$ with respect to $SU(2)\times SU(6)$
yields:%
\begin{align}
E_{6}  &  \supset SU(2)\times SU(6)\\
27  &  \rightarrow(2,\overline{6})+(1,15)\\
78  &  \rightarrow(3,1)+(1,35)+(2,20)\text{.}%
\end{align}
Returning to the maximal subgroups of $SU(6)$ presented in lines
(\ref{SU6one})-(\ref{SU6last}), the list of all possible nested sequences of
maximal subgroups of $E_{6}$ descend to $G_{std}$ as:%
\begin{align}
E_{6}  &  \supset SU(2)\times SU(6)\supset SU(2)\times\lbrack SU(5)\times
U(1)]\\
&  \supset SU(2)\times\lbrack SU(3)\times SU(2)\times\lbrack U(1)]\times
U(1)]\\
E_{6}  &  \supset SU(2)\times SU(6)\supset SU(2)\times\lbrack SU(2)\times
SU(4)\times U(1)]\label{E6SU6two}\\
&  \supset SU(2)\times\lbrack SU(2)\times\lbrack SU(3)\times U(1)]\times
U(1)]\\
E_{6}  &  \supset SU(2)\times SU(6)\supset SU(2)\times\lbrack SU(3)\times
SU(3)\times U(1)]\label{E6SU6three}\\
E_{6}  &  \supset SU(2)\times SU(6)\supset SU(2)\times SU(4)\supset
SU(2)\times\lbrack SU(3)\times U(1)]\label{E6SU6four}\\
E_{6}  &  \supset SU(2)\times SU(6)\supset SU(2)\times USp(6)\supset
SU(2)\times\lbrack SU(3)\times U(1)]\label{E6SU6five}\\
E_{6}  &  \supset SU(2)\times SU(6)\supset SU(2)\times\lbrack SU(2)\times
SU(3)] \label{E6SU6six}\text{.}
\end{align}
In the first two nested sequences the resulting breaking pattern descends to
the same representation content as breaking patterns analyzed previously.
\ For this reason, we confine our analysis to breaking patterns reached via
lines (\ref{E6SU6three})-(\ref{E6SU6six}).

$E_{6}\supset SU(2)\times SU(6)\supset SU(2)\times\lbrack SU(2)\times SU(3)]$

In this case, the representations of $E_{6}$ decompose as:%
\begin{align}
E_{6}  &  \supset SU(2)\times SU(6)\supset SU(2)\times\lbrack SU(2)\times
SU(3)]\\
27  &  \rightarrow(2,2,\overline{3})+(1,1,\overline{6})+(1,3,3)\\
78  &  \rightarrow(3,1,1)+(3,3,1)+(3,3,8)+(2,4,1)+(2,2,8)\text{.}%
\end{align}
Decomposing one of the $SU(2)$ factors with respect to a $U(1)$ subgroup, it
follows that the ratio of $U(1)$ charge for the $Q$- and $E$-fields is
incorrect so that the MSSM\ cannot be obtained via this path.

$E_{6}\supset SU(2)\times SU(6)\supset SU(2)\times USp(6)\supset
SU(2)\times\lbrack SU(3)\times U(1)]$

The representations of $E_{6}$ descend as:%
\begin{align}
E_{6}  &  \supset SU(2)\times SU(6)\supset SU(2)\times USp(6)\supset
SU(2)\times\lbrack SU(3)\times U(1)]\\
27  &  \rightarrow(2,3)_{1}+(2,\overline{3})_{-1}+(1,1)_{0}+(1,3)_{-2}%
+(1,\overline{3})_{2}+(1,8)_{0}\\
78  &  \rightarrow(3,1)_{0}+(1,3)_{-2}+(1,\overline{3})_{2}+(1,8)_{0}%
+(1,1)_{0}+(1,6)_{2}\\
&  +(1,\overline{6})_{-2}+(1,8)_{0}+(2,3)_{1}+(2,\overline{3})_{-1}%
+(2,1)_{3}+(2,1)_{-3}\\
&  +(2,6)_{-1}+(2,\overline{6})_{1}%
\end{align}
Because the $U(1)$ charge assignment is incorrect, we cannot reach the
MSSM\ via this nested sequence either.

$E_{6}\supset SU(2)\times SU(6)\supset SU(2)\times SU(4)\supset SU(2)\times
\lbrack SU(3)\times U(1)]$

Here, the representations of $E_{6}$ descend as:%
\begin{align}
E_{6}  &  \supset SU(2)\times SU(6)\supset SU(2)\times SU(4)\supset
SU(2)\times\lbrack SU(3)\times U(1)]\\
27  &  \rightarrow(2,3)_{2}+(2,\overline{3})_{-2}+(1,1)_{0}+(1,3)_{-4}%
+(1,\overline{3})_{4}+(1,8)_{0}\\
78  &  \rightarrow(3,1)_{0}+(1,1)_{0}+(1,3)_{-4}+(1,\overline{3}%
)_{4}+(1,8)_{0}+(1,\overline{6})_{-4}\\
&  +(1,6)_{4}+(1,8)_{0}+(2,3)_{2}+(2,6)_{-2}+(2,1)_{6}+(2,\overline{3})_{-2}\\
&  +(2,\overline{6})_{2}+(2,1)_{-6}%
\end{align}
which does not contain any candidate $E$-fields.

$E_{6}\supset SU(2)\times SU(6)\supset SU(2)\times\lbrack SU(3)\times
SU(3)\times U(1)]$

All of the breaking patterns in this case have already been classified in our
discussion of breaking patterns for the maximal subgroup $SU(3)\times
SU(3)\times SU(3)$. \ Indeed, this essentially follows from the fact that
$SU(3)$ contains the maximal subgroup $SU(2)\times U(1)$. \ We therefore
proceed to the other rank six bulk gauge groups and their breaking patterns.

\subsubsection{$SU(7)$}

We assume that the matter content of $SU(7)$ descends from the adjoint
representation of $E_{8}$. \ For this reason, we only treat the adjoint, $7$,
$21$, $35$ and complex conjugate representations of $SU(7)$. \ The maximal
subgroups of $SU(7)$ are:%
\begin{align}
SU(7)  &  \supset SU(6)\times U(1)\\
SU(7)  &  \supset SU(2)\times SU(5)\times U(1)\\
SU(7)  &  \supset SU(3)\times SU(4)\times U(1)\\
SU(7)  &  \supset SO(7)
\end{align}
of which only the first three contain $G_{std}$.

\paragraph{$SU(7)\supset SU(6)\times U(1)$}

There are three maximal subgroups of $SU(6)$ which can contain the non-abelian
factor of $G_{std}$ and can be reached via an instanton:%
\begin{align}
SU(7)  &  \supset SU(6)\times U(1)\supset SU(5)\times U(1)\times U(1)\\
SU(7)  &  \supset SU(6)\times U(1)\supset SU(2)\times SU(4)\times U(1)\times
U(1)\\
SU(7)  &  \supset SU(6)\times U(1)\supset SU(3)\times SU(3)\times U(1)\times
U(1)\text{.}%
\end{align}
In this case, in order to preserve an $SU(3)\times SU(2)$ factor, the only
available instanton configuration must generically take values in $U(1)^{3}$
so that all nested sequences of maximal subgroups which can be reached by an
instanton configuration all descend to the group $SU(3)\times SU(2)\times
U(1)\times U(1)\times U(1)$. \ It is therefore enough to consider the breaking
pattern:%
\begin{align}
SU(7)  &  \supset SU(6)\times U(1)\supset SU(5)\times U(1)\times U(1)\\
&  \supset SU(3)\times SU(2)\times U(1)\times U(1)\times U(1)\\
7  &  \rightarrow1_{0,0,6}+1_{0,-5,-1}+(1,2)_{3,1,-1}+(3,1)_{-2,1,-1}\\
21  &  \rightarrow(1,1)_{6,2,-2}+(\overline{3},1)_{-4,2,-2}+(3,2)_{1,2,-2}%
+(3,1)_{-2,2,-2}\\
&  +(1,2)_{3,2,-2}+(1,1)_{0,-5,5}+(3,1)_{-2,1,5}+(1,2)_{3,1,5}\\
35  &  \rightarrow(1,1)_{6,-3,-3}+(\overline{3},1)_{-4,-3,-3}+(3,2)_{1,-3,-3}%
+(1,1)_{-6,3,-3}\\
&  +(3,1)_{4,3,-3}+(\overline{3},2)_{-1,3,-3}+(1,1)_{6,2,4}+(\overline
{3},1)_{-4,2,4}\\
&  +(3,2)_{1,2,4}+(3,1)_{-2,-4,4}+(1,2)_{3,-4,4}\\
48  &  \rightarrow1_{0,0,0}+1_{0,0,0}+(3,1)_{-2,6,0}+(1,2)_{3,6,0}%
+(\overline{3},1)_{2,-6,0}\\
&  +(1,2)_{-3,-6,0}+(1,1)_{0,0,0}+(1,3)_{0,0,0}+(8,1)_{0,0,0}\\
&  +(3,2)_{-5,0,0}+(\overline{3},2)_{5,0,0}+(1,1)_{0,-5,-7}+(1,1)_{0,5,7}\\
&  +(3,1)_{-2,1,-7}+(1,2)_{3,1,-7}+(\overline{3},1)_{2,-1,7}+(1,2)_{-3,-1,7}%
\text{.}%
\end{align}
By inspection, all of the representations of the MSSM are present in the above decompositions.

\paragraph{$SU(7)\supset SU(2)\times SU(5)\times U(1)$}

The representations of $SU(7)$ now decompose as:%
\begin{align}
SU(7)  &  \supset SU(2)\times SU(5)\times U(1)\\
7  &  \rightarrow(2,1)_{5}+(1,5)_{-2}\\
21  &  \rightarrow(1,1)_{10}+(1,10)_{-4}+(2,5)_{3}\\
35  &  \rightarrow(1,5)_{8}+(2,10)_{1}+(1,\overline{10})_{-6}\\
48  &  \rightarrow(3,1)_{0}+(1,24)_{0}+(2,5)_{-7}+(2,\overline{5})_{7}\text{.}%
\end{align}
In order to retain an $SU(3)$ subgroup, an instanton must take values in an
appropriate $U(1)$ or $SU(2)$ subgroup of $SU(5)$. \ As before, a generic
$U(1)^{3}$ instanton will yield the expected MSSM\ spectrum. \ If we instead
consider an $SU(2)\times U(1)$ instanton, it is also immediate that we can
again obtain the desired spectrum of the MSSM. \ This alternative breaking
pattern has the added benefit that it contains one less extraneous $U(1)$ factor.

\paragraph{$SU(7)\supset SU(3)\times SU(4)\times U(1)$}

The representations of $SU(7)$ decompose as:%
\begin{align}
SU(7)  &  \supset SU(3)\times SU(4)\times U(1)\\
7  &  \rightarrow(3,1)_{4}+(1,4)_{-3}\\
21  &  \rightarrow(\overline{3},1)_{8}+(3,4)_{1}+(1,6)_{-6}\\
35  &  \rightarrow(1,1)_{12}+(\overline{3},4)_{5}+(3,6)_{-2}+(1,\overline
{4})_{-9}\\
48  &  \rightarrow(1,1)_{0}+(8,1)_{0}+(1,15)_{0}+(3,\overline{4}%
)_{7}+(3,4)_{-7}\text{.}%
\end{align}
First suppose that the instanton configuration preserves the $SU(3)$ subgroup
of $SU(4)\supset SU(3)\times U(1)$. \ Such an instanton must then also
preserve an $SU(2)$ subgroup of the first $SU(3)$ factor so that the resulting
$U(1)^{3}$ instanton reduces to the generic situation treated previously.

Alternatively, an instanton can preserve all of the first $SU(3)$ factor and
break $SU(4)$ to a smaller subgroup. \ To this end, recall that the maximal
subgroups of $SU(4)$ which can contain an $SU(2)$ subgroup are:%
\begin{align}
SU(4)  &  \supset SU(3)\times U(1)\label{SU4one}\\
SU(4)  &  \supset SU(2)\times SU(2)\times U(1)\\
SU(4)  &  \supset USp(4)\\
SU(4)  &  \supset SU(2)\times SU(2)\text{.} \label{SU44}%
\end{align}
In order to preserve an $SU(2)$ subgroup, the first case necessarily descends
to the previously treated case of a $U(1)^{3}$ instanton. \ We therefore focus
on the remaining cases.

$SU(7)\supset SU(3)\times SU(4)\times U(1)\supset SU(3)\times\lbrack
SU(2)\times SU(2)\times U(1)]\times U(1)$

In this case, we note that the resulting nested sequence of maximal subgroups
descends to the same subgroup as:%
\begin{equation}
SU(7)\supset SU(2)\times SU(5)\times U(1)\supset SU(2)\times\lbrack
SU(3)\times SU(2)\times U(1)]\times U(1)
\end{equation}
whose breaking patterns have already been analyzed.

$SU(7)\supset SU(3)\times SU(4)\times U(1)\supset SU(3)\times USp(4)\times
U(1)$

Under this subgroup, the representations of $SU(7)$ decompose as:%
\begin{align}
SU(7)  &  \supset SU(3)\times SU(4)\times U(1)\supset SU(3)\times USp(4)\times
U(1)\\
7  &  \rightarrow(3,1)_{4}+(1,4)_{-3}\\
21  &  \rightarrow(\overline{3},1)_{8}+(3,4)_{1}+(1,1)_{-6}+(1,5)_{-6}\\
35  &  \rightarrow(1,1)_{12}+(\overline{3},4)_{5}+(3,1)_{-2}+(3,5)_{-2}%
+(1,4)_{-9}\\
48  &  \rightarrow(1,1)_{0}+(8,1)_{0}+(1,5)_{0}+(1,10)_{0}+(3,4)_{7}%
+(\overline{3},4)_{-7}%
\end{align}
there are two possible maximal subgroups of $USp(4)$ which can be reached by a
general breaking pattern:%
\begin{align}
a)  &  :USp(4)\supset SU(2)\times SU(2)\label{USP4a}\\
b)  &  :USp(4)\supset SU(2)\times U(1)\text{.} \label{USP4b}%
\end{align}

We first consider the decomposition with respect to case $a)$:%
\begin{align}
SU(7)  &  \supset SU(3)\times USp(4)\times U(1)\supset SU(3)\times\lbrack
SU(2)\times SU(2)]\times U(1)\label{firstSU7USP}\\
7  &  \rightarrow(3,1,1)_{4}+(1,2,1)_{-3}+(1,1,2)_{-3}\\
21  &  \rightarrow(\overline{3},1,1)_{8}+(3,2,1)_{1}+(3,1,2)_{1}%
+(1,1,1)_{-6}\\
&  +(1,1,1)_{-6}+(1,2,2)_{-6}\\
35  &  \rightarrow(1,1,1)_{12}+(\overline{3},2,1)_{5}+(\overline{3}%
,1,2)_{5}+(3,1,1)_{-2}\\
&  +(3,1,1)_{-2}+(3,2,2)_{-2}+(1,1,2)_{-9}+(1,2,1)_{-9}\\
48  &  \rightarrow(1,1,1)_{0}+(8,1,1)_{0}+(1,1,1)_{0}+(1,2,2)_{0}\\
&  +(1,3,1)_{0}+(1,1,3)_{0}+(1,2,2)_{0}+(3,1,2)_{7}\\
&  +(3,2,1)_{7}+(\overline{3},1,2)_{-7}+(\overline{3},2,1)_{-7}\text{.}%
\end{align}
Without loss of generality, we may consider an instanton which breaks the
first $SU(2)$ factor to either the trivial group, or a $U(1)$ subgroup.
\ Indeed, we find that even when the instanton configuration contains a
non-abelian factor, it is possible to reach the MSSM\ spectrum:%
\begin{equation}%
\begin{tabular}
[c]{|l|l|l|l|l|l|l|}\hline
$Q_{21}$ & $U_{\overline{7}}$ & $D_{\overline{35}}$ & $L_{7}$ & $E_{\overline
{21}}$ & $H_{u\overline{7}}$ & $H_{d7}$\\\hline
$(3,2,1)_{1}$ & $(3,1,1)_{-4}$ & $(\overline{3},1,1)_{2}$ & $(1,2,1)_{-3}$ &
$(1,1,1)_{6}$ & $(1,2,1)_{3}$ & $(1,2,1)_{-3}$\\\hline
\end{tabular}
\text{.} \label{SU7USPcase}%
\end{equation}
Note that in this case, an $SU(2)\times U(1)$ instanton will break $SU(7)$
directly to $G_{std}$ with no extraneous $U(1)$ factors.

Next consider the decomposition with respect to case $b)$:%
\begin{align}
SU(7)  &  \supset SU(3)\times USp(4)\times U(1)\supset SU(3)\times\lbrack
SU(2)\times U(1)]\times U(1)\\
7  &  \rightarrow(3,1)_{0,4}+(1,2)_{1,-3}+(1,2)_{-1,-3}\\
21  &  \rightarrow(\overline{3},1)_{0,8}+(3,2)_{1,1}+(3,2)_{-1,1}%
+(1,1)_{0,-6}\\
&  +(1,1)_{2,-6}+(1,1)_{-2,-6}+(1,3)_{0,-6}\\
35  &  \rightarrow(1,1)_{0,12}+(\overline{3},2)_{1,5}+(\overline{3}%
,2)_{-1,5}+(3,1)_{0,-2}\\
&  +(3,1)_{2,-2}+(3,1)_{-2,-2}+(3,3)_{0,-2}+(1,2)_{1,-9}+(1,2)_{-1,-9}\\
48  &  \rightarrow(1,1)_{0,0}+(8,1)_{0,0}+(1,1)_{2,0}+(1,1)_{2,0}%
+(1,1)_{-2,0}\\
&  +(1,3)_{0,0}+(1,3)_{2,0}+(1,3)_{-2,0}+(3,2)_{1,7}\\
&  +(3,2)_{-1,7}+(\overline{3},2)_{1,-7}+(\overline{3},2)_{-1,-7}\text{.}%
\end{align}
In fact, with respect to the corresponding $U(1)\times U(1)$ instanton, we
find that the matter content again organizes into the precise analogue of line
(\ref{SU7USPcase}) in this case as well. \ We therefore conclude that these
candidate breaking patterns can in principle be used to eliminate additional
$U(1)$ factors.

$SU(7)\supset SU(3)\times SU(4)\times U(1)\supset SU(3)\times SU(2)\times
SU(2)\times U(1)$

The final case of interest proceeds via a different embedding of $SU(2)\times
SU(2)$ in $SU(4)$ such that:%
\begin{align}
SU(7)  &  \supset SU(3)\times SU(4)\times U(1)\supset SU(3)\times\lbrack
SU(2)\times SU(2)]\times U(1)\\
7  &  \rightarrow(3,1,1)_{4}+(1,2,2)_{-3}\\
21  &  \rightarrow(\overline{3},1,1)_{8}+(3,2,2)_{1}+(1,1,3)_{-6}%
+(1,3,1)_{-6}\\
35  &  \rightarrow(1,1,1)_{12}+(\overline{3},2,2)_{5}+(3,1,3)_{-2}%
+(3,3,1)_{-2}+(1,2,2)_{-9}\\
48  &  \rightarrow(1,1,1)_{0}+(8,1,1)_{0}+(1,1,3)_{0}+(1,3,1)_{0}%
+(1,3,3)_{0}\\
&  +(3,2,2)_{7}+(\overline{3},2,2)_{-7}\text{.}%
\end{align}
Although this decomposition is indeed different from that presented below line
(\ref{firstSU7USP}), we note that an $SU(2)$ instanton can generate a very
similar breaking pattern. \ Indeed, under the forgetful homomorphism which
trivializes all representations of the first $SU(2)$ factor, we find that the
two decompositions are in fact identical. \ In particular, this implies that a
similar packaging of the field content of the MSSM\ as in line
(\ref{SU7USPcase}) will hold in this case as well.

\subsubsection{$SO(12)$}

We now proceed to the final rank six bulk gauge group which can occur in a
candidate F-theory GUT model. \ Starting from the adjoint representation of
$E_{8}$, the matter content of the bulk $SO(12)$ theory descends from the
vector $12$, the spinors $32$, $32^{\prime}$ and adjoint $66$. \ The maximal
subgroups of $SO(12)$ are:%
\begin{align}
SO(12)  &  \supset SU(6)\times U(1)\\
SO(12)  &  \supset SU(2)\times SU(2)\times SO(8)\\
SO(12)  &  \supset SU(4)\times SU(4)\\
SO(12)  &  \supset SO(10)\times U(1)\\
SO(12)  &  \supset SO(11)\\
SO(12)  &  \supset SU(2)\times SO(9)\\
SO(12)  &  \supset SU(2)\times USp(6)\\
SO(12)  &  \supset USp(4)\times SO(7)\text{.}\\
SO(12)  &  \supset SU(2)\times SU(2)\times SU(2)
\end{align}
of which all but the last entry contain $G_{std}$. \ As in previous examples,
our expectation is that many distinct nested sequences of maximal subgroups
can describe the breaking pattern of the same instanton configuration.

\paragraph{$SO(12)\supset USp(4)\times SO(7)$}

The decomposition of representations of $SO(12)$ is:%
\begin{align}
SO(12)  &  \supset USp(4)\times SO(7)\\
12  &  \rightarrow(5,1)+(1,7)\\
32,32^{\prime}  &  \rightarrow(4,8)\\
66  &  \rightarrow(10,1)+(1,21)+(5,7)
\end{align}
Of the two simple group factors, only $SO(7)$ contains an $SU(3)$ subgroup.
\ Further, while $G_{2}$ and $SU(4)$ are the two maximal subgroups of $SO(7)$
which contain an $SU(3)$ subgroup, an instanton can only break $SO(7)$ $\ $to
$SU(3)$ via the $SU(4)$ path. \ Further decomposing with respect to the nested
sequence $SO(7)\supset SU(4)\supset SU(3)\times U(1)$ therefore yields:%
\begin{align}
SO(12)  &  \supset USp(4)\times SO(7)\supset USp(4)\times SU(4)\supset
USp(4)\times SU(3)\times U(1)\\
12  &  \rightarrow(5,1)_{0}+(1,1)_{0}+(1,3)_{2}+(1,\overline{3})_{-2}\\
32,32^{\prime}  &  \rightarrow(4,3)_{1}+(4,1)_{-3}+(4,\overline{3}%
)_{-1}+(4,1)_{3}\\
66  &  \rightarrow(10,1)_{0}+(1,1)_{0}+(1,3)_{2}+(1,\overline{3}%
)_{-2}+(1,3)_{2}\\
&  +(1,\overline{3})_{-2}+(1,1)_{0}+(1,3)_{-4}+(1,\overline{3})_{4}%
+(1,8)_{0}\\
&  +(5,1)_{0}+(5,3)_{2}+(5,\overline{3})_{-2}\text{.}%
\end{align}
With conventions as in lines (\ref{USP4a}) and (\ref{USP4b}), we now decompose
$USp(4)$ with respect to the two maximal subgroups which can break to an
$SU(2)$ factor in the presence of an $SU(2)$ factor.

$SO(12)\supset USp(4)\times SO(7)\supset\lbrack SU(2)\times SU(2)]\times
\lbrack SU(3)\times U(1)_{b}]$

First consider the maximal subgroup $USp(4)\supset SU(2)\times SU(2)$: \
\begin{align}
SO(12)  &  \supset USp(4)\times SO(7)\supset\lbrack SU(2)\times SU(2)]\times
\lbrack SU(3)\times U(1)_{b}]\\
12  &  \rightarrow(1,1,1)_{0}+(2,2,1)_{0}+(1,1,1)_{0}+(1,1,3)_{2}%
+(1,1,\overline{3})_{-2}\\
32,32^{\prime}  &  \rightarrow(2,1,3)_{1}+(1,2,3)_{1}+(2,1,1)_{-3}%
+(1,2,1)_{-3}+(2,1,\overline{3})_{-1}\\
&  +(1,2,\overline{3})_{-1}+(2,1,1)_{3}+(1,2,1)_{3}\\
66  &  \rightarrow(3,1,1)_{0}+(1,3,1)_{0}+(2,2,1)_{0}+(1,1,1)_{0}%
+(1,1,3)_{2}\\
&  +(1,1,\overline{3})_{-2}+(1,1,3)_{2}+(1,1,\overline{3})_{-2}+(1,1,1)_{0}%
+(1,1,3)_{-4}\\
&  +(1,1,\overline{3})_{4}+(1,1,8)_{0}+(1,1,1)_{0}+(2,2,1)_{0}+(1,1,3)_{2}\\
&  +(1,1,\overline{3})_{-2}+(2,2,\overline{3})_{-2}\text{.}%
\end{align}
In this case it follows that an $SU(2)$ instanton cannot yield the correct
$U(1)_{Y}$ assignments for the fields of the MSSM. \ If we instead consider a
$U(1)$ instanton which breaks one of the $SU(2)$ factors to $U(1)_{a}$, the
following combinations of representations satisfy the requirements that all
$U(1)_{Y}$ charge assignments are correct and further, that all interaction
terms are consistent with gauge invariance of the parent theory:%
\begin{align}%
\begin{tabular}
[c]{|l|l|l|l|l|}\hline
& $Q$ & $U$ & $D$ & $L$\\\hline
$1$ & $(1_{0},2,3)_{1}$ & $(1_{-1},1,\overline{{\small 3}})_{-1}$ &
$(1_{1},1,\overline{3})_{-1}\text{ }$ & $(1_{0},2,1)_{-3}$\\\hline
$2$ & $(1_{0},2,3)_{1}$ & $(1_{1},1,\overline{3})_{-1}$ & $(1_{-1}%
,1,\overline{3})_{-1}\text{ }$ & $(1_{0},2,1)_{-3}$\\\hline
\end{tabular}
& \\%
\begin{tabular}
[c]{|l|l|l|l|l|}\hline
& $E$ & $H_{u}$ & $H_{d}$ & $(a,b)$\\\hline
$1$ & $(1_{1},1,1)_{3}$ & $(1_{1},2,1)_{0}$ & $(1_{-1},2,1)_{0}$ &
$(3,1)$\\\hline
$2$ & $(1_{-1},1,1)_{3}$ & $(1_{-1},2,1)_{0}$ & $(1_{1},2,1)_{0}$ &
$(-3,1)$\\\hline
\end{tabular}
&  \text{.}%
\end{align}

$SO(12)\supset USp(4)\times SO(7)\supset\lbrack SU(2)\times U(1)]\times\lbrack
SU(3)\times U(1)_{b}]$

Next consider the maximal subgroup $USp(4)\supset SU(2)\times U(1)$:%
\begin{align}
SO(12)  &  \supset USp(4)\times SO(7)\supset\lbrack SU(2)\times U(1)_{a}%
]\times\lbrack SU(3)\times U(1)_{b}]\\
12  &  \rightarrow(1_{2},1_{0})+(1_{-2},1_{0})+(3_{0},1_{0})+(1_{0}%
,1_{0})+(1_{0},3_{2})+(1_{0},\overline{3}_{-2})\\
32,32^{\prime}  &  \rightarrow(2_{1},3_{1})+(2_{-1},3_{1})+(2_{1}%
,1_{-3})+(2_{-1},1_{-3})+(2_{1},\overline{3}_{-1})\\
&  +(2_{-1},\overline{3}_{-1})+(2_{1},1_{3})+(2_{-1},1_{3})\\
66  &  \rightarrow(1_{0},1_{0})+(3_{0},1_{0})+(3_{2},1_{0})+(3_{-2}%
,1_{0})+(1_{0},1_{0})+(1_{0},3_{2})\\
&  +(1_{0},\overline{3}_{-2})+(1_{0},3_{2})+(1_{0},\overline{3}_{-2}%
)+(1_{0},1_{0})+(1_{0},3_{-4})+(1_{0},\overline{3}_{4})\\
&  +(1_{0},8_{0})+(1_{2},1_{0})+(1_{-2},1_{0})+(3_{0},1_{0})+(1_{2}%
,3_{2})+(1_{-2},3_{2})\\
&  +(3_{0},3_{2})+(1_{2},\overline{3}_{-2})+(1_{-2},\overline{3}_{-2}%
)+(3_{0},\overline{3}_{-2})
\end{align}
Listing all possible $Q$-, $U$- and $H_{u}$-fields we find:%
\begin{equation}%
\begin{tabular}
[c]{|l|l|l|}\hline
$Q$ & $U$ & $H_{u}$\\\hline
$(2_{\pm1},3_{1})$ & $(1_{0},\overline{3}_{-2})\text{ or }(1_{0},\overline
{3}_{4})\text{ or }(1_{\pm2},\overline{3}_{-2})$ & $(2_{\pm1},1_{\pm3}%
)$\\\hline
\end{tabular}
\text{.}%
\end{equation}
Note in particular that in this case, it is not possible to form a gauge
invariant $QUH_{u}$, so this path is excluded.

\paragraph{$SO(12)\supset SU(2)\times USp(6)$}

Because there is a unique maximal subgroup of $USp(6)$ which contains an
$SU(3)$ factor, we may perform the unique decomposition:%
\begin{align}
SO(12)  &  \supset SU(2)\times USp(6)\supset SU(2)\times\lbrack SU(3)\times
U(1)]\\
12  &  \rightarrow(2,3)_{1}+(2,\overline{3})_{-1}\\
32  &  \rightarrow(4,1)_{0}+(2,3)_{-2}+(2,\overline{3})_{2}+(2,8)_{0}\\
32^{\prime}  &  \rightarrow(3,3)_{1}+(3,\overline{3})_{-1}+(1,1)_{3}%
+(1,1)_{-3}+(1,6)_{-1}+(1,\overline{6})_{1}\\
66  &  \rightarrow(3,1)_{0}+(1,1)_{0}+(1,6)_{2}+(1,\overline{6})_{-2}%
+(1,8)_{0}+(3,3)_{-2}\\
&  +(3,\overline{3})_{2}+(3,8)_{0}\text{.}%
\end{align}
By inspection, the relative $U(1)_{Y}$ charge assignments for the $E$- and
$Q$-fields are incorrect. \ We therefore conclude that this breaking pattern
is not viable.

\paragraph{$SO(12)\supset SU(2)\times SO(9)$}

The decomposition of $SO(12)$ representations in this case yields:
\begin{align}
SO(12)  &  \supset SU(2)\times SO(9)\\
12  &  \rightarrow(3,1)+(1,9)\\
32,32^{\prime}  &  \rightarrow(2,16)\\
66  &  \rightarrow(3,1)+(1,36)+(3,9)\text{.}%
\end{align}
There are three maximal subgroups of $SO(9)$ which contain an $SU(3)$ factor
via a nested sequence of maximal subgroups:%
\begin{align}
SO(9)  &  \supset SU(2)\times SU(4)\supset SU(2)\times SU(3)\times U(1)\\
SO(9)  &  \supset SO(8)\supset SO(7)\supset SU(4)\supset SU(3)\times U(1)\\
SO(9)  &  \supset SO(8)\supset SU(4)\times U(1)\supset SU(3)\times U(1)\times
U(1)\\
SO(9)  &  \supset SO(7)\times U(1)\supset SU(4)\times U(1)\supset SU(3)\times
U(1)\times U(1)\text{.}%
\end{align}
By inspection, the $U(1)\times U(1)$ valued instanton associated with the last
two nested sequences yield identical breaking patterns.

$SO(12)\supset SU(2)\times SO(9)\supset SU(2)\times SU(2)\times SU(4)$

Decomposing the representations of $SO(12)$ with respect to this breaking
pattern yields:%
\begin{align}
SO(12)  &  \supset SU(2)\times SO(9)\supset SU(2)\times SU(2)\times SU(4)\\
12  &  \rightarrow(3,1,1)+(1,3,1)+(1,1,6)\\
32,32^{\prime}  &  \rightarrow(2,2,4)+(2,2,\overline{4})\\
66  &  \rightarrow(3,1,1)+(1,3,1)+(1,1,15)+(1,3,16)+(3,3,1)+(3,1,6)
\end{align}
In this case, the analysis of breaking patterns is similar to that of the
maximal subgroup $SO(10)\supset SU(2)\times SU(2)\times SU(4)$. \ We therefore
conclude that the appropriate $U(1)\times U(1)$ instanton configuration can
produce the spectrum of the MSSM.

$SO(12)\supset SU(2)\times SO(9)\supset SU(2)\times SO(8)\supset SU(2)\times
SO(7)$

$\supset SU(2)\times SU(4)\supset SU(2)\times\lbrack SU(3)\times U(1)]$

In this case, the decomposition to the appropriate subgroup does not yield a
viable candidate for the $E$-field:%
\begin{align}
SO(12)  &  \supset...\supset SU(2)\times SU(4)\supset SU(2)\times\lbrack
SU(3)\times U(1)]\\
12  &  \rightarrow(3,1)_{0}+(1,1)_{0}+(1,1)_{0}+(1,1)_{0}+(1,3)_{2}%
+(1,\overline{3})_{-2}\\
32,32^{\prime}  &  \rightarrow(2,1)_{3}+(2,3)_{-1}+(2,1)_{-3}+(2,\overline
{3})_{1}+(2,1)_{3}+(2,3)_{-1}\\
&  +(2,1)_{-3}+(2,\overline{3})_{1}\\
66  &  \rightarrow(3,1)_{0}+(1,1)_{0}+(1,1)_{0}+(1,1)_{0}+(1,3)_{2}%
+(1,\overline{3})_{-2}\\
&  +(1,3)_{2}+(1,\overline{3})_{-2}+(1,1)_{0}+(1,3)_{-4}+(1,\overline{3}%
)_{4}+(1,8)_{0}\\
&  +(1,3)_{2}+(1,\overline{3})_{-2}+(3,1)_{0}+(3,1)_{0}+(3,1)_{0}+(3,3)_{2}\\
&  +(3,\overline{3})_{-2}%
\end{align}

$SO(12)\supset SU(2)\times SO(9)\supset SU(2)\times SO(8)$

$\supset SU(2)\times SU(4)\times U(1)\supset SU(2)\times SU(3)\times
U(1)\times U(1)$

The decomposition to $G_{std}$ now yields:%
\begin{align}
SO(12)  &  \supset...\supset SU(2)\times SU(4)\times U(1)\\
&  \supset SU(2)\times SU(3)\times U(1)_{a}\times U(1)b\\
12  &  \rightarrow(3,1)_{0,0}+(1,1)_{0,2}+(1,1)_{0,-2}+(1,1)_{0,0}\\
&  +(1,3)_{2,0}+(1,\overline{3})_{-2,0}\\
32,32^{\prime}  &  \rightarrow(2,1)_{3,1}+(2,3)_{-1,1}+(2,1)_{-3,1}%
+(2,\overline{3})_{1,1}\\
&  +(2,1)_{-3,-1}+(2,\overline{3})_{1,-1}+(2,1)_{3,-1}+(2,3)_{-1,-1}\\
66  &  \rightarrow(3,1)_{0,0}+(1,1)_{0,0}+(1,1)_{0,2}+(1,3)_{2,2}\\
&  +(1,\overline{3})_{-2,2}+(1,1)_{0,-2}+(1,3)_{2,-2}+(1,\overline{3}%
)_{-2,-2}\\
&  +(1,3)_{2,0}+(1,\overline{3})_{-2,0}+(1,1)_{0,0}+(1,3)_{-4,0}\\
&  +(1,\overline{3})_{4,0}+(1,8)_{0,0}+(3,1)_{0,2}+(3,1)_{0,-2}\\
&  +(3,1)_{0,0}+(3,3)_{2,0}+(3,\overline{3})_{-2,0}\text{.}%
\end{align}
In this case, the candidate $E$- and $Q$-fields yield the relations:%
\begin{align}
E  &  :\pm2b=6\\
Q  &  :-a\pm b=1
\end{align}
so that $b=\pm3$ and $a=2$ or $-4$. \ Because the candidate $L$-fields all
descend from $(2,1)_{\pm3,\pm1}$, we further deduce that $a=2$. \ Without loss
of generality, we fix the sign of $b=+3$. \ This in turn implies that the
representation content of the remaining fields is now fixed to be:%
\begin{equation}%
\begin{tabular}
[c]{|l|l|l|l|l|l|l|}\hline
$Q$ & $U$ & $D$ & $L$ & $E$ & $H_{u}$ & $H_{d}$\\\hline
$(2,3)_{-1,-1}$ & $(1,\overline{3})_{-2,0}$ & $(1,\overline{3})_{-2,2}$ &
$(2,1)_{-3,1}$ & $(1,1)_{0,2}$ & $(2,1)_{3,-1}$ & $(2,1)_{-3,1}$\\\hline
\end{tabular}
\ \text{.}%
\end{equation}
Because some of the necessary interaction terms of the MSSM\ are now forbidden
by gauge invariance of the parent theory, we conclude that this does not yield
a viable breaking pattern.

\paragraph{$SO(12)\supset SO(11)$}

In this case, the breaking patterns of $SO(12)$ directly descend to the
analysis of $SO(11)$ breaking patterns previously analyzed. \ Indeed, the
representations of $SO(12)$ descend as:%
\begin{align}
SO(12)  &  \supset SO(11)\\
12  &  \rightarrow1+11\\
32,32^{\prime}  &  \rightarrow32\\
66  &  \rightarrow11+55\text{.}%
\end{align}

\paragraph{$SO(12)\supset SU(6)\times U(1)$}

First recall that the maximal subgroups of $SU(6)$ which contain $SU(3)\times
SU(2)$ are:%
\begin{align}
SU(6)  &  \supset SU(5)\times U(1)\\
SU(6)  &  \supset SU(2)\times SU(4)\times U(1)\\
SU(6)  &  \supset SU(3)\times SU(3)\times U(1)\\
SU(6)  &  \supset SU(2)\times SU(3)\text{.}%
\end{align}
In the first three cases we find that the resulting breaking pattern must
descend to the usual breaking pattern via a $U(1)^{3}$ instanton. \ Finally,
by inspection of the decomposition of $SO(12)\supset SU(6)\times U(1)$, we
note that the resulting integral $U(1)$ charges of each decomposition are
bounded in magnitude by two. \ Hence, only the first three maximal subgroups
can yield a consistent breaking pattern. \ While it would be of interest to
classify all possible ways of packaging the field content of the MSSM\ in
representations of $SO(12)$ in this case, this analysis is not necessary for
the purposes of classifying breaking patterns.

\paragraph{$SO(12)\supset SU(2)\times SU(2)\times SO(8)$}

Decomposing all relevant representations of $SO(12)$ with respect to this
maximal subgroup yields:%
\begin{align}
SO(12)  &  \supset SU(2)\times SU(2)\times SO(8)\\
12  &  \rightarrow(2,2,1)+(1,1,8^{v})\\
32  &  \rightarrow(1,2,8^{s})+(2,1,8^{c})\\
32^{\prime}  &  \rightarrow(1,2,8^{c})+(2,1,8^{s})\\
66  &  \rightarrow(3,1,1)+(1,3,1)+(1,1,28)+(2,2,8^{v})\text{.}%
\end{align}
There are two maximal subgroups of $SO(8)$ which are consistent with a
breaking pattern generated by an instanton configuration:%
\begin{align}
SO(8)  &  \supset SU(4)\times U(1)\\
SO(8)  &  \supset SO(7)\supset SU(4)\text{.}%
\end{align}
We now consider breaking patterns which can descend from both maximal subgroups.

$SO(12)\supset SU(2)\times SU(2)\times SO(8)\supset SU(2)\times SU(2)\times
\lbrack SU(4)\times U(1)]$

Because the only simple group factor which contains an $SU(3)$ subgroup is
$SU(4)$, we may further decompose $SU(4)\supset SU(3)\times U(1)$. \ This
yields:%
\begin{align}
SO(12)  &  \supset SU(2)\times SU(2)\times SO(8)\supset SU(2)\times
SU(2)\times\lbrack SU(4)\times U(1)]\\
&  \supset SU(2)\times SU(2)\times\lbrack SU(3)\times U(1)\times U(1)]\\
12  &  \rightarrow(2,2,1)_{0,0}+(1,1,1)_{0,2}+(1,1,1)_{0,-2}\\
&  +(1,1,3)_{2,0}+(1,1,\overline{3})_{-2,0}\\
32  &  \rightarrow(1,2,1)_{3,1}+(1,2,3)_{-1,1}+(1,2,1)_{-3,-1}\\
&  +(1,2,\overline{3})_{1,-1}+(2,1,1)_{3,-1}+(2,1,3)_{-1,-1}\\
&  +(2,1,1)_{-3,1}+(2,1,\overline{3})_{1,1}\\
32^{\prime}  &  \rightarrow(2,1,1)_{3,1}+(2,1,3)_{-1,1}+(2,1,1)_{-3,-1}\\
&  +(2,1,\overline{3})_{1,-1}+(1,2,1)_{3,-1}+(1,2,3)_{-1,-1}\\
&  +(1,2,1)_{-3,1}+(1,2,\overline{3})_{1,1}\\
66  &  \rightarrow(3,1,1)_{0,0}+(1,3,1)_{0,0}+(1,1,1)_{0,0}\\
&  +(1,1,3)_{2,2}+(1,1,\overline{3})_{-2,2}+(1,1,3)_{2,-2}\\
&  +(1,1,\overline{3})_{-2,-2}+(1,1,1)_{0,0}+(1,1,3)_{-4,0}\\
&  +(1,1,\overline{3})_{4,0}+(1,1,8)_{0,0}+(2,2,1)_{0,2}\\
&  +(2,2,1)_{0,-2}+(2,2,3)_{2,0}+(2,2,\overline{3})_{-2,0}\text{.}
\end{align}

If we now consider a $U(1)$ instanton which breaks one of the $SU(2)$ factor,
we again obtain a $U(1)^{3}$ instanton configuration. \ Indeed, this case is
quite similar to breaking via the maximal subgroup $SU(2)\times SU(2)\times
SU(4)\subset SO(10)$ considered previously.

Next suppose without loss of generality that an instanton configuration takes
values in the first $SU(2)$ factor such that it breaks either to $U(1)$ or
trivial group. \ Because the abelian case is quite similar, we assume that the
non-abelian instanton breaks all of $SU(2)$. \ In this case, the list of
candidate $Q$-, $U$- and $H_{u}$-fields which can yield a gauge invariant
$QUH_{u}$ interaction are:%
\begin{equation}%
\begin{tabular}
[c]{|l|l|l|l|l|}\hline
& $Q$ & $U$ & $H_{u}$ & $(a,b)$\\\hline
$1$ & $(1,2,3)_{-1,1}$ & $(1,1,\overline{3})_{-2,0}$ & $(1,2,1)_{3,-1}$ &
$(2,3)$\\\hline
$2$ & $(1,2,3)_{-1,1}$ & $(2,1,\overline{3})_{1,1}$ & $(2,2,1)_{0,-2}$ &
$(-5/2,-3/2)$\\\hline
$3$ & $(1,2,3)_{-1,1}$ & $(2,1,\overline{3})_{1,-1}$ & $OUT$ & $OUT$\\\hline
$4$ & $(1,2,3)_{-1,-1}$ & $(1,1,\overline{3})_{-2,0}$ & $(1,2,1)_{3,1}$ &
$(2,-3)$\\\hline
$5$ & $(1,2,3)_{-1,-1}$ & $(2,1,\overline{3})_{1,1}$ & $OUT$ & $OUT$\\\hline
$6$ & $(1,2,3)_{-1,-1}$ & $(2,1,\overline{3})_{1,-1}$ & $(2,2,1)_{0,2}$ &
$(-5/2,3/2)$\\\hline
$7$ & $(2,2,3)_{2,0}$ & $(1,1,\overline{3})_{-2,0}$ & $OUT$ & $OUT$\\\hline
$8$ & $(2,2,3)_{2,0}$ & $(2,1,\overline{3})_{1,1}$ & $(1,2,1)_{-3,-1}$ &
$(1/2,-9/2)$\\\hline
$9$ & $(2,2,3)_{2,0}$ & $(2,1,\overline{3})_{1,-1}$ & $(1,2,1)_{-3,1}$ &
$(1/2,9/2)$\\\hline
\end{tabular}
\ \text{.}%
\end{equation}
Restricting to the six viable remaining possibilities, we now find that no
candidate $D$-field reproduces the correct $U(1)_{Y}$ charge assignment. \ We
therefore conclude that only abelian instanton configurations can yield the
spectrum of the MSSM\ in this case.

$SO(12)\supset SU(2)\times SU(2)\times SO(8)\supset SU(2)\times SU(2)\times
SO(7)$

$\supset SU(2)\times SU(2)\times SU(4)\supset SU(2)\times SU(2)\times\lbrack
SU(3)\times U(1)]$

Along this nested sequence of maximal subgroups, the decomposition of the
representations of $SO(12)$ is:
\begin{align}
SO(12)  &  \supset SU(2)\times SU(2)\times SO(8)\supset SU(2)\times
SU(2)\times SO(7)\\
&  \supset SU(2)\times SU(2)\times SU(4)\supset SU(2)\times SU(2)\times\lbrack
SU(3)\times U(1)_{b}]\\
12  &  \rightarrow(2,2,1)_{0}+(1,1,1)_{0}+(1,1,1)_{0}+(1,1,3)_{2}%
+(1,1,\overline{3})_{-2}\\
32,32^{\prime}  &  \rightarrow(1,2,1)_{3}+(1,2,3)_{-1}+(1,2,1)_{-3}%
+(1,2,\overline{3})_{1}+(2,1,1)_{3}\\
&  +(2,1,3)_{-1}+(2,1,1)_{-3}+(2,1,\overline{3})_{1}\\
66  &  \rightarrow(3,1,1)_{0}+(1,3,1)_{0}+(1,1,1)_{0}+(1,1,3)_{2}%
+(1,1,\overline{3})_{-2}\\
&  +(1,1,3)_{2}+(1,1,\overline{3})_{-2}+(1,1,1)_{0}+(1,1,3)_{-4}%
+(1,1,\overline{3})_{4}\\
&  +(1,1,8)_{0}+(2,2,1)_{0}+(2,2,1)_{0}+(2,2,3)_{2}+(2,2,\overline{3}%
)_{-2}\text{.}%
\end{align}

By inspection, the above $U(1)_{b}$ charge assignments do not agree with those
of the MSSM. \ It thus follows that we must further break one of the $SU(2)$
factors to $U(1)$. \ Without loss of generality, we assume that the first
$SU(2)$ factor decomposes further to a maximal $U(1)_{a}$ subgroup. \ The list
of candidate $Q$-, $U$- and $H_{u}$-fields which can yield a gauge invariant
$QUH_{u}$ interaction are therefore:%
\begin{equation}%
\begin{tabular}
[c]{|l|l|l|l|l|}\hline
& $Q$ & $U$ & $H_{u}$ & $(a,b)$\\\hline
$1$ & $(1_{0},2,3)_{-1}$ & $(1_{0},1,\overline{3})_{-2}$ & $(1_{0},2,1)_{3}$ &
$OUT$\\\hline
$2$ & $(1_{0},2,3)_{-1}$ & $(1_{1},1,\overline{3})_{1}$ & $(1_{-1},2,1)_{0}$ &
$(-3,-1)$\\\hline
$3$ & $(1_{0},2,3)_{-1}$ & $(1_{-1},1,\overline{3})_{1}$ & $(1_{1},2,1)_{0}$ &
$(3,-1)$\\\hline
$4$ & $(1_{0},2,3)_{-1}$ & $(1_{0},1,\overline{3})_{4}$ & $(1_{0},2,1)_{-3}$ &
$(a,-1)$\\\hline
$5$ & $(1_{1},2,3)_{2}$ & $(1_{0},1,\overline{3})_{-2}$ & $(1_{-1},2,1)_{0}$ &
$(-3,2)$\\\hline
$6$ & $(1_{1},2,3)_{2}$ & $(1_{1},1,\overline{3})_{1}$ & $OUT$ & $OUT$\\\hline
$7$ & $(1_{1},2,3)_{2}$ & $(1_{-1},1,\overline{3})_{1}$ & $(1_{0},2,1)_{-3}$ &
$(3,-1)$\\\hline
$8$ & $(1_{1},2,3)_{2}$ & $(1_{0},1,\overline{3})_{4}$ & $OUT$ & $OUT$\\\hline
$9$ & $(1_{-1},2,3)_{2}$ & $(1_{0},1,\overline{3})_{-2}$ & $(1_{1},2,1)_{0}$ &
$(3,2)$\\\hline
$10$ & $(1_{-1},2,3)_{2}$ & $(1_{1},1,\overline{3})_{1}$ & $(1_{0},2,1)_{-3}$
& $(-3,-1)$\\\hline
$11$ & $(1_{-1},2,3)_{2}$ & $(1_{-1},1,\overline{3})_{1}$ & $OUT$ &
$OUT$\\\hline
$12$ & $(1_{-1},2,3)_{2}$ & $(1_{0},1,\overline{3})_{4}$ & $OUT$ &
$OUT$\\\hline
\end{tabular}
\text{.}%
\end{equation}
Next, we list all candidate $D$- and $H_{d}$-fields which can yield a gauge
invariant $QDH_{d}$ interaction term:%
\begin{equation}%
\begin{tabular}
[c]{|l|l|l|l|l|}\hline
& $Q$ & $D$ & $H_{d}$ & $(a,b)$\\\hline
$2a$ & $(1_{0},2,3)_{-1}$ & $(1_{0},1,\overline{3})_{-2}$ & $(1_{0},2,1)_{3}$
& $(-3,-1)$\\\hline
$2b$ & $(1_{0},2,3)_{-1}$ & $(1_{-1},1,\overline{3})_{1}$ & $(1_{1},2,1)_{0}$
& $(-3,-1)$\\\hline
$3a$ & $(1_{0},2,3)_{-1}$ & $(1_{0},1,\overline{3})_{-2}$ & $(1_{0},2,1)_{3}$
& $(3,-1)$\\\hline
$3b$ & $(1_{0},2,3)_{-1}$ & $(1_{1},1,\overline{3})_{1}$ & $(1_{-1},2,1)_{0}$
& $(3,-1)$\\\hline
$4a$ & $(1_{0},2,3)_{-1}$ & $(1_{0},1,\overline{3})_{-2}$ & $(1_{0},2,1)_{3}$
& $(a,-1)$\\\hline
$4b$ & $(1_{0},2,3)_{-1}$ & $(1_{1},1,\overline{3})_{1}$ & $(1_{-1},2,1)_{0}$
& $(3,-1)$\\\hline
$4c$ & $(1_{0},2,3)_{-1}$ & $(1_{-1},1,\overline{3})_{1}$ & $(1_{1},2,1)_{0}$
& $(-3,-1)$\\\hline
$5$ & $(1_{1},2,3)_{2}$ & $OUT$ & $OUT$ & $(-3,2)$\\\hline
$7a$ & $(1_{1},2,3)_{2}$ & $(1_{0},1,\overline{3})_{-2}$ & $(1_{-1},2,1)_{0}$
& $(3,-1)$\\\hline
$7b$ & $(1_{1},2,3)_{2}$ & $(1_{1},1,\overline{3})_{1}$ & $OUT$ &
$(3,-1)$\\\hline
$9$ & $(1_{-1},2,3)_{2}$ & $(1_{1},1,\overline{3})_{1}$ & $OUT$ &
$(3,2)$\\\hline
$10a$ & $(1_{-1},2,3)_{2}$ & $(1_{0},1,\overline{3})_{-2}$ & $(1_{1},2,1)_{0}$
& $(-3,-1)$\\\hline
$10b$ & $(1_{-1},2,3)_{2}$ & $(1_{-1},1,\overline{3})_{1}$ & $OUT$ &
$(-3,-1)$\\\hline
\end{tabular}
\text{.}%
\end{equation}
Of these remaining possibilities, we now determine all possible candidate $L$-
and $E$-fields which can yield the gauge invariant interaction term $LEH_{d}$:%
\begin{equation}%
\begin{tabular}
[c]{|l|l|l|l|l|}\hline
& $L$ & $E$ & $H_{d}$ & $(a,b)$\\\hline
$2a$ & $(1_{1},2,1)_{0}$ & $(1_{-1},1,1)_{-3}$ & $(1_{0},2,1)_{3}$ &
$(-3,-1)$\\\hline
$2b$ & $(1_{0},2,1)_{3}$ & $(1_{-1},1,1)_{-3}$ & $(1_{1},2,1)_{0}$ &
$(-3,-1)$\\\hline
$2b^{\prime}$ & $(1_{1},2,1)_{0}$ & $(1_{-2},1,1)_{0}$ & $(1_{1},2,1)_{0}$ &
$(-3,-1)$\\\hline
$3a$ & $(1_{-1},2,1)_{0}$ & $(1_{1},1,1)_{-3}$ & $(1_{0},2,1)_{3}$ &
$(3,-1)$\\\hline
$3b$ & $(1_{-1},2,1)_{0}$ & $(1_{2},1,1)_{0}$ & $(1_{-1},2,1)_{0}$ &
$(3,-1)$\\\hline
$3b^{\prime}$ & $(1_{0},2,1)_{3}$ & $(1_{1},1,1)_{-3}$ & $(1_{-1},2,1)_{0}$ &
$(3,-1)$\\\hline
$4a$ & $(1_{\pm1},2,1)_{0}$ & $(1_{\mp1},1,1)_{-3}$ & $(1_{0},2,1)_{3}$ &
$(\mp3,-1)$\\\hline
$4b$ & $(1_{-1},2,1)_{0}$ & $(1_{2},1,1)_{0}$ & $(1_{-1},2,1)_{0}$ &
$(3,-1)$\\\hline
$4b^{\prime}$ & $(1_{0},2,1)_{3}$ & $(1_{1},1,1)_{-3}$ & $(1_{-1},2,1)_{0}$ &
$(3,-1)$\\\hline
$4c$ & $(1_{1},2,1)_{0}$ & $(1_{-2},1,1)_{0}$ & $(1_{1},2,1)_{0}$ &
$(-3,-1)$\\\hline
$4c^{\prime}$ & $(1_{0},2,1)_{3}$ & $(1_{-1},1,1)_{-3}$ & $(1_{1},2,1)_{0}$ &
$(-3,-1)$\\\hline
$7a$ & $(1_{-1},2,1)_{0}$ & $(1_{2},1,1)_{0}$ & $(1_{-1},2,1)_{0}$ &
$(3,-1)$\\\hline
$7a^{\prime}$ & $(1_{0},2,1)_{3}$ & $(1_{1},1,1)_{-3}$ & $(1_{-1},2,1)_{0}$ &
$(3,-1)$\\\hline
$10a$ & $(1_{1},2,1)_{0}$ & $(1_{-2},1,1)_{0}$ & $(1_{1},2,1)_{0}$ &
$(-3,-1)$\\\hline
$10a^{\prime}$ & $(1_{0},2,1)_{3}$ & $(1_{-1},1,1)_{-3}$ & $(1_{1},2,1)_{0}$ &
$(-3,-1)$\\\hline
\end{tabular}
\text{.}%
\end{equation}
Note that in this case, there are many distinct ways to package the field
content of the MSSM\ such that $SO(12)$ breaks to $SU(3)\times SU(2)\times
U(1)\times U(1)$ via a $U(1)^{2}$ instanton configuration.

\paragraph{$SO(12)\supset SU(4)\times SU(4)$}

Decomposing representations of $SO(12)$ with respect to the maximal subgroup
$SU(4)\times SU(4)$ yields:%
\begin{align}
SO(12)  &  \supset SU(4)\times SU(4)\\
12  &  \rightarrow(6,1)+(1,6)\\
12  &  \rightarrow(6,1)+(1,6)\\
32^{\prime}  &  \rightarrow(4,\overline{4})+(\overline{4},4)\\
66  &  \rightarrow(15,1)+(1,15)+(6,6)\text{.}%
\end{align}
Without loss of generality, we assume that the first $SU(4)$ factor further
breaks to $SU(3)\times U(1)$. \ The remaining nested sequences of maximal
subgroups which can yield the Standard Model gauge group are:%
\begin{align}
SU(4)  &  \supset SU(2)\times SU(2)\times U(1)\\
SU(4)  &  \supset USp(4)\supset SU(2)\times SU(2)\\
SU(4)  &  \supset USp(4)\supset SU(2)\times U(1)\\
SU(4)  &  \supset SU(2)\times SU(2)\text{.}%
\end{align}

$SO(12)\supset SU(4)\times SU(4)\supset\lbrack SU(3)\times U(1)_{a}]\times
SU(2)\times SU(2)\times U(1)$

In this case, it follows at once from the local isomorphisms $SU(4)\simeq
SO(6)$ and $SU(2)\times SU(2)\simeq SO(4)$ that the endpoint of this breaking
pattern is identical to the endpoint of the nested sequence of maximal
subgroups:%
\begin{align}
SO(12)  &  \supset SU(2)\times SU(2)\times SO(8)\supset SU(2)\times
SU(2)\times SU(4)\times U(1)\\
&  \supset SU(2)\times SU(2)\times\lbrack SU(3)\times U(1)]\times U(1)\text{.}%
\end{align}
We therefore conclude that all breaking patterns via instantons have in this
case been catalogued.

$SO(12)\supset SU(4)\times SU(4)\supset\lbrack SU(3)\times U(1)_{a}]\times
USp(4)\supset\lbrack SU(3)\times U(1)_{a}]\times\lbrack SU(2)\times SU(2)]$

The decomposition of the representations of $SO(12)$ with respected to this
sequence of maximal subgroups is:%
\begin{align}
SO(12)  &  \supset SU(4)\times SU(4)\supset\lbrack SU(3)\times U(1)_{a}]\times
USp(4)\\
&  \supset\lbrack SU(3)\times U(1)_{a}]\times\lbrack SU(2)\times SU(2)]\\
12  &  \rightarrow(3,1,1)_{2}+(\overline{3},1,1)_{-2}+(1,1,1)_{0}%
+(1,1,1)_{0}+(1,2,2)_{0}\\
32,32^{\prime}  &  \rightarrow(1,2,1)_{3}+(1,1,2)_{3}+(3,2,1)_{-1}%
+(3,1,2)_{-1}+(1,2,1)_{-3}\\
&  +(1,1,2)_{-3}+(\overline{3},2,1)_{1}+(\overline{3},1,2)_{1}\\
66  &  \rightarrow(1,1,1)_{0}+(3,1,1)_{-4}+(\overline{3},1,1)_{4}%
+(8,1,1)_{0}+(1,1,1)_{0}\\
&  +(1,2,2)_{0}+(1,3,1)_{0}+(1,1,3)_{0}+(1,2,2)_{0}+(3,1,1)_{2}\\
&  +(\overline{3},1,1)_{-2}+(3,1,1)_{2}+(3,2,2)_{2}+(\overline{3}%
,1,1)_{-2}+(\overline{3},2,2)_{-2}\text{.}%
\end{align}

By inspection of the above representation content, we note that while an
$SU(2)$ instanton which breaks either of the $SU(2)$ factors could yield the
correct gauge group, the resulting $U(1)_{Y}$ charge assignments of the fields
would be incorrect. \ It is therefore enough to consider abelian instanton
configurations which break one of the $SU(2)$ factors to $U(1)_{b}$. \ Due to
the symmetry between the two $SU(2)$ factors, we assume without loss of
generality that the instanton preserves the first $SU(2)$ factor. \ We begin
by listing the candidate representations for the $Q$-, $U$- and $H_{u}$-
fields which can yield the interaction term $QUH_{u}$ as well as the correct
$U(1)_{Y}$ charge assignments:%
\begin{equation}%
\begin{tabular}
[c]{|l|l|l|l|l|}\hline
& $Q$ & $U$ & $H_{u}$ & $(a,b)$\\\hline
$1$ & $(3,2,1_{0})_{-1}$ & $(\overline{3},1,1_{\pm1})_{1}$ & $(1,2,1_{\mp
1})_{0}$ & $(\mp3,-1)$\\\hline
$2$ & $(3,2,1_{0})_{-1}$ & $(\overline{3},1,1_{0})_{4}$ & $(1,2,1_{0})_{-3}$ &
$(a,-1)$\\\hline
$3$ & $(3,2,1_{\pm1})_{2}$ & $(\overline{3},1,1_{0})_{-2}$ & $(1,2,1_{\mp
1})_{0}$ & $(\mp3,2)$\\\hline
$4$ & $(3,2,1_{\pm1})_{2}$ & $(\overline{3},1,1_{\mp1})_{1}$ & $(1,2,1_{0}%
)_{-3}$ & $(\pm3,-1)$\\\hline
\end{tabular}
\end{equation}
where in the above, all $\pm$'s of a given row are correlated. \ Of these four
possibilities, we now list all candidate representations for the $D$- and
$H_{d}$-fields which can yield the interaction term $QDH_{d}$:%
\begin{equation}%
\begin{tabular}
[c]{|l|l|l|l|l|}\hline
& $Q$ & $D$ & $H_{d}$ & $(a,b)$\\\hline
$1a$ & $(3,2,1_{0})_{-1}$ & $(\overline{3},1,1_{\mp1})_{1}$ & $(1,2,1_{\pm
1})_{0}$ & $(\mp3,-1)$\\\hline
$1b$ & $(3,2,1_{0})_{-1}$ & $(\overline{3},1,1_{0})_{-2}$ & $(1,2,1_{0})_{3}$
& $(\mp3,-1)$\\\hline
$2a$ & $(3,2,1_{0})_{-1}$ & $(\overline{3},1,1_{\mp1})_{1}$ & $(1,2,1_{\pm
1})_{0}$ & $(\mp3,-1)$\\\hline
$2b$ & $(3,2,1_{0})_{-1}$ & $(\overline{3},1,1_{0})_{-2}$ & $(1,2,1_{0})_{3}$
& $(a,-1)$\\\hline
$4$ & $(3,2,1_{\pm1})_{2}$ & $(\overline{3},1,1_{0})_{-2}$ & $(1,2,1_{\mp
1})_{0}$ & $(\pm3,-1)$\\\hline
\end{tabular}
\text{.}%
\end{equation}
Finally, we list all candidate $E$- and $L$- fields which can yield the term
$ELH_{d}$:%
\begin{equation}%
\begin{tabular}
[c]{|l|l|l|l|l|}\hline
& $E$ & $L$ & $H_{d}$ & $(a,b)$\\\hline
$1a$ & $(1,1,1_{\mp1})_{-3}$ & $(1,2,1_{0})_{3}$ & $(1,2,1_{\pm1})_{0}$ &
$(\mp3,-1)$\\\hline
$1a^{\prime}$ & $(1,1,1_{\mp2})_{0}$ & $(1,2,1_{\pm1})_{0}$ & $(1,2,1_{\pm
1})_{0}$ & $(\mp3,-1)$\\\hline
$1b$ & $(1,1,1_{\mp1})_{-3}$ & $(1,2,1_{\pm1})_{0}$ & $(1,2,1_{0})_{3}$ &
$(\mp3,-1)$\\\hline
$2a$ & $(1,1,1_{\mp1})_{-3}$ & $(1,2,1_{0})_{3}$ & $(1,2,1_{\pm1})_{0}$ &
$(\mp3,-1)$\\\hline
$2a^{\prime}$ & $(1,1,1_{\mp2})_{0}$ & $(1,2,1_{\pm1})_{0}$ & $(1,2,1_{\pm
1})_{0}$ & $(\mp3,-1)$\\\hline
$2b$ & $(1,1,1_{\mp1})_{-3}$ & $(1,2,1_{\pm1})_{0}$ & $(1,2,1_{0})_{3}$ &
$(\mp3,-1)$\\\hline
$4$ & $(1,1,1_{\pm1})_{-3}$ & $(1,2,1_{0})_{3}$ & $(1,2,1_{\mp1})_{0}$ &
$(\pm3,-1)$\\\hline
$4^{\prime}$ & $(1,1,1_{\pm2})_{0}$ & $(1,2,1_{\mp1})_{0}$ & $(1,2,1_{\mp
1})_{0}$ & $(\pm3,-1)$\\\hline
\end{tabular}
\text{.}%
\end{equation}
We note that in this case, while there are only two linear combinations of the
two $U(1)$ factors which can yield $U(1)_{Y}$, there are different ways to
package the fields of the MSSM\ in representations of $SO(12)$.

$SO(12)\supset SU(4)\times SU(4)\supset\lbrack SU(3)\times U(1)_{a}]\times
USp(4)\supset\lbrack SU(3)\times U(1)_{a}]\times\lbrack SU(2)\times U(1)_{b}]$

In this case, the decomposition of representations of $SO(12)$ yields:%
\begin{align}
SO(12)  &  \supset SU(4)\times SU(4)\supset\lbrack SU(3)\times U(1)_{a}]\times
USp(4)\\
&  \supset\lbrack SU(3)\times U(1)_{a}]\times\lbrack SU(2)\times U(1)_{b}]\\
12  &  \rightarrow(3_{2},1_{0})+(\overline{3}_{-2},1_{0})+(1_{0},1_{0}%
)+(1_{0},1_{2})+(1_{0},1_{-2})+(1_{0},3_{0})\\
32,32^{\prime}  &  \rightarrow(1_{3},2_{1})+(1_{3},2_{-1})+(3_{-1}%
,2_{1})+(3_{-1},2_{-1})\\
&  +(1_{-3},2_{1})+(1_{-3},2_{-1})+(\overline{3}_{1},2_{1})+(\overline{3}%
_{1},2_{-1})\\
66  &  \rightarrow(1_{0},1_{0})+(3_{-4},1_{0})+(\overline{3}_{4},1_{0}%
)+(8_{0},1_{0})+(1_{0},1_{2})+(1_{0},1_{-2})\\
&  +(1_{0},3_{0})+(1_{0},1_{0})+(1_{0},3_{0})+(1_{0},3_{2})+(1_{0},3_{-2})\\
&  +(3_{2},1_{0})+(\overline{3}_{-2},1_{0})+(3_{2},1_{2})+(3_{2}%
,1_{-2})+(3_{2},3_{0})\\
&  +(\overline{3}_{-2},1_{2})+(\overline{3}_{-2},1_{-2})+(\overline{3}%
_{-2},3_{0})\text{.}%
\end{align}
We note in passing that this indeed yields a distinct decomposition from the
previous breaking pattern. \ By inspection, the only candidate $E$-fields are
$(1_{0},1_{\pm2})$ so that $b=\pm3$. \ Listing all $Q$-, $U$- and $H_{u}%
$-fields which can yield a gauge invariant interaction term $QUH_{u}$ such
that $b=\pm3$ is indeed a solution, we find:%
\begin{equation}%
\begin{tabular}
[c]{|l|l|l|l|}\hline
$Q$ & $U$ & $H_{u}$ & $(a,b)$\\\hline
$(3_{-1},2_{\pm1})$ & $(\overline{3}_{-2},1_{0})$ & $(1_{3},2_{\mp1})$ &
$(2,\pm3)$\\\hline
\end{tabular}
\end{equation}
where all $\pm$'s in a given row are correlated. \ Listing all $Q$-, $D$- and
$H_{d}$- fields which can yield the term $QDH_{d}$, we find:%
\begin{equation}%
\begin{tabular}
[c]{|l|l|l|l|}\hline
$Q$ & $D$ & $H_{d}$ & $(a,b)$\\\hline
$(3_{-1},2_{\pm1})$ & $(\overline{3}_{-2},1_{\mp2})$ & $(1_{-3},2_{\pm1})$ &
$(2,\pm3)$\\\hline
\end{tabular}
\ \text{.}%
\end{equation}
Now, we find that in this case, the only candidate $L$- and $H_{d}$-fields are
$(1_{-3},2_{\pm1})$. \ In particular, it follows that the purported $ELH_{d}$
interaction will violate $U(1)_{a}$ because the only candidate $E$-field is
neutral under $U(1)_{a}$ so that this breaking pattern cannot yield the
spectrum of the MSSM.

$SO(12)\supset SU(4)\times SU(4)\supset\lbrack SU(3)\times U(1)_{a}%
]\times\lbrack SU(2)\times SU(2)]$

In this case, the decomposition of the representations of $SO(12)$ is given
by:%
\begin{align}
SO(12)  &  \supset SU(4)\times SU(4)\supset\lbrack SU(3)\times U(1)_{a}%
]\times\lbrack SU(2)\times SU(2)]\\
12  &  \rightarrow(3_{2},1,1)+(\overline{3}_{-2},1,1)+(1_{0},3,1)+(1_{0}%
,1,3)\\
32,32^{\prime}  &  \rightarrow(1_{3},2,2)+(3_{-1},2,2)+(1_{-3},2,2)+(\overline
{3}_{1},2,2)\\
66  &  \rightarrow(1_{0},1,1)+(3_{-4},1,1)+(\overline{3}_{4},1,1)+(8_{0}%
,1,1)+(1_{0},1,3)\\
&  +(1_{0},3,1)+(1_{0},3,3)+(3_{2},1,3)+(3_{2},3,1)+(\overline{3}_{-2},1,3)\\
&  +(\overline{3}_{-2},3,1)\text{.}%
\end{align}
By inspection, we must consider an abelian instanton configuration which
breaks one of the $SU(2)$ factors to a $U(1)_{b}$ subgroup. \ Without loss of
generality, we assume that the instanton preserves the first $SU(2)$ factor.
\ In this case, the resulting candidate $E$-fields are all of the form
$(1_{0},1,1_{\pm2})$ so that $b=\pm3$. \ Listing all candidate $Q$-, $U$- and
$H_{u}$-fields which can yield a gauge invariant term of the form $QUH_{u}$,
we find:%
\begin{equation}%
\begin{tabular}
[c]{|l|l|l|l|}\hline
$Q$ & $U$ & $H_{u}$ & $(a,b)$\\\hline
$(3_{-1},2,1_{\pm1})$ & $(\overline{3}_{-2},1,1_{0})$ & $(1_{3},2_{\mp1})$ &
$(2,\pm3)$\\\hline
\end{tabular}
\end{equation}
where all $\pm$'s are correlated in the above. \ This in turn implies that
there is a unique candidate $H_{d}$-field given by $(1_{-3},2_{\pm1})$. \ This
in turn requires that in order to obtain a non-zero $QDH_{d}$ interaction
term, a candidate $D$-field must have representation content $(\overline
{3}_{4},1,1_{\mp2})$ which is not present in the given decomposition described
above. \ We therefore conclude that this breaking pattern cannot yield the
spectrum of the MSSM.

\paragraph{$SO(12)\supset SO(10)\times U(1)$}

This is the final maximal subgroup of $SO(12)$ which can in principle contain
$G_{std}$. \ The representation content of $SO(12)$ decomposes under this
maximal subgroup as:%
\begin{align}
SO(12)  &  \supset SO(10)\times U(1)\\
12  &  \rightarrow1_{2}+1_{-2}+10_{0}\\
32  &  \rightarrow16_{1}+\overline{16}_{-1}\\
32^{\prime}  &  \rightarrow\overline{16}_{1}+16_{-1}\\
66  &  \rightarrow1_{0}+10_{2}+10_{-2}+45_{0}\text{.}%
\end{align}

Recall that the maximal subgroups of $SO(10)$ are listed in lines
(\ref{SO10one})-(\ref{SO10last}), of which only lines (\ref{SO10one}%
)-(\ref{SO10four}) contain an $SU(3)\times SU(2)$ subgroup. \ In the present
context, we wish to determine whether the presence of the additional $U(1)$
factor can yield a new breaking pattern distinct from those already treated
for $G_{S}=SO(10)$. \ Moreover, while it is in principle of interest to
classify all ways of packaging the fields of the MSSM\ into $SO(12)$
representations, our primary interest is in the classification of all possible
breaking patterns. \ For this reason, we again confine our classification to
this more narrow question.

$SO(12)\supset SO(10)\times U(1)\supset SU(5)\times U(1)\times U(1)$

In this case, there is a unique way in which the $SU(5)$ factor can further
break to $G_{std}$. \ Indeed, this is the natural extension of the analogous
breaking pattern of $SO(10)$ analyzed previously. \ We thus conclude that in
this case the abelian $U(1)^{3}$ instanton breaks $SO(12)$ to $G_{std}\times
U(1)\times U(1)$.

$SO(12)\supset SO(10)\times U(1)\supset\lbrack SU(2)\times SU(2)\times
SU(4)]\times U(1)$

Under this nested sequence of maximal subgroups, $SU(4)$ is the only factor
which contains an $SU(3)$ subgroup. \ The representation content of $SO(12)$
therefore must decompose as:%
\begin{align}
SO(12)  &  \supset SO(10)\times U(1)\supset\lbrack SU(2)\times SU(2)\times
SU(4)]\times U(1)_{b}\\
&  \supset\lbrack SU(2)\times SU(2)\times\lbrack SU(3)\times U(1)_{a}]]\times
U(1)_{b}\\
12  &  \rightarrow(1,1,1)_{0,2}+(1,1,1)_{0,-2}+(2,2,1)_{0,0}+(1,1,3)_{2,0}%
+(1,1,\overline{3})_{-2,0}\\
32  &  \rightarrow(2,1,1)_{3,1}+(2,1,3)_{-1,1}+(1,2,1)_{-3,1}+(1,2,\overline
{3})_{1,1}\\
&  +(2,1,1)_{-3,-1}+(2,1,\overline{3})_{1,-1}+(1,2,1)_{3,-1}+(1,2,3)_{-1,-1}\\
32^{\prime}  &  \rightarrow(1,2,1)_{3,1}+(1,2,3)_{-1,1}+(2,1,1)_{-3,1}%
+(2,1,\overline{3})_{1,1}\\
&  +(1,2,1)_{-3,-1}+(1,2,\overline{3})_{1,-1}+(2,1,1)_{3,-1}+(2,1,3)_{-1,-1}\\
66  &  \rightarrow(1,1,1)_{0,0}+(2,2,1)_{0,2}+(1,1,3)_{2,2}+(1,1,\overline
{3})_{-2,2}\\
&  +(2,2,1)_{0,-2}+(1,1,3)_{2,-2}+(1,1,\overline{3})_{-2,-2}+(3,1,1)_{0,0}\\
&  +(1,3,1)_{0,0}+(1,1,1)_{0,0}+(1,1,3)_{-4,0}+(1,1,\overline{3})_{4,0}\\
&  +(1,1,8)_{0,0}+(2,2,3)_{2,0}+(2,2,\overline{3})_{-2,0}\text{.}%
\end{align}

In the present context, breaking one of the $SU(2)$ factors to a $U(1)$
subgroup yields a breaking pattern identical to that already studied in the
context of the sequence of maximal subgroups $SO(12)\supset SO(10)\times
U(1)\supset SU(5)\times U(1)\times U(1)\supset SU(3)\times SU(2)\times
U(1)\times U(1)\times U(1)$. \ In order to classify all candidate breaking
patterns, it is therefore enough to restrict to cases where one of the $SU(2)$
factors is completely broken. \ Without loss of generality, we assume that the
candidate non-abelian instanton preserves the second $SU(2)$ factor. \ Listing
all candidate $Q$-, $U$- and $H_{u}$-fields which can yield the gauge
invariant interaction term $QUH_{u}$, we find:%
\begin{equation}%
\begin{tabular}
[c]{|l|l|l|l|l|}\hline
& $Q$ & $U$ & $H_{u}$ & $(a,b)$\\\hline
$1$ & $(1,2,3)_{-1,\pm1}$ & $(1,1,\overline{3})_{-2,0}$ & $(1,2,1)_{3,\mp1}$ &
$(2,\pm3)$\\\hline
$2$ & $(1,2,3)_{-1,\pm1}$ & $(1,1,\overline{3})_{4,0}$ & $(1,2,1)_{-3,\mp1}$ &
$(-1,0)$\\\hline
$3$ & $(1,2,3)_{-1,\pm1}$ & $(1,1,\overline{3})_{-2,\mp2}$ & $(1,2,1)_{3,\pm
1}$ & $(1/2,\pm3/2)$\\\hline
$4$ & $(1,2,3)_{-1,\pm1}$ & $(2,1,\overline{3})_{1,\pm1}$ & $(2,2,1)_{0,\mp2}$
& $(-5/2,\mp3/2)$\\\hline
$5$ & $(2,2,3)_{2,0}$ & $(2,1,\overline{3})_{1,\pm1}$ & $(1,2,1)_{-3,\mp1}$ &
$(1/2,\mp9/2)$\\\hline
$6$ & $(2,2,3)_{2,0}$ & $(1,1,\overline{3})_{-2,\mp2}$ & $(2,2,1)_{0,\pm2}$ &
$(1/2,\pm3/2)$\\\hline
\end{tabular}
\text{.}%
\end{equation}
Listing all choices of representations for candidate $D$- and $H_{d}$-fields
which also admit the gauge invariant interaction term $QDH_{d}$, we find:%
\begin{equation}%
\begin{tabular}
[c]{|l|l|l|l|l|}\hline
& $Q$ & $D$ & $H_{d}$ & $(a,b)$\\\hline
$2a$ & $(1,2,3)_{-1,\pm1}$ & $(1,1,\overline{3})_{-2,0}$ & $(1,2,1)_{3,\mp1}$
& $(-1,0)$\\\hline
$2b$ & $(1,2,3)_{-1,\pm1}$ & $(1,1,\overline{3})_{-2,\mp2}$ & $(1,2,1)_{3,\pm
1}$ & $(-1,0)$\\\hline
$3a$ & $(1,2,3)_{-1,\pm1}$ & $(1,1,\overline{3})_{4,0}$ & $(1,2,1)_{-3,\mp1}$
& $(1/2,\pm3/2)$\\\hline
$3b$ & $(1,2,3)_{-1,\pm1}$ & $(2,1,\overline{3})_{1,\pm1}$ & $(2,2,1)_{0,\mp
2}$ & $(1/2,\pm3/2)$\\\hline
$6a$ & $(2,2,3)_{2,0}$ & $(1,1,\overline{3})_{-2,\pm2}$ & $(2,2,1)_{0,\mp2}$ &
$(1/2,\pm3/2)$\\\hline
$6b$ & $(2,2,3)_{2,0}$ & $(2,1,\overline{3})_{1,\pm1}$ & $(1,2,1)_{-3,\mp1}$ &
$(1/2,\pm3/2)$\\\hline
\end{tabular}
\text{.}%
\end{equation}
Because the only candidate $E$-fields are given by $(1,1,1)_{0,\pm2}$ or
$(1,2,1)_{\pm3,\pm1}$, we now observe that all consistent choices of
$U(1)_{Y}$ given previously cannot yield the correct value for the $E$-fields.
\ Hence, an instanton configuration must break one of the $SU(2)$ factors to a
$U(1)$ subgroup in order to reproduce the spectrum of the MSSM.

$SO(12)\supset SO(10)\times U(1)\supset SO(9)\times U(1)$

In order to obtain an $SU(3)\times SU(2)$ subgroup along this nested sequence
of maximal subgroups, the $SO(9)$ factor must also contain such a subgroup.
\ Returning to lines (\ref{firstSO9})-(\ref{lastSO9}), we again conclude that
the only maximal subgroup of $SO(9)$ satisfying this criterion is $SU(2)\times
SU(4)$. \ Further decomposing the $SU(4)$ factor to the maximal subgroup
$SU(3)\times U(1)$, the decomposition of representations of $SO(12)$ now
descends to:%
\begin{align}
SO(12)  &  \supset SO(10)\times U(1)_{b}\supset SO(9)\times U(1)_{b}%
\label{SO12SO10SO9first}\\
&  \supset\lbrack SU(2)\times SU(4)]\times U(1)_{b}\\
&  \supset\lbrack SU(2)\times\lbrack SU(3)\times U(1)_{a}]]\times U(1)_{b}\\
12  &  \rightarrow(1,1)_{0,2}+(1,1)_{0,-2}+(1,1)_{0,0}+(3,1)_{0,0}\\
&  +(1,3)_{2,0}+(1,\overline{3})_{-2,0}\\
32,32^{\prime}  &  \rightarrow(2,1)_{3,1}+(2,3)_{-1,1}+(2,1)_{-3,1}%
+(2,\overline{3})_{1,1}\\
&  +(2,1)_{-3,-1}+(2,\overline{3})_{1,-1}+(2,1)_{3,-1}+(2,3)_{-1,-1}\\
66  &  \rightarrow(1,1)_{0,0}+(1,1)_{0,2}+(1,1)_{0,-2}+(3,1)_{0,2}\\
&  +(3,1)_{0,-2}+(3,1)_{0,0}+(1,3)_{2,2}+(1,\overline{3})_{-2,2}\\
&  +(1,3)_{2,-2}+(1,\overline{3})_{-2,-2}+(3,1)_{0,0}+(1,1)_{0,0}\\
&  +(1,3)_{2,0}+(1,\overline{3})_{-2,0}+(1,3)_{-4,0}+(1,\overline{3})_{4,0}\\
&  +(1,8)_{0,0}+(3,3)_{2,0}+(3,\overline{3})_{-2,0}\text{.}
\label{SO12SO10SO9last}%
\end{align}
Listing all $Q$-, $U$- and $H_{u}$- fields which can yield the term $QUH_{u}$,
we find:%
\begin{equation}%
\begin{tabular}
[c]{|l|l|l|l|l|}\hline
& $Q$ & $U$ & $H_{u}$ & $(a,b)$\\\hline
$1$ & $(2,3)_{-1,\pm1}$ & $(1,\overline{3})_{-2,0}$ & $(2,1)_{3,\mp1}$ &
$(2,\pm3)$\\\hline
$2$ & $(2,3)_{-1,\pm1}$ & $(1,\overline{3})_{-2,\mp2}$ & $(2,1)_{3,\pm1}$ &
$(1/2,\pm3/2)$\\\hline
$3$ & $(2,3)_{-1,\pm1}$ & $(1,\overline{3})_{4,0}$ & $(2,1)_{-3,\mp1}$ &
$(-1,0)$\\\hline
\end{tabular}
\ \text{.}%
\end{equation}
Because the candidate $E$-fields all descend from the representation
$(1,1)_{0,\pm2}$, it follows that $b=\pm3$ so that the second and third cases
are ruled out. \ Restricting to this case, the candidate $D$-fields are
therefore $(1,\overline{3})_{-2,\pm2}$, where the $\pm$ sign is correlated
with that given in the first case. \ In order to obtain a gauge invariant
$QDH_{d}$ interaction term, the resulting $H_{d}$-field must transform in the
representation $(2,1)_{3,\mp3}$, which does not descend from a representation
of $SO(12)$. \ We therefore conclude that this breaking pattern cannot yield
the spectrum of the MSSM.

$SO(12)\supset SO(10)\times U(1)\supset\lbrack SU(2)\times SO(7)]\times U(1)$

In this final case, $SU(4)$ and $G_{2}$ or the only maximal subgroups of
$SO(7)$ which contains an $SU(3)$ subgroup. \ Of these two possibilities, an
instanton can only break the former case to $SU(3)$. \ Decomposing the
representations of $SO(12)$ under the corresponding nested sequence of maximal
subgroups yields:%
\begin{align}
SO(12)  &  \supset SO(10)\times U(1)_{v}\supset\lbrack SU(2)\times
SO(7)]\times U(1)_{b}\supset\lbrack SU(2)\times SU(4)]\times U(1)_{b}\\
&  \supset\lbrack SU(2)\times\lbrack SU(3)\times U(1)_{a}]]\times U(1)_{b}\\
12  &  \rightarrow(1,1)_{0,2}+(1,1)_{0,-2}+(1,1)_{0,0}+(3,1)_{0,0}\\
&  +(1,3)_{2,0}+(1,\overline{3})_{-2,0}\\
32,32^{\prime}  &  \rightarrow(2,1)_{3,1}+(2,3)_{-1,1}+(2,1)_{-3,1}%
+(2,\overline{3})_{1,1}\\
&  +(2,1)_{-3,-1}+(2,\overline{3})_{1,-1}+(2,1)_{3,-1}+(2,3)_{-1,-1}\\
66  &  \rightarrow(1,1)_{0,0}+(1,1)_{0,2}+(1,1)_{0,-2}+(3,1)_{0,2}\\
&  +(3,1)_{0,-2}+(3,1)_{0,0}+(1,3)_{2,2}+(1,\overline{3})_{-2,2}\\
&  +(1,3)_{2,-2}+(1,\overline{3})_{-2,-2}+(3,1)_{0,0}+(1,1)_{0,0}\\
&  +(1,3)_{2,0}+(1,\overline{3})_{-2,0}+(1,3)_{-4,0}+(1,\overline{3})_{4,0}\\
&  +(1,8)_{0,0}+(3,3)_{2,0}+(3,\overline{3})_{-2,0}\text{.}%
\end{align}
In fact, this decomposition is identical to that given for the previously
considered nested sequence of maximal subgroups described by lines
(\ref{SO12SO10SO9first})-(\ref{SO12SO10SO9last}). \ We therefore conclude that
just as in that case, this breaking pattern cannot yield the spectrum of the
MSSM. \newpage
\bibliographystyle{ssg}
\bibliography{fgutsii}

\end{document}